\documentclass[numberedappendix]{emulateapj}
\usepackage{longtable}

\shorttitle{Cold Molecular Gas in Merger Remnants}
\shortauthors{Ueda et al.}

\begin{document}

\title{Cold Molecular Gas in Merger Remnants: I. Formation of Molecular Gas Disks}
\author{
Junko Ueda\altaffilmark{1,2,3}, Daisuke Iono\altaffilmark{1,4}, Min S. Yun\altaffilmark{5}, 
Alison F. Crocker\altaffilmark{6}, Desika Narayanan\altaffilmark{7}, Shinya Komugi\altaffilmark{1}, 
Daniel Espada\altaffilmark{1,4,8}, Bunyo Hatsukade\altaffilmark{1}, Hiroyuki Kaneko\altaffilmark{9}, 
Yuichi Matsuda\altaffilmark{1,4}, Yoichi Tamura\altaffilmark{10}, David J. Wilner\altaffilmark{3}, 
Ryohei Kawabe\altaffilmark{1}, Hsi-An Pan\altaffilmark{4,11,12}}
\altaffiltext{1}{National Astronomical Observatory of Japan, 2-21-1 Osawa, Mitaka,Tokyo, 181-8588, Japan}
\altaffiltext{2}{Department of Astronomy, School of Science, The University of Tokyo, 7-3-1 Hongo, Bunkyo-ku, Tokyo 133-0033, Japan}
\altaffiltext{3}{Harvard-Smithsonian Center for Astrophysics, 60 Garden Street, Cambridge, MA 02138, USA}
\altaffiltext{4}{Department of Astronomical Science, The Graduate University for Advanced Studies (SOKENDAI), 2-21-1 Osawa, Mitaka, Tokyo 181-8588, Japan}
\altaffiltext{5}{Department of Astronomy, University of Massachusetts, Amherst, MA 01003, USA}
\altaffiltext{6}{Ritter Astrophysical Research Center, University of Toledo, Toledo, OH 43606, USA}
\altaffiltext{7}{Department of Astronomy, Haverford College, 370 Lancaster Ave, Haverford, PA 19041, USA}
\altaffiltext{8}{Joint ALMA Observatory, Alonso de C\'ordova 3107, Vitacura, Santiago 763-0355, Chile}
\altaffiltext{9}{Graduate School of Pure and Applied Sciences, University of Tsukuba, 1-1-1 Tennodai, Tsukuba, Ibaraki 305-8577, Japan}
\altaffiltext{10}{Institute of Astronomy, The University of Tokyo, 2-21-1 Osawa, Mitaka,Tokyo, 181-0015, Japan}
\altaffiltext{11}{Nobeyama Radio Observatory, National Astronomical Observatory of Japan, 462-2 Minamimaki, Minamisaku, Nagano 384-1305, Japan}
\altaffiltext{12}{Department of Physics, Faculty of Science, Hokkaido University, Kita-ku, Sapporo 060-0810, Japan}

\begin{abstract}
We present $\lesssim$ 1~kpc resolution $^{12}$CO imaging study 
of 37 optically-selected local merger remnants 
using new and archival interferometric maps 
obtained with ALMA, CARMA, SMA and PdBI.  
We supplement a sub-sample with single-dish measurements 
obtained at the NRO 45~m telescope 
for estimating the molecular gas mass (10$^{7-11}$~M$_{\sun}$), 
and evaluating the missing flux of the interferometric measurements.  
Among the sources with robust CO detections, 
we find that 80~\% (24/30) of the sample show 
kinematical signatures of rotating molecular gas disks 
(including nuclear rings) in their velocity fields, 
and the sizes of these disks vary significantly from 1.1~kpc to 9.3~kpc.  
The size of the molecular gas disks in 54~\% of the sources 
is more compact than the $K$-band effective radius.  
These small gas disks may have formed from a past gas inflow 
that was triggered by a dynamical instability during a potential merging event.  
On the other hand, the rest (46~\%) of the sources have gas disks 
which are extended relative to the stellar component, 
possibly forming a late-type galaxy with a central stellar bulge.  
Our new compilation of observational data suggests that  
nuclear and extended molecular gas disks are common in the final stages of mergers.  
This finding is consistent with recent major-merger simulations of gas rich progenitor disks.  
Finally, we suggest that some of the rotation-supported turbulent disks observed at high redshifts 
may result from galaxies that have experienced a recent major merger.
\end{abstract}


\section{Introduction}
Galaxy interactions and mergers play important roles 
in the formation and evolution of galaxies, 
as illustrated by the increasing galaxy merger rate 
at higher redshifts \citep[e.g.,][]{Lin04, Bridge10}.  
Gravitational interactions between galaxies have significant effects 
on their morphology and physical states.  
The degree of these effects depends on a wide variety of parameters 
such as collision orbit, collision speed, and relative size.  
Major mergers, which can lead to significant dynamical and morphological disturbances, 
are strongly linked to starbursts and AGNs \citep[e.g.,][]{Sanders96}.  
Since the 1970's, it has been long predicted from numerical simulations 
that the major merger of two disk galaxies results 
in the formation of a spheroid-dominated early-type galaxy 
\citep{Toomre77, Barnes92, Naab03}.  
This classical scenario is supported by observations 
of the stellar structures of merger remnants that resemble elliptical galaxies 
\citep[e.g.,][]{Schweizer82, Rothberg06a} 
and signatures of past mergers such as shells and tidal features around elliptical galaxies 
\citep[e.g.,][]{Schweizer92, Schweizer96}.

Contrary to the classical scenario of merger evolution, 
recent high-resolution simulations that include more realistic gas physics 
have shown that not all of the major mergers will become an early-type galaxy, 
but some will reemerge as a disk-dominated late-type galaxy 
\citep{Barnes02, Robertson08, Hopkins09, Xu10}.  
During a merger, any gas that does not lose significant angular momentum 
through stellar torques will survive the collision, 
reforming a gaseous (and subsequently a stellar) disk \citep{Springel05}, 
while gas that falls to the galaxy center will contribute 
to a nuclear starburst and subsequent formation of the spheroidal component.  
In galaxy mergers, the chance of disk survival during the merging event depends 
on orbital parameters, mass ratio, and gas mass fraction of the progenitors \citep{Hopkins09}.  
The same simulations also suggest that increasing the gas mass fraction leads 
to a more efficient disk survival as there are fewer stars to lose angular momentum to.  

In order to check this theoretical prediction and 
to investigate the evolution of galaxies after a merging event, 
we have conducted a $^{12}$CO imaging study of merger remnants in the local Universe 
using millimeter/submillimeter interferometric measurements.  
This paper is the first of a series of our study on merger remnants, 
which includes a summary of the data and focuses on 
the properties of the molecular gas disk in merger remnants.  
The focus of the second paper is to understand the evolution
of merger remnants by comparing the properties of molecular
gas and stellar component with early-type galaxies 
\citep[i.e. Atlas3D Project;][]{Cappellari11, Alatalo13, Davis13} 
and late-type galaxies \citep[i.e. BIMA-SONG;][]{Regan01, Helfer03}.  
We will also investigate the kinematics of diffuse gas 
using \ion{H}{1} line measurements in a subsequent paper.  

Merger remnants are completely merged galaxies 
that still have tidal tails, shells, and loops, which indicate past dynamical interactions.  
Most previous studies on merger remnants 
\citep[e.g.,][]{Lake86, Shier98, James99, Genzel01, 
Rothberg04, Dasyra06, Rothberg06a, Rothberg06b} 
focused on the stellar properties.  
This is because the classical scenario, 
in which a major merger between two disk galaxies 
forms a new elliptical galaxy, received higher attention.  
The main purpose of this study is to characterize 
the cold molecular disks emerging from mergers 
by analyzing new high-resolution interferometric CO maps 
for 27 optically-selected merger remnants 
obtained using the \textit{Atacama Large Millimeter/submillimeter Array} (ALMA), 
\textit{Combined Array for Research in Millimeter-wave Astronomy} (CARMA), 
and \textit{Submillimeter Array} (SMA).
We further obtained published and archival interferometric maps 
from ALMA, \textit{Plateau de Bure Interferometer} (PdBI), and SMA, for 10 galaxies 
to supplement our analysis, forming a grand sample of 37 galaxies.  
In addition, new \textit{Nobeyama Radio Observatory} (NRO) 45~m data
and published single-dish measurement 
from the \textit{Five College Radio Astronomical Observatory } (FCRAO) 14~m, 
\textit{Institut de Radioastronomie and Millim\'{e}trique} (IRAM) 30~m, 
\textit{National Radio Astronomy Observatory } (NRAO) 12~m, 
and \textit{Swedish-ESO Submillimeter Telescope} (SEST) 15~m telescopes 
were obtained to analyze the total molecular gas content for 32 galaxies, 
including six galaxies which do not overlap with the sample for interferometric observations.  
The low critical density of low $J$ CO transitions enable tracing diffuse molecular gas, 
which is distributed in a more extended fashion compared to high $J$ CO transitions.  
Thus low $J$ CO lines are the best tracer for 
investigating the distribution and kinematics of diffuse molecular gas.

This paper is organized as follows.  
We present the properties of our merger remnant sample in Section~2, 
observational details and data reduction process in Section~3, 
and new results in Section~4.
In Section~5, we characterize the size and the kinematics
of the molecular gas by fitting a simple rotating
disk model to the interferometric maps.
We summarize this paper in Section~6.  
All values are calculated based on a $\Lambda$CDM model 
with $\Omega_{\rm M}$ = 0.3, $\Omega_{\rm \Lambda}$ = 0.7, 
and $H_{0}$ = 73 km s$^{-1}$ Mpc$^{-1}$.

\section{Sample Sources}
Our sample is drawn from the merger remnant sample 
studied by \citet[][RJ04 hereafter]{Rothberg04}, 
which is a large-scale observational study of 51 merger remnants 
in the local Universe ($<$ 200~Mpc).  
The sample galaxies in RJ04 are selected solely based on optical morphology 
that suggests an advanced merger stage: 
(1) tidal tails, loops and shells, (2) a single nucleus, 
and (3) absence of nearby companions, 
regardless of star formation or AGN activities.  
The stellar components in a majority of their sample have undergone violent relaxation, 
though seven sources show evidence for incomplete phase mixing (RJ04).  
Their analysis also shows the presence of ``excess light'' 
in the surface brightness profiles of nearly one-third of their sample, 
suggesting the effect of central starbursts.  

We have obtained new and published interferometric CO maps 
of 37 merger remnants from this parent sample.  
In addition, we have obtained new and published single-dish CO measurements 
of 26/37 merger remnants to accurately estimate the total CO flux.  
We note that the exact fraction of these sources 
that result from a major as opposed to a minor merger is unknown 
since it is difficult to reverse the chronology 
and disentangle the exact mass and morphology of the progenitors.  
The $K$-band images (RJ04) of 37 merger remnants 
are presented in Figure~\ref{fig:f1} (left)  and  Figure~\ref{fig:f2}, 
and the basic properties of these sources are summarized in Table~\ref{tab:t1}.  
The $K$-band images show single-nuclei in all of the sample sources.  
However a double nucleus, which suggests an incomplete merger, 
was found in a few sources in the radio continuum maps 
\citep[e.g., NGC3256;][]{Norris95}.  

We estimate the far-infrared (FIR) luminosities of our sample 
using the $IRAS$ data at 60~$\mu$m and 100~$\mu$m 
\citep{Beichman88, Moshir92, Moshir08}.  
The FIR luminosities of our sample range 
from 10$^{9}$~$L_{\sun}$ to 10$^{12}$~$L_{\sun}$, 
but nine galaxies are not detected by $IRAS$ 
and we estimate upper limits of the FIR luminosities.  
One galaxy (UGC~8058) is classified as an Ultra Luminous Infrared Galaxy 
(ULIRG; $L_{\rm FIR} \geq 10^{12}$~$L_{\sun}$), 
and 12 galaxies are classified as LIRGs 
($L_{\rm FIR} \geq 10^{11}$~$L_{\sun}$).  

RJ04 estimate the S\'{e}rsic index ($n$) of the merger remnant from the S\'{e}rsic profile, 
which is a generalized form of the de Vaucouleurs profile 
to probe the stellar surface brightness.  
The S\'{e}rsic index is $n$ = 1 for an exponential disk, 
whereas $n$ = 4 for a de Vaucouleurs profile.  
The upper limit of the S\'{e}rsic fits is $n$ = 10, 
thus a fit resulting in $n$ = 10 may signify inaccurate fits.
In our sample of 37, there are 12 galaxies with $n <$ 3, 
12 galaxies with 3 $\leq n <$ 5, three galaxies with 5 $\leq n <$ 10, 
and 10 galaxies with $n$ = 10.

\section{Observations and Archival Data}
\subsection{Interferometric Observations with ALMA}
The $^{12}$CO~($J$ = 1--0) observations for 20 galaxies (see Table~\ref{tab:t2}) 
were carried out during the ALMA Cycle~0 period.  
All galaxies except for NGC~455 are located in the southern sky (Decl.~$<$~0\arcdeg).  
We used the compact configuration for the three nearby galaxies 
(Arp~230, AM~0956-282, and NGC~7727) and 
the extended configuration for the other 17 galaxies 
to achieve $\lesssim$ 1~kpc spatial resolution.  
The number of 12~m antennas was 16 -- 24 depending on observation.
The primary beam of the 12~m antenna is $\sim$ 50\arcsec at 115~GHz.  
The correlator was configured to have four spectral windows, 
each with 1.875~GHz bandwidth and 488~kHz frequency resolution.  

We imaged the calibrated visibility data 
using the Common Astronomy Software Applications package (CASA).  
The basic properties of the data are summarized in Table~\ref{tab:t2}.  
The synthesized beam sizes range between 1\farcs2 and 4\farcs1 
by adopting briggs weighting of the visibilities (robust = 0.5).  
This weighting is recommended, as it offers a good compromise 
between sensitivity and resolution.  
The achieved rms noise levels in the velocity resolution of 20 km s$^{-1}$  
are 1.78 -- 5.26 mJy beam$^{-1}$.  
In addition, we made the continuum maps using the line free channels.
The uncertainty of flux calibration is 5~\%.

\subsection{Interferometric Observations with SMA}
The $^{12}$CO~($J$ = 2--1) observations for five galaxies 
(NGC~828, UGC~2238, UGC~9829, NGC 6052, and UGC~10675) 
were carried out using the SMA in June and October 2011.  
The data were obtained using the compact configuration. 
The primary beam is $\sim$ 50\arcsec at 230~GHz.  
The correlator was configured to have two sidebands, 
each with 2~GHz bandwidth and 3.25~MHz frequency resolution.  

Data inspection was carried out using the IDL-based SMA calibration tool MIR 
and imaging was done using the MIRIAD package.  
The basic properties of the data are summarized in Table~\ref{tab:t2}.  
The synthesized beam sizes range between 2\farcs6 and 4\farcs2 
by adopting natural weighting of the visibilities, which gives maximum sensitivity.  
The achieved rms noise levels in the velocity resolution of 20 km~s$^{-1}$ 
are 18.2 -- 27.3 mJy~beam$^{-1}$.
The uncertainty of flux calibration is 20~\%.  

\subsection{Interferometric Observations with CARMA}
The $^{12}$CO~($J$ = 1--0) observations for two galaxies (UGC~6 and UGC~4079)
were carried out using the CARMA in January and February 2011.  
The data were obtained using the C configuration.  
The primary beam of the 10.4~m antenna is $\sim$60\arcsec at 115~GHz.  
The correlator was composed of 10~bands, 
and each band consisted of 15~channels with 31.25~MHz frequency resolution. 

Data inspection and imaging were carried out using the MIRIAD package.  
The synthesized beam sizes are 1\farcs93~$\times$~1\farcs44 for UGC~6 
and 1\farcs74~$\times$~1\farcs53 for UGC~4079 
by adopting natural weighting of the visibilities.  
The achieved rms noise levels in the velocity resolution of 100 km~s$^{-1}$ 
are 3.72 mJy~beam$^{-1}$ for UGC~6 and 2.05 mJy~beam$^{-1}$ for UGC~4079.  
The uncertainty of flux calibration is 20~\%.  

\subsection{Archival Interferometric Data}
We further obtained published and archival interferometric CO maps for 10 galaxies, 
out of which seven galaxies (see Table~\ref{tab:t3}) 
were observed using the SMA \citep[e.g.,][]{Wilson08}, 
two galaxies (NGC~2782 and NGC~4441) were observed 
using the PdBI \citep{Hunt08, Jutte10}, 
and one galaxy (NGC~3256) was observed using ALMA for Science Verification.  
The basic properties of the data are summarized in Table~\ref{tab:t3}.  
The observed CO transitions are $J$ = 1--0 for three galaxies, 
$J$ = 2--1 for five galaxies, and $J$ = 3--2 for two galaxies.  
The synthesized beam sizes range between $0\farcs7$ and $7\farcs6$, 
which correspond to $\lesssim$ 1~kpc for each source.  
The achieved rms noise levels in the velocity resolution of 20 km~s$^{-1}$ 
are 1.37 -- 37.2 mJy beam$^{-1}$.  
The uncertainty of flux calibration is 5~\% for NGC~3256, 
which was observed using ALMA, and 20~\% for the rest.  

\subsection{Single-dish Observations and Archival Single-dish Data}
The $^{12}$CO~($J$ = 1--0) observations were carried out using the NRO 45~m telescope 
for 10 galaxies (see Table~\ref{tab:t5}) during  2011 -- 2012.  
The single-beam, 2-Sideband, and dual-polarization receiver 
\citep[T100;][]{Nakajima08} was used as the receiver front end.  
The analog signal from T100 is downconverted to 2--4~GHz and digitized to 3-bits 
before being transferred to the digital FX-type spectrometer SAM45.  
The 488~kHz resolution mode (2~GHz bandwidth) of SAM45 
was used for the frequency resolution.  
Typical system temperatures were 150 -- 200~K.  

Data reduction was carried out 
using the AIPS-based NRO calibration tool NEWSTAR.  
After flagging scans with large ripples, 
the final spectra were made by fitting a first order baseline.  
The antennae temperature ($T_{\rm A}^{*}$) was converted 
to the main beam temperature ($T_{\rm mb}$) 
using a main beam efficiency of $\eta_{\rm mb} = 0.4$ 
and $T_{\rm mb} = T_{\rm A}^{*}/\eta_{\rm mb}$.  
The final spectra are converted to Jansky 
using a Kelvin--to--Jansky conversion factor of 2.4 Jy K$^{-1}$.  
The primary beam size is $\sim$15\arcsec at 115~GHz.  
The rms noise levels in the velocity resolution of 30 km~s$^{-1}$ 
range between 1~mK and 13~mK.  

In addition, we gather archival single-dish CO~(1--0) data for 16 galaxies.  
The basic properties of the archival data are summarized in Table~\ref{tab:t5}.  
Eight galaxies were observed 
using the SEST 15~m telescope 
\citep{Mirabel90, Casoli92, Huchtmeier92, Andreani95, Wiklind95, Horellou97}.  
Five galaxies were observed using the NRAO 12~m telescope \citep{Sanders91, Maiolino97, Evans05}.  
Two galaxies were obtained with the IRAM 30~m telescope \citep{Zhu99, Bertram06}, 
and one galaxy was obtained with the FCRAO 14~m telescope \citep{Young95}.

\section{Results}
\subsection{Distribution and Kinematics of the CO emission}
The CO emission was newly detected in the interferometric maps of 20 out of 27 galaxies.  
The final number of our sample is 37 including seven galaxies 
after combining with archival data.  
The integrated line intensities are summarized in Table~\ref{tab:t4}.  
The integrated intensity maps and velocity fields of 30 galaxies 
are presented in Figure~\ref{fig:f1}.  
They were made by smoothing and clipping the intensity 
using the AIPS task, MOMNT.  
The cleaned image cube was smoothed both spatially and in velocity 
and then this cube was clipped at the 1.5~$\sigma$ level per channel 
for 20 galaxies observed with ALMA (Cycle~0), 
5~$\sigma$ level per channel for NGC~3256 
because of high side-lobe levels which cannot be eliminated, 
and 2~$\sigma$ level per channel for the others.
The molecular gas in 20 out of 30 galaxies detected in the CO line 
is distributed in the nuclear regions and 
the CO emission peaks roughly correspond to the peak in the $K$-band images (RJ04).  
In the other 10 galaxies, the CO distribution show multiple components
such as arms, bars, and other structures which are not associated to the main body.  
Detailed description of the individual galaxies are provided in the Appendix.  

Six galaxies (NGC~455, NGC~1210, AM~0318-230, 
UGC~4079, NGC ~5018, and NGC~7585) 
were not detected in the CO~(1--0) line.  
The observations of all these galaxies except for UGC~4079 
have been conducted at ALMA.  
The 3~$\sigma$ upper limits of the molecular gas mass 
are $2 \times 10^{7-8}$ $M_{\sun}$, 
which are estimated from the integrated intensity maps 
and assuming a velocity width of 300 km~s$^{-1}$.  
AM~1419-263 is tentatively detected in the CO (1--0) line, 
but there is no robust emission ($>$ 1.5~$\sigma$).  
Thus we classify AM~1419-263 as a non-detection 
because the emission is too faint to discuss 
the distribution and kinematics of the molecular gas.  
The CO~(1--0) emission is not seen 
in the interferometric map of UGC~4079 due to limited sensitivity, 
while the CO~(1--0) spectra was clearly obtained 
using the single-dish measurements (Figure~\ref{fig:f3}).  
The rms noise level in the interferometric maps is 
about 10 times larger than that in the single-dish spectra, 
assuming a beam filling factor of unity and 
a point source unresolved with the beam size.  
There are two similarities among these seven galaxies.  
One similarity is low FIR luminosities 
($L_{\rm FIR} < 3 \times 10^{10} L_{\sun}$).  
The FIR luminosities of five galaxies are upper limits 
because they are not detected in the $IRAS$ maps.  
The other is that the S\'{e}rsic indices estimated by RJ04 
are distributed around $n$ = 4, 
which suggests a de Vaucouleurs profile.  
The $K$-band images (RJ04) of these galaxies 
are presented in Figure~\ref{fig:f2}.  
These galaxies show relatively smooth and featureless 
stellar spheroidal structures.

\subsection{The 3~mm Continuum Maps}
The 3~mm continuum emission was detected 
in 14 out of 21 galaxies observed using ALMA.  
While the CO emission was not detected in 
NGC~1210, NGC~5018, and AM~1419-263, 
the unresolved 3~mm continuum emission 
was clearly seen in the nuclear regions.  
The 3~mm continuum contour maps of 11 galaxies 
detected in both CO~(1--0) line and 3~mm continuum emission 
are presented in Figure~\ref{fig:f1} (middle) using red color contours.  
The total flux of the 3~mm continuum is summarized in Table \ref{tab:t4} 
and ranges from 0.54~mJy to 27.5~mJy.  
The total flux was estimated by integrating the extended emission 
if the 3~mm continuum emission was resolved.  
If not, the total flux corresponds to the peak flux of the 3~mm continuum.  
For seven galaxies undetected in the 3~mm continuum, 
we estimated 3~$\sigma$ upper limits, 
assuming an unresolved point source of the nucleus.  
The 3~$\sigma$ upper limits are below 0.4~mJy.
The continuum emission was unresolved 
in nine galaxies and associated with the nuclei defined by the $K$-band images.  
In the rest, the continuum emission is resolved.  
We remark the 3~mm continuum emission detected on the east of AM~0956-282.  
The peak flux is 0.24~mJy ($>$~3~$\sigma$ detection).  
The 3~mm continuum emission is very faint, but this is a robust detection.
Detailed description of the individual galaxies are provided in the Appendix.  

\subsection{Single-dish CO~(1--0) Spectra and Integrated Intensity}
The new CO~(1--0) spectra of 16 galaxies 
observed using the NRO 45~m telescope are presented in Figure~\ref{fig:f3} 
and the properties of the data are summarized in Table~\ref{tab:t5}.  
We report the first detections of the CO line in seven galaxies.  
The integrated line intensity of Arp~230 is 4.1$\pm$0.8 K~km~s$^{-1}$, 
which corresponds to 10 \% of the previous measurement 
using the NRAO 12~m telescope \citep{Galletta97}, 
the difference of which is likely attributed to the large CO extent.  
The integrated line intensities of NGC~2623 and NGC~2782 are 
37.4$\pm$7.5 K~km~s$^{-1}$ and 23.5$\pm$4.7 K~km~s$^{-1}$, respectively.  
These correspond to 53 \% and 25 \% of the previous measurements 
using the FCRAO 14~m telescope \citep{Young95}.  
The integrated line intensities of UGC~5101, NGC~3656, NGC~4194, 
UGC~8058, and NGC~6052 are 17.4$\pm$3.5 K~km~s$^{-1}$, 
51.6$\pm$10.3 K~km~s$^{-1}$, 47.7$\pm$9.5 K~km~s$^{-1}$, 
13.0$\pm$2.6 K~km~s$^{-1}$, and 16.9$\pm$3.4 K~km~s$^{-1}$, respectively.  
These five galaxies were observed using the IRAM 30~m telescope, and 
the integrated line intensities are 15.6$\pm$3.1 K~km~s$^{-1}$ for UGC~5101 
and 22.0$\pm$4.4 K~km~s$^{-1}$ for UGC~8058 \citep{Solomon97}, 
23.4 K~km~s$^{-1}$ for NGC~3656 \citep{Wiklind95},  
29.26$\pm$0.26 K~km~s$^{-1}$ for NGC~4194 \citep{Albrecht04}, 
and 16.4 K~km~s$^{-1}$ for NGC~6052 \citep{Albrecht07}.  
Using the Kelvin--to--Jansky conversion factor of 4.95 for the IRAM telescope, 
the ratios of the intensities of the NRO to the IRAM single-dish measurements are 
60 \% for UGC~5101, 118\% for NGC~3656, 
87\% for NGC~4194, 32 \% for UGC~8058, and 50 \% for NGC~6052.  
The ratios are quite different, which may result from multiple factors 
such as pointing accuracy.  
The total flux of NGC~4441 is 65.5$\pm$13.1 Jy~km~s$^{-1}$, 
which is 1.6 larger than that estimated from the previous observations 
using the \textit{Onsala Space Observatory} (OSO) 20~m telescope \citep{Jutte10}.  
The total flux of the OSO single-dish measurements 
is estimated to be $\sim$40 Jy~km~s$^{-1}$ 
from the molecular gas mass reported by \citet{Jutte10}.  
We also use literature data of single-dish CO~(1--0) measurements for 16 galaxies 
to discuss the molecular gas mass in a later section
and the CO properties of these sources are summarized in Table~\ref{tab:t5}.  

We estimate the recovered flux for 18 galaxies, 
comparing the flux of the same CO transition in the same region 
measured from the interferometric maps with the single-dish spectra.  
For six galaxies which were observed 
in the CO (2--1) and CO (3--2) lines with interferometers, 
we obtain the CO~(2--1) and CO~(3--2) single-dish data from literature.  
The integrated line intensity $I_{\rm  CO}$ in K km s$^{-1}$ is converted 
into the integrated flux $S_{\rm CO} \Delta v$ in Jy km s$^{-1}$ 
using Kelvin--to--Jansky conversion factors of 
23.1 Jy K$^{-1}$ for the FCRAO 14~m telescope, 
4.95 Jy K$^{-1}$ for the IRAM 30~m telescope, 
30.4 Jy K$^{-1}$ for the NRAO 12~m telescope, 
2.4 Jy K$^{-1}$ for the NRO 45~m telescope, 
and 27 Jy K$^{-1}$ for the SEST.  
The recovered flux is shown in Table~\ref{tab:t4}.

\subsection{Molecular Gas Mass}
The molecular gas mass ($M_{\rm H_{2}}$) is estimated from the CO integrated intensity 
using the following two equations \citep{Carilli13, Casey14};
\begin{eqnarray}
\frac{L^{\prime}_{\rm CO}}{\rm K~km~s^{-1}} &=& 3.25 \times 10^{7}\Bigl(\frac{S_{\rm CO}\Delta v}{\rm Jy~km~s^{-1}}\Bigr) \nonumber \\
&\times&\Bigl(\frac{\nu_{\rm obs}}{\rm GHz}\Bigr)^{-2}\Bigl(\frac{D_{\rm L}}{\rm Mpc}\Bigr)^{2}~(1+z)^{-3},
\label{eq:eq1}
\end{eqnarray}
where $L_{\rm CO}$ is the CO luminosity, 
$S_{\rm CO}\Delta v$ is the CO integrated intensity, 
$\nu_{obs}$ is the observing frequency, 
$D_{L}$ is the luminosity distance, and $z$ is the redshift.  
Then the molecular gas mass is derived by;
\begin{eqnarray}
\frac{M_{\rm H_{2}}}{ \rm M_{\sun}} &=& \alpha_{\rm CO} \frac{L_{\rm CO}}{ \rm L_{\sun}},
\label{eq:eq2}
\end{eqnarray}
where $\alpha_{\rm CO}$ is the CO luminosity-to-H$_{2}$ mass conversion factor 
in M$_{\sun}$~pc$^{-2}$~(K~km~s$^{-1}$)$^{-1}$.  
The conversion factor has been a controversial issue for a long time.  
\citet{Solomon91} find that the conversion factor is roughly constant 
in the Galaxy and nearby spiral galaxies 
and suggest that the conversion factor is 
$X_{\rm CO}$ = 2.2 $\times$ 10$^{20}$ cm$^{-2}$ (K~km~s$^{-1}$)$^{-1}$, 
corresponding to $\alpha_{\rm CO}$ = 4.8 M$_{\sun}$~pc$^{-2}$~(K~km~s$^{-1}$)$^{-1}$.  
However, the conversion factor, in gas-rich galaxies at high redshift and local U/LIRGs, 
is lower than the standard value in the Galaxy 
\citep[e.g., $\alpha_{\rm CO}$~$\sim$~0.6--0.8;][]{Downes98, Papadopoulos12a}.  
It is suggested that the conversion factor in mergers, on average, 
are lower values than those in high-redshift disk galaxies 
which may have gravitationally unstable clumps \citep{Narayanan12}.  
Indeed, the low $\alpha_{\rm CO}$ values are found in mergers 
such as the Antennae galaxies \citep[$\alpha_{\rm CO} \sim$ 0.2--1;][]{Zhu03} 
and Arp~299 \citep[$\alpha_{\rm CO} \sim$ 0.2--0.6;][]{Sliwa12}.

We estimate the molecular gas mass of a merger remnant 
with a FIR luminosity less than 10$^{11}$~$L_{\sun}$ 
using the conversion factor of $\alpha_{\rm CO}$ = 4.8 \citep{Solomon91}, 
and we use $\alpha_{\rm CO}$ = 0.8 \citep{Downes98} for the 14 U/LIRGs.  
In estimating the molecular gas mass from the CO~(3--2) and the CO~(2--1) luminosities, 
we convert the high-$J$ CO luminosity into the CO~(1--0) luminosity, 
assuming that the CO~(3--2)/CO~(1--0) and the CO~(2--1)/CO~(1--0) 
brightness temperature ratio is 0.5, 
which is similar to the average CO~(3--2)/CO~(1--0) ratio 
in local U/LIRGs estimated by \citet{Iono09}.  
We estimate the molecular gas mass 
using the single-dish CO data ($M_{\rm H_{2}, SD}$)
and also using the interferometric CO maps ($M_{\rm H_{2}, INT}$), 
and summarize $M_{\rm H_{2}, SD}$ and $M_{\rm H_{2}, INT}$ 
in Table~\ref{tab:t4} and Table~\ref{tab:t5}, respectively.  
The molecular gas mass in the sample of merger remnants 
ranges between 10$^{7}$~$M_{\sun}$ and 10$^{11}$~$M_{\sun}$.  

Although $M_{\rm H_{2}, SD}$ should in general be larger than $M_{\rm H_{2}, INT}$
as it retrieves extended flux that is not recovered by the interferometer, 
some sources in Table~\ref{tab:t4} and Table~\ref{tab:t5}
have $M_{\rm H_{2}, INT}$ that is larger than the single dish flux.  
Several factors may contribute to this, such as the uncertainty in calibration, 
and possibilities that the single dish beam did not cover all of the CO distribution for extended sources.  
The most likely explanation is the assumption of our CO line ratio 
where we use CO~(3--2) or CO~(2--1) for interferometric data, 
and CO~(1-0) for single dish.  
We use a uniform line ratio for all sources, 
but there is an uncertainty in this assumed line ratio 
because the line ratios vary among the sources \citep[e.g.][]{Iono09} 
due to the different physical conditions in the interstellar matter.  
We note, however, that the uncertainty in the relative mass 
between that obtained from interferometric and single dish data 
does not affect the results presented in this paper.  
A full discussion of the molecular mass will be provided 
in a forthcoming paper (Ueda et al., in prep.; Paper II).

\section{Formation of the Molecular Gas Disk} 
In this section, we characterize the size and the kinematics of the molecular gas 
by fitting a simple rotating disk model to the interferometric maps.  
The main motivation for this work is to evaluate the size and 
the frequency of the occurrence of molecular disks and 
to investigate this in the context of recent numerical simulations 
that suggest the formation of an extended gas disk 
as a final product of a major merger \citep{Springel05}.  
We further estimate the relative size of the molecular gas distribution 
to the stellar component traced in $K$-band emission.  
Finally, we end this subsection with an implication to high-redshift galaxies.  

\subsection{Molecular Gas Disks in Merger Remnants}
It is clear from Figure~\ref{fig:f1} that most of the merger remnants 
show disk-like rotation in their CO velocity fields.  
In order to quantify the kinematics and the geometry of the rotating molecular gas, 
we apply a fitting program (AIPS's task GAL) 
to the CO velocity fields of the 30 galaxies detected in the CO line.  
The galactic center is defined by the emission peak of the $K$-band image (RJ04), and 
we assume that the molecular gas rotates around the galactic center.  
We fit tilted concentric ring models to the velocity field 
using least-squares fitting routine for the kinematical parameters 
and a default convergence condition.  

The estimated position angle, inclination, and systemic velocity 
are summarized in Table~\ref{tab:t6}.  
The fitting routine converged for 24 out of the 30 galaxies.  
The fitting procedure failed for the remaining six galaxies 
(AM~1158-333, NGC~4194, AM~1255-430, 
UGC~9829, NGC~6052, and NGC~7135) 
due to the clumpy CO distribution or the complex velocity distribution 
that is not characterized by a simple gas disk model.  
The CO distribution is strongly distorted in these six galaxies, 
and the CO emission is not associated with the galactic center 
in NGC~6052 and NGC~7135.   

In addition, we investigate the Position-Velocity (PV) diagrams 
along the kinematical major axes of 24 galaxies with molecular gas disks  
(Figure~\ref{fig:f4}).  
The kinematical major axes are estimated by fitting the CO velocity fields.  
The PV diagrams reveal that the emission peaks are clearly shifted from 
the galactic centers in 18/24 galaxies (see Table~\ref{tab:t6}), 
suggesting the presence of a ring and/or a bar.  
The PV diagrams of NGC~828, UGC~2238, and NGC~7252 show 
a double peak located at the transition point between rigid rotation and flat rotation, 
which may signify a spiral and/or a bar structure rather than a ring structure.  

\subsection{The Extent of the Molecular Gas}
In order to quantify the extent of the molecular gas disk, 
we estimate two types of radii ($R_{\rm CO}$ and $R_{80}$) 
using the integrated intensity maps.  
$R_{\rm CO}$ is the radius from the galactic center 
that encloses the maximum extent of the molecular gas disk, 
and $R_{80}$ is the radius from the galactic center 
which contains 80~\% of the total CO flux.  
We chose this method because most of our attempts 
to fit an exponential disk were unsuccessful. 
In particular, the radial profiles of molecular gas deviate from 
an exponential profile in the outer regions. 
For the 24 galaxies with robust disk models in Section 5.1, 
we take into account the derived position angle and 
the inclination of the gas disk (Table~6).  
We define these 24 galaxies as ``Type~A''.  
For six galaxies in which the CO velocity field cannot be modeled by circular motion, 
we define them as ``Type~B'' and estimate $R_{\rm CO}$ and $R_{80}$ 
without correcting for the geometry of the galaxy.  
The $R_{\rm CO}$ and $R_{80}$ of 30 galaxies 
are summarized in Table~\ref{tab:t6}.  
The histogram of $R_{\rm CO}$ is shown in Figure~\ref{fig:f5} (left).  
The dark- and light-gray bars show Type~A and Type~B, respectively.  
For three galaxies (NGC~2782, NGC~3256, and NGC~4441), 
we exclude the CO extensions which clearly exist outside 
the molecular gas disk (see Figure~\ref{fig:f1}).  

The histogram of $R_{80}$ is shown in Figure~\ref{fig:f5} (right).  
The representation of the color is the same as in Figure~\ref{fig:f5} (left).  
In order to examine the effect of the detection limit on the estimated radii, 
we investigate the relation between 1~$\sigma$ mass sensitivity and $R_{\rm 80}$.
The absence of an obvious systematic correlation allows us 
to argue that the detection limit does not strongly affect 
the observed size of the molecular gas.  
In addition, there is no significant effect of the different low $J$ rotational transitions 
($J$ = 1--0, 2--1, 3--2) on the estimated radii, 
because the distributions of the CO~(1--0), CO~(2--1), and CO~(3--2) 
are not different in local galaxies such as M51 \citep{Vlahakis13}, 
although the distribution may vary in different population of galaxies 
with extreme CO excitation conditions.  
The average $R_{\rm CO}$ of Type~A galaxies is 3.5$\pm$2.3~kpc and 
the average $R_{80}$ of  the molecular gas disks is 1.5$\pm$0.8~kpc.  
Arp 187 has a molecular gas disk 
with the largest $R_{80}$ of 3.5$\pm$0.3~kpc.  
Its maximum extent is also the largest in our sample.  
On the other hand, AM~0956-282 has a molecular gas disk 
with the smallest $R_{80}$ of 0.4$\pm$0.1~kpc.  
The $R_{\rm CO}$ of AM~0956-282 is smaller than the average, 
but it is not the smallest in our sample.  
We estimate $R_{\rm 80, MW}$ of the molecular gas disk in the Milky Way 
using the radially averaged surface brightness \citep{Nakanishi06}.  
As a result, we find that $R_{\rm 80, MW}$ is 8.5$\pm$0.5~kpc, 
which is about five times larger than the average $R_{80}$ in our sample.  
There is no merger remnant in the present sample 
which has a less concentrated molecular gas disk than the gas disk in the Milky Way.  

\subsection{Comparison of the Sizes of the Molecular Gas Disk with the Stellar Component}
We estimate the ratio of $R_{80}$ to the $K$-band effective radius 
($R_{\rm eff}$; Table~\ref{tab:t1}) for our sample of merger remnants 
to investigate the relative size of the molecular gas disk to the stellar component.  
$R_{\rm eff}$ is the radius of the isophote 
containing half of the total $K$-band luminosity.  
The ratio between $R_{80}$ and $R_{\rm eff}$ 
($R_{\rm ratio}$, hereafter) is summarized in Table~\ref{tab:t6}.  
The histogram of $R_{\rm ratio}$ of the 29 sources except for UGC~8058 
is shown in Figure~\ref{fig:f6}. 
The $R_{\rm ratio}$ of UGC~8058 is extremely large ($R_{\rm ratio}$ = 16).  
UGC~8058 is the most distant source ($D_{\rm L} \simeq$ 180~Mpc) in our sample, 
and the CO distribution is only marginally resolved and 
$R_{80}$ is likely overestimated due to the large beam, 
leading to an overestimation of $R_{\rm ratio}$.  
Another possibility of the large ratio is that $R_{\rm eff}$ is underestimated 
due to an excess light in the galactic center.

The $R_{\rm ratio}$ of Type~A galaxies except for UGC~8058 
range between 0.08 and 4.5.  
The source with the largest $R_{\rm ratio}$ is 
NGC~34 ($R_{\rm ratio}$ = 4.5$\pm$0.9), 
followed by UGC~5101 ($R_{\rm ratio}$ = 4.0$\pm$0.8) 
and UGC~10675 ($R_{\rm ratio}$ = 3$\pm$1).  
These three sources have relatively high FIR luminosities ($L_{\rm FIR} >$ 10$^{11}~L_{\sun}$), 
suggesting that the molecular gas may eventually become the stellar disk.  
However, the timescale for stellar disk formation 
is related to the complex interplay among 
gravitational compression, residual gas inflow, shear motion, stellar feedback, 
and the gas dynamics in the non-axisymmetric potential, 
and it may not be easy to determine.  
In order to confirm the formation of stellar disks, 
we will have to observe these sources 
using dense gas and ionized gas tracers at high angular resolution.  
In addition, there is no relation between the relative size of the molecular gas disk 
and FIR luminosity for the whole sample (Figure~\ref{fig:f7}).  
While half of the U/LIRGs in our sample have compact disks 
which are consistent with previous studies \citep{Downes98}, 
the rest of U/LIRGs have extended molecular gas disks.
Therefore the formation of extended molecular gas disks is not related to 
the physical activity that increases the FIR luminosity, namely the starburst/AGN.

We find that the molecular gas disks in 54~\% (13/24) of the sources  
are more compact than the $K$-band effective radius.  
These gas disks may have formed by past gas inflow 
that was triggered by dynamical instability during the merging event \citep{Barnes02} 
or dry mergers.  
Since the molecular gas is already concentrated in the nuclear regions 
as seen in present-day early-type galaxies \citep[e.g.,][]{Crocker11, Alatalo13, Davis13}, 
these sources are candidates which could become early-type galaxies in the future.  
On the other hand, the molecular gas disks in 46~\% (11/24) of the sources 
are larger than the $K$-band effective radius.  
The frequent occurrence of large gas disks suggest the possibility 
that merging galaxies can evolve into bulge-dominated late-type galaxies, 
unless there are further mechanisms to transport the molecular gas toward the central region 
\citep[e.g., nuclear bar;][]{Bournaud02} and decrease the size of the molecular gas disk.  
The ubiquitous presence of both compact and extended gas disks 
is consistent with the numerical predictions 
that mergers of gas-rich galaxies can produce 
both early- and late-type galaxy population \citep{Springel05}.  

\subsection{Implications for Galaxies at  $z$ = 1 to 3}
Understanding the process of disk formation is particularly important 
at high-redshift when morphological segregation into different Hubble types begins.  
However, the exact formation mechanism of disk galaxies is not well understood.   
Two commonly adopted scenarios are, 
major mergers of gas rich progenitor disks \citep[e.g.][]{Robertson08}, 
or cosmological inflow of low-angular momentum cold gas 
that eventually settles onto the galactic disks \citep[e.g.][]{Keres05, Dekel09}. 
Recent high-sensitivity submillimeter observations 
and optical integral field spectroscopy reveal 
the presence of molecular and ionized gas disks 
in star-forming galaxies at high-redshift 
\citep[e.g.,][]{Forster-Schreiber06, Tacconi13}.  
\citet{Tacconi13} classify 50--75~\% of their sample of $z$ = 1--3 galaxies 
as disk/spiral galaxies based on the CO kinematics and stellar morphology.  

An important finding from our CO study of merger remnants 
is the high occurrence rate of molecular gas disks with various sizes, 
some of which are approaching the size of the Milky Way disk.  
While the surrounding environment and the properties of the progenitor disks 
may be different at low and high redshifts, 
this suggests the possibility that a good fraction of what are identified 
as disk galaxies (i.e. favoring the cold accretion model) 
at high redshifts through rotational kinematics 
may in fact contain a substantial subset of merger remnants.  
Deep and high resolution imaging of the rest-frame 
$K$-band emission is needed 
in order to understand the true nature of these sources.  
Finally, we note that the initial collision of our sample sources probably occurred 
a few billion years ago ($\sim$ typical merger timescale), 
and the observational evidence of high ($>$ 30\%) gas mass fraction 
even at z = 0.3 (3~Gyrs ago) \citep[e.g.,][]{Daddi10, Combes13} 
is consistent with the theoretical prediction that disks with higher gas mass fraction 
are more likely to survive the collision \citep[see discussion in Introduction, and in][]{Hopkins09}.  

\section{Summary}
We have conducted a CO imaging study of 
optically-selected merger remnants in the local Universe 
using millimeter/submillimeter interferometers including ALMA, CARMA, and SMA 
in order to investigate the properties of molecular gas in merger remnants.  
We also obtained archival interferometric data from ALMA, IRAM PdBI, and SMA, 
for 10 further galaxies, forming a grand sample of 37 merger remnants.  
In addition, new NRO 45~m telescope data and earlier single-dish measurements 
from the IRAM 30~m, SEST 15~m, FCRAO 14~m, and NRAO 12~m telescopes 
were obtained to analyze the total CO flux.

We investigate interferometric CO maps of 37 merger remnants, 
including seven galaxies which were undetected in the CO line, 
and find that 80~\% (24/30) of the sample with robust CO detections 
show kinematical signatures of rotating molecular gas disks.  
We also find that the emission peaks in a PV diagram are clearly shifted 
from the galactic centers in 75~\% (18/24) sources with molecular gas disks, 
suggesting the presence of a ring and/or a bar.  
The CO distribution of the remaining six galaxies 
is not characterized by simple gas disk model.  
These sources show clumpy CO distributions and complex velocity fields, 
indicating that the molecular gas is still strongly disturbed 
and has not settled in the galactic plane.  
The molecular gas masses range between 
10$^{7}$ $M_{\sun}$and 10$^{11}$ $M_{\sun}$, 
adopting standard CO luminosity-to-H$_{2}$ conversion factors.  

The sizes of the molecular gas disks ($R_{\rm CO}$) vary significantly 
from 1.1~kpc to 9.3~kpc with an average of 3.5$\pm$2.3~kpc.  
We also estimate the size ratio of the molecular gas disk 
to the stellar component for 24 merger remnants 
in order to investigate whether cold molecular gas disks are in extended form, 
as predicted from recent numerical simulations.  
The size ratios for 54~\% (13/24) of the sample are less than unity, 
hence these sources have compact molecular gas disks 
as seen in present-day early-type galaxies.  
These disks may have formed by past gas inflow 
that was triggered by dynamical instability following the merging.  
On the other hand, 46~\% (11/24) of the sample have gas disks 
which are extended relative to the stellar component.  
We suggest the possibility that these extended gas disks 
may be rebuilding stellar disks and forming a late-type galaxy.  
\\

The CO data products of 17 merger remnants which we observed 
will be publicly released on the website (\url{http://alma-intweb.mtk.nao.ac.jp/\~jueda/data/mr/index.html}).  
The products include data cubes, integrated intensity maps, 
and velocity fields in FITS format.  

\acknowledgments
We thank L. K. Hunt, S. Garc{\'{\i}}a-Burillo, E. Juette, B. Rothberg, and C. D. Wilson 
for kindly providing their published data.
We also thank Chihomi Hara for her help on preparing the CARMA observing scripts.
We would like to express our gratitude to the EA-ARC, SMA and CARMA staff members 
for making the new observations possible.

This paper has made use of the following ALMA data: ADS/JAO.ALMA\#2011.0.00099.S
and ALMA Science Verification data: ADS/JAO.ALMA\#2011.0.00002.SV.
ALMA is a partnership of ESO (representing its member states), NSF (USA) and NINS (Japan), 
together with NRC (Canada) and NSC and ASIAA (Taiwan), in cooperation with the Republic of Chile.  
The Joint ALMA Observatory is operated by ESO, AUI/NRAO and NAOJ.

The Submillimeter Array is a joint project between the Smithsonian Astrophysical Observatory 
and the Academia Sinica Institute of Astronomy and Astrophysics 
and is funded by the Smithsonian Institution and the Academia Sinica.

Support for CARMA construction was derived from the Gordon and Betty Moore Foundation, 
the Kenneth T. and Eileen L. Norris Foundation, the James S. McDonnell Foundation, 
the Associates of the California Institute of Technology, the University of Chicago, 
the states of California, Illinois, and Maryland, and the National Science Foundation. 
Ongoing CARMA development and operations are supported by the National Science Foundation 
under a cooperative agreement, and by the CARMA partner universities. 

This research has made use of the NASA/IPAC Extragalactic Database (NED) 
which is operated by the Jet Propulsion Laboratory, California Institute of Technology, 
under contract with the National Aeronautics and Space Administration. 

Finally, J.U. is financially supported by a Research Fellowship 
from the Japan Society for the Promotion of Science for Young Scientists
and the Sasakawa Scientific Research Grant from The Japan Science Society.  
D.I. is supported by JSPS KAKENHI Grant Number 2580016.  
D.N. acknowledges support from the US NSF via grant AST-144560.  

\appendix
\section{Notes on individual galaxies}
This section provides background information on each of the sources
revealed in previous observations and numerical simulations.  
Here we also give a short description of the morphological
and kinematical features of the molecular gas
seen in the integrated intensity maps, velocity fields,
channel maps, and PV diagrams along the kinematic major axis.  
The rotational velocity of the molecular gas,
which we describe in the following subsections, 
is corrected for inclination using the
estimation given in Table~\ref{tab:t6}.

\subsection{UGC~6 (VV~806, Mrk~334)}
Several loops and shells are identified in this galaxy in optical wavelengths.  
The \textit{Hubble Space Telescope} ($HST$)/Wide Field and Planetary Camera 2 (WFPC2)
$V$-band image clearly shows circumnuclear spiral structures \citep{Martini03}.
The H$\alpha$ velocity field reveals a rotating disk
whose rotational velocity is almost constant ($V_{\rm rot}$ = 210--220 km~s$^{-1}$)
over a radial distance range of 4\arcsec -- 12\arcsec \citep{Smirnova10}.  
Optical spectroscopy classifies the galaxy
as a Seyfert~1.8 \citep[e.g.,][]{Dahari88, Osterbrock93}.
The \ion{H}{1} spectra show a double-horn profile \citep{Mirabel88},
which is characteristic of a rotating disk.

The CO~(1--0) emitting gas is distributed 
within a radius of 3\arcsec ($\simeq$ 1.3~kpc) from the galactic center.  
The detailed distribution and kinematics is not clear 
due to the limited velocity resolution ($\Delta V$ = 100 km~s$^{-1}$), 
but the velocity field reveals the presence of a rotating disk.  
The maximum rotational velocity is estimated to be $\sim$280 km~s$^{-1}$.
This molecular gas disk is about four times smaller than the ionized gas disk 
studied by \citet{Smirnova10}.

\subsection{NGC~34 (NGC~17, VV~850, Mrk~938)}
This galaxy is classified as a LIRG.  
The galaxy has several ripple-like patterns and two linear tidal tails 
extending towards the northeast and the southwest seen in the optical.  
The projected length of the longer tail is over 20~kpc \citep{Schweizer07}.  
The $HST$/Advanced Camera for Surveys (ACS) $B$-band image 
clearly shows a bluish central disk with spirals 
and its optical light is dominated by a $\sim$400~Myr old post-starburst population \citep{Schweizer07}.  
The nature of the activity in this galaxy based on the optical spectra is controversial.  
Some authors classify the galaxy as a Seyfert~2 \citep[e.g.,][]{Veilleux95, Yuan10, Brightman11}, 
while others suggest that the galaxy is classified as a \ion{H}{2} galaxy 
rather than a Seyfert galaxy \citep[e.g.,][]{Mulchaey96, Goncalves99}.  
\citet{Fernandez10} find two diffuse radio lobes, spanning 390~kpc, 
which could be a signature of an AGN or a starburst-driven superwind.  
\citet{Fernandez10} also suggest that an atomic gas disk is forming from the gas of the northern tail.  
\citet{Xu14} present ALMA Cycle~0 observations of the CO~(6--5) line 
and find a compact gas disk with a size of 200~pc.

This galaxy has a molecular gas disk extended 
in comparison with the stellar structure ($R_{80}/R_{\rm eff}$ = 4.54), 
but the absolute size of the molecular gas disk ($R_{\rm CO}$ $\simeq$ 3 kpc) 
is approximately equal to the average size in our sample.  
The PV diagram shows nearly symmetric isophotes with respect to the systemic velocity 
and two emission peaks on opposite sides of the nucleus, 
indicating the presence of a ring or a disk with non-uniform distribution.  

\subsection{Arp~230 (IC~51)}
This galaxy is well known as a shell galaxy and 
listed as a candidate of a polar ring galaxy \citep{Whitmore90}.  
The $HST$/WFPC2 $V$-band image shows strong dust lanes along the polar ring.  
Stellar shells are clearly visible in the northeast and southeast. 
\citet{McGaugh90} estimate the timescale for shell formation 
based on the optical color of the shells.  
The color implies an age of 1--2 Gyr for the outermost shell, 
an age of 0.5--0.7~Gyr for the intermediate shell, and 
a young age ($\leq$ 0.3~Gyr) for the inner shells.  
The majority of the 6~cm radio continuum and \ion{H}{1} emission 
is aligned with the optical dust lane \citep{Cox04, Schiminovich13}.  
From the \ion{H}{1} velocity field, \citet{Schiminovich13} find 
the presence of an outer \ion{H}{1} disk and a dense inner \ion{H}{1} disk/ring.

The CO~(1--0) emitting gas is seen along the optical dust lane 
and distributed in three main components.  
One component is associated with the nucleus and 
the others are located at the tangential points of the polar ring.  
The velocity field shows a shallow velocity gradient of $\sim$75 km~s$^{-1}$~kpc$^{-1}$, 
which has been corrected using the estimated inclination of 65\arcdeg.  
These signatures suggest the presence of a slow-rotating ring/disk of the molecular gas, 
which appears to coincide with the inner \ion{H}{1} disk/ring \citep{Schiminovich13}.  
In addition, we find three 3~mm continuum components, 
whose peaks roughly correspond to the CO emission peak of the three main components.
 
\subsection{NGC~455 (Arp~164, UGC~815)}
There are few studies focusing on this galaxy.  
Optical imaging reveals two diffuse tails
extending towards the northwest and the southeast.  
The galaxy is undetected in the $IRAS$ Sky Survey Atlas,
but detected in the $AKARI$ All-Sky Survey.  
We estimate a FIR luminosity of 3.0 $\times$ 10$^{9}$~$L_{\sun}$ 
using $AKARI$ data at 65$\mu$m, 90$\mu$m, and 140$\mu$m \citep{Yamamura10}
and the formula defined by \citet{Takeuchi10}.

This galaxy is undetected in the CO~(1--0) line.  
The 3~$\sigma$ upper limit on the molecular gas mass 
is $3.3 \times 10^{7}$~$M_{\sun}$.

\subsection{NGC~828 (UGC~1655)}
This galaxy is classified as a LIRG.  
The NIR image shows several filaments surrounding the main body 
and spiral features within 10\arcsec~from the galactic center \citep{Smith96}.  
The small 3.3~$\mu$m polycyclic aromatic hydrocarbon (PAH) equivalent width (EW $\leq$ 20~nm) 
suggests that a powerful buried AGN is present in the galaxy \citep{Imanishi06}.
In the H$\alpha$ emission line, there are three sources 
with comparable surface brightness \citep{Hattori04}.  
One diffuse source is associated with the nucleus 
and the others are located nearly symmetrically along the major axis of the galaxy.  
The rotation curve measured from the H$\alpha$ emission 
shows flat rotation with a constant velocity ($V_{\rm rot}~$sin~$i$) of 200 km~s$^{-1}$ 
outside $\ga$ 8\arcsec~from the galactic center \citep{Marquez02}.
Multiple CO line spectra show double horn profiles, 
suggesting the presence of a rotating disk
\citep{Sanders91, Casoli92, Narayanan05}.
Previous CO~(1--0) interferometric observations provide 
a rotation curve of the molecular gas disk \citep{Wang91}, 
but the observed disk size was limited to a 3.5~kpc radius.  

Our new CO~(2--1) observations confirm the presence of 
a large molecular gas disk with a radius of 8~kpc.  
The rotation curve of the molecular gas disk shows flat rotation 
with a velocity of $\sim$270 km~s$^{-1}$ outside a radius of 8\arcsec.
This rotational velocity is larger than that measured form the H$\alpha$ emission 
\citep{Marquez02} by 70 km~s$^{-1}$.  
The PV diagram shows a symmetric distribution and a double peak 
located at the transition point between rigid rotation and flat rotation, 
which may signify a spiral and/or a bar structure rather than a ring structure.

\subsection{UGC~2238}
This galaxy is classified as a LIRG.  
The H$\alpha$ emission is distributed
across the galaxy in several knots \citep{Hattori04}.
According to optical spectroscopy, \citet{Veilleux95} classify the galaxy
as a Low Ionization Nuclear Emission Region galaxy (LINER),
while \citet{Yuan10} suggest that this is a starburst--AGN composite.  
However, there is no sign of an AGN or an obscured AGN
in the NIR spectra \citep{Imanishi10}.  
The radio continuum emission is elongated in the northeast and southwest,
corresponding to the optical morphology \citep{Condon90}.  

We find a large molecular gas disk in this galaxy 
for the first time with a radius of 8.5~kpc.
The CO~(2--1) emitting gas is elongated in the direction of 
the morphological major axis of the stellar component.  
The rotation curve of the molecular gas disk shows flat rotation 
with a velocity of $\sim$200 km~s$^{-1}$
at $\ga$ 6\arcsec~from the galactic center.  
The PV diagram reveals nearly symmetric isophotes and a double peak 
located at the transition point between rigid rotation and flat rotation.  

\subsection{NGC~1210 (AM~0304--255)}
This galaxy presents shells in optical images \citep{Malin83}.  
The $K$-band surface brightness is approximately a de Vaucouleurs profile 
with a the S\'{e}rsic index of 4.08 (RJ04).  
The UV imaging reveals a tidal tail extending towards the north 
and a diffuse debris structure on the south region \citep{Marino09}.  
Furthermore, several UV knots with a luminosity 
similar to that of the nucleus are seen along the tidal tail.  
The \ion{H}{1} emission observed by \citet{Schiminovich01} 
corresponds to the tidal tail visible in the UV images.  
The PV diagram measured from the [\ion{O}{2}] line at 4959~\AA~shows 
rigid rotation, suggesting a rotating ionized gas component \citep{Longhetti98}.

This galaxy is undetected in the CO~(1--0) line. 
The 3~$\sigma$ upper limit on the molecular gas mass 
is $1.7 \times 10^{7}$~$M_{\sun}$.

\subsection{AM~0318--230}
There are few studies focusing on this galaxy.
The $K$-band isophotal shape clearly shows a featureless rectangular core,
even though the overall shape is dominated by disky isophotes \citep[][RJ04]{Chitre02}.

This galaxy is undetected in the CO~(1--0) line.  
The 3~$\sigma$ upper limit on the molecular gas mass 
is $2.2 \times 10^{8}$~$M_{\sun}$.

\subsection{NGC~1614 (Arp~186, Mrk~617)}
This galaxy is classified as a LIRG.  
Optical imaging shows a bright center with two spiral arms at scales of few kpc,
a liner tail extending over 20~kpc towards the southwest \citep{Mullan11},
and a large curved extension to the southeast of the nucleus.  
The \ion{H}{1} map shows a single tail \citep{Hibbard96},
which does not follow the optical tail.  
Several knots and condensations traced by the H$\alpha$ and the P$\alpha$ emission
are distributed along the eastern spiral arm \citep{Alonso-Herrero01, Rodriguez-Zaurin11}.  
A star-forming ring with a radius of $\sim$300~pc is observed
at the P$\alpha$ emission \citep{Alonso-Herrero01} and radio \citep[e.g.][]{Neff90, Olsson10}.  
\citet{Haan11} suggest that the galaxy built up a central stellar cusp due to nuclear starbursts,
judging from the significant core light excess in the $H$-band image and relatively large 6.2~$\mu$m PAH EW.  
\citet{Soifer01} find that 87~\% of the 12~$\mu$m flux come from
within a radius of 4\arcsec (1.2~kpc) from the nucleus.  
This galaxy is classified as a starburst-dominated galaxy with no AGN signature
through optical and infrared spectroscopy \citep[e.g.,][]{Veilleux95, Imanishi10}.  
Small HCN/HCO$^{+}$ line ratios are further evidence for dominance of starburst activity
\citep{Costagliola11, Imanishi13}.
High-resolution CO~(2--1) observations at 0\farcs5 resolution reveals
a nuclear ring of molecular gas \citep{Konig13} fed by dust lanes.  

We use the CO~(2--1) data obtained by \citet{Wilson08}.  
The PV diagram shows rigid rotation and a double peak
located at a distance of 300--400~pc from the galactic center,
which is consistent with the CO~(1--0) data
obtained at the \textit{Owens Valley Radio Observatory} (OVRO) \citep{Olsson10}.  
This CO~(2--1) data is unable to spatially resolve
the nuclear molecular ring presented by \citet{Konig13}. 

\subsection{Arp~187}
This is a well known radio galaxy.  
Many radio continuum surveys including Arp~187 have been conducted, 
but there are few studies focusing on this galaxy using all available wavelengths.  
Optical imaging shows two diffuse tails running in the north and south directions.  
\citet{Evans05} suggest that the star formation rate is low 
because the galaxy has a low $L_{\rm IR}/L_{\rm CO}$, 
which is similar to the global $L_{\rm IR}/L_{\rm CO}$ of local spiral galaxies, 
and the dust temperature is lower (30--45~K) than the typical value of radio galaxies.  
Assuming that the IR luminosity is entirely due to star formation, 
the star formation rate (SFR) derived from the FIR luminosity is a few solar masses per year.  

The CO~(1--0) emitting gas appears to form a disk overall, 
but the distribution is disturbed and warped in the nuclear region.  
We detect two 3~mm continuum components 
located at both sides of the nucleus, spanning 4~kpc.  
The direction connecting these two components is different 
from the kinematical major axis of the molecular gas.  
These components can be small radio lobes, if an AGN exists.  
In order to check for the presence of an AGN, 
we estimate the $q$-value of the radio--FIR correlation \citep{Helou85} 
using the 1.4~GHz radio continuum data \citep{Condon98}.  
Arp~187 shows a radio-excess ($q$ = -0.1), suggesting an AGN activity.  
We also investigate the radio spectral index at a few GHz 
using the 5~GHz and 8.4~GHz radio data from the literature \citep{Wright92}.  
We find a spectral index of $\alpha$ = -1.0, which also indicates the presence of an AGN.  

\subsection{AM~0612--373}
There are few studies focusing on this galaxy.  
Optical imaging shows a long tail extending over 30~kpc towards the south 
and a dust lane running along the morphological minor axis of the stellar body \citep{Smith91b}.  
The rotation curve measured from the \ion{Ca}{2} triplet absorption line 
($\lambda$ $\sim$ 0.85~$\mu$m) 
reveals a velocity gradient \citep[$V_{\rm rot}~$sin~$i$ = $\pm$~160 km~s$^{-1}$;][]{Rothberg10}, 
which cannot be fully explained by either rigid or flat rotation.

Most of the CO~(1--0) emitting gas is distributed 
within a radius of 2.5~kpc from the galactic center.  
We note that the kinematical major axis of the CO distribution is 
perpendicular to the apparent morphological major axis of the stellar body.  
The PV diagram shows rigid rotation and a multi-peak.   
The maximum rotational velocity is estimated to be $\sim$380 km~s$^{-1}$.  

\subsection{UGC~4079 (Mrk~84)}
There are few studies focusing on this galaxy.  
The $HST$/WFPC2 $B$-band imaging shows clumpy structures in the main body 
and a faint tail with several blobs, extending towards the northeast.
The \ion{H}{1} spectra show a double-horn profile 
with a full-width at half maximum (FWHM) of 234 km~s$^{-1}$ \citep{Springob05}.  

The new CO~(1--0) spectra obtained using the NRO 45~m telescope 
show a double-horn profile with a FWHM of 240 km~s$^{-1}$, 
suggesting the presence of a rotating gas disk.  
However, the galaxy is undetected in the interferometric maps 
due to the limited sensitivity.  
Thus the rotating molecular gas disk is not confirmed.

\subsection{NGC~2623 (Arp~243, UGC~4509, VV~79)}
This galaxy is classified as a LIRG.  
Optical image reveals two prominent tails extending over 120\arcsec
($\sim$43~kpc; 1\arcsec = 0.36~kpc) towards the northeast and the southwest \citep{Soifer01}.  
High-resolution $HST$/ACS images reveal $\sim$100 star clusters in the southern region of the nucleus.  
 Optical colors of the clusters are consistent with ages of $\sim$1--100~Myr \citep{Evans08}.  
\citet{Soifer01} find that 80~\% of the 12~$\mu$m flux comes from
within a radius of 2\arcsec ($\sim$0.7~kpc) from the galactic center.   
The rotation curves measured form the H$\alpha$ and the [\ion{N}{2}] emission show 
flat rotation with a constant velocity ($V_{\rm rot}$~sin~$i$) of $\sim$100$\pm$50 km~s$^{-1}$ \citep{Smith96}.
While \citet{Risaliti00} find no X-ray evidence for the presence of an AGN,
later work shows the signature of a Compton-thick AGN based on the X-ray data \citep{Maiolino03}.  
A high HCN/HCO$^{+}$ line ratio is further evidence for an AGN 
\citep[HCN/HCO$^{+}$(1--0) = 1.5;][]{Imanishi09}.
The \ion{H}{1} spectra show both emission and absorption,
and two distinct absorption features are observed at velocities below and above
the systemic velocity of the galaxy \citep[e.g.][]{Biermann78, Mirabel88}.  
The VLA \ion{H}{1} map shows two tidal tail, which follow the optical tails \citep{Hibbard96}.

We use the CO~(2--1) data obtained by \citet{Wilson08}.  
The CO~(2--1) emitting gas is distributed within a radius of 3\arcsec
($\simeq$ 1.1~kpc) from the galactic center.
This $R_{\rm CO}$ is the smallest of our sample sources with molecular gas disks.  
The CO~(2--1) maps are similar to the CO~(1--0) map obtained at the OVRO by \citet{Bryant99}.  
The PV diagram shows rigid rotation and 
the maximum rotational velocity is estimated to be $V_{\rm rot}$~sin~$i$ = 180 km~s$^{-1}$.  

\subsection{NGC~2782 (Arp~215, UGC~4862)}
This galaxy has a long \ion{H}{1} tail ($\sim$5\arcmin) extending towards the northwest
and a short \ion{H}{1} tail ($\sim$2\arcmin) extending toward the east,
which is associated with a stellar tail \citep{Smith91a}.  
The \ion{H}{1} velocity field shows the presence of an inner disk counter-rotating
with respect to the gas motions in the outer region \citep{Smith91a}.
\citet{Smith94} conclude using a three-body dynamical model
that the galaxy is a merger of two disk galaxies with a mass ratio of $\sim$4:1.  
The $HST$/WFPC2 images reveals a nuclear stellar bar within a radius of 7\farcs5,
which appears to be fueling gas into the nuclear starburst \citep{Jogee99}.  
\citet{Mullan11} identify 87 star cluster candidates in the eastern tail
and 10 candidates in the northwestern tail
using the $HST$WFPC2 $V$-and $I$-band images.  
Previous studies considered its nuclear activity 
to be dominated by a nuclear starburst \citep[e.g.,][]{Kinney84, Boer92},
although recent radio and X-ray observations
suggest the presence of a hidden AGN \citep{Braatz04, Zhang06, Krips07}.

\citet{Hunt08} investigate the CO~(1--0) and CO~(2--1) distribution and kinematics
using high spatial and velocity resolution maps obtained at the PdBI.  
The CO emission is aligned along the stellar nuclear bar with two emission peaks
on opposite side of the nucleus, spanning $\sim$6\arcsec,
and distributed in an elongated structure with two spiral arms within a radius of 8\arcsec.  
We estimate the rotation curve of the molecular gas disk
and find flat rotation with a velocity of $\sim$260 km s$^{-1}$ outside a radius of 4\arcsec.  
In addition, there are two more extended spiral features extending towards the north and the south.  
Additional CO interferometric maps were obtained using the OVRO by \citet{Jogee99}.  
They find that the bar-like gaseous feature is offset in a leading sense
with respect to the stellar bar and shows non-circular motion,
which suggests gas inflow into the nucleus.

\subsection{UGC~5101}
This galaxy is classified as a LIRG or an ULIRG
depending on the assumed distance,
and well-known as an object with a buried AGN
based on multi-wavelength observations.  
\citet{Imanishi03} detect an absorbed hard X-ray component
with a 6.4 keV Fe K$\alpha$ line, while
 \citet{Armus04} detect the 14.3~$\mu$m [\ion{Ne}{5}] line.  
These detections indicate the presence of a buried AGN.  
A high HCN/HCO$^{+}$ line ratio (= 2.5) is further evidence
for dominance of AGN activity \citep{Imanishi06}.
Optical imaging shows a 38~kpc linear tail extending toward the west
and a large ring \citep{Surace00}.
\citet{Windhorst02} suggest the presence of an inclined dusty disk
using the $HST$/WFPC2 $I$-band image.
The rotation curves measured from the H$\alpha$ and [\ion{N}{2}] emission
show flat rotation with a constant velocity ($V_{\rm rot}$~sin~$i$)
of $\sim$200 km~s$^{-1}$ \citep{Smith96}.

We use the CO~(2--1) data obtained by \citet{Wilson08}.
The PV diagram shows shows rigid rotation and a double peak 
located at a distance of 400--500~pc from the galactic center.  
The PdBI CO~(1--0) map \citep{Genzel98} is
similar to the CO~(2--1) map \citep{Wilson08},
but the CO~(1--0) emission is more extended than the CO~(2--1).  
Thus the radius of the molecular gas disk which we estimate in this study
is underestimated.

\subsection{AM~0956--282 (VV~592)}
There are few studies focusing on this galaxy.  
Optical imaging shows a diffuse core with patchy structures.  
RJ04 could not determine structural parameters 
of the $K$-band surface brightness due to its disturbed structure.  
It is possible that the uncertainty of the position of the nucleus 
defined in the $K$-band peak is a large.  
This galaxy is detected in the $AKARI$ All-Sky Survey, 
and the FIR luminosity is estimated to be 7.0 $\times$ 10$^{8}$~$L_{\sun}$ 
using the 3-bands \citep[65$\mu$m, 90$\mu$m, and 140$\mu$m;][]{Yamamura10}.  

The CO (1--0) emitting gas is distributed in three main components.  
The molecular gas masses of these components are 
4.8$\times$10$^{7}$~$M_{\sun}$, 4.2$\times$10$^{6}$~$M_{\sun}$, 
and 1.3$\times$10$^{6}$~$M_{\sun}$.  
The largest mass component is associated with the main body of the galaxy.  
The position angle of the kinematic major axis is almost 0\arcdeg, 
which is not aligned with the apparent isophotal major axis of the stellar body.  
The PV diagram reveals a velocity gradient, 
which is steeper in the northern region than the southern region.  
The intermediate mass component located in the southeast of the largest component 
shares the same velocity distribution.  
A diffuse optical counterpart, possibly a short tidal tail is seen in the same region.  
Thus this component is likely to have been ejected 
from the main body due to the interaction.  
The smallest mass component is located in the west of the main body.  
This component has a stellar counterpart, 
with a weak 3~mm continuum emission associated to it.  
This component may be a star cluster or a dwarf galaxy.

\subsection{NGC~3256 (AM~1025--433, VV~65)}
This galaxy is classified as a LIRG.  
The galaxy has two prominent tidal tails
seen in the optical and the \ion{H}{1} emission \citep{English03}.  
The projected length of the tails are longer than 40~kpc \citep{Mullan11}.  
The $K$-band image (RJ04) shows a single nucleus,
but a double nucleus separated by $\sim$5\arcsec ($\sim$0.9~kpc)
along the north--south direction has been detected
in the radio, NIR, and X-ray \citep{Norris95, Kotilainen96, Lira02}.  
The SMA CO~(2--1) observations also reveal
the presence of the double nucleus \citep{Sakamoto06}.  
The northern nucleus has been identified as a intense star-forming region
with a powerful outflowing galactic wind \citep{Lipari00, Lira02}.  
While \citet{Neff03} suggest that the both nuclei may host low-luminosity AGNs
using radio and X-ray observations,
\citet{Jenkins04} find no strong evidence for active galactic nuclei
based on \textit{XMM-Newton} X-ray observations.
The $HST$/WFPC2 images shows spiral structures \citep{Laine03}.  
\citet{Zepf99} find more than 1000 young stellar clusters
with ages ranging from a few to several hundred Myrs
in the inner region (7~kpc $\times$ 7~kpc) using the $HST$/WFPC2 images.  

We use the ALMA Science Verification (SV) data.  
The velocity field and PV diagram reveals the presence of a disturbed disk.  
The rotation curve of the molecular gas disk shows
rigid rotation within a radius of 20\arcsec.    
The strongest 3~mm continuum emission in our sample is detected in this galaxy.
The SV data was published by \citet{Sakamoto14} which, 
combined with their own higher spatial resolution Cycle~0 data, 
show several arm-like structures \citep{Sakamoto14}.  
The SV data by itself is unable to spatially resolve the double nucleus
presented by \citet{Sakamoto06, Sakamoto14}.

\subsection{NGC3597 (AM~1112--232)}
Optical imaging shows plumes elongated in the west of the main body \citep{Lutz91}.  
\citet{vanDriel91} detected two radio components at opposite sides of the galactic center.  
Optical spectroscopy classifies the galaxy as a \ion{H}{2} galaxy \citep[e.g.,][]{ Veilleux95, Kewley00}.  
\citet{Lutz91} find a population of 14 blue stellar clusters
and later work by \citet{Carlson99} also find $\sim$700 compact objects
using the $HST$/WFPC2 images,
which suggests that the progenitor galaxies were gas-rich.

The CO~(1--0) integrated intensity map shows a bar-like structure.  
The velocity field shows a smooth velocity gradient and
the rotation curve of the molecular gas disk shows flat rotation
with a velocity of $\sim$155 km s$^{-1}$ outside a radius of 1~kpc.  
The 3~mm continuum is roughly distributed along the bar-like CO structure.  
The strongest peak of the 3~mm continuum is located at a distance of 3\farcs7
from the galactic center and corresponds to one of the CO peaks.  

\subsection{AM~1158--333}
There are few studies focusing on this galaxy.  
Optical imaging shows a diffuse extension towards the south of the main stellar body.  
The main body is elongated in the northeast and southwest.  
The brightest peak, which we define as the nucleus, is located at the northeast edge.  
\citet{Arp87} classify the galaxy as an elliptical galaxy with jets.  
The \ion{H}{1} spectra show a double peaked profile 
with a FWHM of $\sim$~100 km~s$^{-1}$ \citep{Theureau98}.  

We detect two molecular gas components traced by the CO~(1--0) line.  
One component is associated with the nucleus.  
This cannot be resolved due to the limited spatial resolution.  
The other is distributed below the main body blue-shifted 80 km~s$^{-1}$ 
from the systemic velocity ($V_{\rm sys}$ = 3027 km s$^{-1}$).  
The molecular gas mass of the component associated with the nucleus 
is 1.2 $\times$ 10$^{7}$ $M_{\sun}$, which is 2.8 times smaller than 
that of the other component.  

\subsection{NGC~4194 (Arp~160, UGC~7241, VV~261, Mrk~201)}
This galaxy is known as the ``Medusa Merger''.
The optical morphology is characterized by a diffuse (``hair-like'') tail
extending over 10~kpc towards the north of the main body.  
The \ion{H}{1} map shows a single $\sim$50~kpc tail
extending towards the opposite side of the optical tail \citep{Manthey08}.  
This galaxy is a candidate of a minor merger
between a large elliptical and a small spiral galaxy \citep{Manthey08}.  
\citet{Weistrop04} identify 48 UV-bright knots,
two-thirds of which are younger than 20~Myr,
and suggest that the UV knots will become globular clusters in several Gyr.  
The HCN/HCO$^{+}$(1--0) line ratio is less than unity,
suggesting a starburst-dominated galaxy \citep{Costagliola11}.
The ionized gas traced by the 12.8~$\mu$m [\ion{N}{2}] line shows
a smooth velocity gradient in the north-south direction
through the galactic enter \citep{Beck14}.  

The galaxy is unusual in that 
the CO~(2--1) emission is distributed along the optical tail.  
As seen in the velocity field, the diffuse gas forms three tails
in the northern region of the main body.  
In addition, the CO~(2--1) emitting gas appears
to be forming a rotating disk in the main body,
although the kinematical parameters
for the molecular gas cannot be determined
because it is difficult to exclude the gaseous tails
from the component associated with the main body.  
These kinematic features are consistent with
the OVRO CO~(1--0) map \citep{Aalto00}. 

\subsection{NGC~4441 (UGC~7572)}
Optical imaging shows one tidal tail extending towards the north 
and two shells to the southwest of the main body.  
This galaxy is a candidate of a merger remnant  
between a spiral and an elliptical galaxy \citep{Manthey08, Jutte10}.  
The large-scale \ion{H}{1} distribution and kinematics is studied by \citet{Manthey08} 
using the Westerbork Radio Synthesis Telescope.  
The total mass of the atomic gas is 1.46$\times$10$^{9}$~$M_{\sun}$.  
The \ion{H}{1} distribution shows two prominent tidal tails 
extending more than 40~kpc towards the north and south.  
The northern tail follows the optical tail closely.  
The \ion{H}{1} velocity field shows simple rotation in the inner region, 
indicating that the gas has settle down.  
Its velocity gradient continues into the tidal tails.

\citet{Jutte10} investigate the CO~(1--0) distribution and kinematics 
using high spatial and velocity resolution maps obtained at the PdBI.  
Most of the CO emission comes from a central rotating disk 
with a radius of about 2~kpc.  
The kinematic major axis of the molecular gas disk is nearly perpendicular to 
the rotation of large-scale \ion{H}{1} structure \citep{Manthey08}, 
indicating a kinematically decoupled core.  
The integrated intensity map we obtained is slightly different from \citet{Jutte10} 
due to different clipping levels.  
In our map, there are CO extensions on the outer side of the molecular gas disk, 
which appear to be kinematically decoupled from the disk.

\subsection{UGC~8058 (Mrk~231)}
This galaxy is the only object in our sample to be classified as a ULIRG,
and well-known as an object which harbours both intense starburst and powerful AGN activities.  
The fractional contribution of the AGN and the starburst to the bolometric IR luminosity is uncertain.
\citet{Farrah03} suggest that the AGN contributes $\sim$30~\% to the IR luminosity,  
and \citet{Veilleux09} estimate that the AGN contribution is $\sim$70~\%.  
This galaxy is also characterized by jets and kpc-scale outflows,
which is likely a result of the negative feedback from the AGN.
The jets and outflows have been detected at radio wavelengths
\citep[e.g.,][]{Carilli98, Lonsdale03},
and recently detected at millimeter wavelengths \citep[e.g.,][]{Feruglio10, Aalto12}.  
The outflow rate of molecular gas is estimated to be $\sim$700 M$_{\sun}$~yr$^{-1}$,
which exceeds the ongoing SFR in the host galaxy \citep{Feruglio10}.  
Optical imaging shows two tidal tails extending towards the north and south,
whose combined projected length is over 70~kpc \citep{Koda09}.  
Multi-wavelength observations (at NIR, millimeter, and radio)
suggest the presence of a nearly face-on disk \citep{Carilli98, Downes98, Taylor99}.

We use the CO~(3--2) data obtained by \citet{Wilson08}.   
The CO~(3--2) map reveals the presence of a molecular gas disk
as previously presented by \citet{Downes98}.  
The maximum extent of the gas disk estimated using the CO~(3--2) map
is similar to that seen in the PdBI CO~(1--0) map \citep{Downes98}.  
The PV diagram shows a main component which appears to be a gas disk
and a high-velocity component located along the line-of-sight in the directoin of the galactic center.  
This high-velocity component may be part of molecular outflows.

\subsection{AM~1255--430}
There are few studies focusing on this galaxy.  
The optical morphology is X-shaped with tails and plumes,   
the longest of which extends over 20~kpc towards the southeast \citep{Smith91b}.  

The distribution of the CO~(1--0) emitting gas is significantly disturbed.  
The kinematical parameters for the molecular gas cannot be determined.  
The channel map ($V$ = 8560 -- 8840 km~s$^{-1}$) shows 
an arc-like distribution in the northeast region.  
The molecular gas moves along the arc and 
merges into a gas component associated with the nucleus.  

\subsection{AM~1300--233}
This galaxy is classified as a LIRG.  
The $HST$/ACS $B$-band image show a disrupted nuclear region,
a small tail extending towards the northeast, and
a bright region located at $\sim$4~kpc southwest of the galactic center.  
Although this region is identified as a secondary nucleus \citep{Miralles-Caballero11},
it is likely to be a massive star-forming region \citep{Monreal-Ibero10, Rodriguez-Zaurin11}.
\citet{Monreal-Ibero10} suggest that the mechanism of ionization
in the nuclear region is shocks (100--150 km~s$^{-1}$),
using the [\ion{O}{1}]$\lambda$6300/H$\alpha$ ratio,
which is a good tracer of shock-induced ionization.
Optical spectroscopy classify the galaxy as a LINER \citet{Corbett03}.  
There is no sign of either an AGN or an obscured AGN in the NIR spectra \citep{Imanishi10}.

The CO~(1--0) emitting gas appears to form a disk overall,
but the distribution is disturbed and warped in the nuclear region.  
The kinematic major axis of the molecular gas is inconsistent with
the morphological major axis of the stellar body.  
The 3~mm continuum emission is elongated
along the kinematic minor axis of the molecular gas.  

\subsection{NGC~5018}
The galaxy has several shells seen 
in the NUV \citep{Rampazzo07} and the optical \citep{Malin83}, 
and it presents a prominent linear dust lane \citep{Fort86}.  
\citet{Goudfrooij94} find extended H$\alpha$+[\ion{N}{2}] emission.  
The projected size of the H$\alpha$+[\ion{N}{2}] emitting region 
is larger than 40\arcsec ($\simeq$ 7.3~kpc).  
The rotation curves measured from the \ion{Ca}{2} triplet absorption line 
($\lambda \sim$ 0.85~$\mu$m) and the CO stellar absorption line 
($\lambda$ = 2.29~$\mu$m) show flat rotation \citep{Rothberg10}, 
suggesting the presence of a rotating disk.  
The \ion{H}{1} emission is not associated with the nuclear region, 
however, it is distributed along a filamentary structure 
connecting between NGC~5018 and its two neighbors, 
NGC~5022 and MCG~03-34-013 \citep{Kim88}.  
This \ion{H}{1} filament is a sign of interaction between these galaxies.

This galaxy is undetected in the CO~(1--0) line.  
The 3~$\sigma$ upper limit on the molecular gas mass 
is 2.3 $\times$ 10$^{7}$~$M_{\sun}$.
 \citet{Lees91} and \citet{Huchtmeier92} also report a non-detection 
in their single-dish CO measurements.  

\subsection{Arp~193 (UGC~8387, VV~821, IC~883)}
This galaxy is classified as a LIRG.  
The $HST$/ACS $B$-band image shows two straight tails
emerging almost at 90\arcdeg angles,
and dust lanes across the nuclear region.  
There are two emission peaks in the nuclear region, spanning $\sim$340~pc,
traced by the IR \citep{Scoville00} and the radio \citep{Condon91},
but the $K$-band image (RJ04) shows a single peak.  
The rotation curves measured form the H$\alpha$ and the [\ion{N}{2}] emission
show flat rotation with a constant velocity ($V_{\rm rot}$~sin~$i$) of 150 km~s$^{-1}$ \citep{Smith96}.
The soft X-ray emission is elongated along the morphological minor axis of the stellar body,
which is likely associated with outflow driven by a nuclear starburst or AGN \citep{Iwasawa11}.
Optical spectroscopy classify the galaxy as a LINER \citep{Veilleux95},
while \citet{Yuan10} suggest that this is a starburst-AGN composite.
However, the small HCN/HCO$^{+}$(1--0) line ratio of 0.9 and the strong 3.3~$\mu$m
are evidence for starburst activity with no significant AGN contribution \citep{Imanishi09}.  

Three CO interferometric maps were published \citep{Downes98, Bryant99, Wilson08}.  
We use the CO~(3--2) data obtained by \citet{Wilson08}.  
The PV diagram shows an asymmetric distribution and the presence of a rotating disk.  
Other CO~(1--0) and CO~(2--1) interferometric data also shows
a nuclear ring or disk of molecular gas \citep{Downes98, Bryant99}.
The CO~(3--2) emission is detected outside the extent of the CO~(1--0) and CO~(2--1) emission,
due to the high sensitivity in CO~(3--2) measurements.  

\subsection{AM~1419--263}
There are few studies focusing on this galaxy.  
Optical imaging shows a $\sim$20~kpc plume 
elongated in the southwest and northeast 
and a 4~kpc tail extending towards the south \citep{Smith91b}.  
Even though the optical structure is strongly distorted, 
the $K$-band surface brightness is almost perfectly 
fit by the de Vaucouleurs profile (RJ04).  
The rotation curve measured from the \ion{Ca}{2} triplet absorption line 
($\lambda$ $\sim$ 0.85~$\mu$m) 
shows a shallow velocity gradient \citep{Rothberg10}, 
which cannot be fully explained by rigid rotation or flat rotation.  

This galaxy is tentatively detected in the CO~(1--0) line, 
but there is no robust emission ($>$ 1.5~$\sigma$) continuing in velocity.   
The detailed distribution and kinematics of the molecular gas is not clear 
due to the limited sensitivity.  
The 3~mm continuum emission is clearly detected 
at the $\sim$20~$\sigma$ level in the nucleus.

\subsection{UGC~9829 (VV~847)}
There are few studies focusing on this galaxy.  
Optical imaging shows two long tidal tails running north-south.  
The projected length of the northern tail is longer than 30~kpc.  
The $K$-band image shows a bar structure across the nucleus (RJ04).

This galaxy is an unusual case in that most of the CO~(2--1) emission comes from  
the root of the northern tail, with only weak emission seen in the nucleus.  
The molecular gas shows a kinematic signature of gas streaming along the tail, 
but it does not appear to inflow into the nuclear region 
because the velocity of the molecular gas in the nucleus is 
about 250 km~s$^{-1}$ smaller than that in the tail leading to the nucleus.

\subsection{NGC~6052 (Arp~209, UGC~10182, VV~86, Mrk~297)}
The optical morphology is dominated by a large number of clumpy features,
which appear to be young massive clusters
according to their color and brightness \citep{Holtzman96}.  
The \ion{H}{1} map shows a tidal tail extending towards the south \citep{Garland07}.  
\citet{Alloin79} interpret the galaxy as a merger of two late-type galaxies,
which might have produced the burst of star formation in clumps.
\citet{Taniguchi91} suggest from numerical $N$-body simulations of the collision of two disk galaxies 
that the evolutionary phase of this galaxy
corresponds to about 150~Myr after the first impact.  
The peak of the radio continuum emission is shifted
from the emission peak in optical images,
suggesting the presence of off-nuclear starbursts \citep[e.g.,][]{Condon90, Deeg93}.
The OVRO CO~(1--0) emission is barely resolved and
the size of the CO~(1--0) source is 7\farcs5 $\times$ 4\farcs8 \citep{Garland07}.

The CO (2--1) emitting gas comes from the northern part of the galaxy
and is not associated with the nucleus nor bright knots in the optical.  
The velocity field of the molecular gas as well as the ionized gas \citep{Garcia-Lorenzo08} 
are significantly disturbed and shows complex structures.  
The majority of the molecular gas appears not to rotate in circular motion around the nucleus.  
It is likely that the galaxy has not undergone violent relaxation,
according to the kinematics of the molecular gas and the apparent morphology.

\subsection{UGC~10675 (VV~805, Mrk~700)}
Optical imaging shows a plume extending towards the southwest of the main body.  
Optical spectroscopy classifies the galaxy as a Seyfert~1 \citep{ Denisyuk76, Gallego96}.  
However, \citet{Goncalves99} suggest that the galaxy is a LINER 
because the line strengths are weaker than those in classical AGNs.  
Furthermore, there is no evidence of an AGN from the radio--FIR correlation \citep{Ji00}.  

The CO~(2--1) emitting gas appears to form a disk overall, but it is extremely disturbed.  
The PV diagram reveals three peaks, two of which overlap along the line of sight.  
There are only three resolution elements along the radial direction 
due to the limited sensitivity and the low angular resolution.  
Thus it is possible that derived radial properties have large uncertainties.

\subsection{AM~2038--382}
Optical imaging shows a plume extending towards the east 
and a prominent southwest loop.  
Optical spectroscopy reveals the dominance of A--F type stars, 
while lacking strong \ion{H}{2} region type emission lines, 
suggesting that the burst of star formation ended 
less than a few $\times$ 10$^{8}$~yr ago \citep{Sekiguchi93}.  
The \ion{H}{1} spectrum is roughly a double-horn profile \citep{Richter94}.  
The rotation curve measured from the CO stellar absorption line at 2.29~$\mu$m, 
however, shows no sign of a rotating disk \citep{Rothberg10}.

The CO~(2--1) integrated intensity map is shaped like a peanut.  
A peanut-shaped gas distribution is often seen in barred galaxies 
because gas piles up at the bar ends.  
The size of the northern concentration is larger than the southern one.  
The PV diagram shows a velocity gradient of $\sim$260 km~s$^{-1}$~kpc$^{-1}$ 
and two emission peaks, which are offset by 2\arcsec ($\sim$0.8~kpc) from the galactic center.  
These signatures imply the presence of a ring or a bar, 
but a bar structure is not identified in optical images.

\subsection{AM~2055--425}
This galaxy is classified as a LIRG or an ULIRG depending on the assumed distance.  
The $HST$/ACS $B$-band image shows two tidal tails 
and knotty structures in the main body.  
The velocity field of the H$\alpha$ emission shows rotational motion 
with a velocity width of 150 km~s$^{-1}$ without inclination correction \citep{Mihos98}.  
Optical spectroscopy classifies the galaxy 
as a starburst-AGN composite \citep{Yuan10, Brightman11}.  
The IR spectroscopy reveals the presence of 
a buried AGN \citep{Risaliti06, Imanishi10} 
and the X-ray emission is clearly dominated 
by a hidden AGN \citep{Franceschini03}.

Most of the CO~(1--0) emission comes from 
the southeast region of the main body, 
but the emission peaks seen in all channel maps 
are located at the nucleus.  
The PV diagram also shows gas concentration in the nucleus.  
In addition, an extension of the molecular gas is distributed 
in the southwest region of the main body,
which does not correspond to the stellar arm structure.  
The velocity field in the molecular gas disk shows 
a less distinct velocity gradient 
due to the low inclination of the molecular gas disk ($i$ = 10\arcdeg) 
and non-circular motion.  
The 3~mm continuum emission is associated with the nucleus 
and the peak corresponds to the CO peak seen in the integrated intensity map.  

\subsection{NGC~7135 (AM~2146--350, IC~5136)}
This galaxy has shell structures seen in the optical \citep{Malin83} 
as well as in the UV \citep{Rampazzo07}.  
The distribution of warm ionized gas shows an asymmetric structure 
relative to the stellar body, elongated in the southwest direction \citep{Rampazzo03}, 
which corresponds to the FUV distribution.  
The velocity field of warm ionized gas indicates rotational motion 
with a maximum rotation velocity ($V_{\rm rot}~$sin~$i$) of 78 km~s$^{-1}$ \citep{Rampazzo03}.  
\citet{Rampazzo07} suggest, using the relation between the optical line-strength indices 
and UV colors, that the galaxy experienced recent nuclear star formation 
likely triggered by the interaction/accretion that formed the shells.  
The \textit{Very Large Array} (VLA) \ion{H}{1} emission is mostly detected 
outside the main body \citep{Schiminovich01}.  

The CO~(1--0) emitting gas is distributed in the southern region of the main body 
and not associated with the nucleus.  
The detailed structure and kinematics of the molecular gas 
are not clear due to the weak CO emission.  
The 3~mm continuum emission is robustly detected and 
the peak of the continuum emission corresponds to the nucleus.

\subsection{NGC~7252 (Arp~226, AM~2217-245)}
This galaxy is a well-known merger, called the ``Atoms for Peace''.  
The galaxy has remarkable loops and two long tidal tails seen in the optical.  
High-resolution optical imaging carried out using the $HST$/WFPC
reveals spiral arms within a radius of 3\farcs5 from the galactic center  
and weak spiral features out to about 9\arcsec~\citep{Whitmore93}.  
\citet{Miller97} detect 499 cluster candidates,
whose mean age is estimated to be 0.6$\pm$0.2~Gyr by \citet{Whitmore97}.  
The \ion{H}{1} emission is associated with the optical tidal tails,
whose projected length are 520\arcsec and 270\arcsec \citep{Hibbard94},
corresponding to 160~kpc and 82~kpc, respectively.  
\citet{Schweizer13} suggest that the atomic gas is falling back to the nucleus.  
Previous CO (1--0) interferometric observations show  
a rotating molecular gas disk \citep{Wang92},
but the observed disk size was limited to a $\sim$1.2~kpc radius.

Numerical simulations focusing on this galaxy
have been conducted \citep[e.g.,][]{Borne91, Hibbard95, Chien10}.
They conclude that the galaxy is a merger
between two disk galaxies and resulting in an elliptical galaxy.  
The merger age is estimated to be about 0.5--1~Gyr
from numerical simulations \citep{Hibbard95, Chien10}
as well as observations \citep{Schweizer82}.  
This is in good agreement with the mean age of the star clusters \citep{Whitmore97}.
\citet{Mihos93} suggest that the SFR has fallen to one third of the value
that it would have been at its peak according their simulation.

Our new CO~(1--0) observations confirm the presence of
a molecular gas disk two times larger ($R_{\rm CO}$ = 2.6~kpc)
than the previous results \citep{Wang92}.  
The velocity field shows a smooth velocity gradient,
and the rotation curve of the molecular gas disk shows
flat rotation with a velocity of $\sim$260 km~s$^{-1}$ outside a radius of 1~kpc.  
The PV diagram shows a double peak located at the transition point
between rigid rotation and flat rotation.  
The 3~mm continuum emission is detected around the nucleus at the 4~$\sigma$ level.  

\subsection{AM~2246--490}
This galaxy is classified as a LIRG or an ULIRG depending on the assumed distance.  
The $HST$/ACS $B$-band image shows a single nucleus with long prominent tails.  
\citet{Haan11} suggest that the galaxy built up a central stellar cusp due to nuclear starbursts, 
judging from the significant core light excess in the $H$-band image 
and relatively large 6.2~$\mu$m PAH EW.  
While optical spectroscopy classifies the galaxy as a Seyfert~2 \citep{Yuan10}, 
there is no signature of an AGN according to a selection 
by hard X-ray color and the 6.4~keV iron line \citep{Iwasawa11}.

The velocity field shows the presence of a warped gas disk.
The PV diagram reveals rigid rotation and a double peak located off-center, 
suggesting the possibility of a ring structure, 
but a non-uniform distribution of molecular gas may also explain the observed diagram.  
The 3~mm continuum emission is associated with the nucleus 
and the peak corresponds to the CO peak seen in the integrated intensity map.

\subsection{NGC~7585 (Arp~223)}
This galaxy has shells seen in the optical \citep{Malin83}.  
Multicolor optical imaging reveals that the shells 
are bluer than the main body, 
and that the timescale from the interaction, by which the shells were formed, 
is typically $\sim$1~Gyr \citep{McGaugh90}.  
The upper limit of the atomic gas mass is 
7.5 $\times$ 10$^{9}$ $M_{\sun}$ \citep{Balick76}.

The galaxy is undetected in the CO~(1--0) line.  
The 3~$\sigma$ upper limit on the molecular gas mass 
is 3.1 $\times$ 10$^{7}$ $M_{\sun}$.  
\citet{Georgakakis01} also report non-detection 
in their CO measurements obtained at the OSO 20~m telescope.

\subsection{NGC~7727 (Arp~222, VV~67)}
The $HST$/WFPC2 $B$-band image reveals spiral structures \citep{Peeples06},
while the $K$-band surface brightness profile roughly follows
the de Vaucouleurs law \citep[][RJ04]{Chitre02}.  
The H$\alpha$ emission is associated with the nuclear region \citep{Knapen05},
where the mean stellar color is slightly bluer than the outer envelopes \citep{Schombert90}.  
The rotation curve measured from the Mg$_{b}$ absorption line 
($\lambda \sim$ 0.52~$\mu$m) reveals a velocity gradient 
\citep[$V_{\rm rot}~$sin~$i$ = $\pm$~150~km~s$^{-1}$;][]{Simien97}.  
This galaxy is detected in the $AKARI$ All-Sky Survey,
and the FIR luminosity is estimated to be 4.5$\times$10$^{8}$ $L_{\sun}$
using three bands (65$\mu$m, 90$\mu$m, and 140$\mu$m; Yamamura et al. 2010).

The PV digram shows rigid rotation, suggesting the presence of a rotating gas disk.  
The maximum rotational velocity is estimated to be $\sim$150 km~s$^{-1}$.  
A close inspection of the channel maps reveals two different components.  
The low-velocity component ($V$ = 1665--1735 km~s$^{-1}$)
is associated with the nucleus and
the high-velocity component ($V$ = 1740--1990 km~s$^{-1}$)
is distributed along the optical dust lane.  
The molecular gas mass in the high-velocity component
is nearly eight times larger than that in the low-velocity component.  
The 3~mm continuum emission is associated with the nucleus
and the peak corresponds to the CO peak seen in the integrated intensity map.

\clearpage
\begin{figure}[htbp]
 \begin{minipage}{1.0\hsize}
  \begin{center}
   \includegraphics[scale=1.3, trim=3.5cm 21.1cm 0cm 4.5cm, clip]{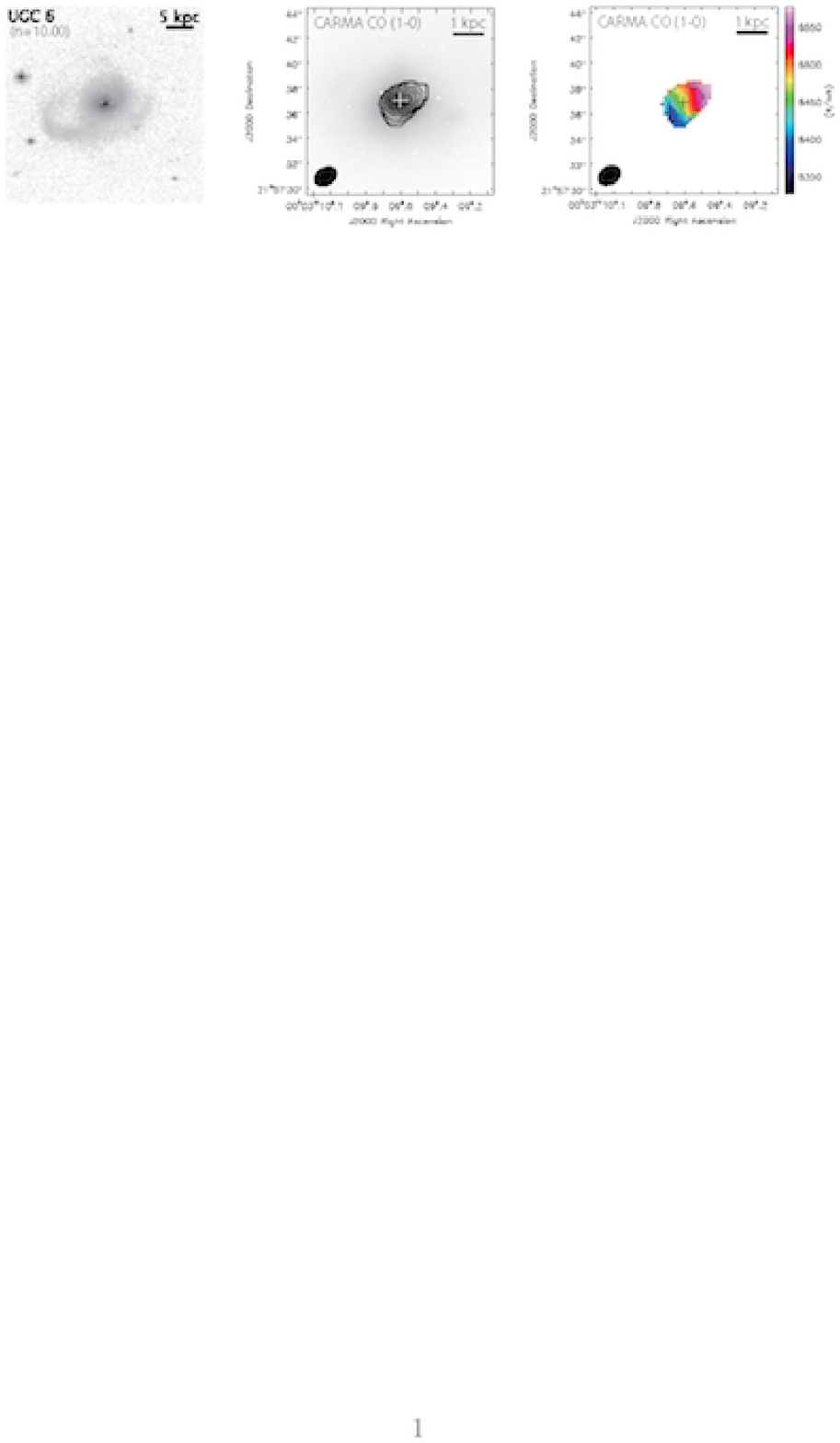}
  \end{center}
 \end{minipage}
 
 \begin{minipage}{1.0\hsize}
  \begin{center}
   \includegraphics[scale=1.3, trim=3.5cm 21.1cm 0cm 4.8cm, clip]{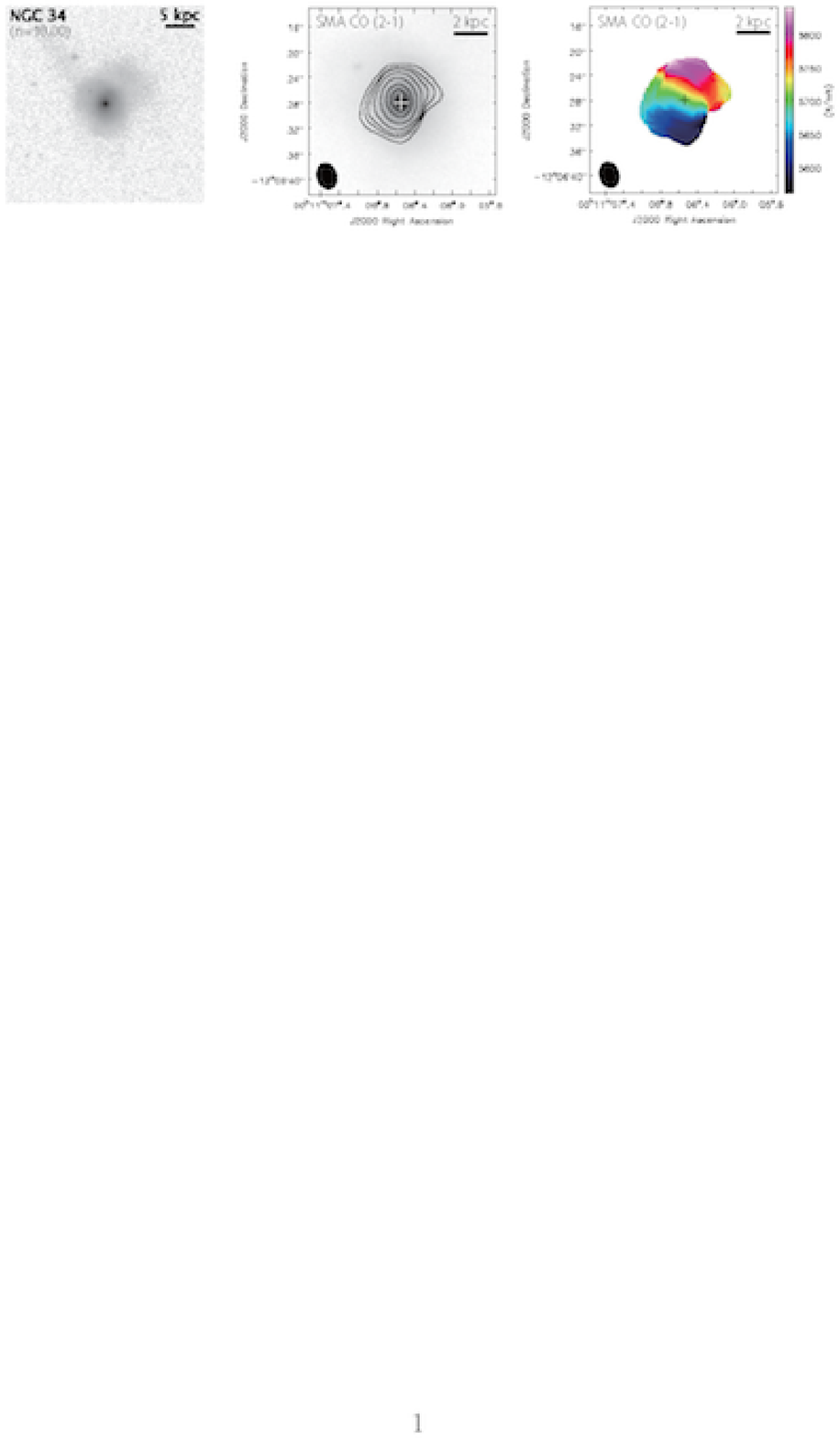}
  \end{center}
 \end{minipage}
 
 \begin{minipage}{1.0\hsize}
  \begin{center}
   \includegraphics[scale=1.3, trim=3.5cm 21.1cm 0cm 4.8cm, clip]{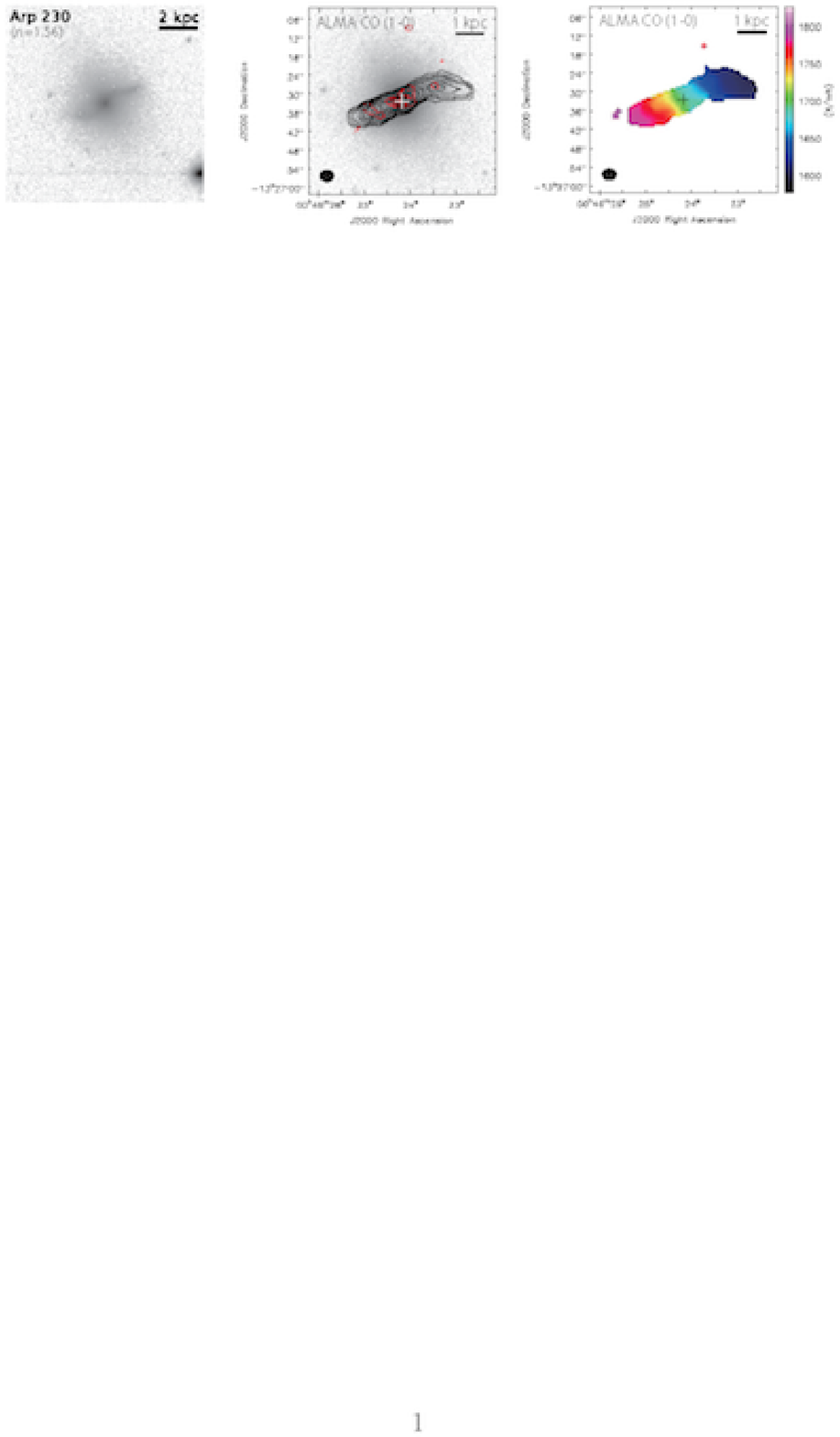}
  \end{center}
 \end{minipage}
 
 \begin{minipage}{1.0\hsize}
  \begin{center}
   \includegraphics[scale=1.3, trim=3.5cm 21.1cm 0cm 4.8cm, clip]{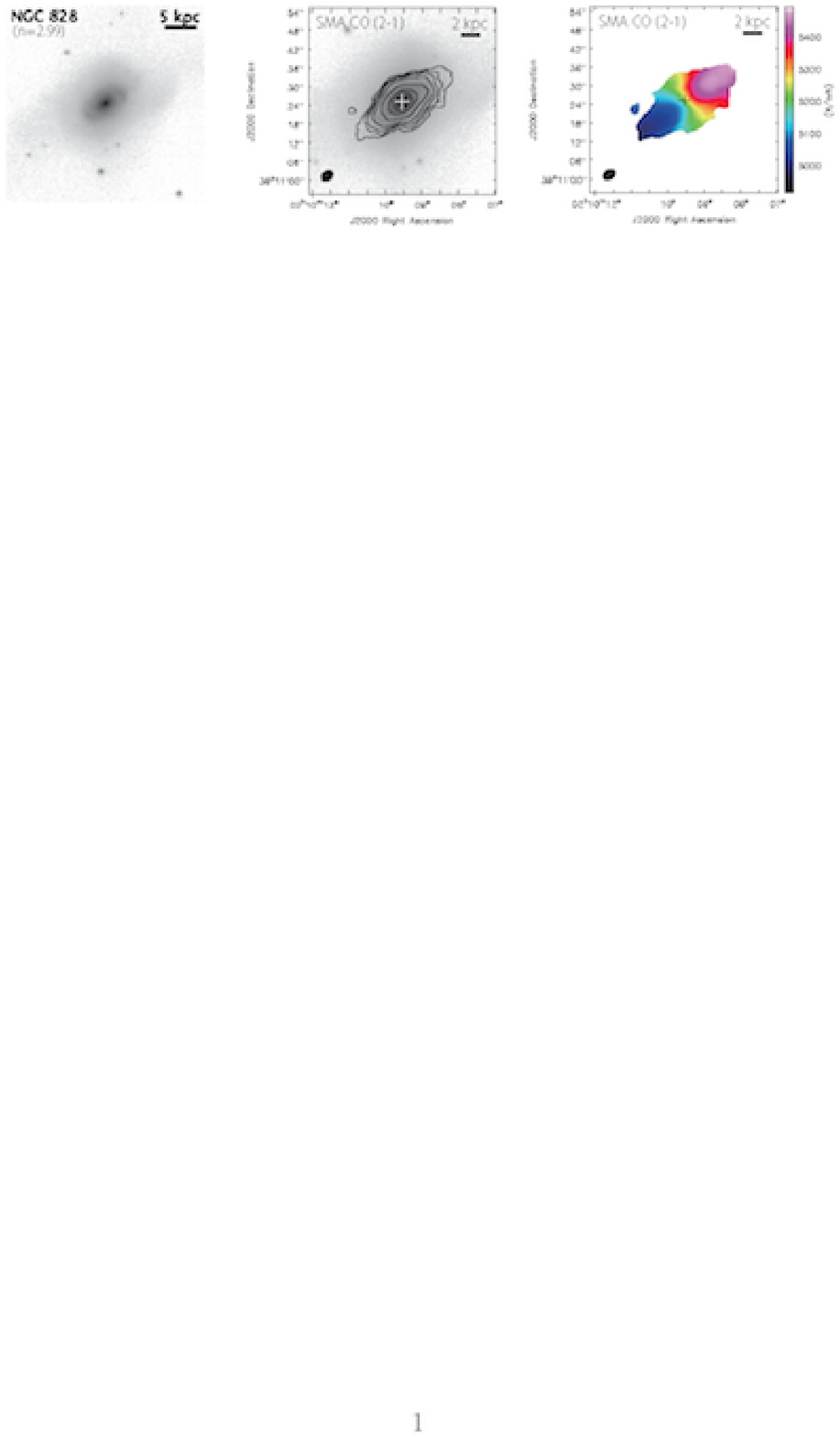}
  \end{center}
 \end{minipage}
 \caption{
 (left) The $K$-band images (RJ04) of UGC~6, NGC~34, Arp~230, and NGC~828.  
 The length of each side corresponds to 90\arcsec.  The magnification is selected individually 
 for each galaxy to clearly illustrate its individual morphological structures.  
 The number in the upper-left corner is the S\'{e}rsic index estimated by \citet{Rothberg04}.  
 (middle) The black contour maps of the CO integrated intensity (moment~0 maps) 
 and red contour maps of the 3~mm continuum overlaid on the $K$-band images.  
 The black contours are 1.2 Jy~km~s$^{-1} \times$ (1, 2, 3, 4, 5, 6, 7) for UGC~6, 
 5.6 Jy~km~s$^{-1} \times$ (1, 2, 3, 5, 10, 20, 30, 40) for NGC~34, 
 0.23 Jy~km s$^{-1} \times$ (1, 2, 3, 5, 7, 9, 11, 13) for Arp~230, 
 4.3 Jy~km s$^{-1} \times$ (1, 2, 3, 5, 10, 20, 30, 40) for NGC~828.  
 The red contours are 68 $\mu$Jy $\times$ (3, 4, 5) for Arp~230.  
 (right) The CO velocity fields (moment~1 maps).  
 The moment~0 and moment~1 maps are clipped using the AIPS task, MOMNT.
 The plus signs in the center of the moment maps show 
 the galactic centers defined by the $K$-band images 
 and the ellipses in the bottom-left corner of the moment maps show the beam sizes.  
 }
 \label{fig:f1}
\end{figure}

\addtocounter{figure}{-1}
\begin{figure}[htbp]
 \begin{minipage}{1.0\hsize}
  \begin{center}
   \includegraphics[scale=1.3, trim=3.5cm 21.1cm 0cm 4.5cm, clip]{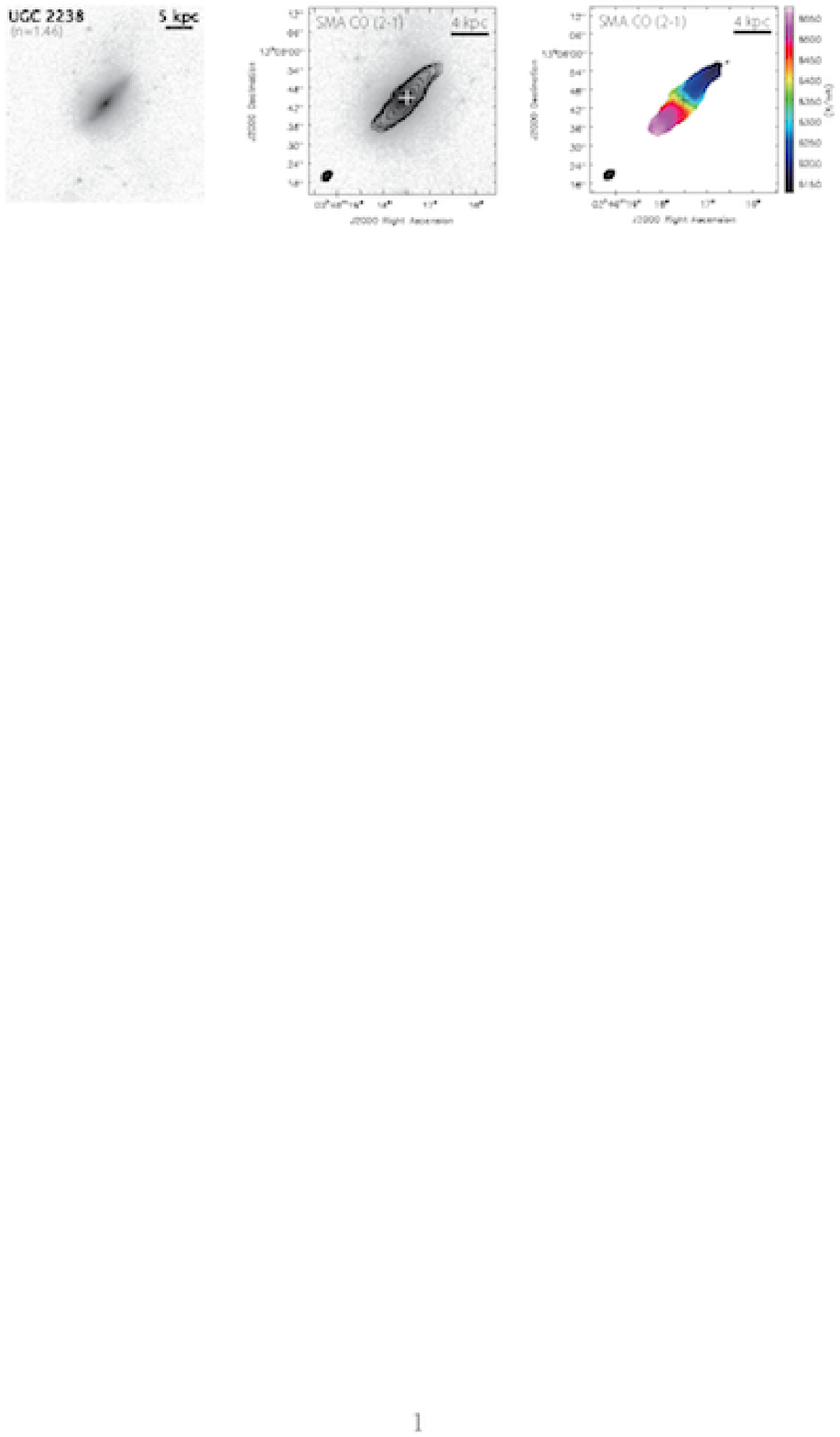}
  \end{center}
 \end{minipage}
 
 \begin{minipage}{1.0\hsize}
  \begin{center}
   \includegraphics[scale=1.3, trim=3.5cm 21.1cm 0cm 4.8cm, clip]{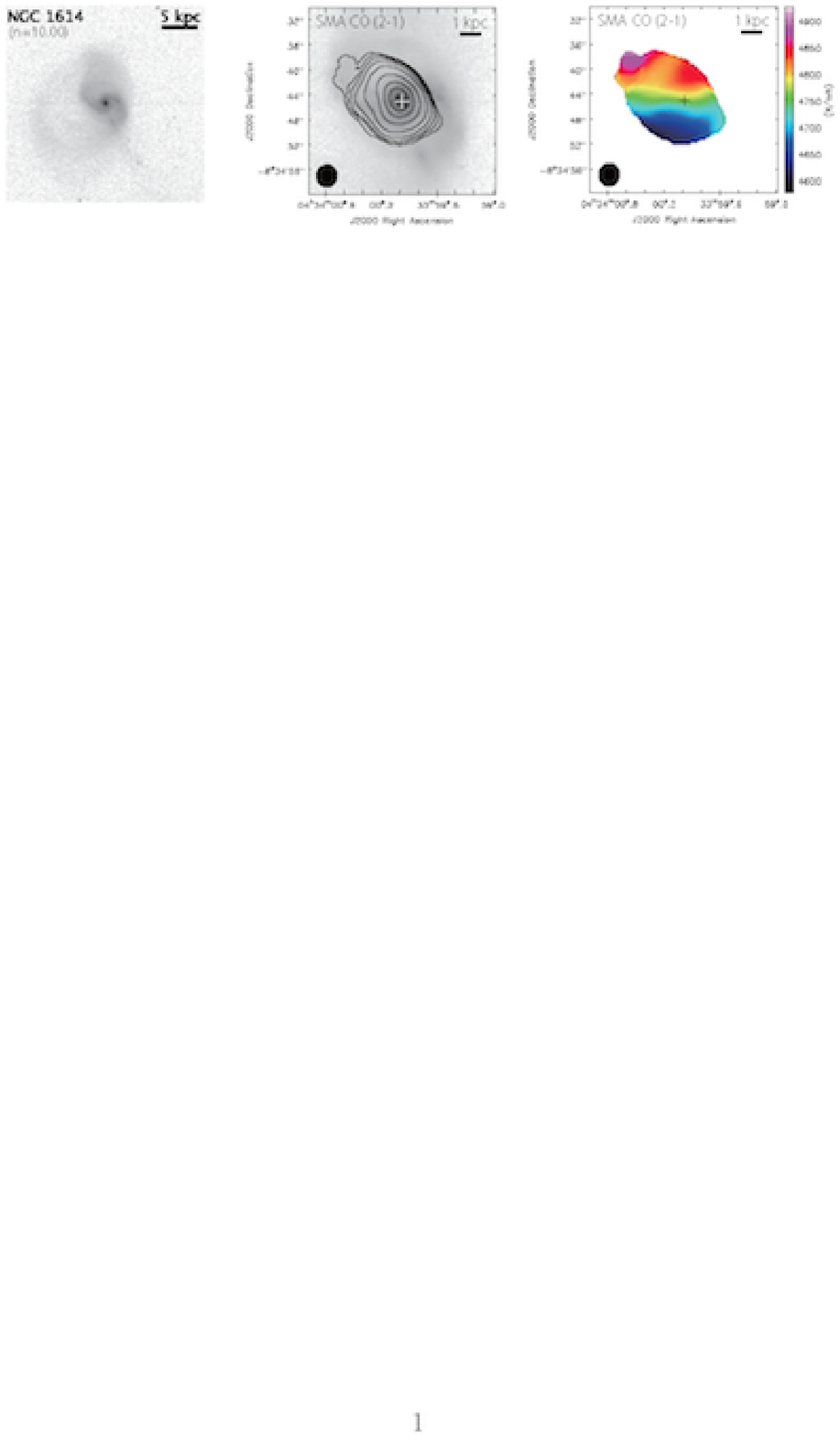}
  \end{center}
 \end{minipage}
 
 \begin{minipage}{1.0\hsize}
  \begin{center}
   \includegraphics[scale=1.3, trim=3.5cm 21.1cm 0cm 4.8cm, clip]{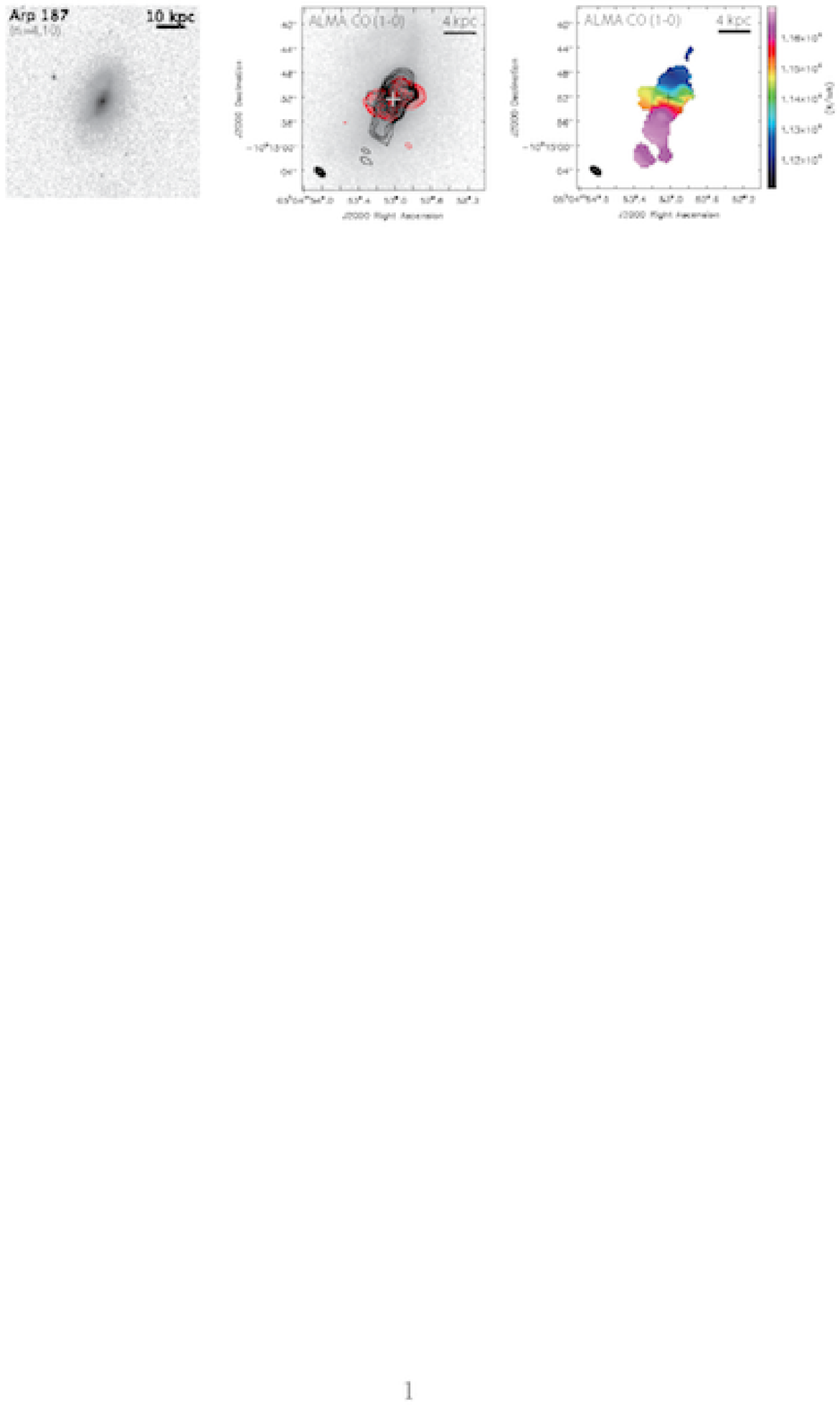}
  \end{center}
 \end{minipage}
 
 \begin{minipage}{1.0\hsize}
  \begin{center}
   \includegraphics[scale=1.3, trim=3.5cm 21.1cm 0cm 4.8cm, clip]{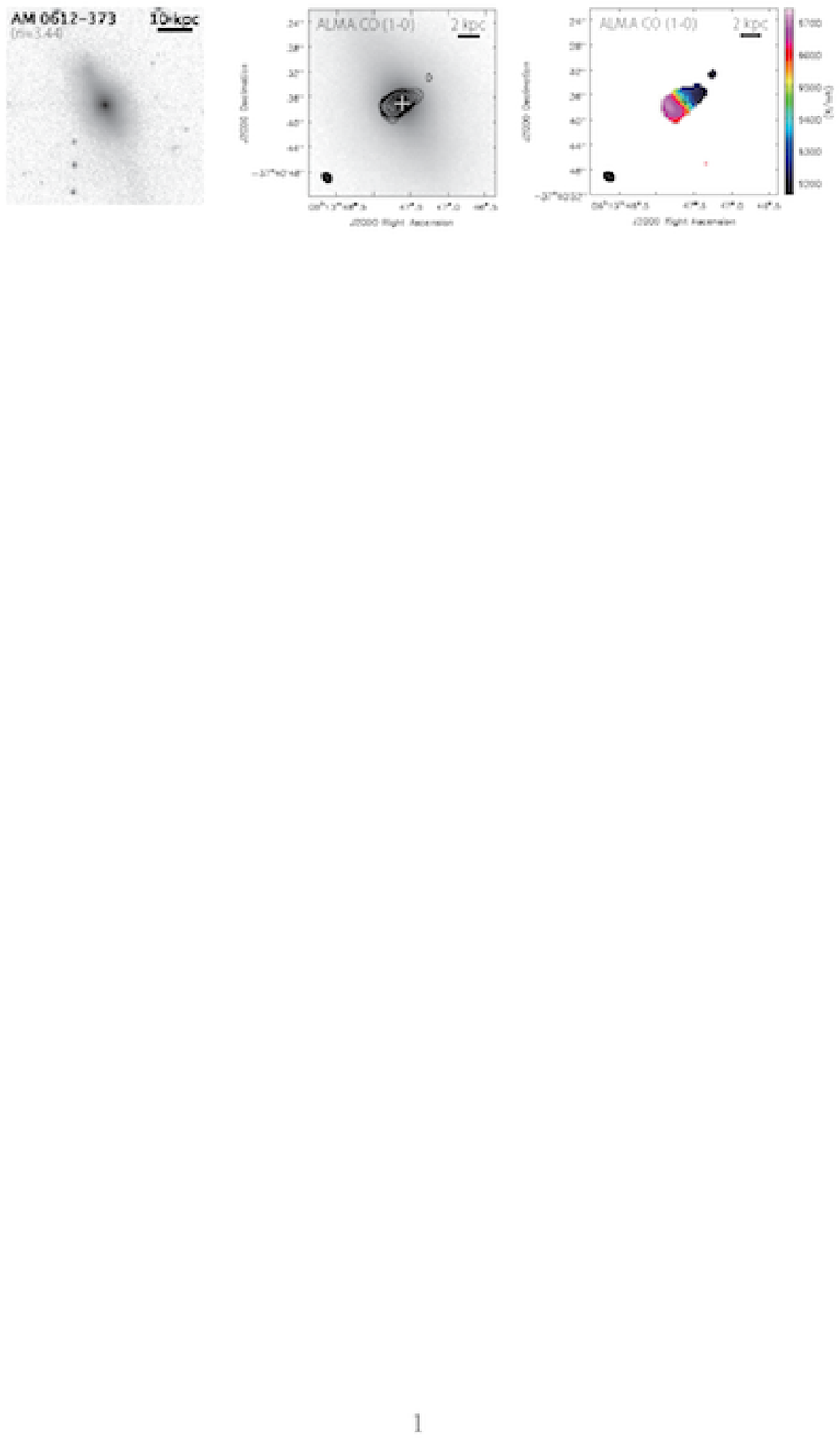}
  \end{center}
 \end{minipage}
 \caption{
 (Continued) The same as the previous figures 
 but for UGC~2238, NGC~1614, Arp~187, and AM~0612-373.  
 The black contours of the CO integrated intensity are 
 3.0 Jy~km~s$^{-1} \times$ (1, 2, 3, 5, 10, 20, 40, 60) for UGC~2238, 
 3.0 Jy~km~s$^{-1} \times$ (1, 2, 3, 5, 10, 25, 50, 75, 100) for NGC~1614, 
 0.46 Jy~km~s$^{-1} \times$ (1, 2, 3, 5, 7, 9, 11, 13, 15) for Arp~187, 
 and 0.38 Jy~km~s$^{-1} \times$ (1, 2, 3, 5, 10, 15, 20) for AM~0612-373.  
 The red contours of the 3~mm continuum are 
 110 $\mu$Jy $\times$ (3, 5, 10, 20, 40, 60) for Arp~187.
 }
 \label{fig:f1}
\end{figure}

\addtocounter{figure}{-1}
\begin{figure}[htbp]
 \begin{minipage}{1.0\hsize}
  \begin{center}
   \includegraphics[scale=1.3, trim=3.5cm 21.1cm 0cm 4.5cm, clip]{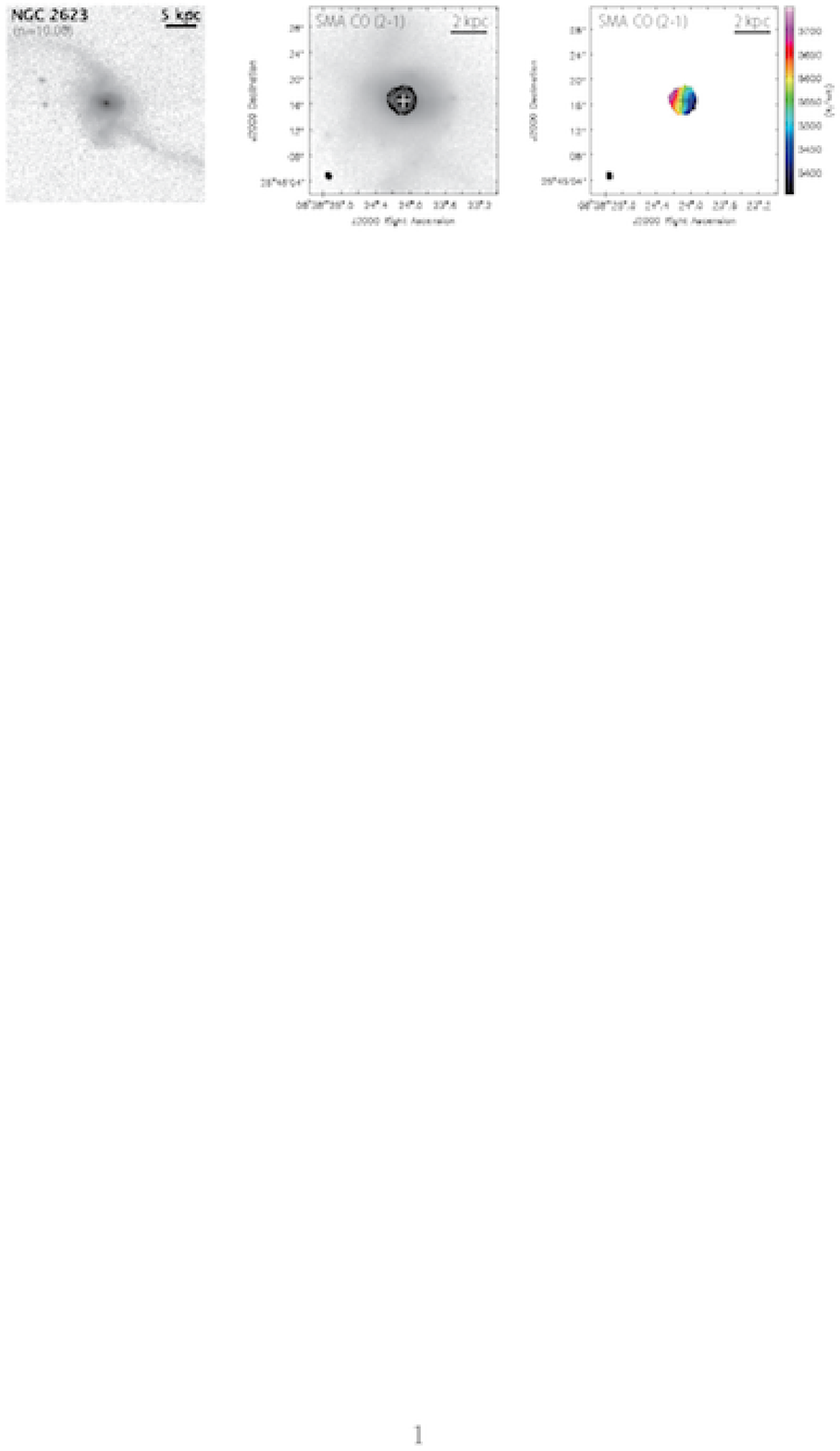}
  \end{center}
 \end{minipage}
 
 \begin{minipage}{1.0\hsize}
  \begin{center}
   \includegraphics[scale=1.3, trim=3.5cm 21.1cm 0cm 4.8cm, clip]{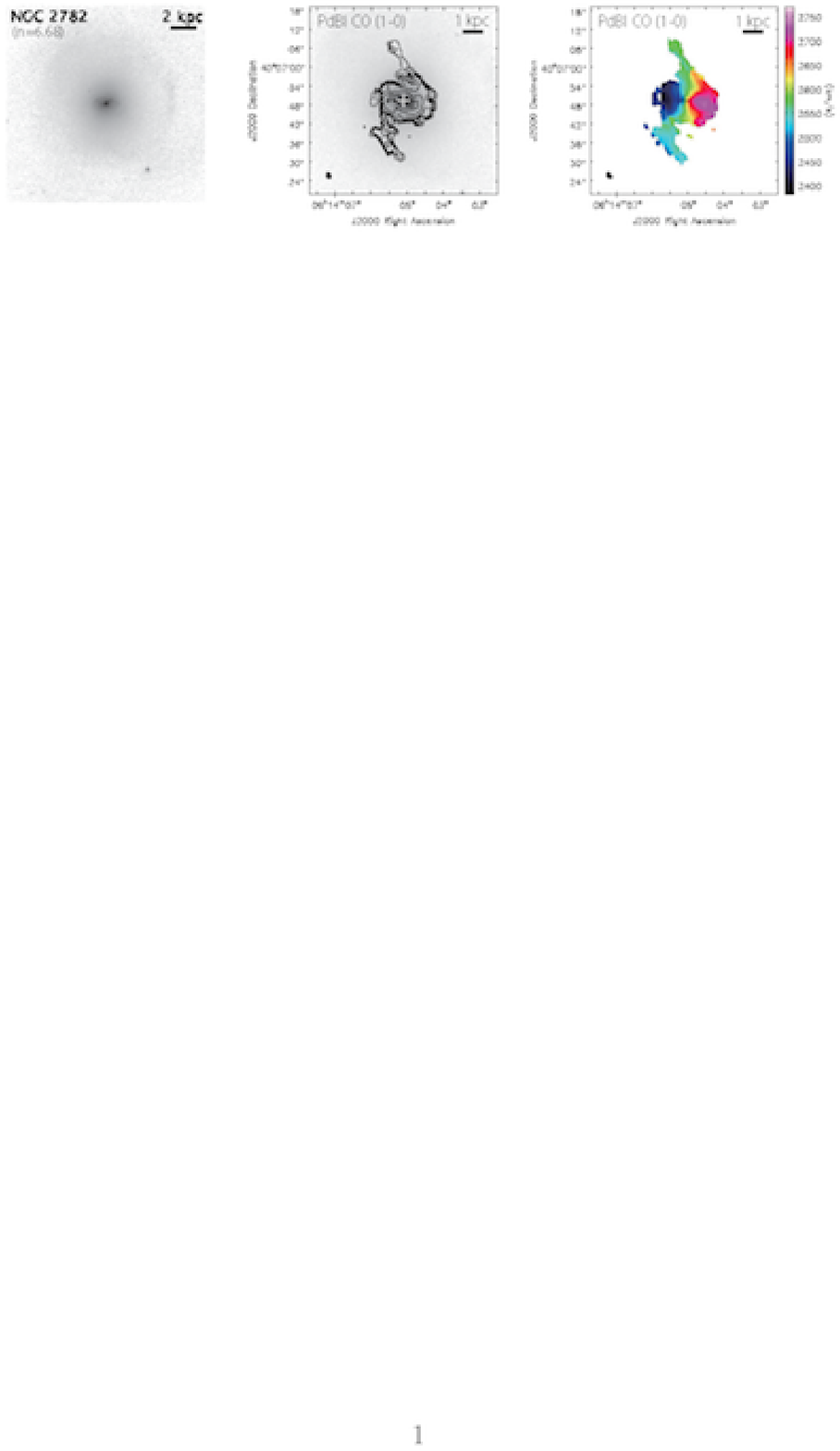}
  \end{center}
 \end{minipage}

 \begin{minipage}{1.0\hsize}
  \begin{center}
   \includegraphics[scale=1.3, trim=3.5cm 21.1cm 0cm 4.8cm, clip]{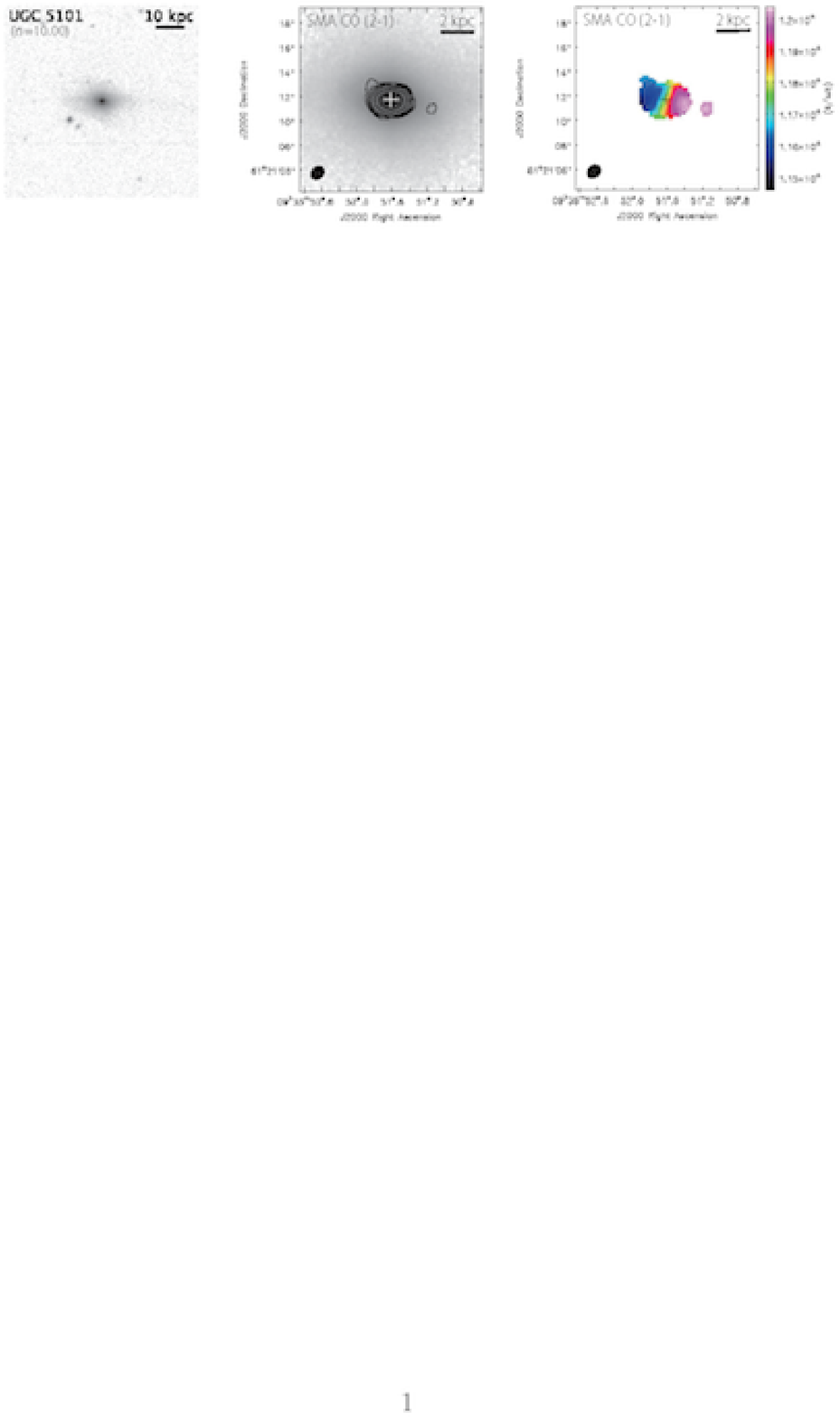}
  \end{center}
 \end{minipage}
 
 \begin{minipage}{1.0\hsize}
  \begin{center}
   \includegraphics[scale=1.3, trim=3.5cm 21.1cm 0cm 4.8cm, clip]{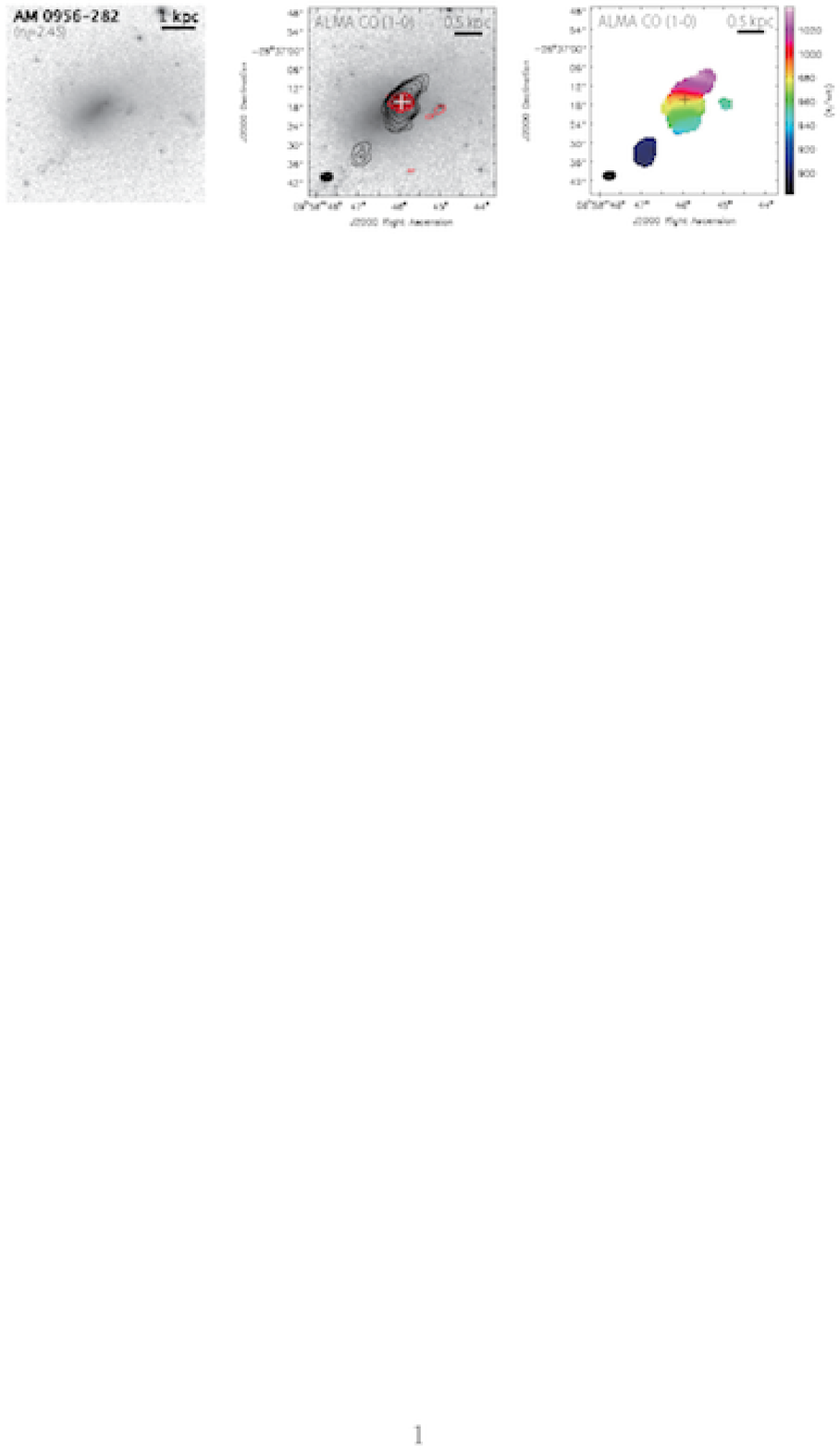}
  \end{center}
 \end{minipage}
 \caption{
 (Continued) The same as the previous figures 
 but for NGC~2623, NGC~2782, UGC~5101, and AM~0956-282.  
 The black contours of the CO integrated intensity are 
 2.0 Jy~km s$^{-1} \times$ (1, 2, 3, 5, 10, 20, 30, 40) for NGC~2323, 
 0.19 Jy~km s$^{-1} \times$ (1, 2, 3, 5, 10, 20, 30, 40) for NGC~2782, 
 2.3 Jy~km s$^{-1} \times$ (1, 2, 3, 5, 10, 25, 50) for UGC~5101, and 
 0.22 Jy~km s$^{-1} \times$ (1, 2, 3, 5, 10, 15, 20, 25) for AM~0956-282.  
 The red contours of the 3~mm continuum are 
 60 $\mu$Jy $\times$ (3, 5, 10, 15, 20, 25) for AM~0956-282.
 }
 \label{fig:f1}
\end{figure}

\addtocounter{figure}{-1}
\begin{figure}[htbp]
 \begin{minipage}{1.0\hsize}
  \begin{center}
   \includegraphics[scale=1.3, trim=3.5cm 21.1cm 0cm 4.5cm, clip]{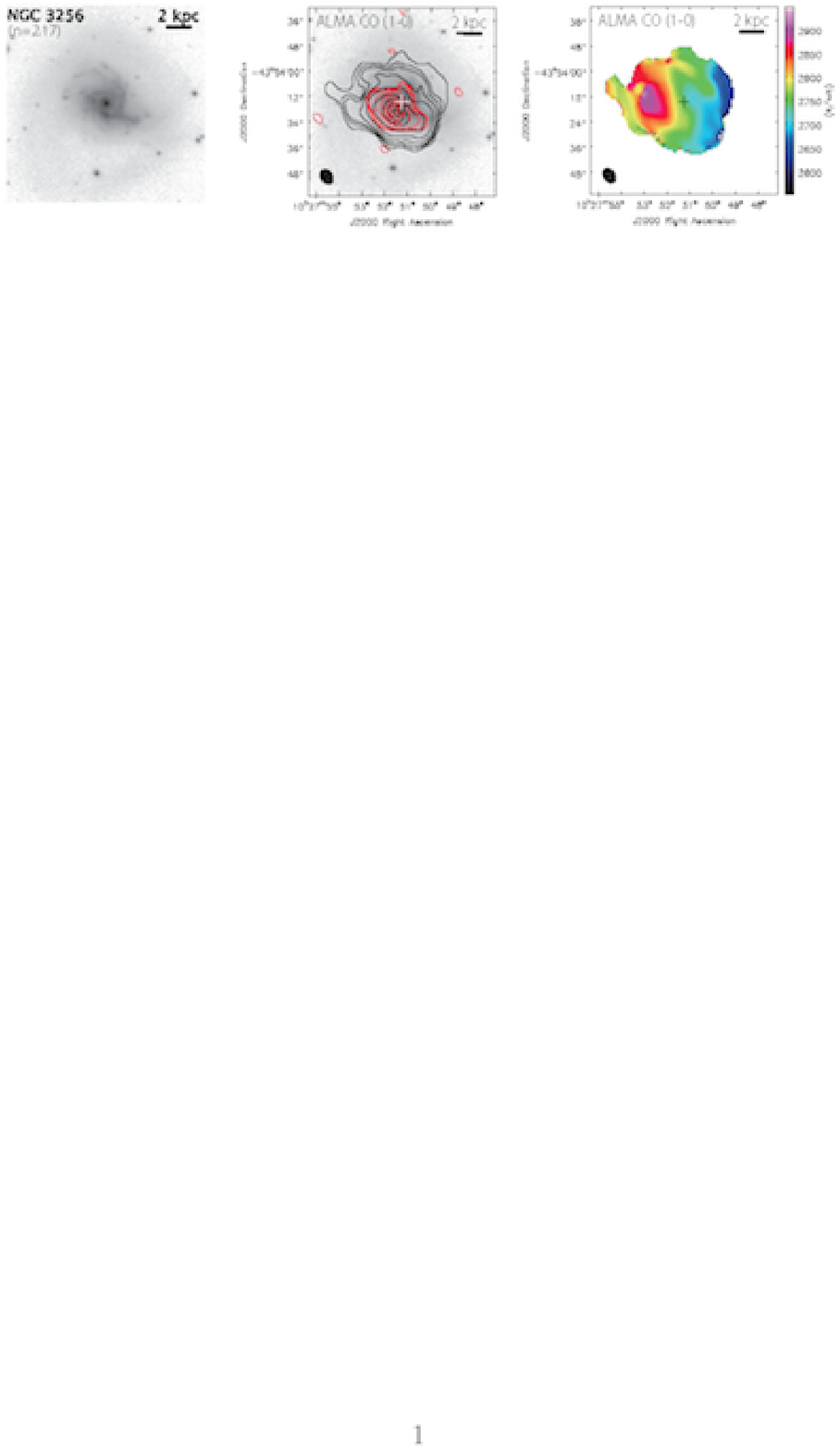}
  \end{center}
 \end{minipage}
 
 \begin{minipage}{1.0\hsize}
  \begin{center}
   \includegraphics[scale=1.3, trim=3.5cm 21.1cm 0cm 4.8cm, clip]{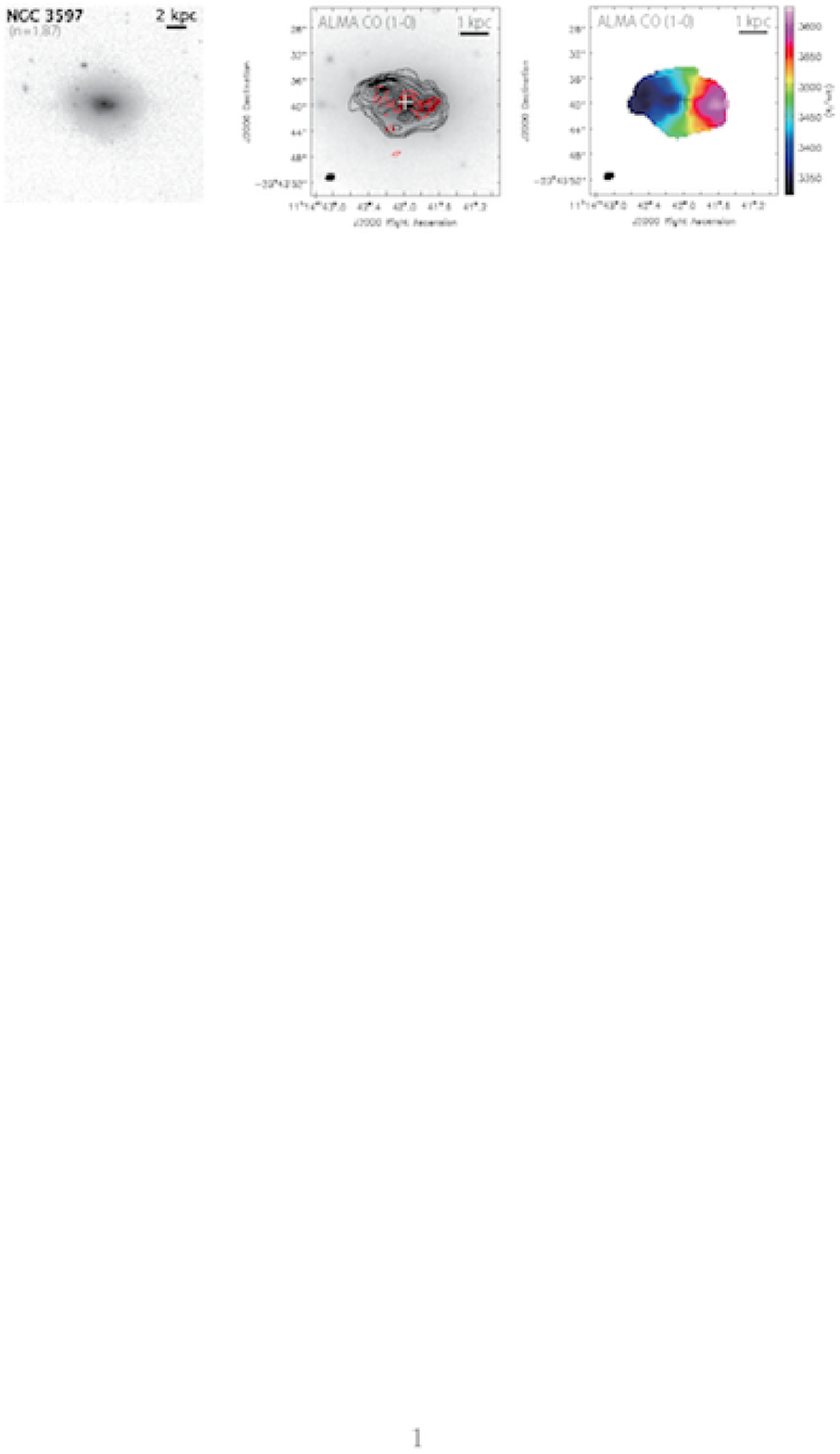}
  \end{center}
 \end{minipage}
 
 \begin{minipage}{1.0\hsize}
  \begin{center}
   \includegraphics[scale=1.3, trim=3.5cm 21.1cm 0cm 4.8cm, clip]{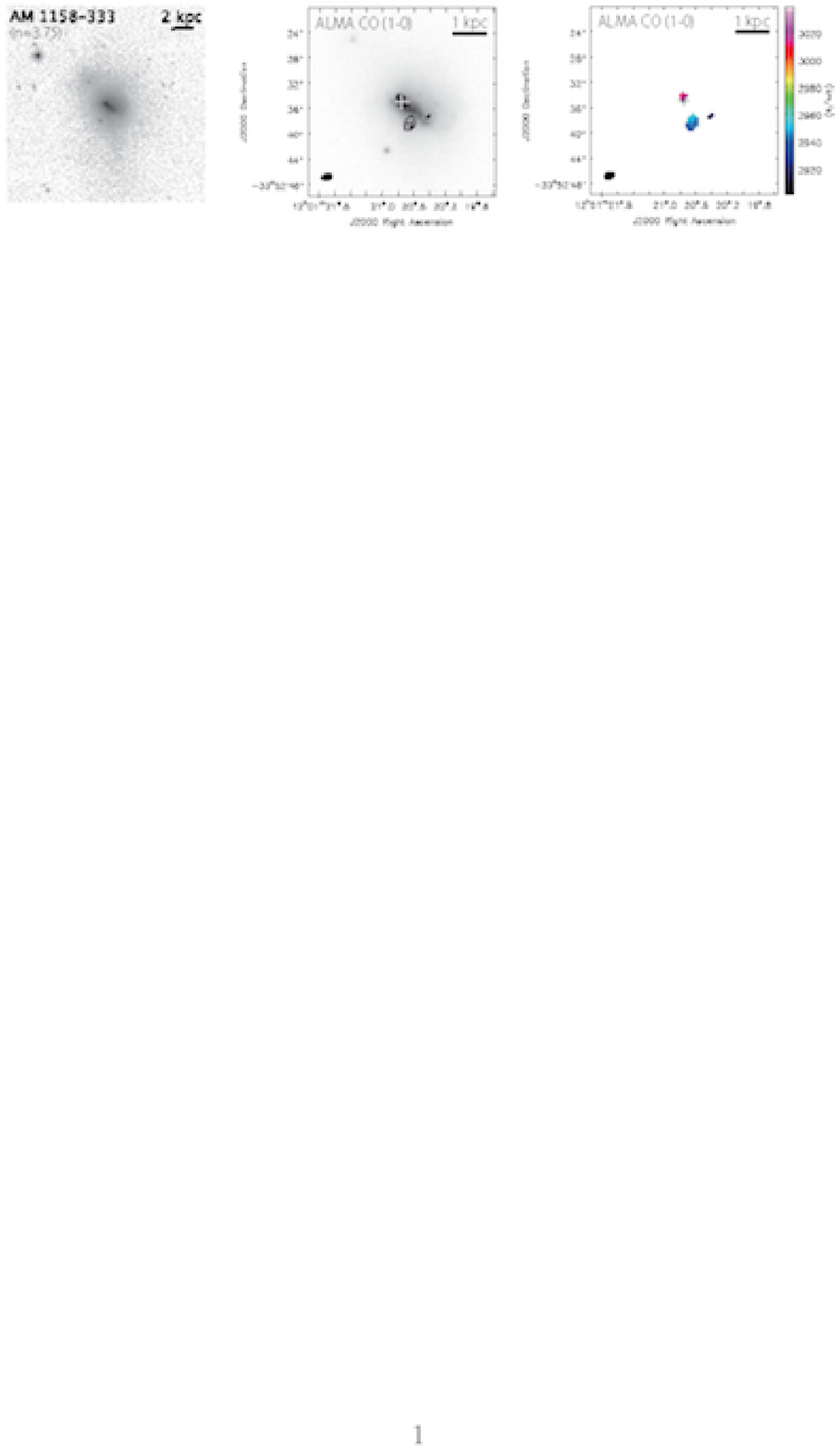}
  \end{center}
 \end{minipage}
 
 \begin{minipage}{1.0\hsize}
  \begin{center}
   \includegraphics[scale=1.3, trim=3.5cm 21.1cm 0cm 4.8cm, clip]{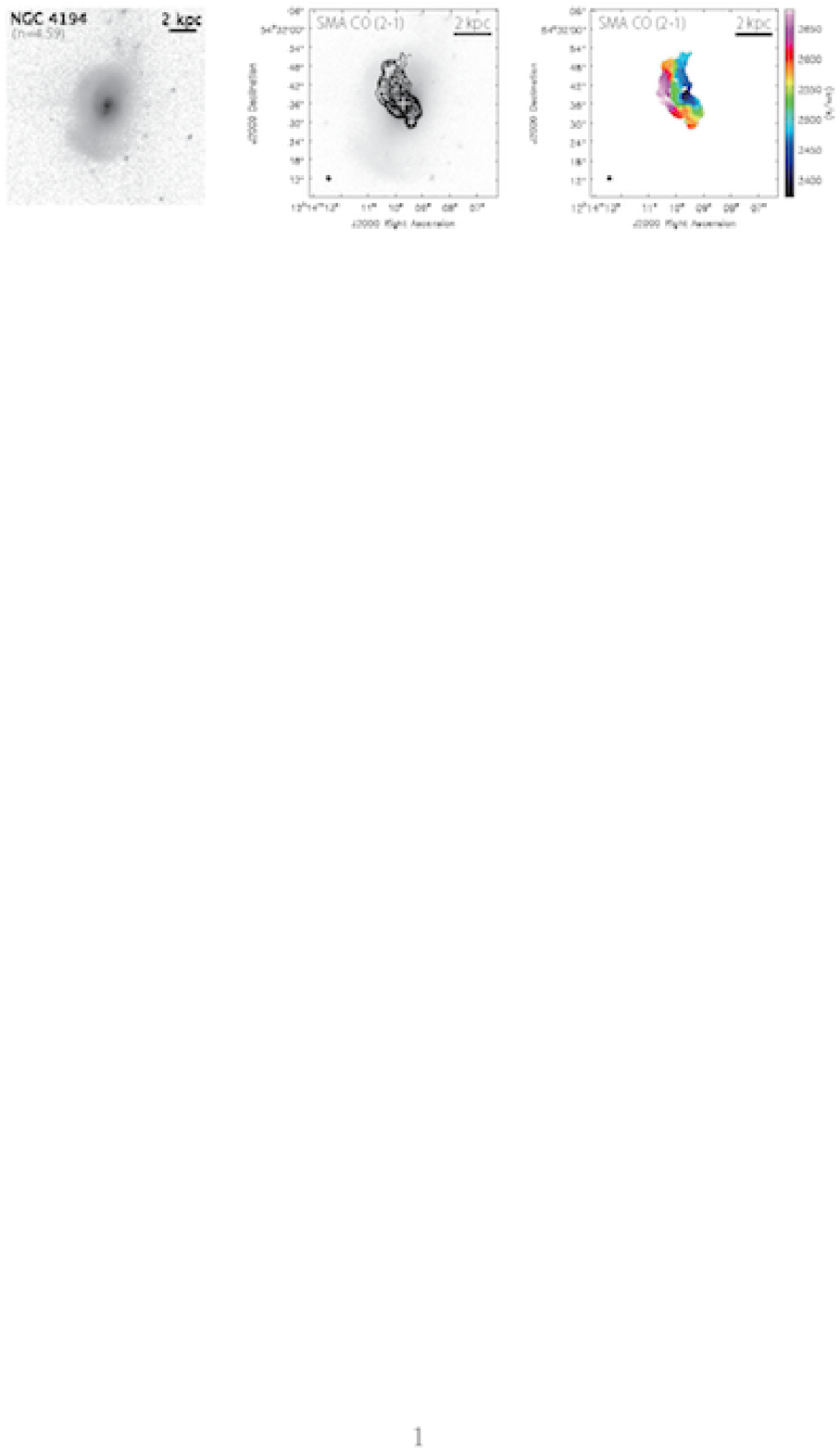}
  \end{center}
 \end{minipage}
 \caption{
 (Continued) The same as the previous figures 
 but for NGC~3256, NGC~3597, AM~1158-333, and NGC~4194.  
 The black contours of the CO integrated intensity are 
 2.0 Jy~km s$^{-1} \times$ (1, 2, 3, 5, 10, 25, 50, 75, 100) for NGC~3256, 
 0.46 Jy~km s$^{-1} \times$ (1, 2, 3, 5, 7, 9, 11, 13, 15) for NGC~3597, 
 0.37 Jy~km s$^{-1} \times$ (1, 2, 3) for AM~1158-333, and 
 0.84 Jy~km s$^{-1} \times$ (1, 2, 3, 5, 10, 20, 40, 60), for NGC~4194.  
 The red contours of the 3~mm continuum are 
 86 $\mu$Jy $\times$ (3, 5, 10, 25, 50, 75, 100, 125) for NGC~3256 
 and 87 $\mu$Jy $\times$ (3, 5, 7, 9) for NGC~3597.
 }
 \label{fig:f1}
\end{figure}

\addtocounter{figure}{-1}
\begin{figure}[htbp]
 \begin{minipage}{1.0\hsize}
  \begin{center}
   \includegraphics[scale=1.3, trim=3.5cm 21.1cm 0cm 4.5cm, clip]{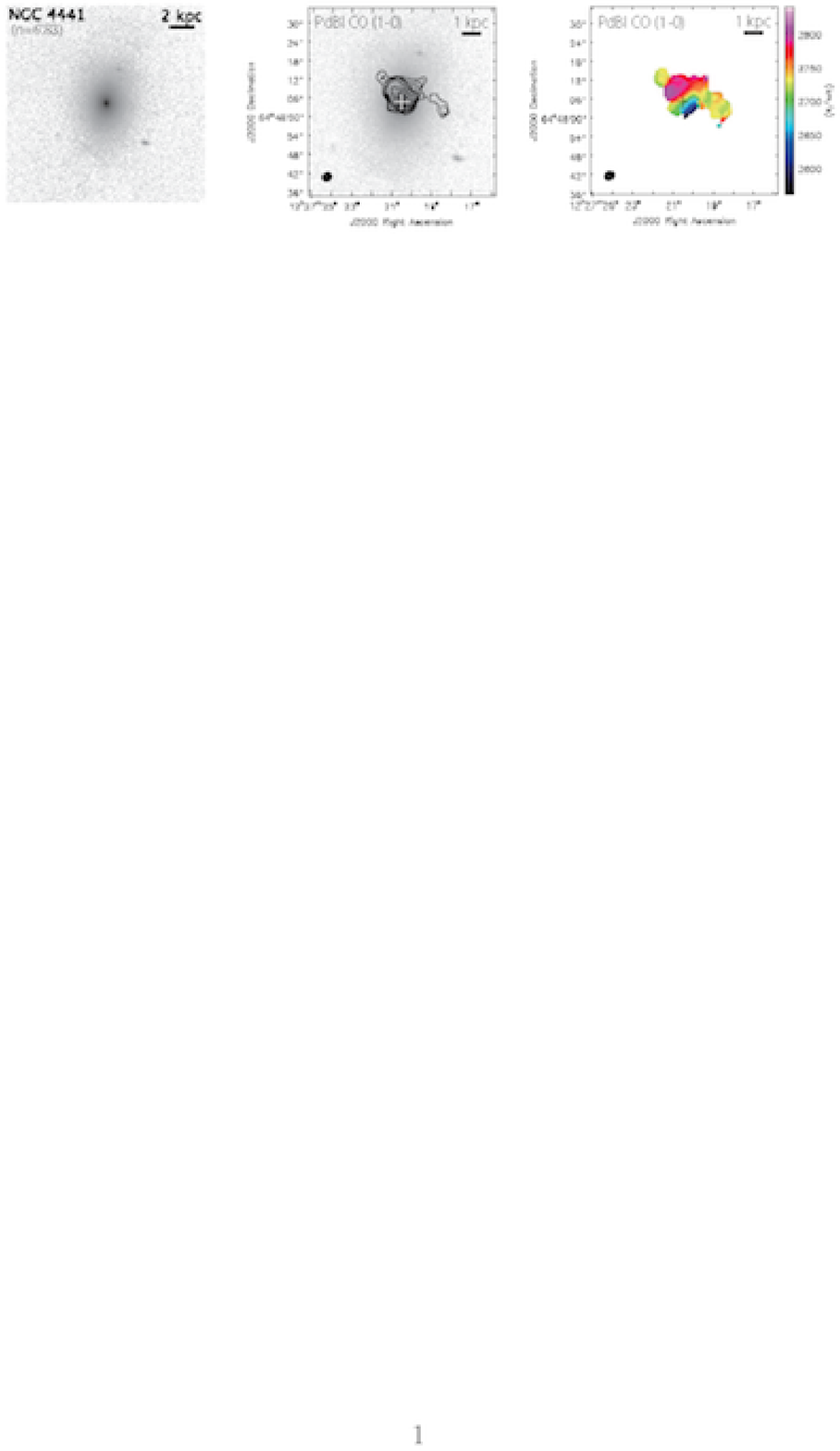}
  \end{center}
 \end{minipage}
  
 \begin{minipage}{1.0\hsize}
  \begin{center}
   \includegraphics[scale=1.3, trim=3.5cm 21.1cm 0cm 4.8cm, clip]{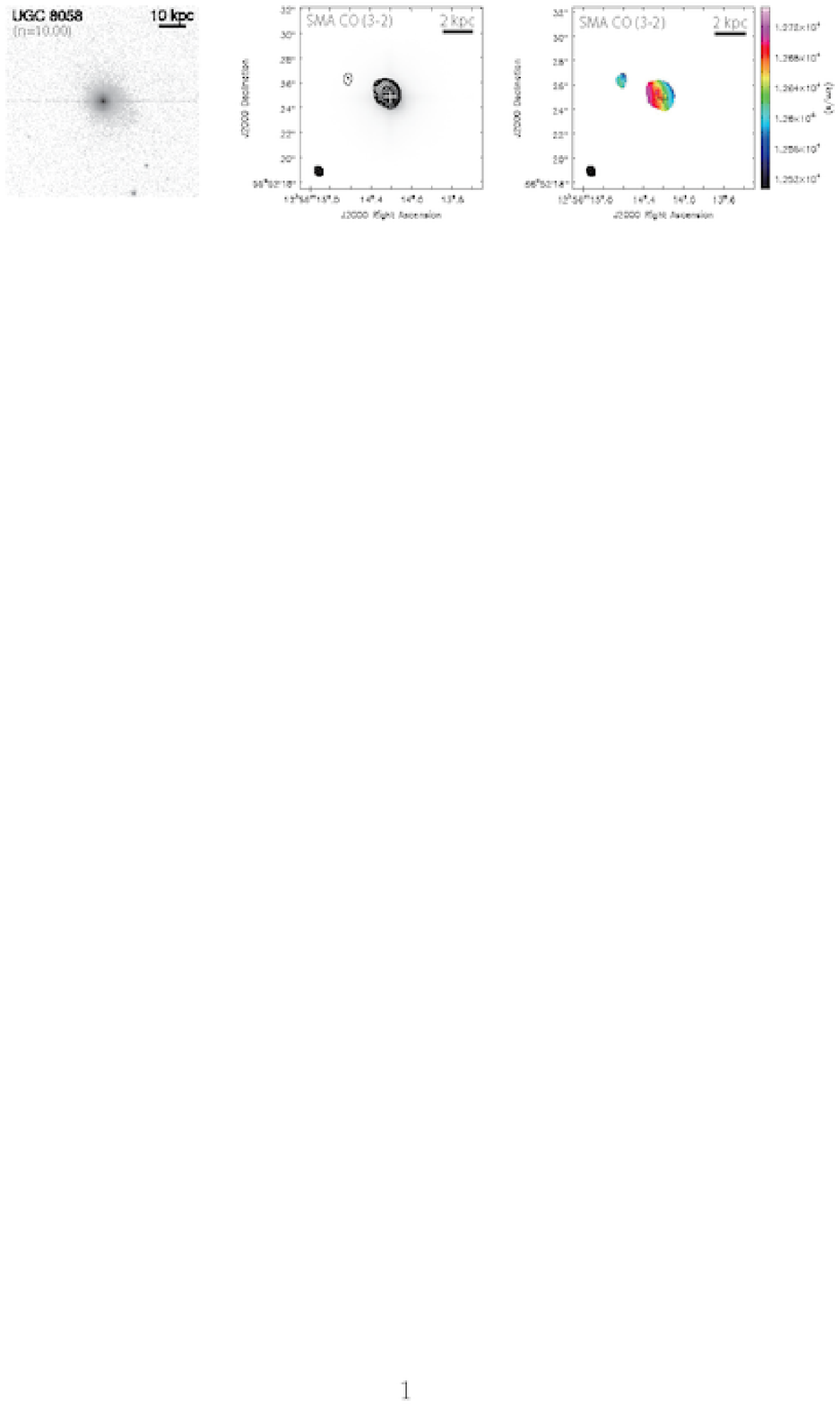}
  \end{center}
 \end{minipage}
 
 \begin{minipage}{1.0\hsize}
  \begin{center}
   \includegraphics[scale=1.3, trim=3.5cm 21.1cm 0cm 4.8cm, clip]{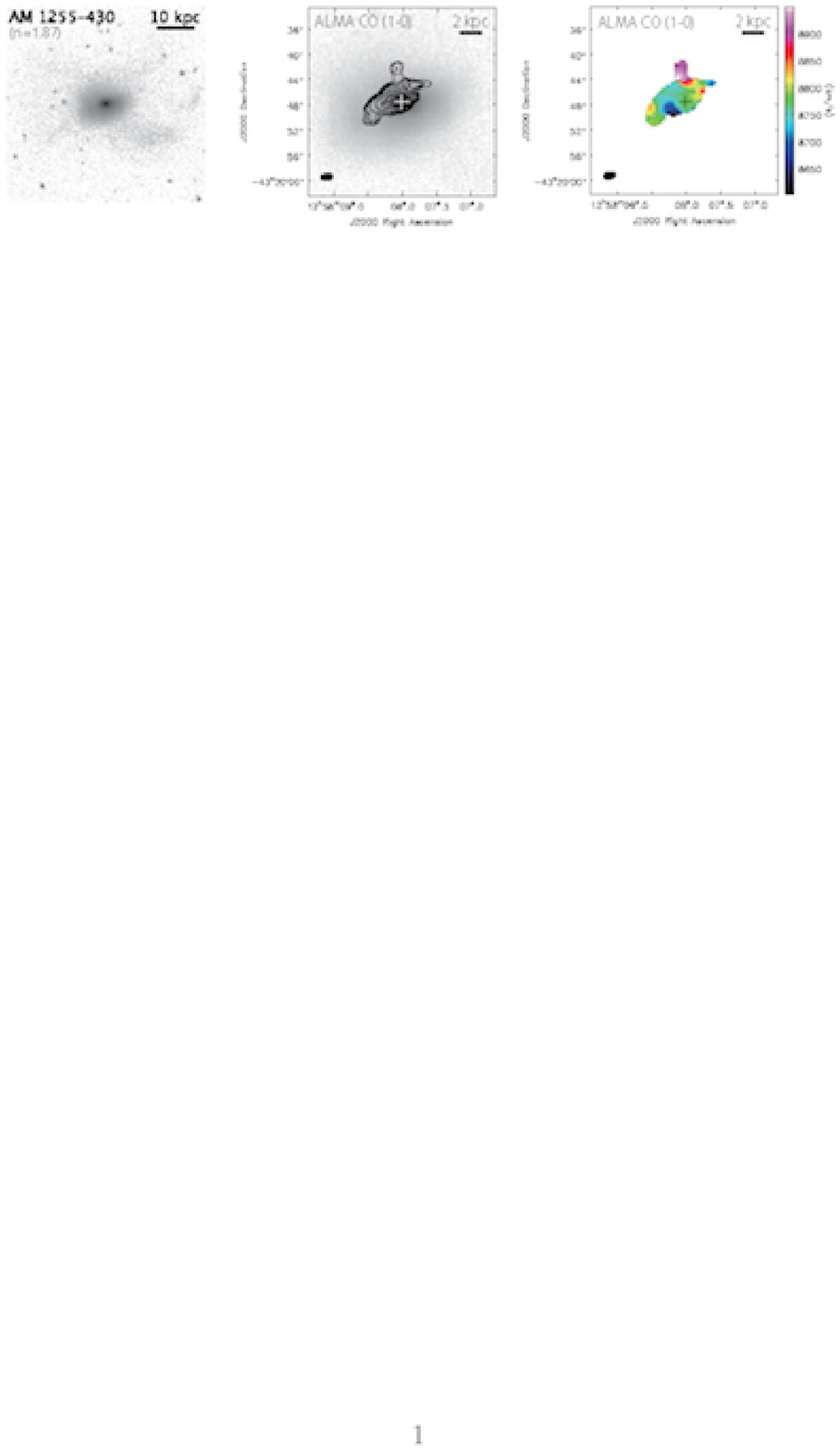}
  \end{center}
 \end{minipage}
  
 \begin{minipage}{1.0\hsize}
  \begin{center}
   \includegraphics[scale=1.3, trim=3.5cm 21.1cm 0cm 4.8cm, clip]{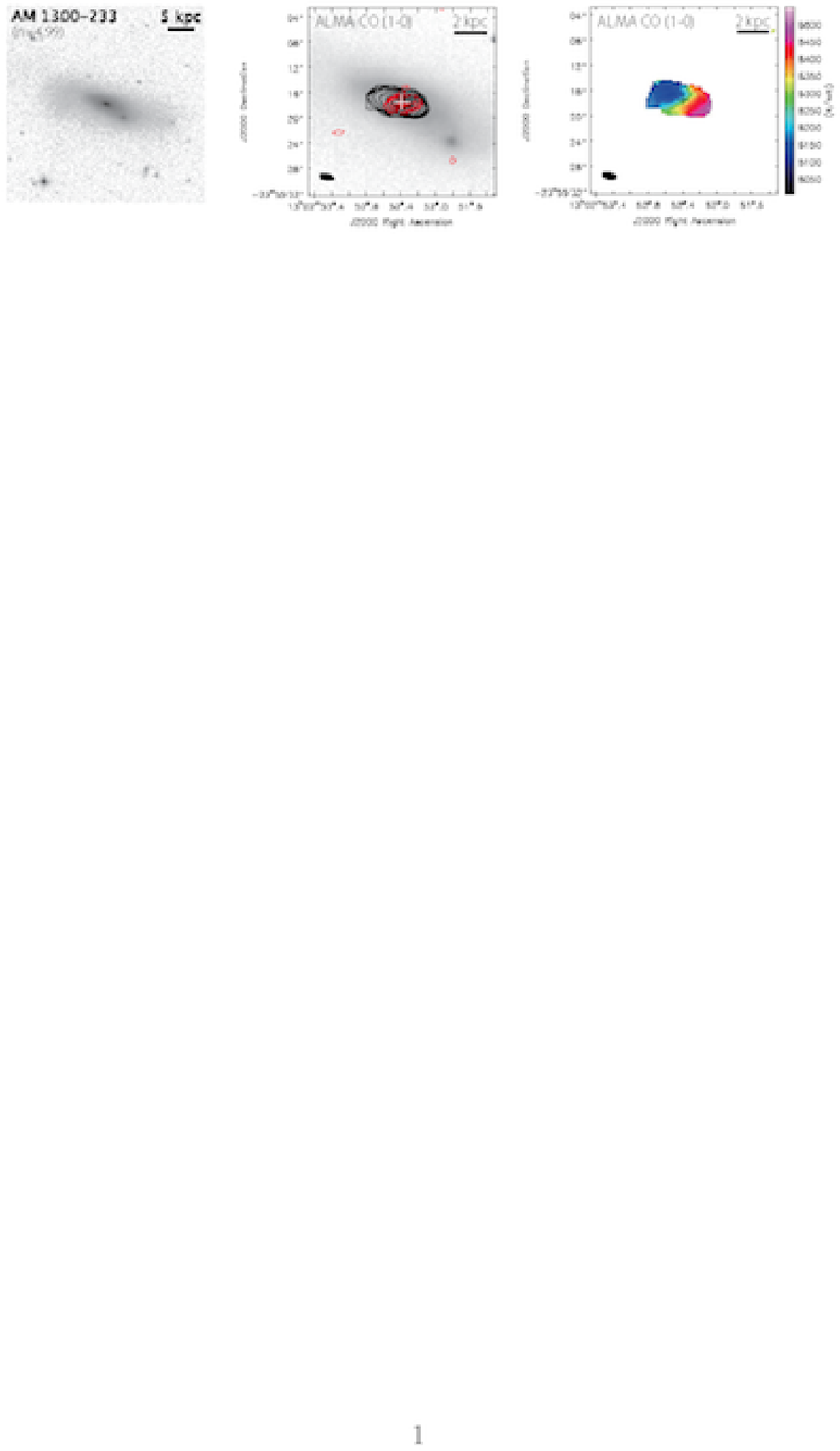}
  \end{center}
 \end{minipage}
 \caption{
 (Continued) The same as the previous figures 
 but for NGC~4441, UGC~8058,  AM~1255-430, and AM~1300-233.  
 The black contours of the CO integrated intensity are 
 0.28 Jy~km s$^{-1} \times$ (1, 2, 3, 5, 10, 25, 50) for NGC~4441, 
 5.8 Jy~km s$^{-1} \times$ (1, 2, 3, 5, 10, 20, 30, 40) for UGC~8058, 
 0.22 Jy~km s$^{-1} \times$ (1, 2, 3, 5, 10, 15, 20, 25) for AM~1255-430, and 
 0.39 Jy~km s$^{-1} \times$ (1, 2, 3, 5, 10, 20, 40, 60) for AM~1300-233.  
 The red contours of the 3~mm continuum are 
 62 $\mu$Jy $\times$ (3, 5, 10, 25, 50, 75, 100) for AM~1300-233.
 }
 \label{fig:f1}
\end{figure}

\addtocounter{figure}{-1}
\begin{figure}[htbp]
 \begin{minipage}{1.0\hsize}
  \begin{center}
   \includegraphics[scale=1.3, trim=3.5cm 21.1cm 0cm 4.5cm, clip]{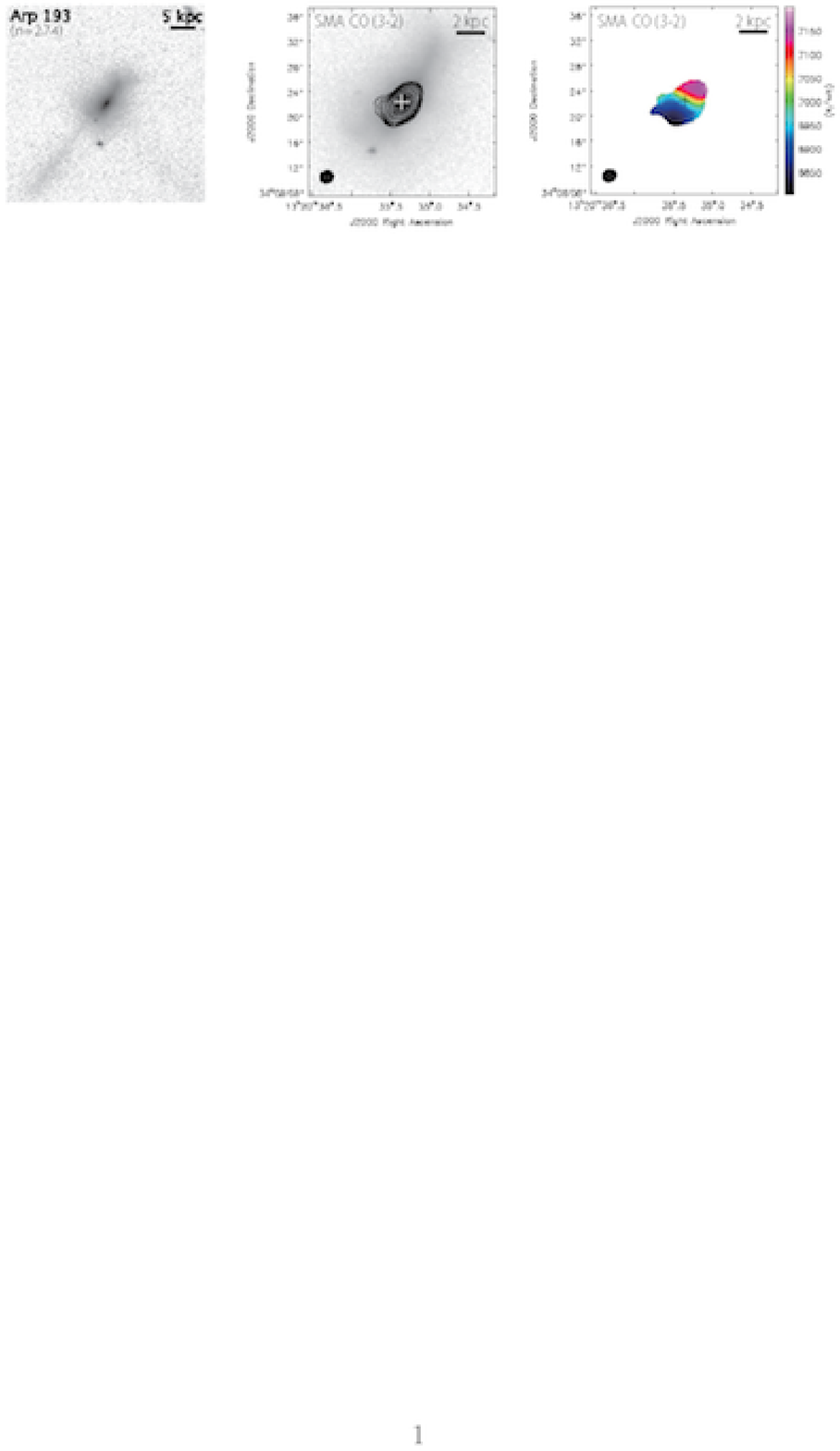}
  \end{center}
 \end{minipage}
 
 \begin{minipage}{1.0\hsize}
  \begin{center}
   \includegraphics[scale=1.3, trim=3.5cm 21.1cm 0cm 4.8cm, clip]{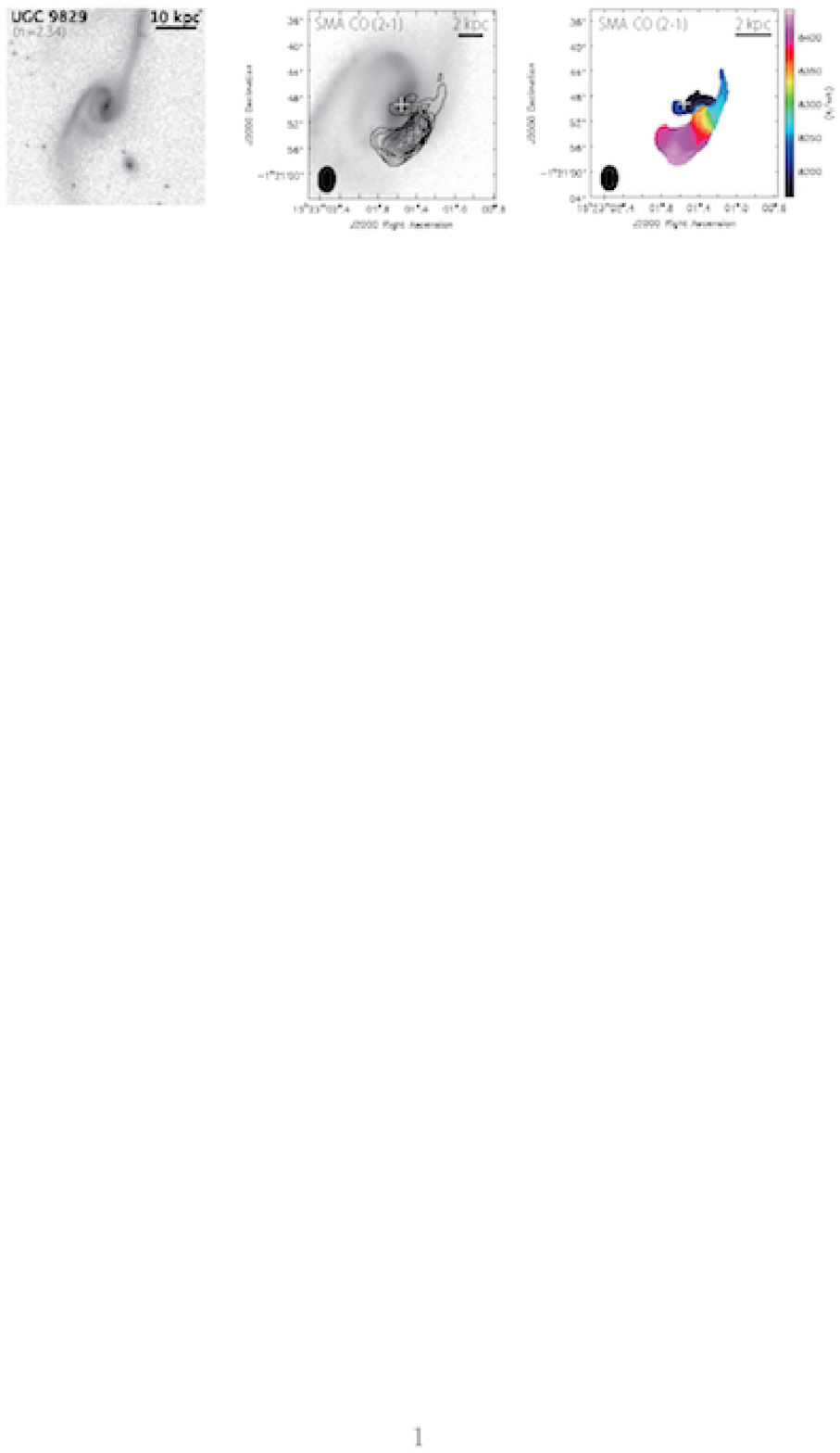}
  \end{center}
 \end{minipage}
 
 \begin{minipage}{1.0\hsize}
  \begin{center}
   \includegraphics[scale=1.3, trim=3.5cm 21.1cm 0cm 4.8cm, clip]{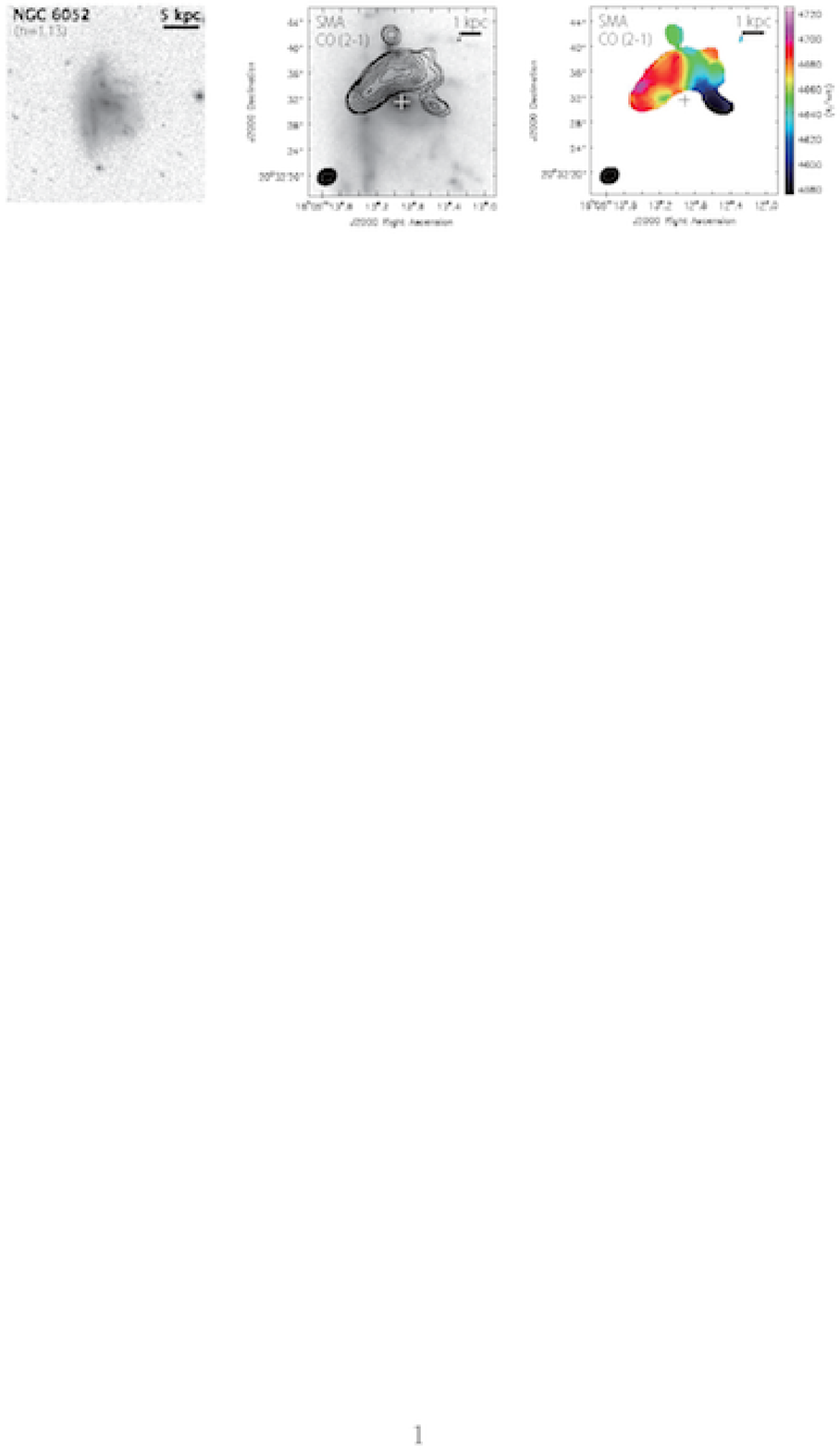}
  \end{center}
 \end{minipage} 
 
 \begin{minipage}{1.0\hsize}
  \begin{center}
   \includegraphics[scale=1.3, trim=3.5cm 21.1cm 0cm 4.8cm, clip]{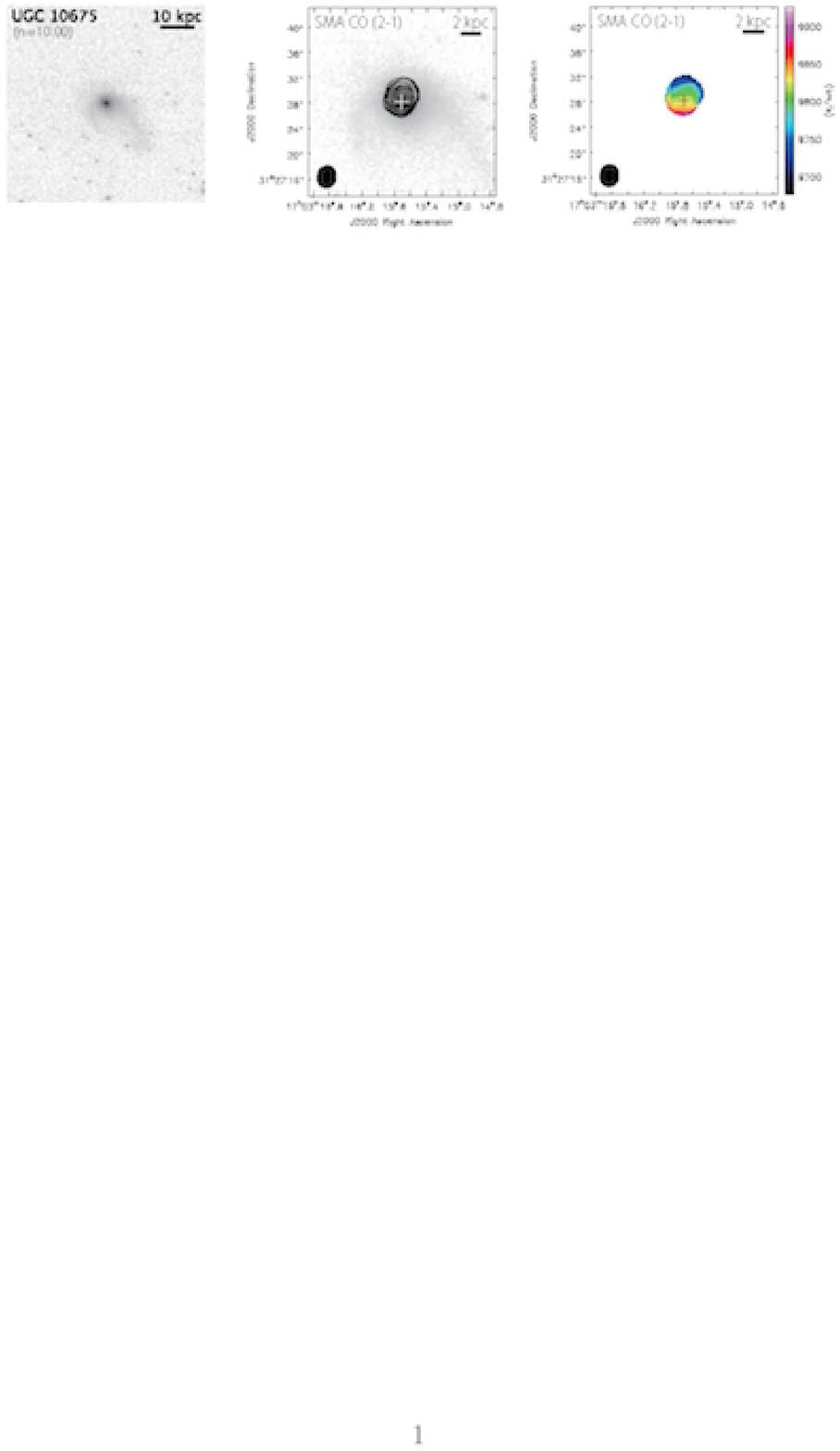}
  \end{center}
 \end{minipage} 
 \caption{
 (Continued) The same as the previous figures 
 but for Arp~193, UGC~9829, NGC~6052, and UGC~10675.
 The black contours of the CO integrated intensity are 
 10 Jy~km s$^{-1} \times$ (1, 2, 3, 5, 10, 20, 30, 40) for Arp~193, 
 2.4 Jy~km s$^{-1} \times$ (1, 2, 3, 4, 5, 6, 7, 8) for UGC~9829, 
 1.7 Jy~km s$^{-1} \times$ (1, 2, 3, 5, 7, 9, 11, 13) for NGC~6052, and 
 2.4 Jy~km s$^{-1} \times$ (1, 2, 3, 5, 7, 9, 11, 13, 15) for UGC~10675.
 }
 \label{fig:f1}
\end{figure}

\addtocounter{figure}{-1}
\begin{figure}[htbp]
 \begin{minipage}{1.0\hsize}
  \begin{center}
   \includegraphics[scale=1.3, trim=3.5cm 21.1cm 0cm 4.5cm, clip]{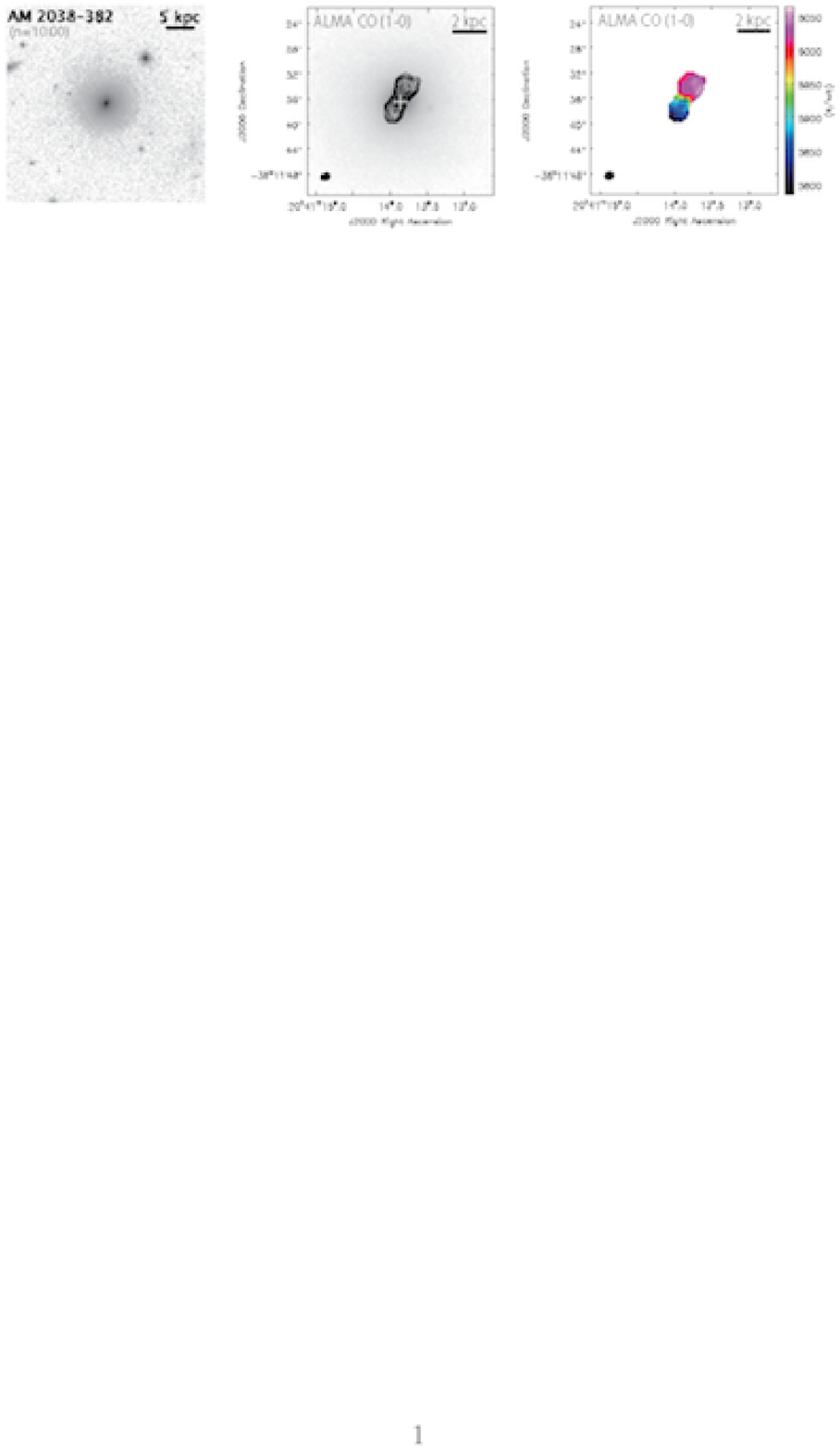}
  \end{center}
 \end{minipage}
 
 \begin{minipage}{1.0\hsize}
  \begin{center}
   \includegraphics[scale=1.3, trim=3.5cm 21.1cm 0cm 4.8cm, clip]{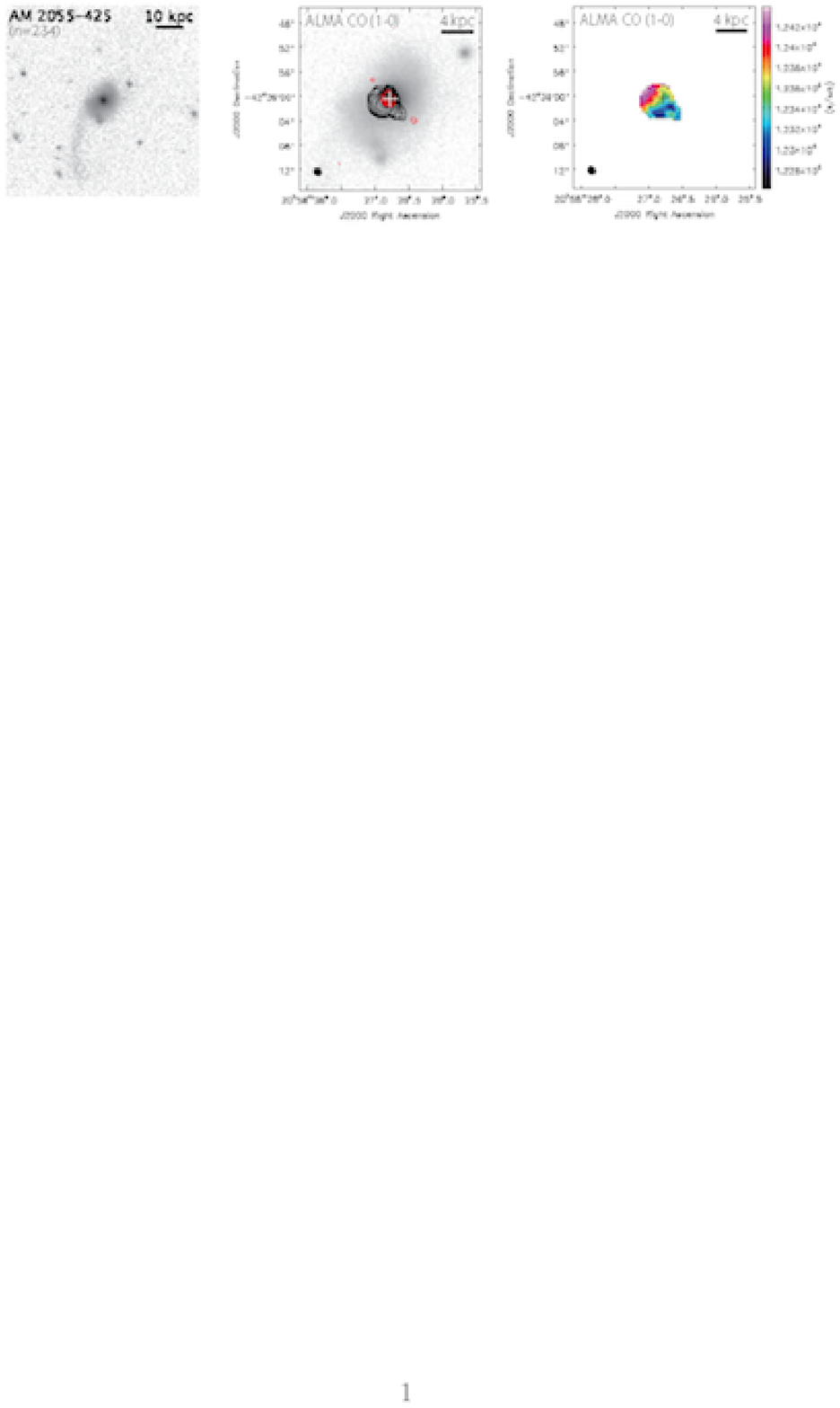}
  \end{center}
 \end{minipage} 

 \begin{minipage}{1.0\hsize}
  \begin{center}
   \includegraphics[scale=1.3, trim=3.5cm 21.1cm 0cm 4.8cm, clip]{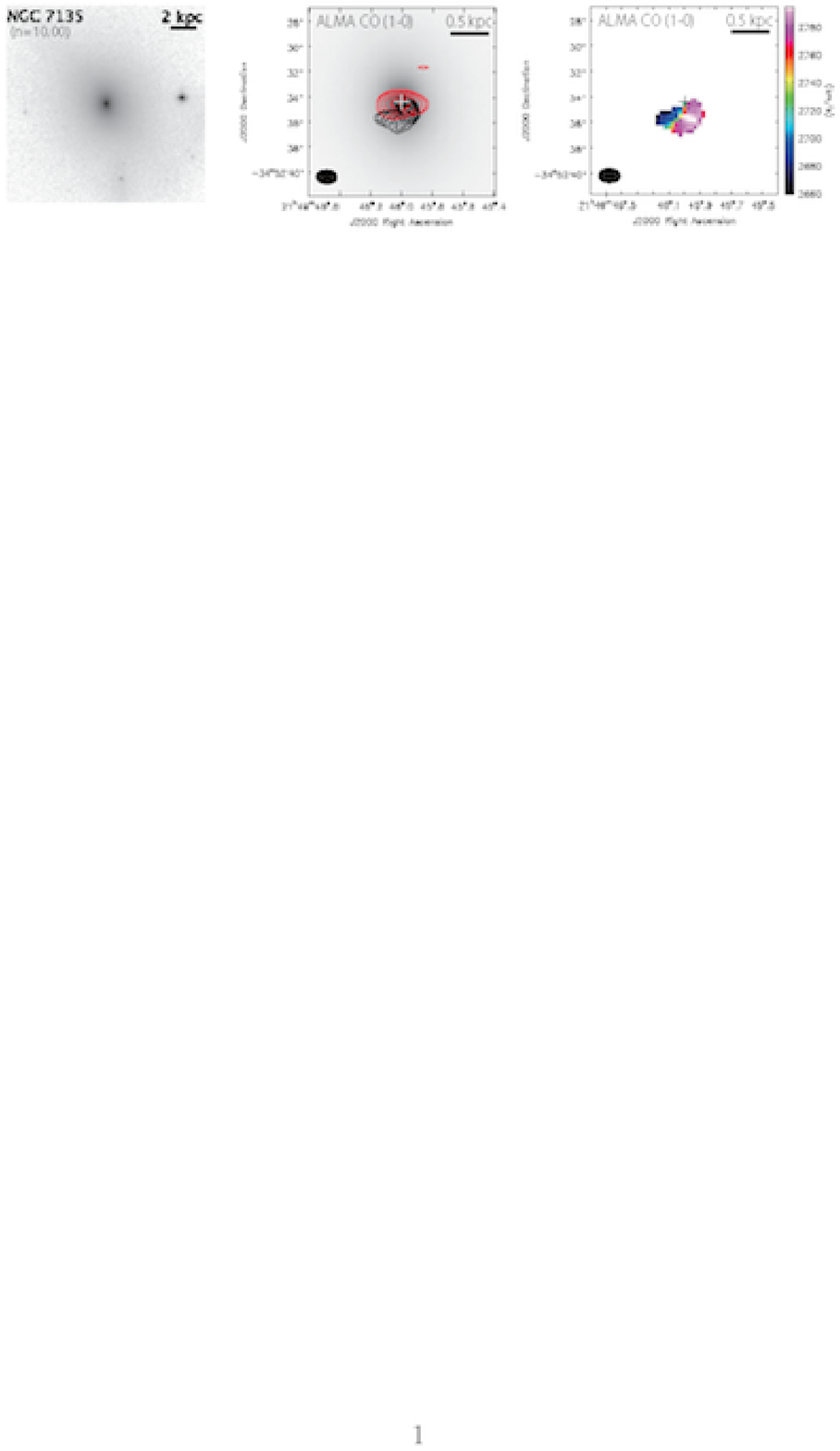}
  \end{center}
 \end{minipage} 

 \begin{minipage}{1.0\hsize}
  \begin{center}
   \includegraphics[scale=1.3, trim=3.5cm 21.1cm 0cm 4.8cm, clip]{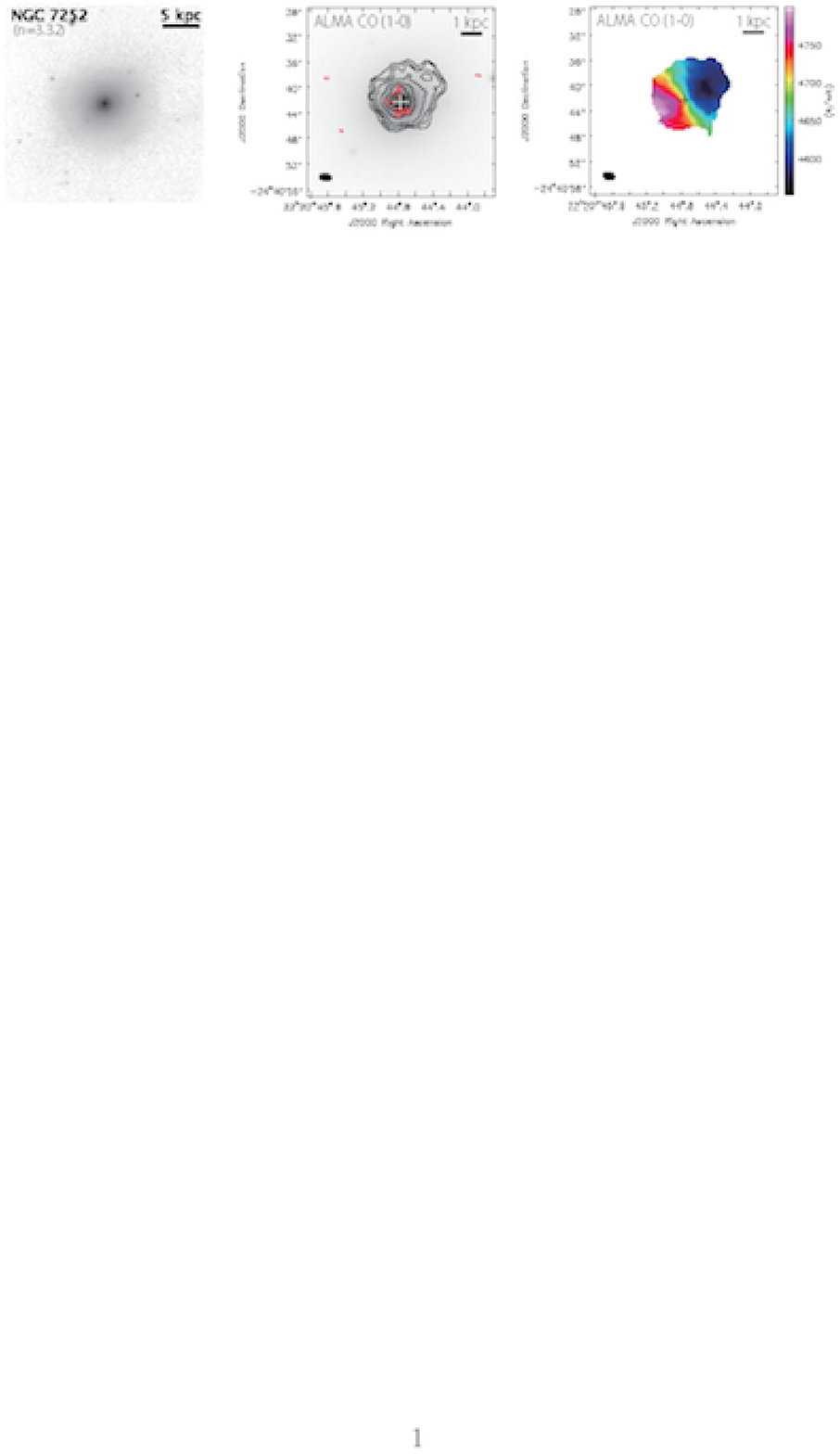}
  \end{center}
 \end{minipage} 
 \caption{
 (Continued) The same as the previous figures 
 but for AM~2038-382, AM~2055-425, NGC~7135, and NGC~7252.  
 The black contours of the CO integrated intensity are 
 0.37 Jy~km s$^{-1} \times$ (1, 2, 3, 5, 10, 15, 20) for AM~2038-382, 
 0.31 Jy~km s$^{-1} \times$ (1, 2, 3, 5, 10, 25, 50) for AM~2055-425, 
 0.37 Jy~km s$^{-1} \times$ (1, 1.5, 2, 2.5, 3, 3.5) for NGC~7135, and 
 0.37 Jy~km s$^{-1} \times$ (1, 2, 3, 5, 10, 15, 20, 25, 30) for NGC~7252.  
 The red contours of the 3~mm continuum are 
 72 $\mu$Jy $\times$ (3, 5, 10, 20, 40) for AM~2055-425, 
 115 $\mu$Jy $\times$ (3, 5, 10, 20, 40, 60) for NGC~7135, and 
 85 $\mu$Jy $\times$ (3, 4) for NGC~7252.
 }
 \label{fig:f1}
\end{figure}

\addtocounter{figure}{-1}
\begin{figure}[htbp]
 \begin{minipage}{1.0\hsize}
  \begin{center}
   \includegraphics[scale=1.3, trim=3.5cm 21.1cm 0cm 4.5cm, clip]{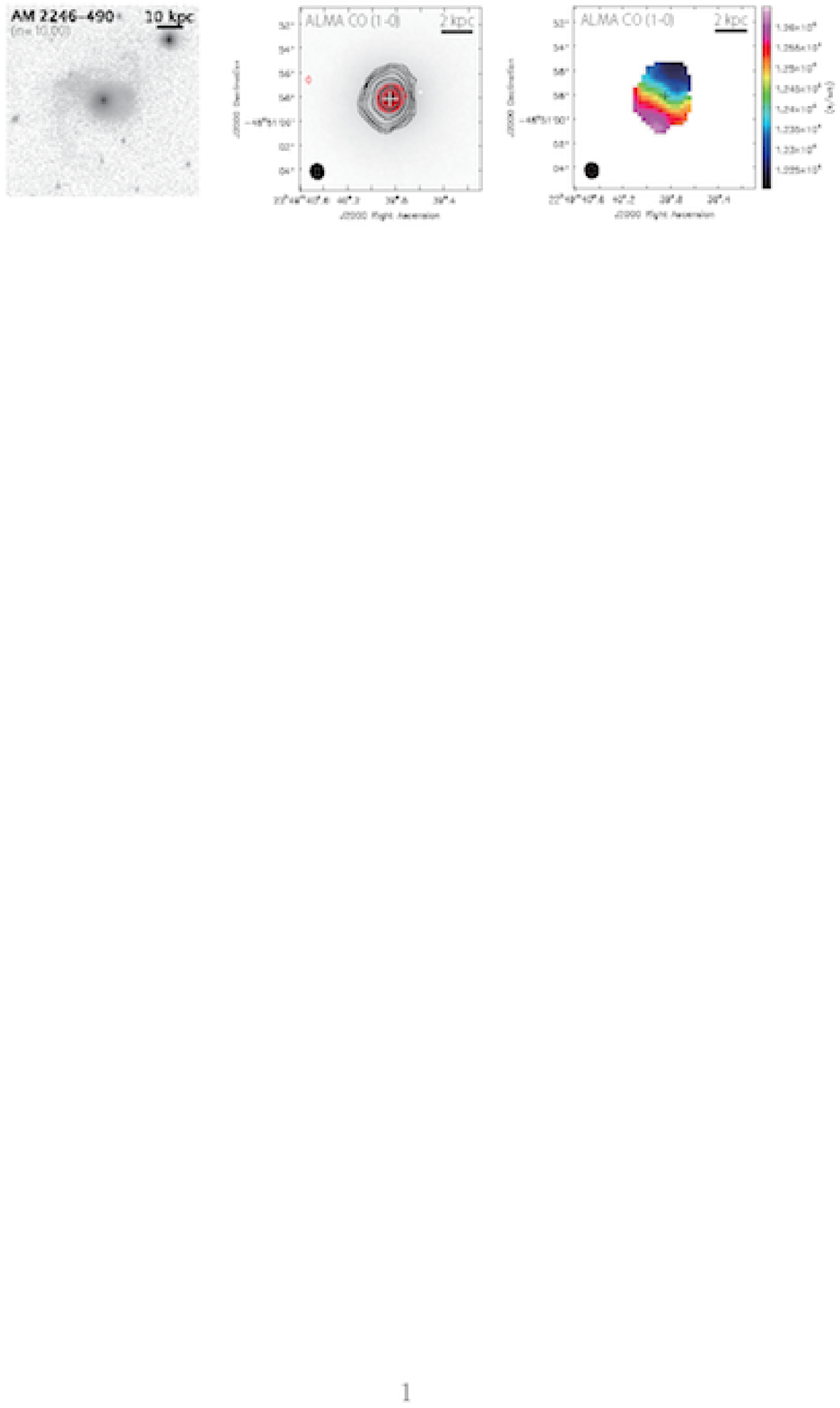}
  \end{center}
 \end{minipage}
 
 \begin{minipage}{1.0\hsize}
  \begin{center}
   \includegraphics[scale=1.3, trim=3.5cm 21.1cm 0cm 4.8cm, clip]{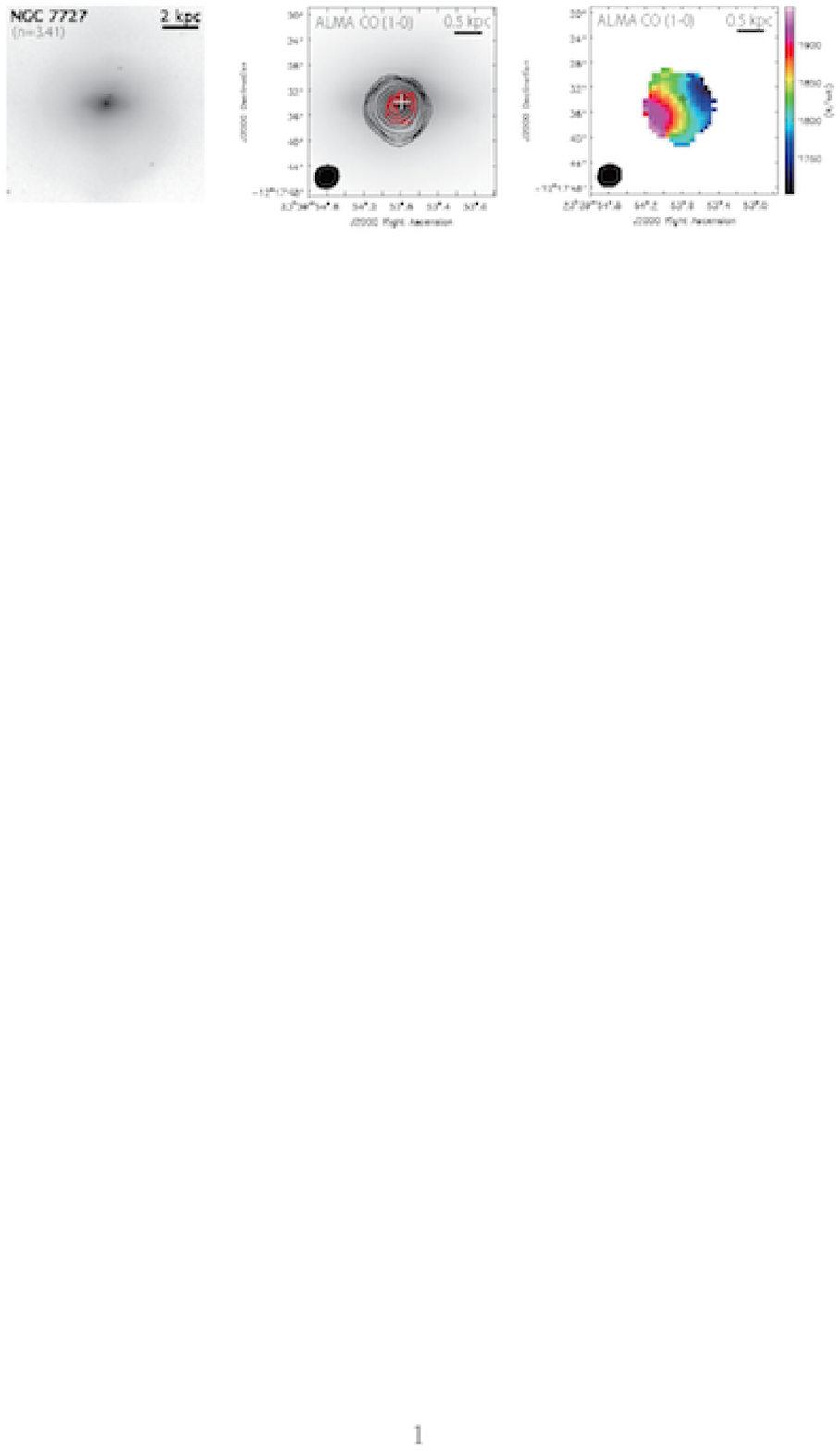}
  \end{center}
 \end{minipage} 
 \caption{
 (Continued) The same as the previous figures 
 but for AM~2246-490 and NGC~7727.  
 The black contours of the CO integrated intensity are 
 0.30 Jy~km s$^{-1} \times$ (1, 2, 3, 5, 10, 25, 50) for AM~2246-490 
 and 0.22 Jy~km s$^{-1} \times$ (1, 2, 3, 5, 10, 15, 20, 25, 30) for NGC~7727.  
 The red contours of the 3~mm continuum are 
 75 $\mu$Jy $\times$ (3, 5, 10, 15, 20) for AM~2246-490 and 
 77 $\mu$Jy $\times$ (3, 4, 5, 6) for NGC~7727.
 }
 \label{fig:f1}
\end{figure}

\begin{figure}[htbp] 
 \begin{minipage}{0.24\hsize}
  \begin{center}
   \includegraphics[scale=0.5, trim=4cm 0.5cm 3cm 0.5cm, clip]{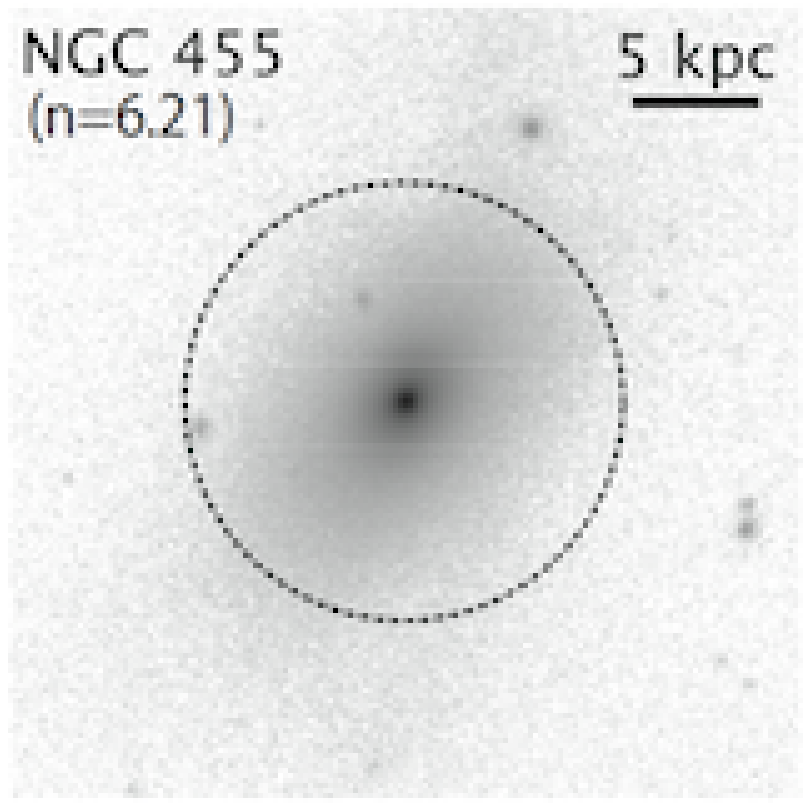}
  \end{center}
 \end{minipage}
 \begin{minipage}{0.24\hsize}
  \begin{center}
   \includegraphics[scale=0.5, trim=4cm 0.5cm 3cm 0.5cm, clip]{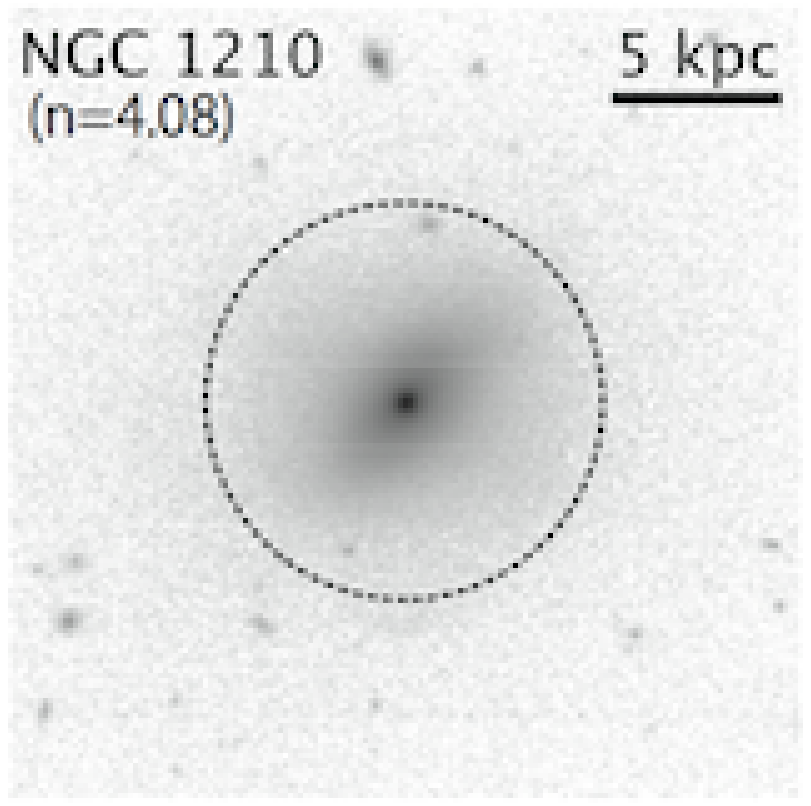}
  \end{center}
 \end{minipage}
 \begin{minipage}{0.24\hsize}
  \begin{center}
   \includegraphics[scale=0.5, trim=4cm 0.5cm 3cm 0.5cm, clip]{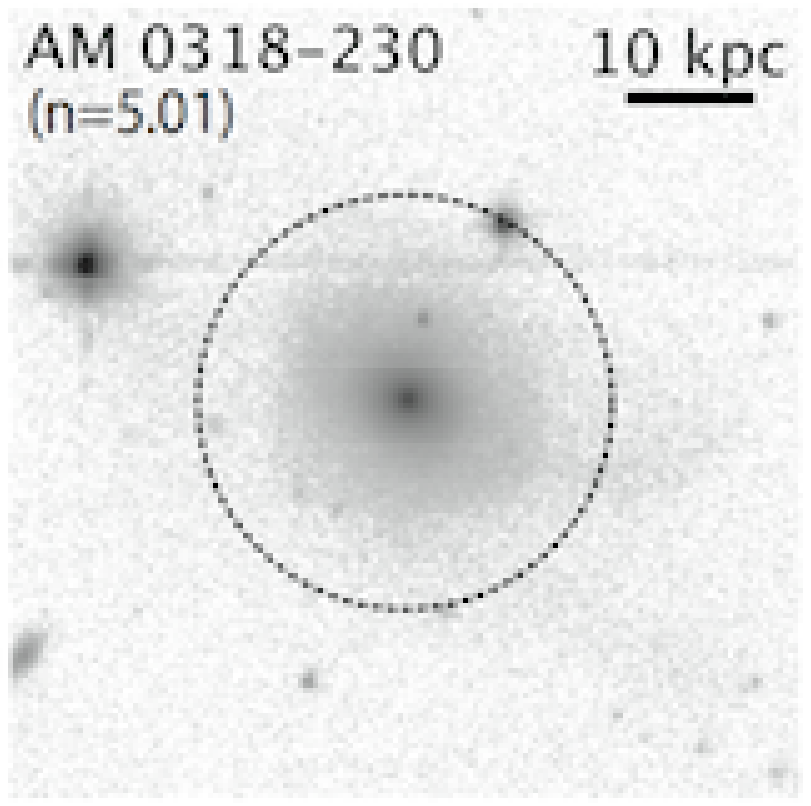}
  \end{center}
 \end{minipage}
 \begin{minipage}{0.24\hsize}
  \begin{center}
   \includegraphics[scale=0.5, trim=4cm 0.5cm 3cm 0.5cm, clip]{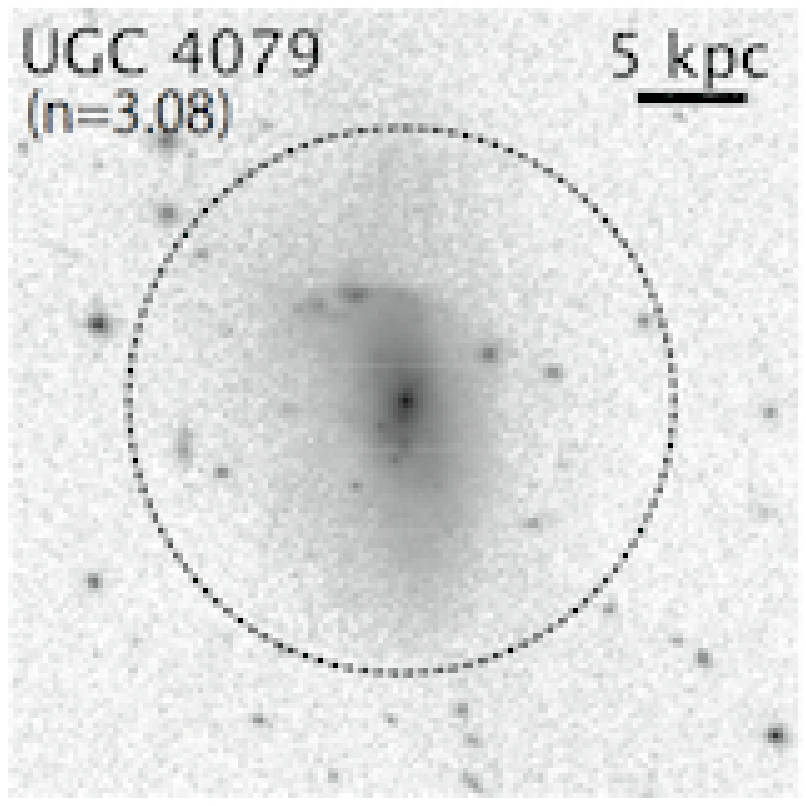}
  \end{center}
 \end{minipage}
 \begin{minipage}{0.24\hsize}
  \begin{center}
   \includegraphics[scale=0.5, trim=4cm 0.5cm 3cm 0.5cm, clip]{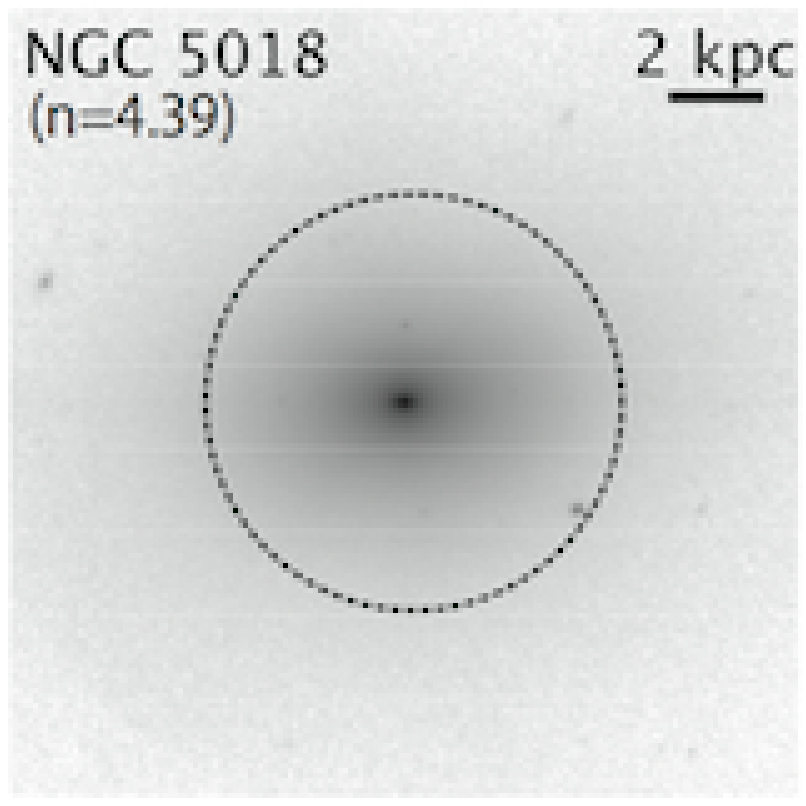}
  \end{center}
 \end{minipage}
 \begin{minipage}{0.24\hsize}
  \begin{center}
   \includegraphics[scale=0.5, trim=4cm 0.5cm 3cm 0.5cm, clip]{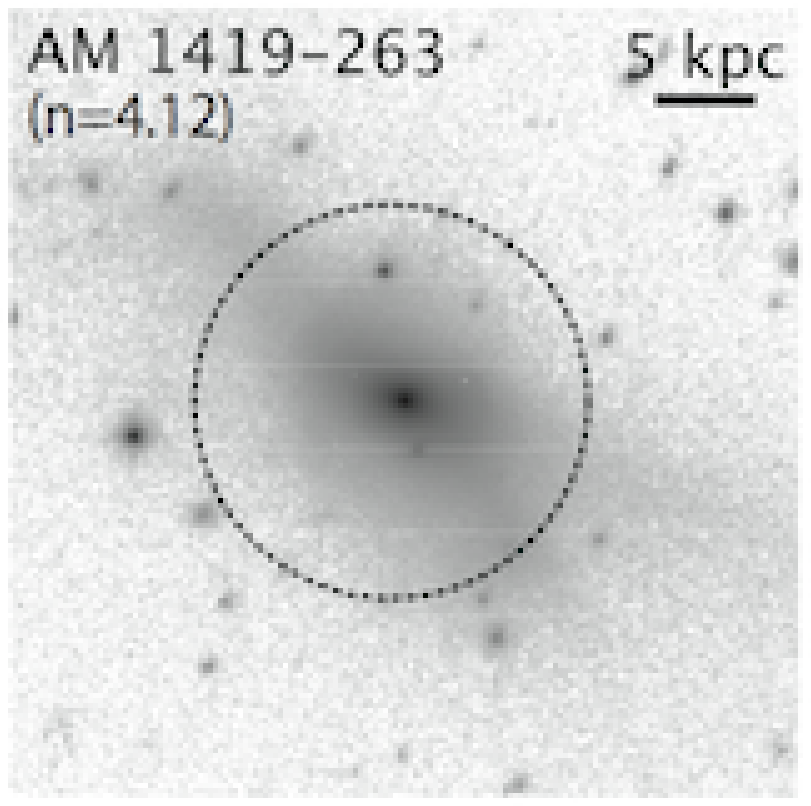}
  \end{center}
 \end{minipage}
  \begin{minipage}{0.24\hsize}
  \begin{center}
   \includegraphics[scale=0.5, trim=4cm 0.5cm 3cm 0.5cm, clip]{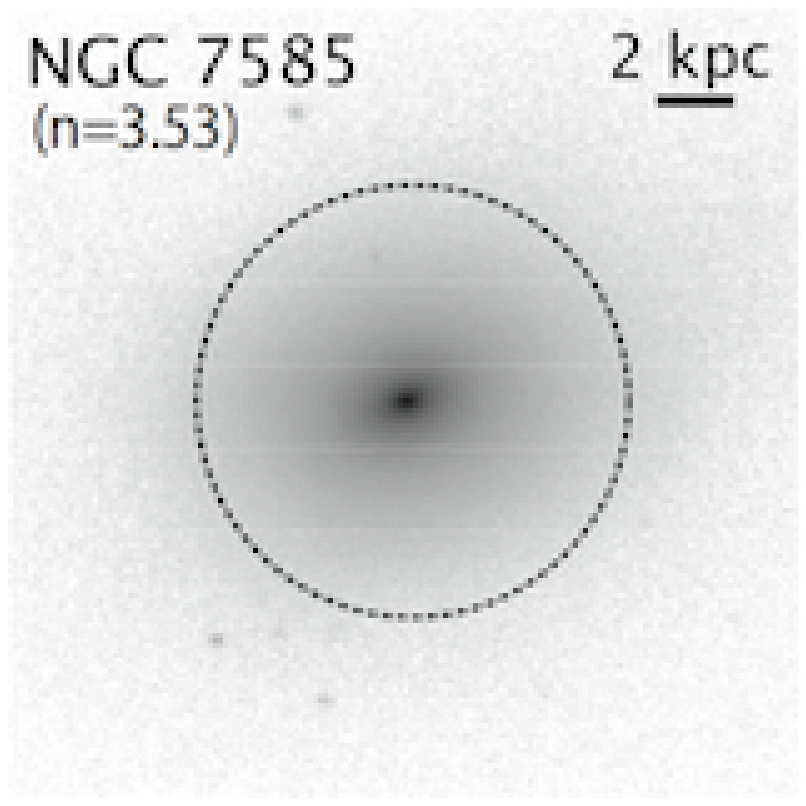}
  \end{center}
 \end{minipage}
 \caption{
 The $K$-band images (RJ04) of seven galaxies 
 undetected in their CO interferometric maps.  
 The length of each side corresponds to 90\arcsec.  
 The magnification is selected individually for each galaxy 
 to clearly illustrate its individual morphological structures.  
 The number in the upper-left corner is the S\'{e}rsic index estimated by (RJ04).  
 The black dotted circle shows the primary beam of interferometric observations for each galaxy.  
 UGC~4079 is undetected in the interferometric map due to lack of sensitivity, 
 but the CO~(1--0) spectra was clearly detected using the single-dish measurements.
 }
 \label{fig:f2}
\end{figure}

\begin{figure}[htbp]
 \begin{minipage}{0.24\hsize}
  \begin{center}
   \includegraphics[scale=0.75, trim=1cm 0cm 1cm 0cm, clip]{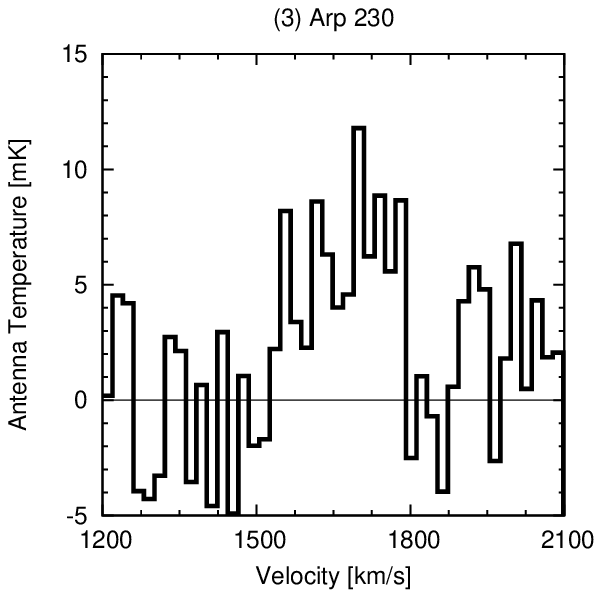}
  \end{center}
 \end{minipage}
 \begin{minipage}{0.24\hsize}
  \begin{center}
   \includegraphics[scale=0.75, trim=1cm 0cm 1cm 0cm, clip]{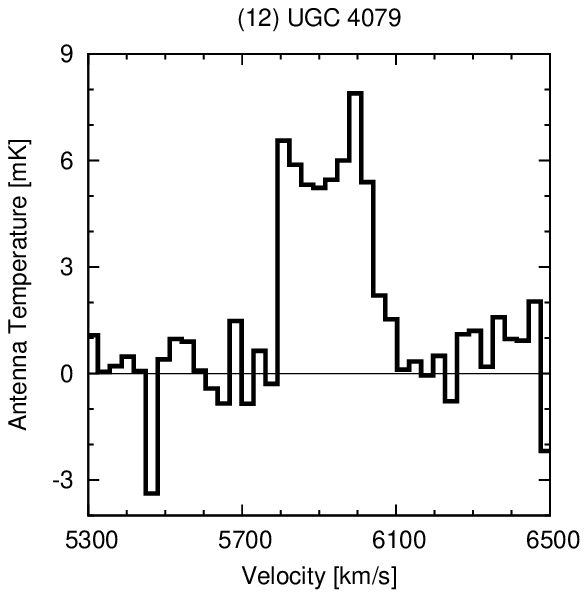}
  \end{center}
 \end{minipage}
 \begin{minipage}{0.24\hsize}
  \begin{center}
   \includegraphics[scale=0.75, trim=1cm 0cm 1cm 0cm, clip]{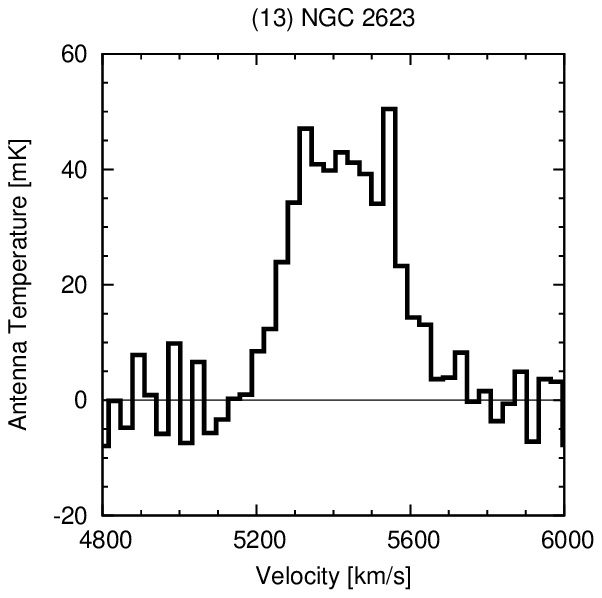}
  \end{center}
 \end{minipage}
  \begin{minipage}{0.24\hsize}
  \begin{center}
   \includegraphics[scale=0.75, trim=1cm 0cm 1cm 0cm, clip]{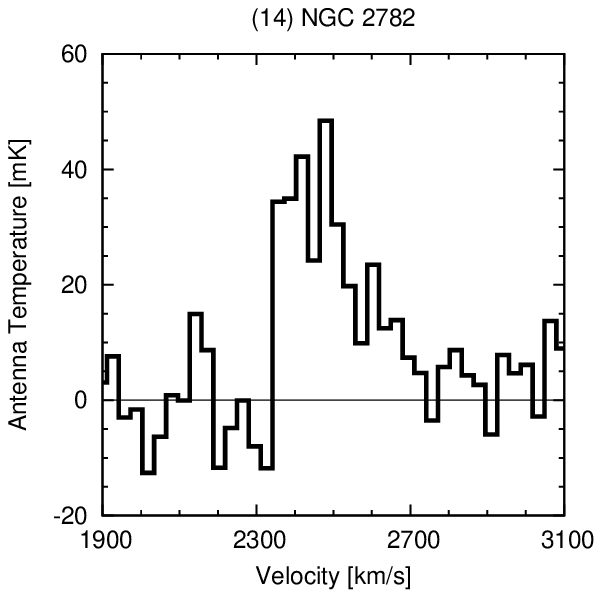}
  \end{center}
 \end{minipage}

 \begin{minipage}{0.24\hsize}
  \begin{center}
   \includegraphics[scale=0.75, trim=1cm 0cm 1cm 0cm, clip]{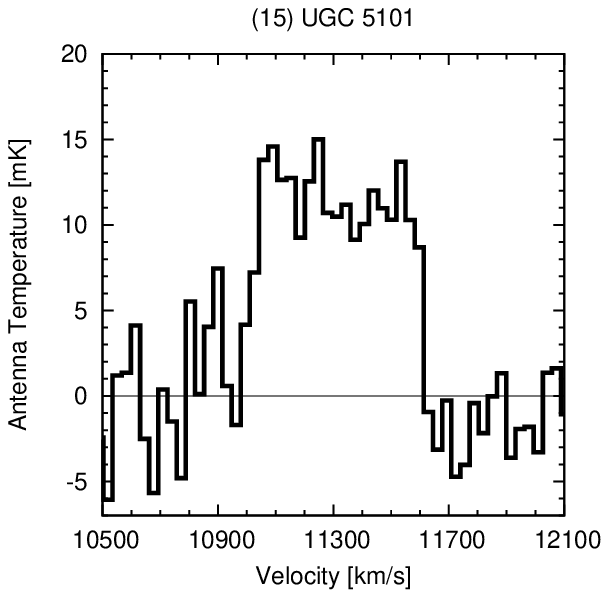}
  \end{center}
 \end{minipage}
 \begin{minipage}{0.24\hsize}
  \begin{center}
   \includegraphics[scale=0.75, trim=1cm 0cm 1cm 0cm, clip]{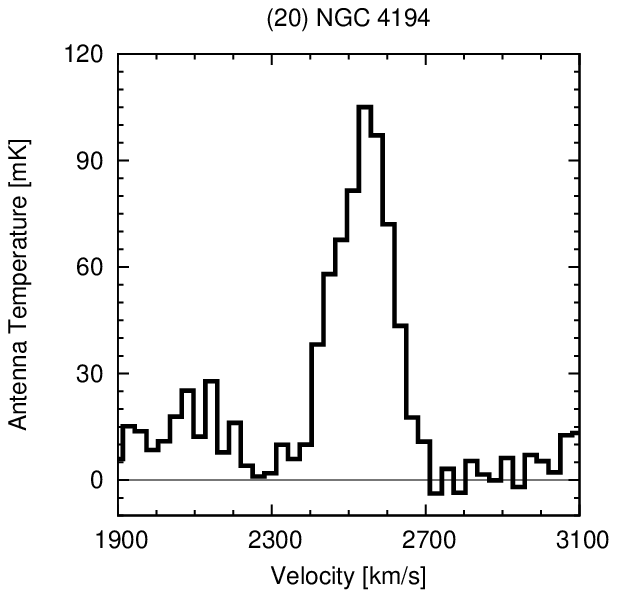}
  \end{center}
 \end{minipage}
 \begin{minipage}{0.24\hsize}
  \begin{center}
   \includegraphics[scale=0.75, trim=1cm 0cm 1cm 0cm, clip]{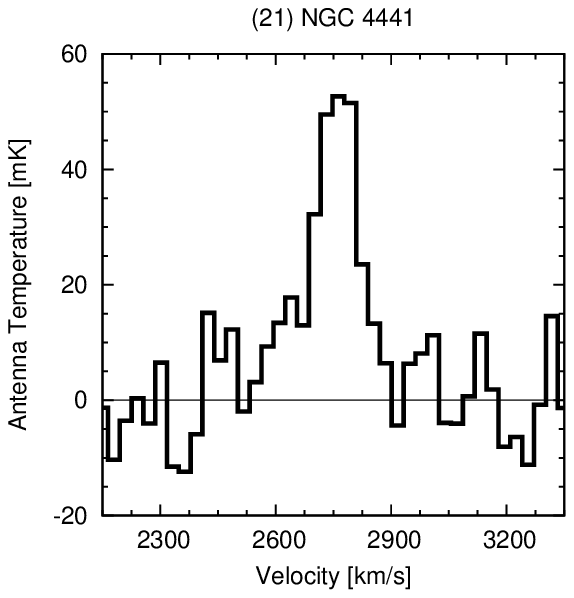}
  \end{center}
 \end{minipage}
  \begin{minipage}{0.24\hsize}
  \begin{center}
   \includegraphics[scale=0.75, trim=1cm 0cm 1cm 0cm, clip]{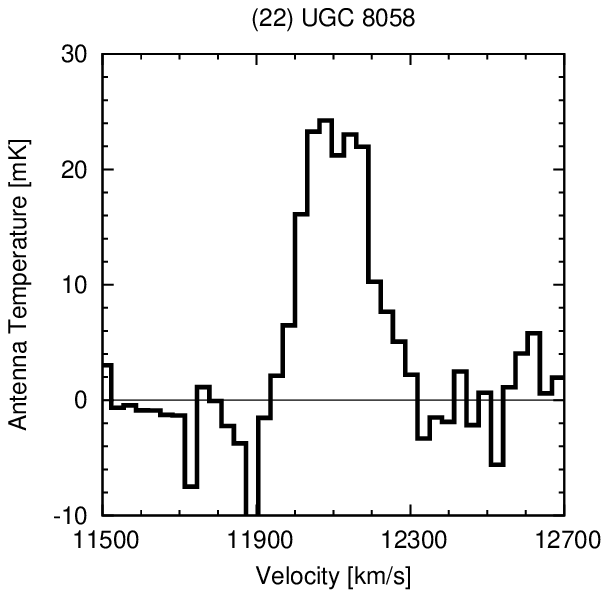}
  \end{center}
 \end{minipage}
 
 \begin{minipage}{0.24\hsize}
  \begin{center}
   \includegraphics[scale=0.75, trim=1cm 0cm 1cm 0cm, clip]{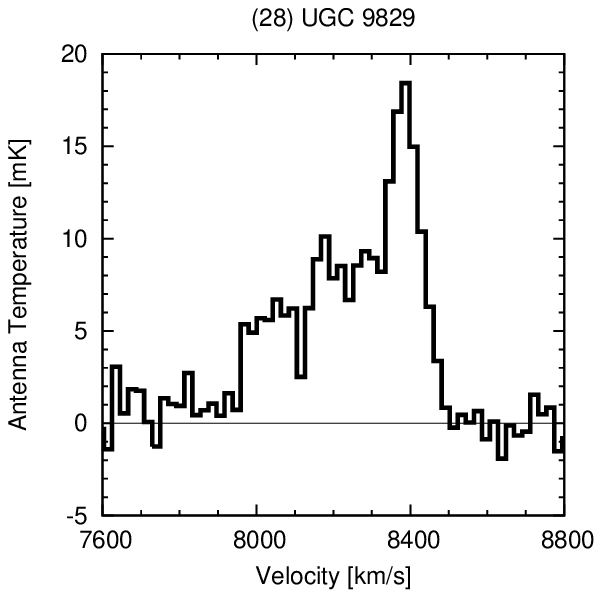}
  \end{center}
 \end{minipage}
 \begin{minipage}{0.24\hsize}
  \begin{center}
   \includegraphics[scale=0.75, trim=1cm 0cm 1cm 0cm, clip]{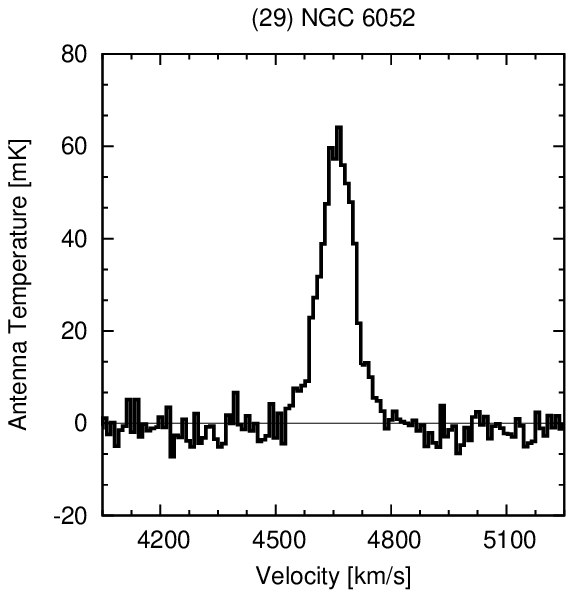}
  \end{center}
 \end{minipage}
 \begin{minipage}{0.24\hsize}
  \begin{center}
   \includegraphics[scale=0.75, trim=1cm 0cm 1cm 0cm, clip]{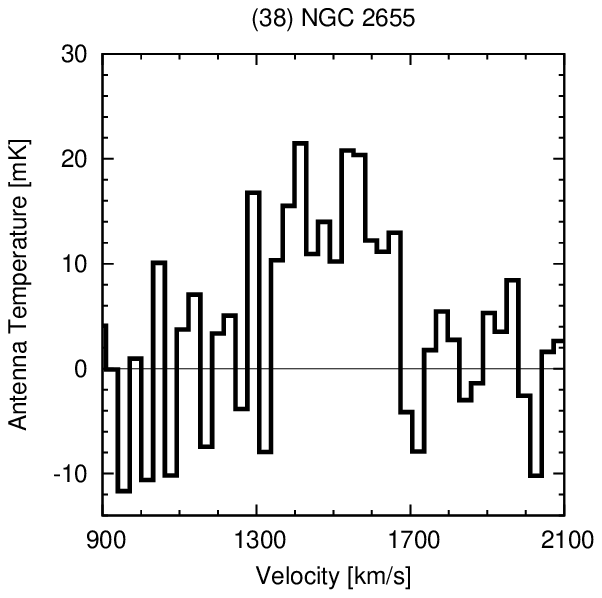}
  \end{center}
 \end{minipage}
  \begin{minipage}{0.24\hsize}
  \begin{center}
   \includegraphics[scale=0.75, trim=1cm 0cm 1cm 0cm, clip]{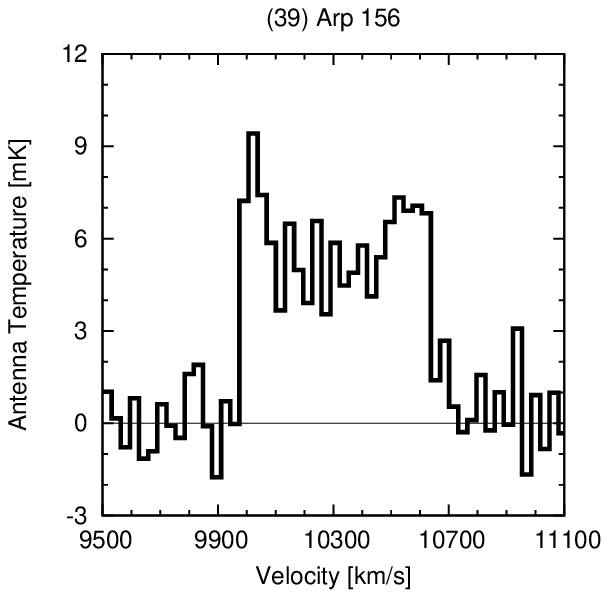}
  \end{center}
 \end{minipage}

 \begin{minipage}{0.24\hsize}
  \begin{center}
   \includegraphics[scale=0.75, trim=1cm 0cm 1cm 0cm, clip]{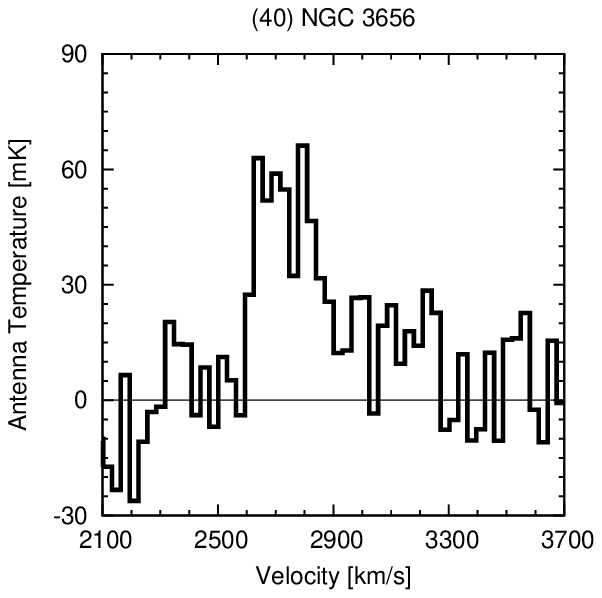}
  \end{center}
 \end{minipage}
 \begin{minipage}{0.24\hsize}
  \begin{center}
   \includegraphics[scale=0.75, trim=1cm 0cm 1cm 0cm, clip]{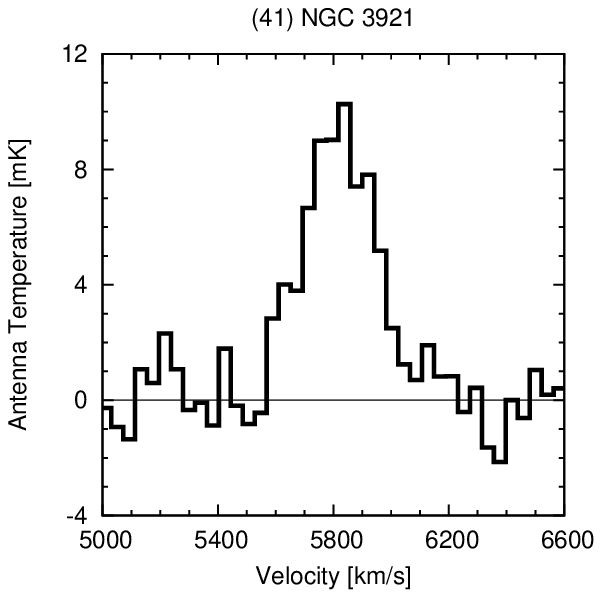}
  \end{center}
 \end{minipage}
 \begin{minipage}{0.24\hsize}
  \begin{center}
   \includegraphics[scale=0.75, trim=1cm 0cm 1cm 0cm, clip]{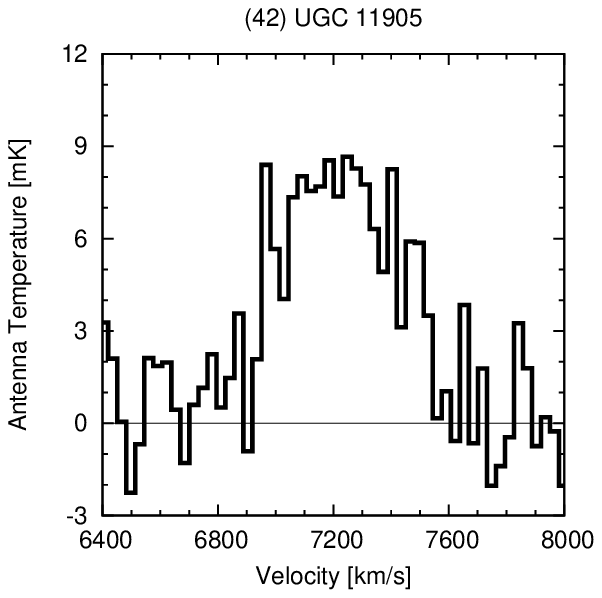}
  \end{center}
 \end{minipage}
  \begin{minipage}{0.24\hsize}
  \begin{center}
   \includegraphics[scale=0.75, trim=1cm 0cm 1cm 0cm, clip]{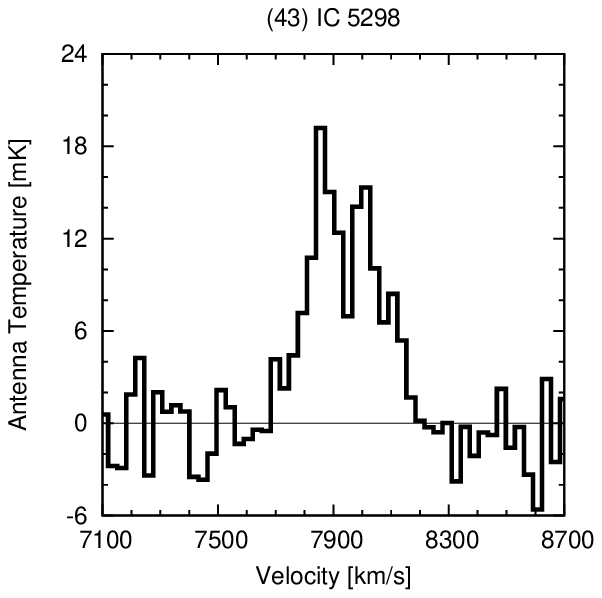}
  \end{center}
 \end{minipage}
 \caption{
 The CO~(1--0) spectra of 16 sources obtained with the NRO 45~m telescope.  
 The primary beam size is about 15\arcsec at 115~GHz.  
 The velocity resolution is different for each galaxy.
 }
 \label{fig:f3}
\end{figure}

\begin{figure}[htbp]
 \begin{minipage}{0.24\hsize}
  \begin{center}
   \includegraphics[scale=0.22]{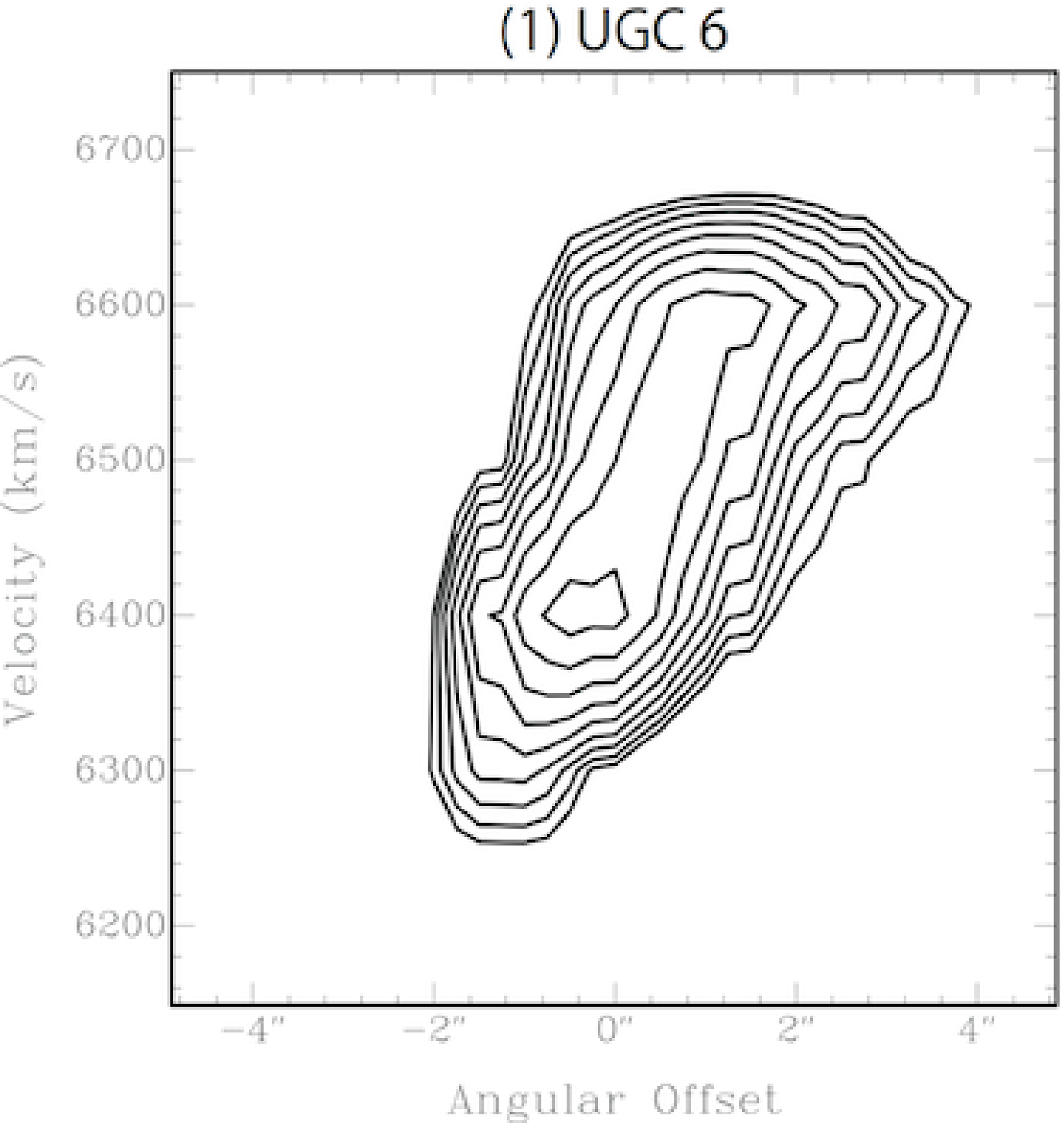}
  \end{center}
 \end{minipage}
 \begin{minipage}{0.24\hsize}
  \begin{center}
   \includegraphics[scale=0.22]{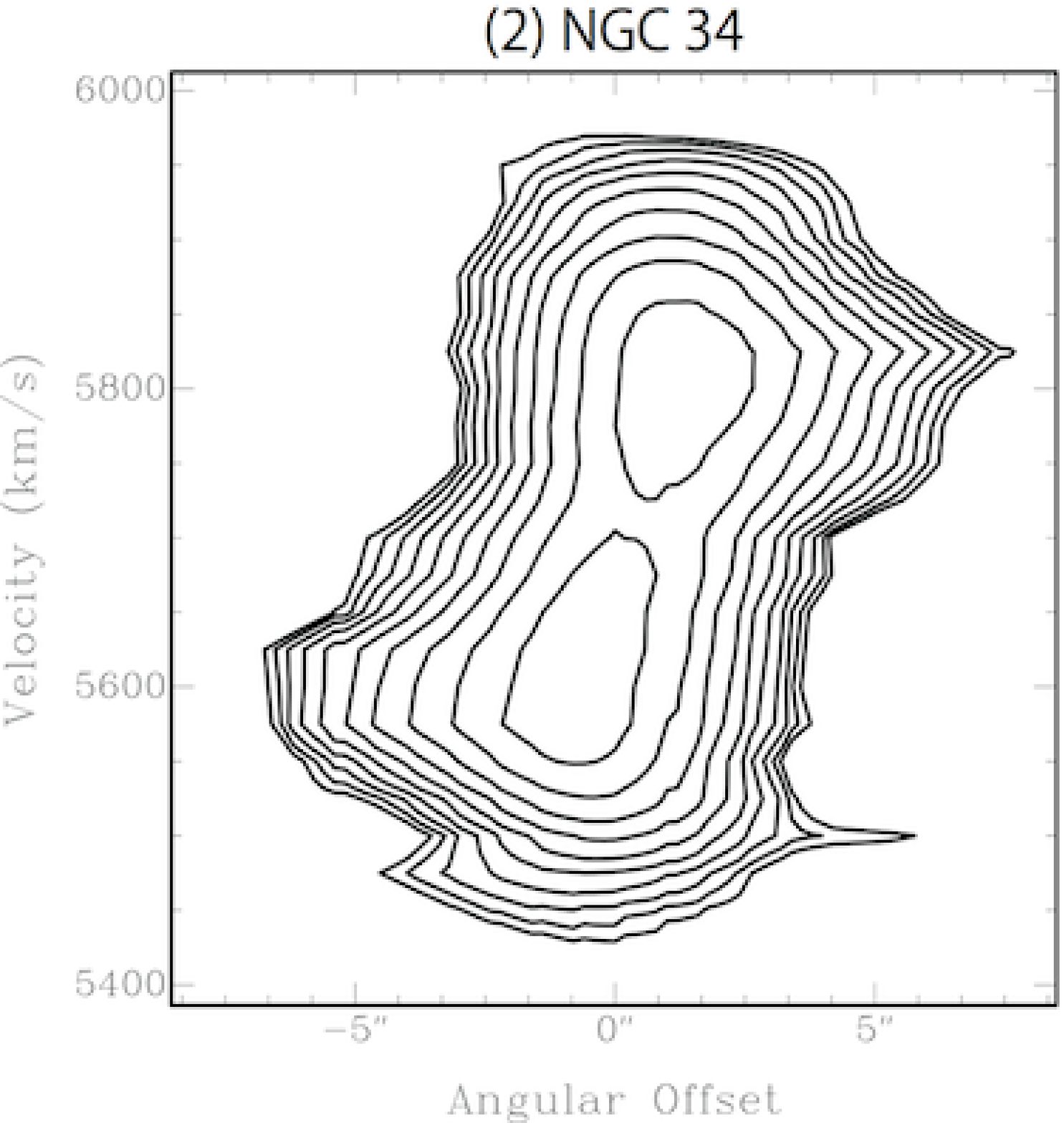}
  \end{center}
 \end{minipage}
 \begin{minipage}{0.24\hsize}
  \begin{center}
   \includegraphics[scale=0.22]{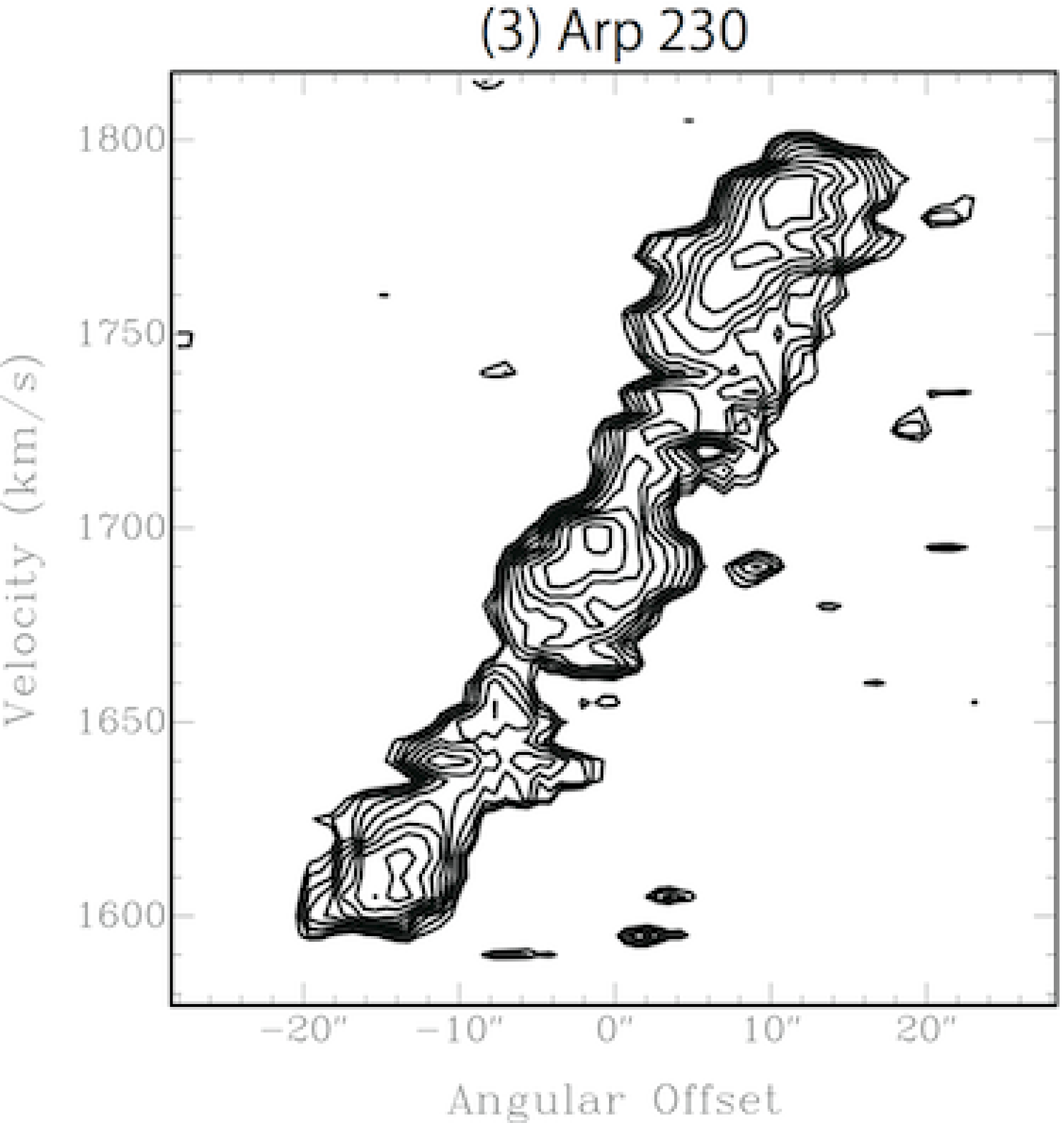}
  \end{center}
 \end{minipage}
  \begin{minipage}{0.24\hsize}
  \begin{center}
   \includegraphics[scale=0.22]{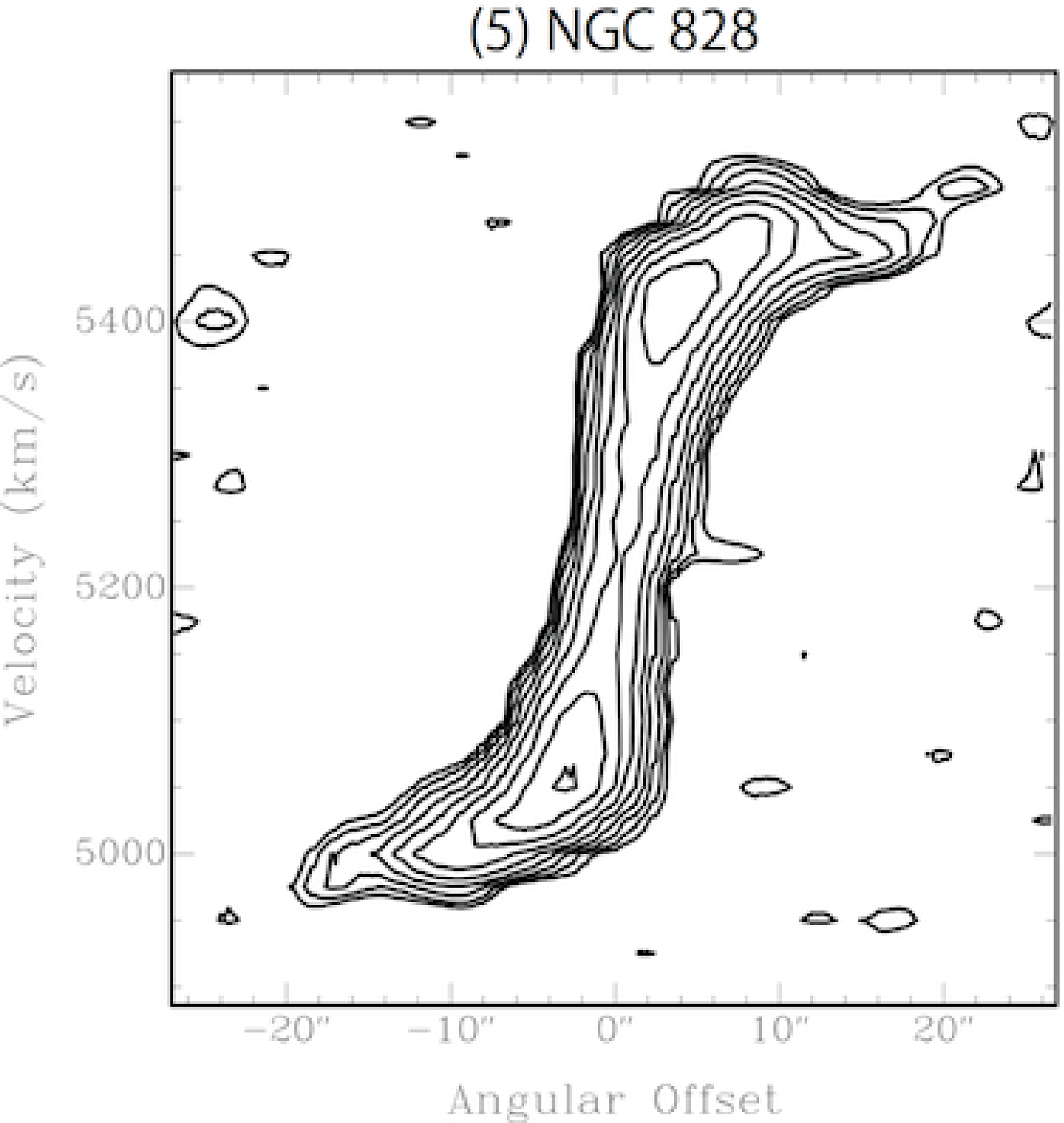}
  \end{center}
 \end{minipage}

 \begin{minipage}{0.24\hsize}
  \begin{center}
   \includegraphics[scale=0.22]{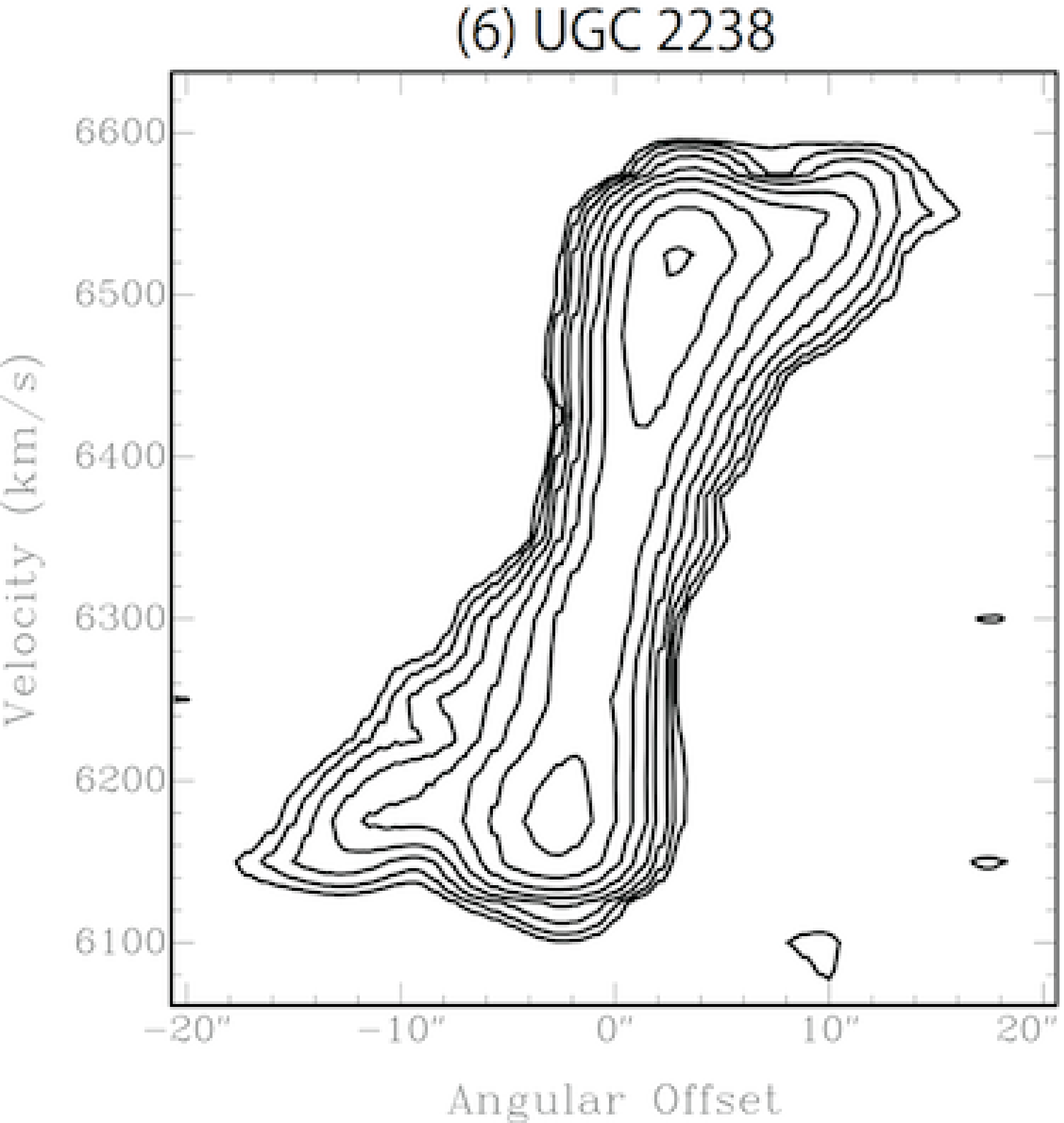}
  \end{center}
 \end{minipage}
 \begin{minipage}{0.24\hsize}
  \begin{center}
   \includegraphics[scale=0.22]{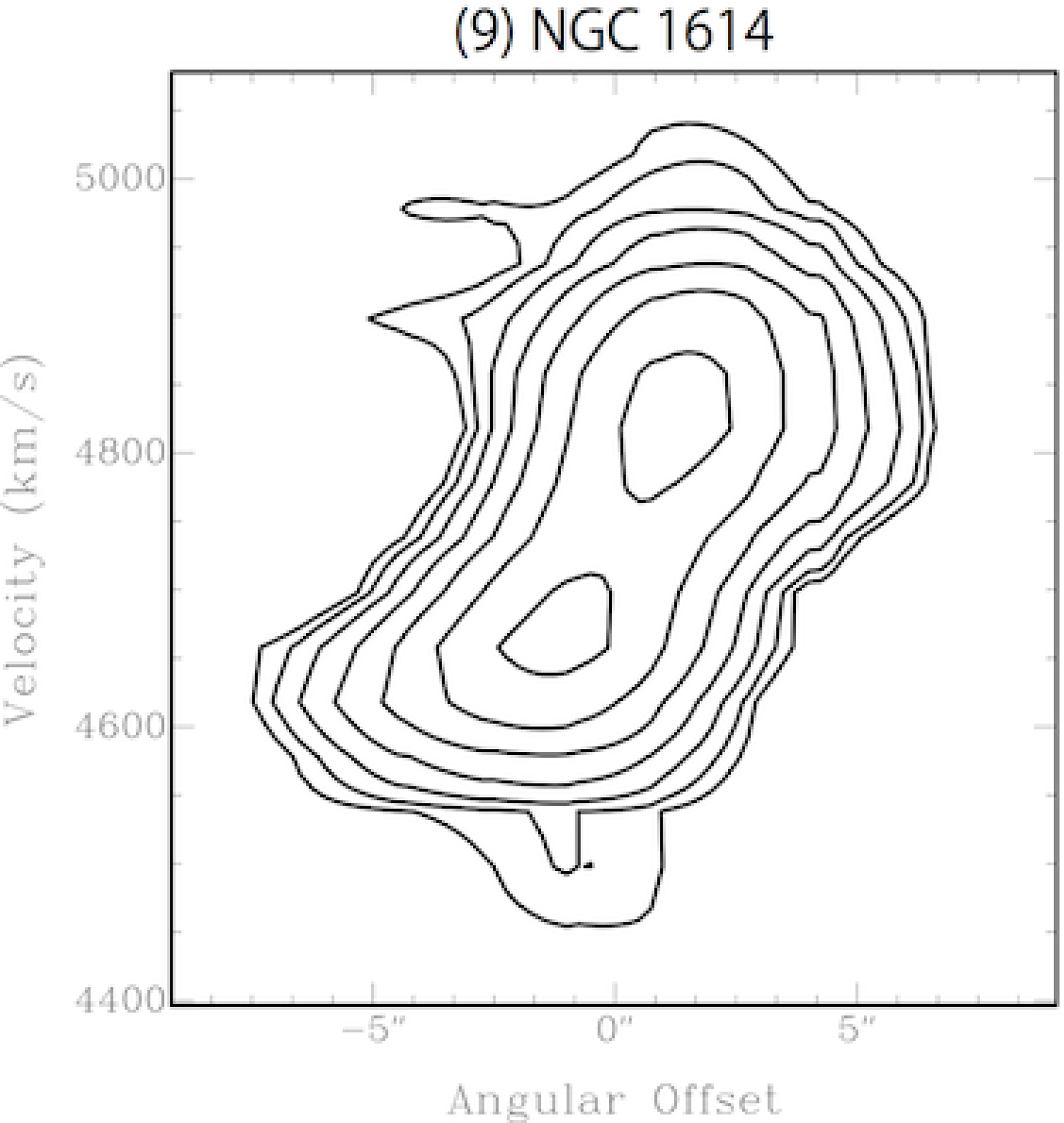}
  \end{center}
 \end{minipage}
 \begin{minipage}{0.24\hsize}
  \begin{center}
   \includegraphics[scale=0.22]{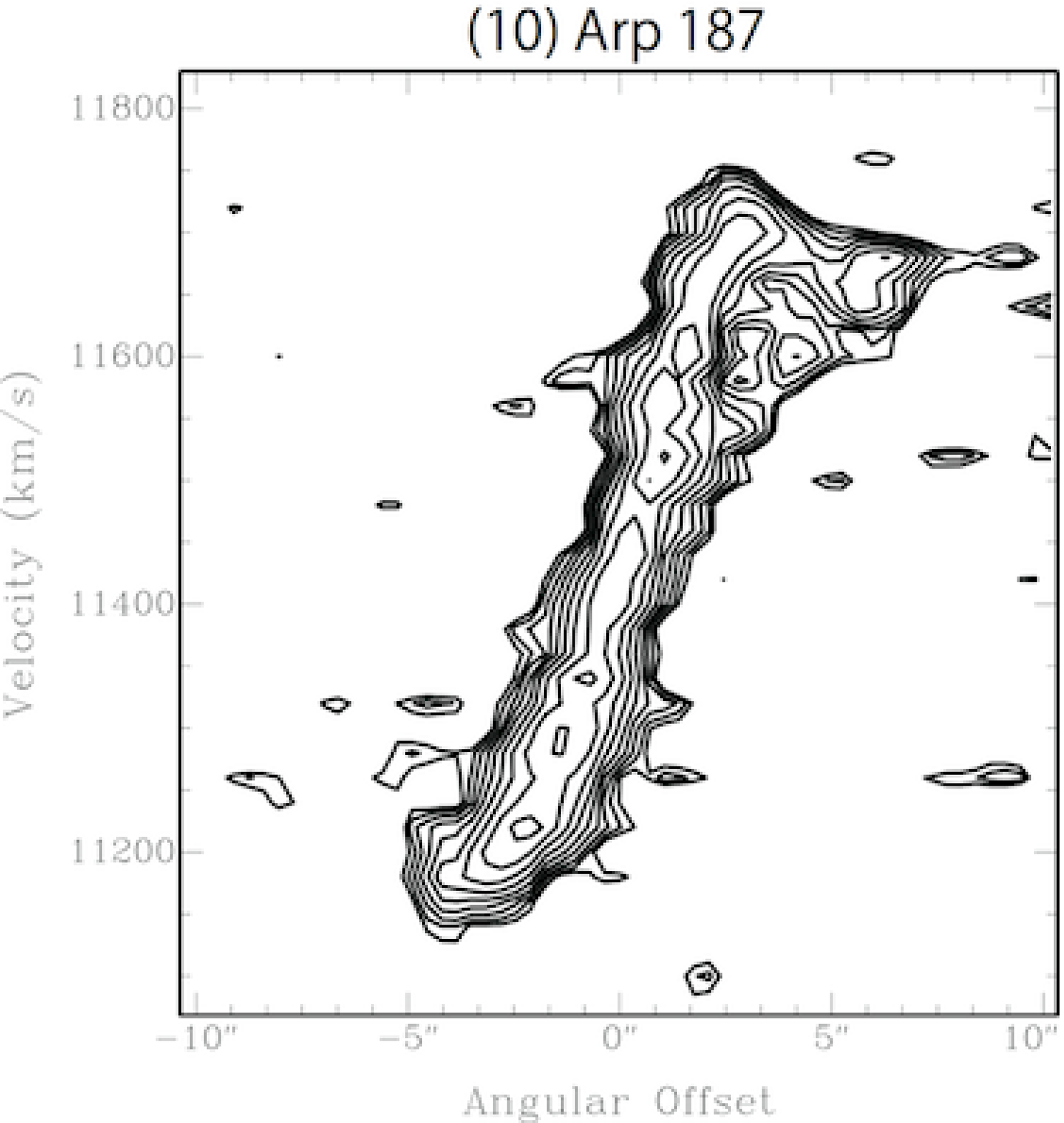}
  \end{center}
 \end{minipage}
  \begin{minipage}{0.24\hsize}
  \begin{center}
   \includegraphics[scale=0.22]{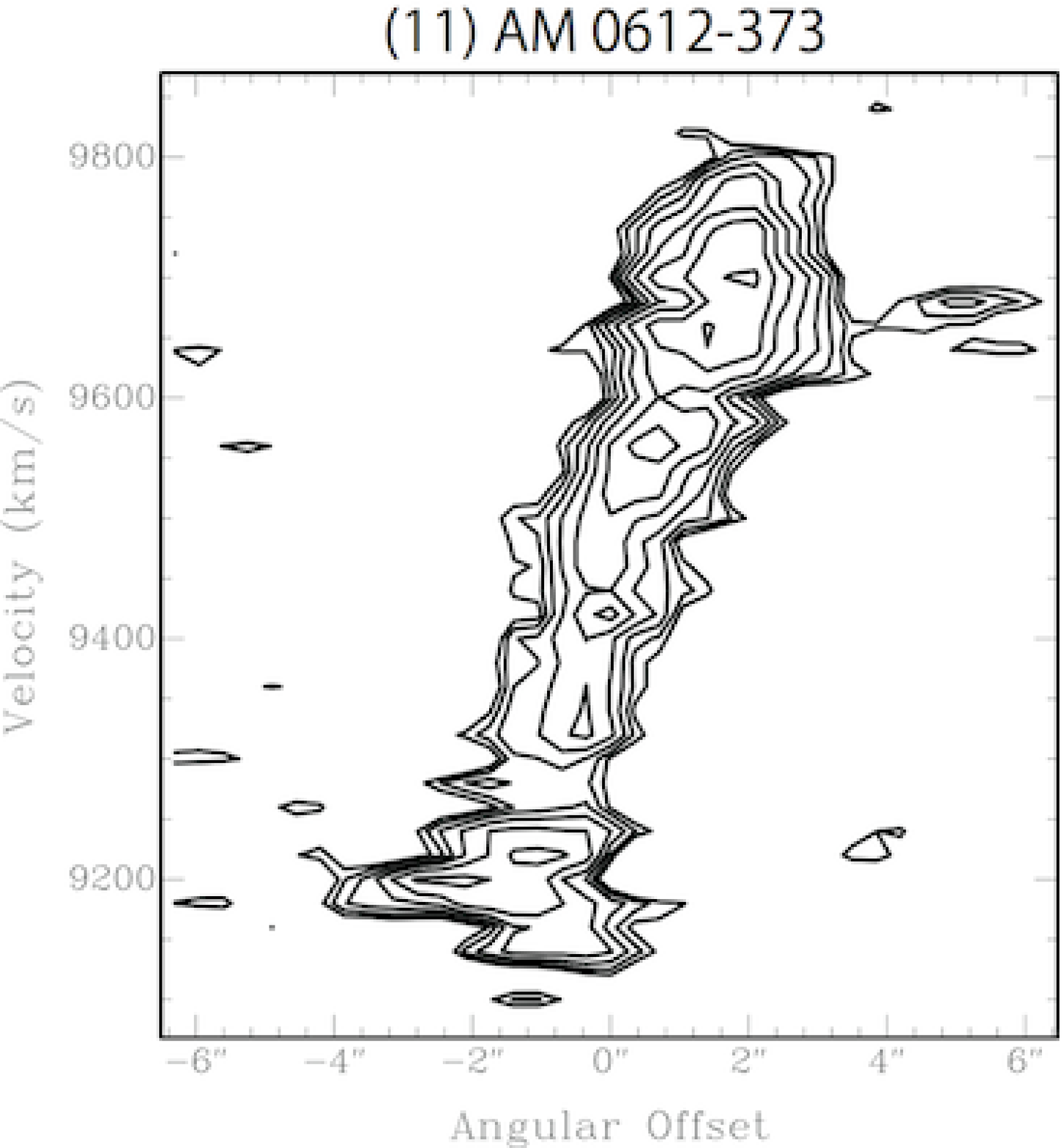}
  \end{center}
 \end{minipage}
 
 \begin{minipage}{0.24\hsize}
  \begin{center}
   \includegraphics[scale=0.22]{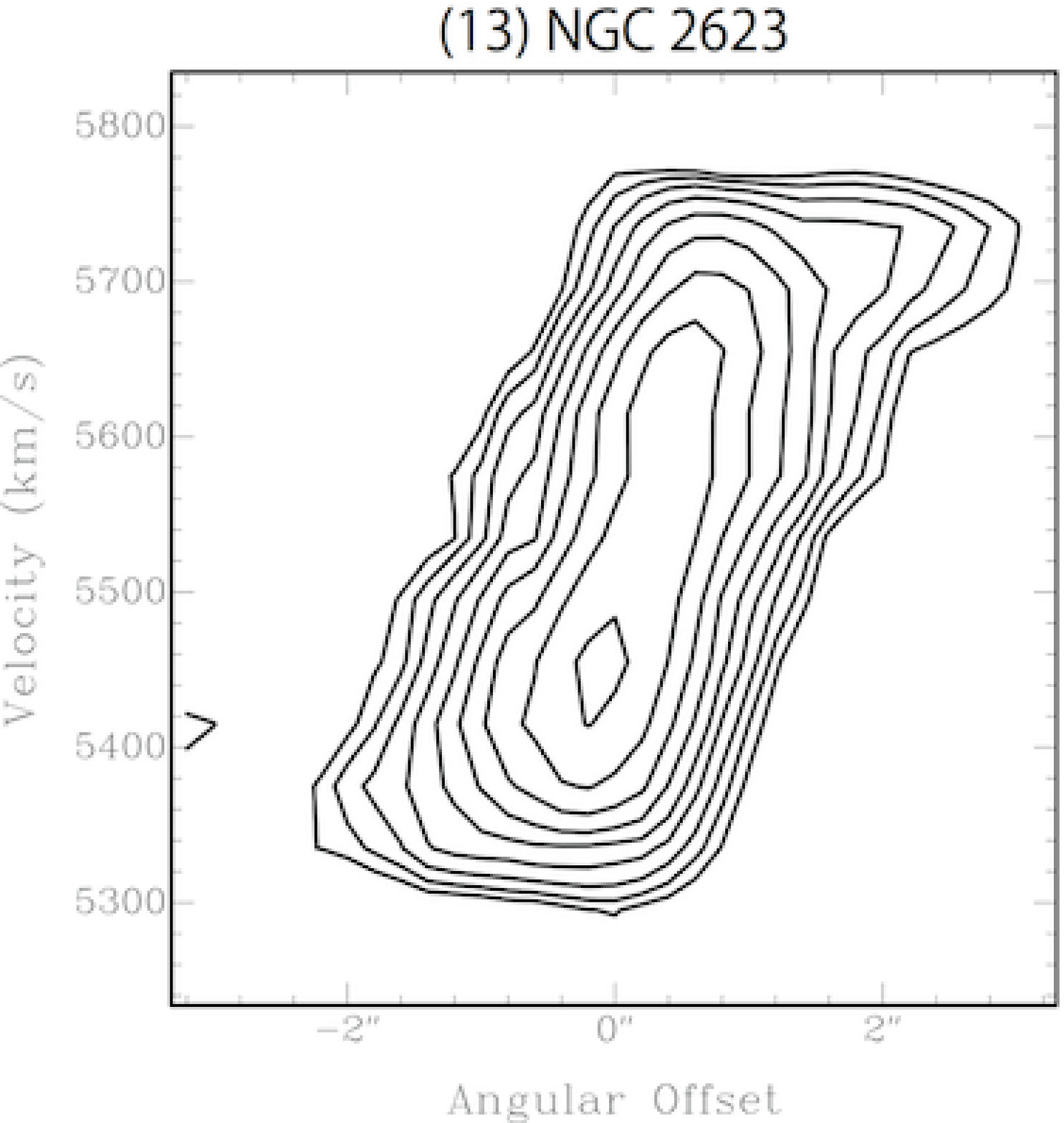}
  \end{center}
 \end{minipage}
 \begin{minipage}{0.24\hsize}
  \begin{center}
   \includegraphics[scale=0.22]{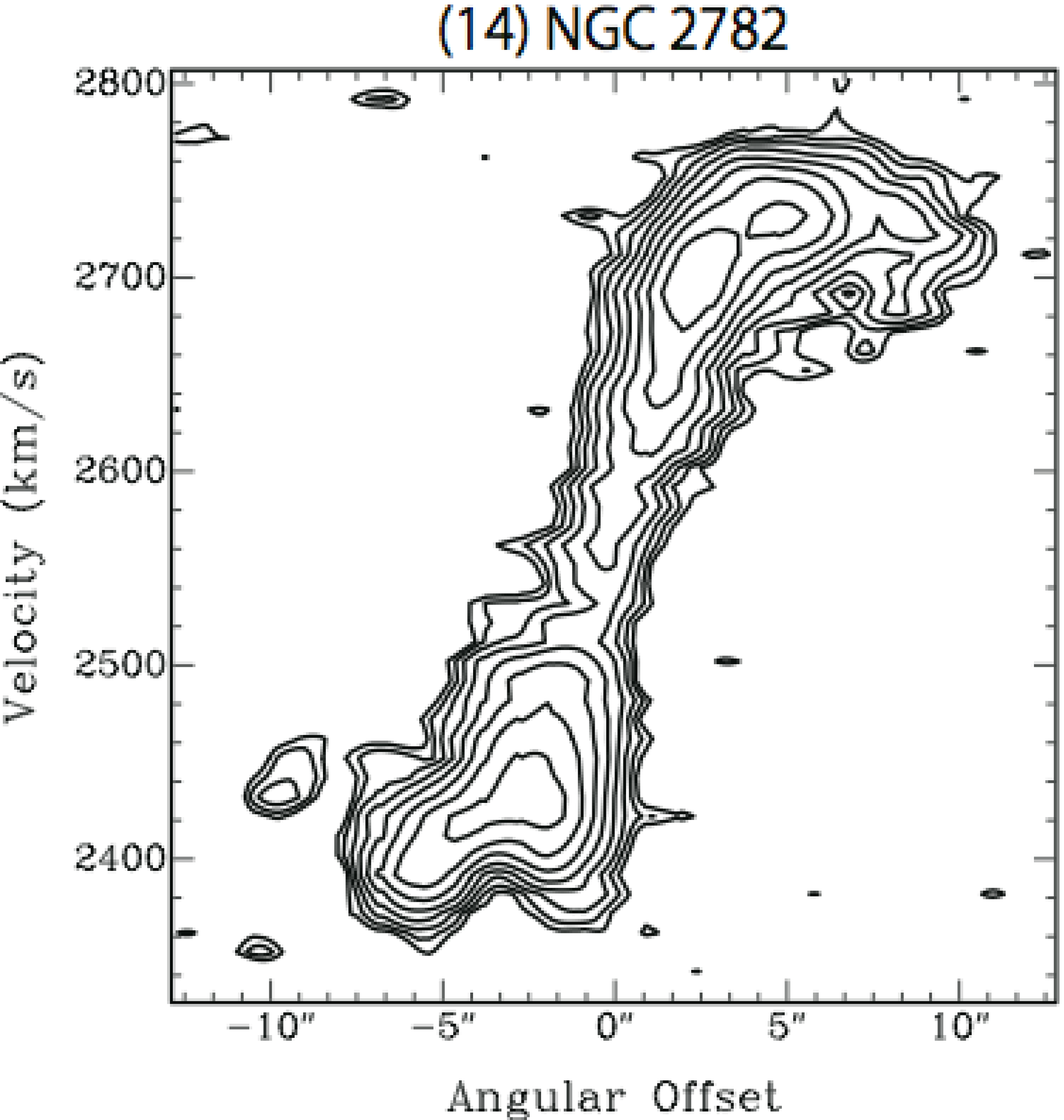}
  \end{center}
 \end{minipage}
 \begin{minipage}{0.24\hsize}
  \begin{center}
   \includegraphics[scale=0.22]{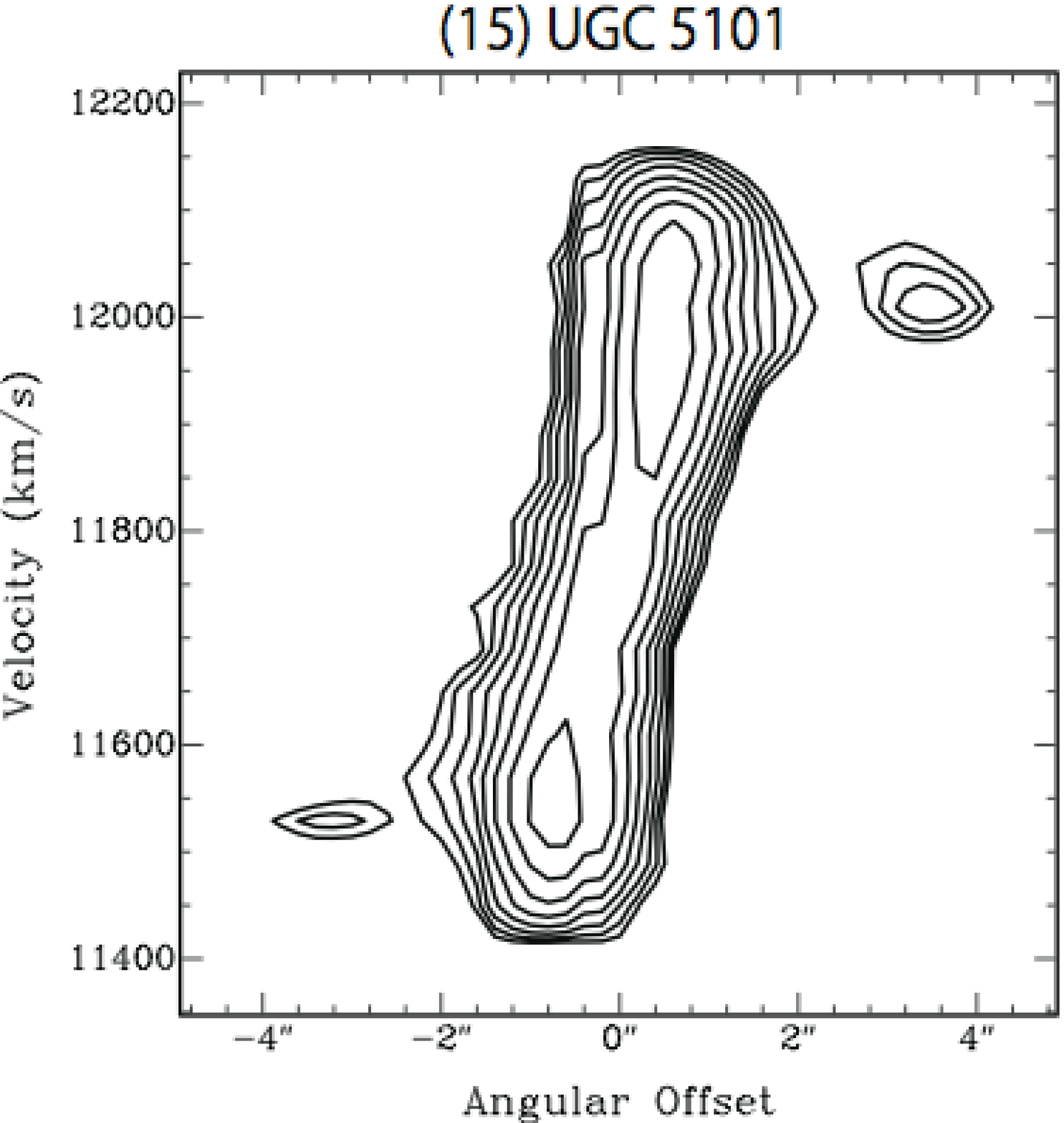}
  \end{center}
 \end{minipage}
  \begin{minipage}{0.24\hsize}
  \begin{center}
   \includegraphics[scale=0.22]{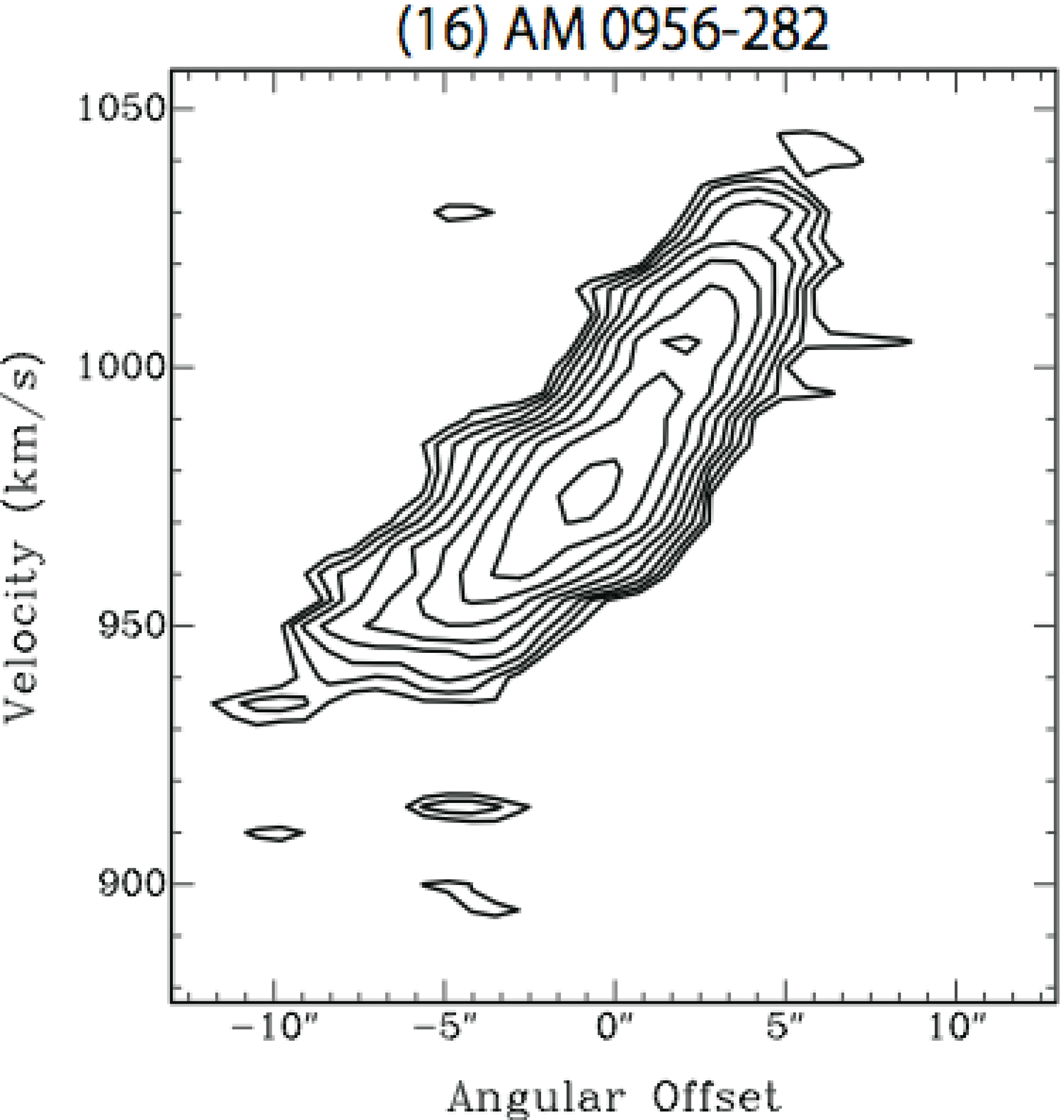}
  \end{center}
 \end{minipage}

 \begin{minipage}{0.24\hsize}
  \begin{center}
   \includegraphics[scale=0.22]{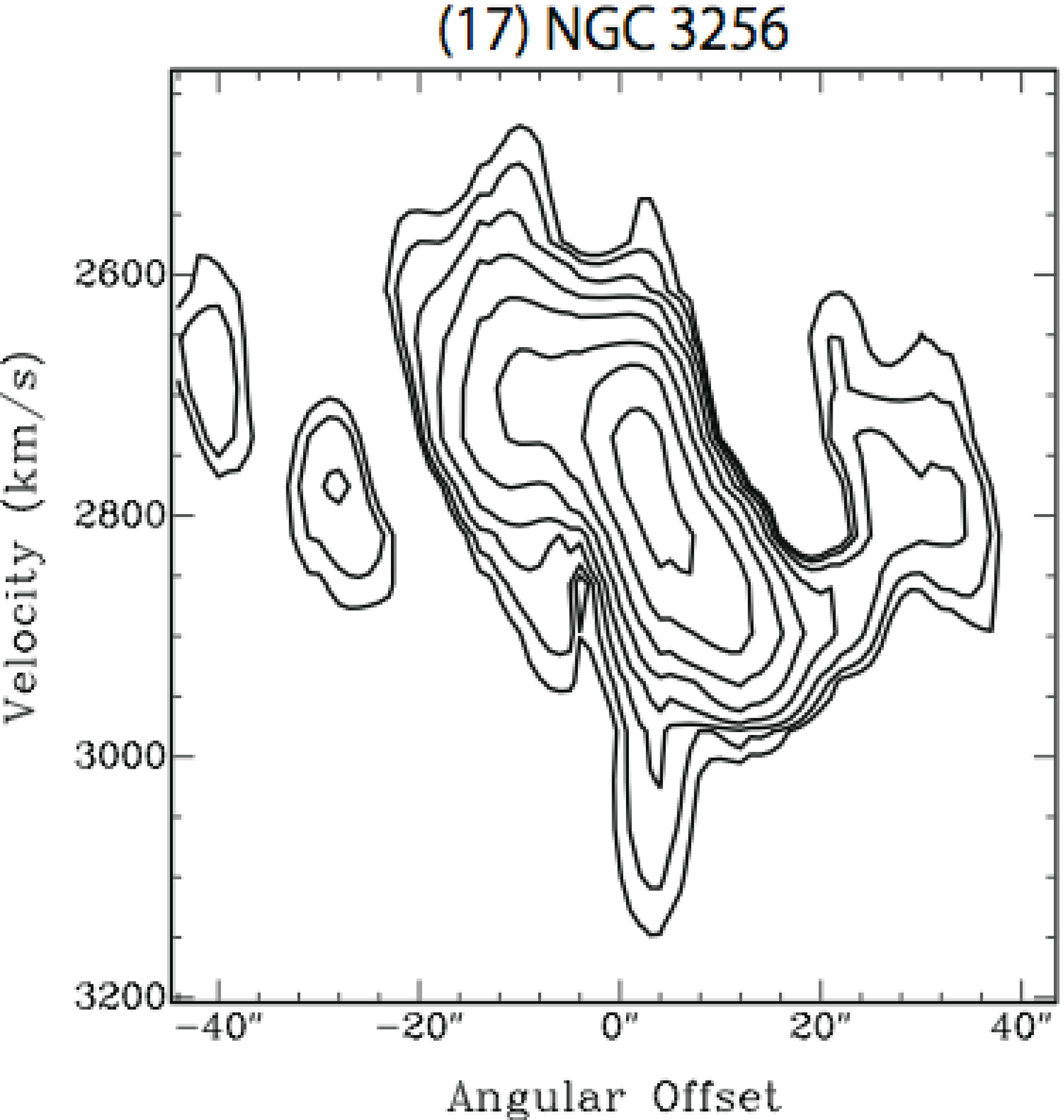}
  \end{center}
 \end{minipage}
 \begin{minipage}{0.24\hsize}
  \begin{center}
   \includegraphics[scale=0.22]{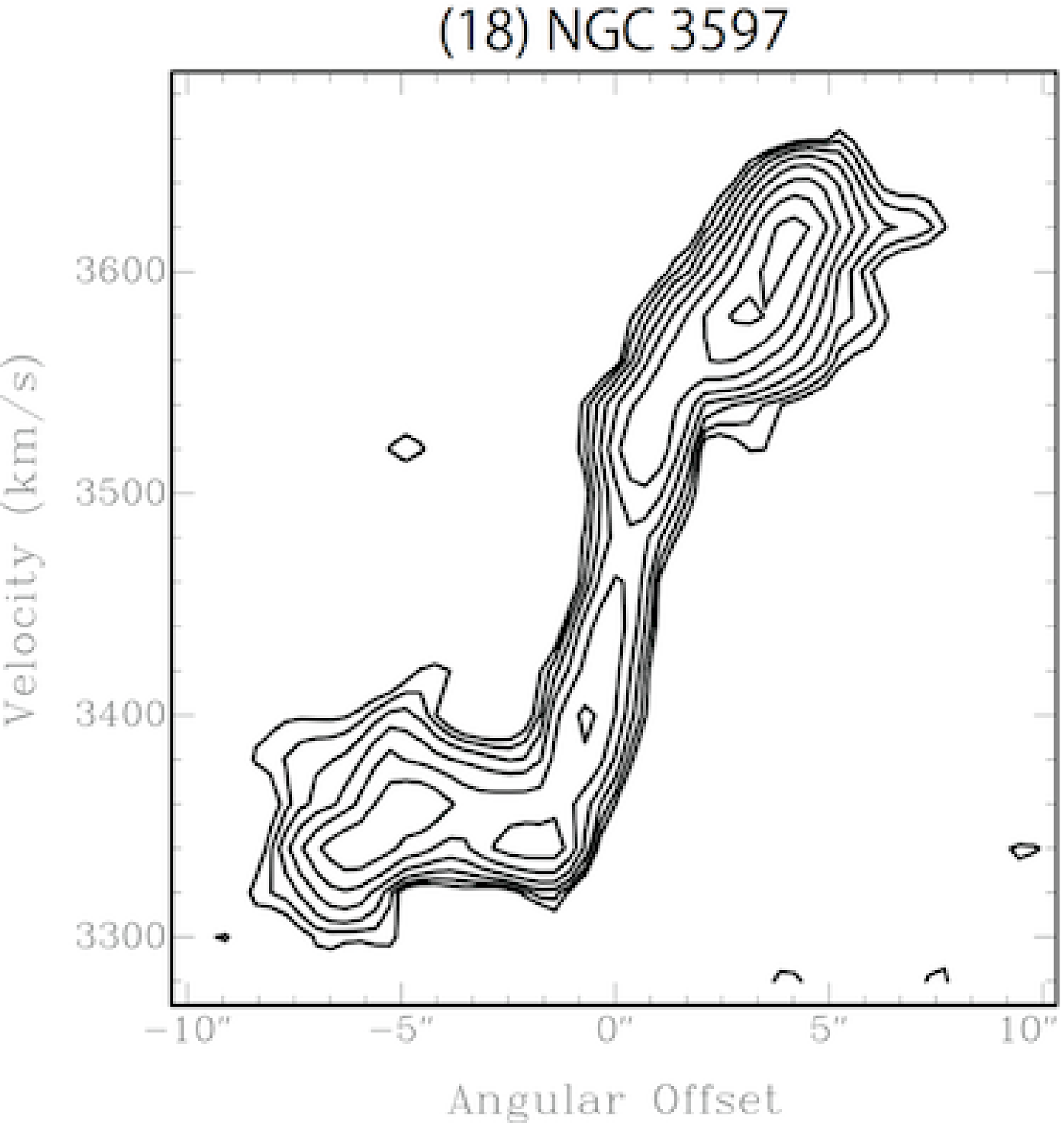}
  \end{center}
 \end{minipage}
 \begin{minipage}{0.24\hsize}
  \begin{center}
   \includegraphics[scale=0.22]{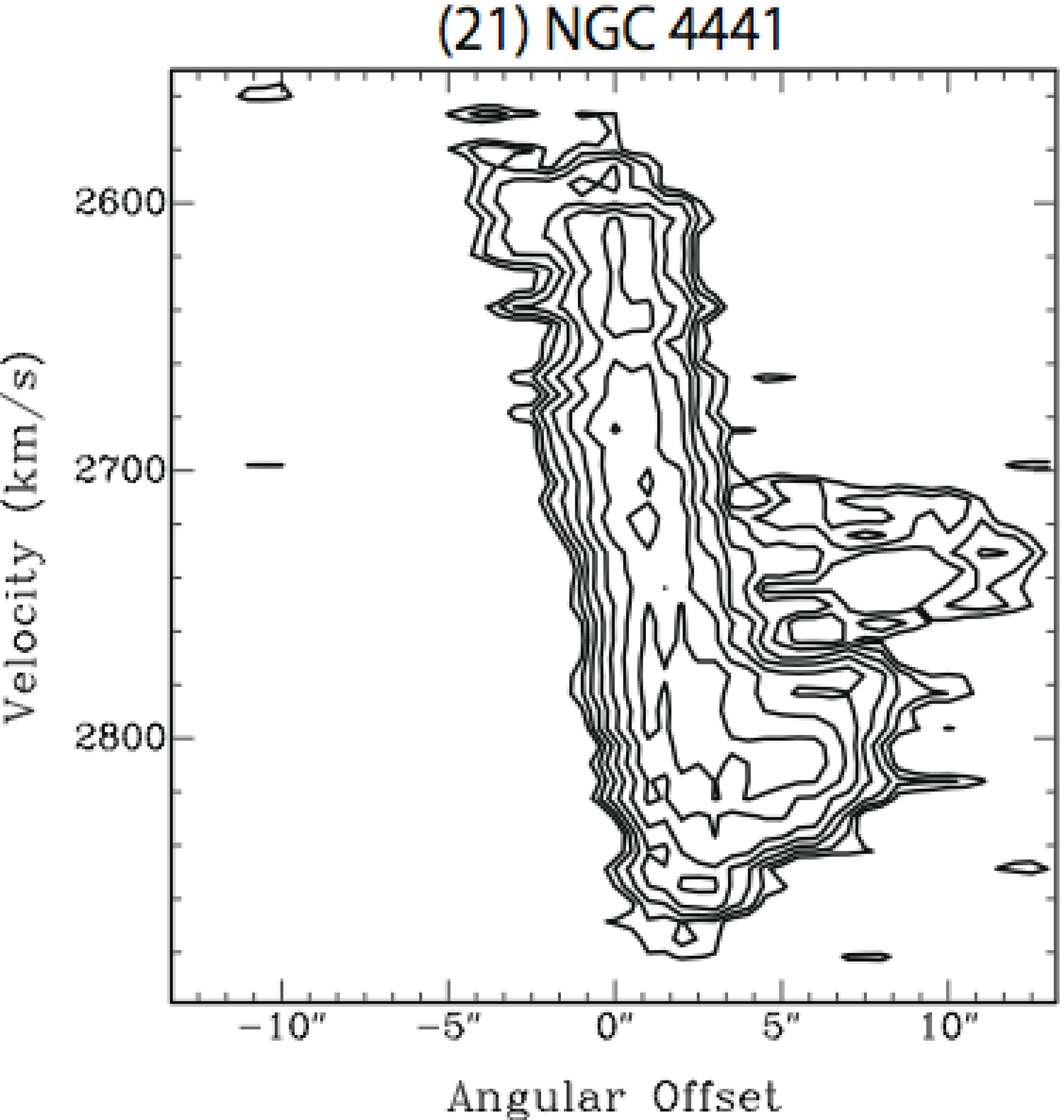}
  \end{center}
 \end{minipage}
  \begin{minipage}{0.24\hsize}
  \begin{center}
   \includegraphics[scale=0.22]{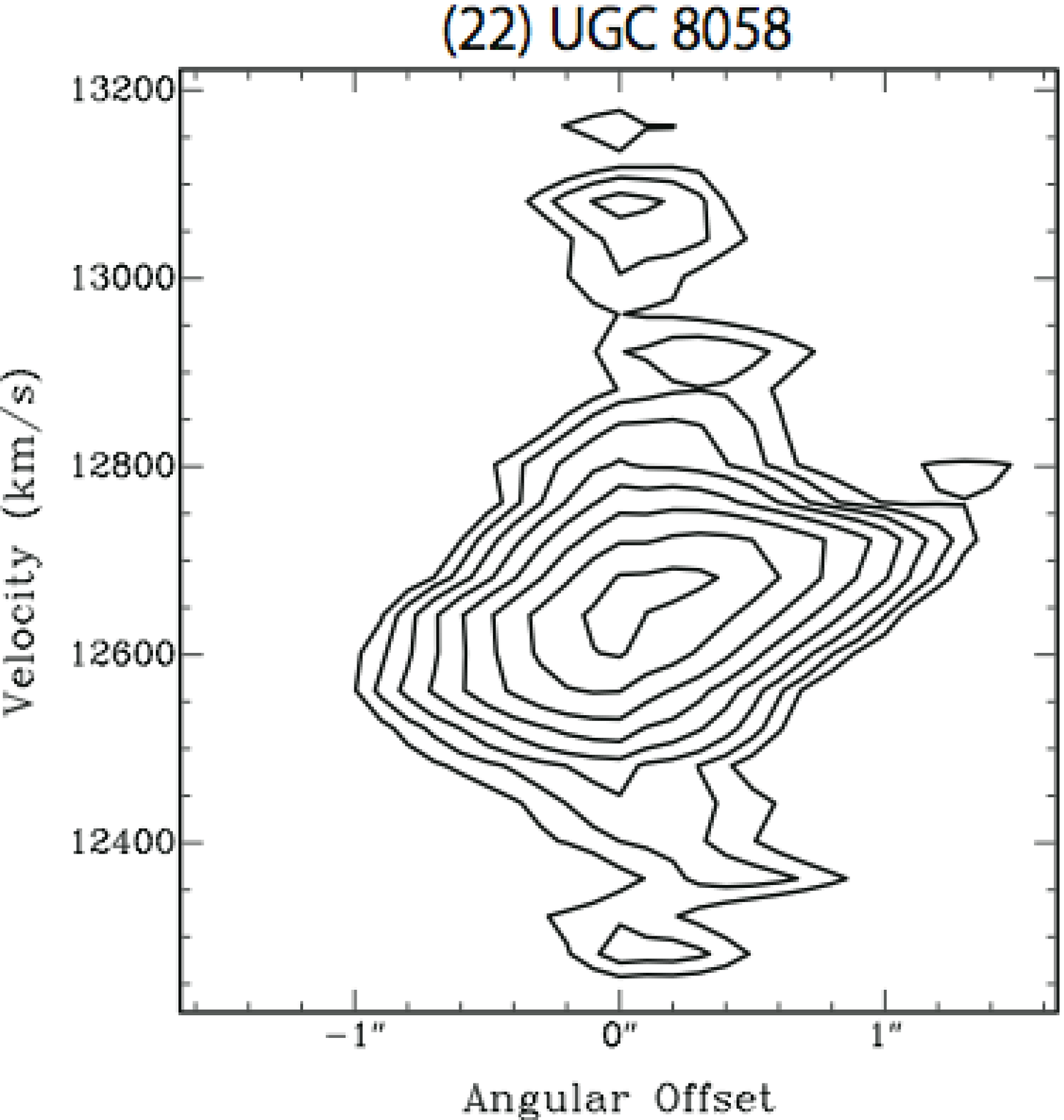}
  \end{center}
 \end{minipage}
 \caption{
 Position-Velocity diagrams along the kinematical major axes 
 of 24 galaxies with molecular gas disks.  
 The $n$th contours are defined by $cp^{n}$ Jy~beam$^{-1}$, 
 where $c$ is the 5~$\sigma$ and 2~$\sigma$ level 
 for NGC3256 and other sources, respectively.  
 The parameters are: ($c$, $p$) = 
 (7.44, 1.2), (27.6, 1.4), (11.7, 1.2), (43.6, 1.5), 
 (35.4, 1.5), (22.4, 1.9), (5.04, 1.3), (5.50, 1.3), 
 (22.4, 1.4), (3.88, 1.5), (23.8, 1.4), (11.4, 1.4), 
 (6.45, 2.0), (7.96, 1.4), (5.84, 1.4), (40.4, 1.5), 
 respectively, from top-left to bottom-right. 
 }
 \label{fig:f4}
\end{figure}

\addtocounter{figure}{-1}
\begin{figure}[htbp]
 \begin{minipage}{0.24\hsize}
  \begin{center}
   \includegraphics[scale=0.22]{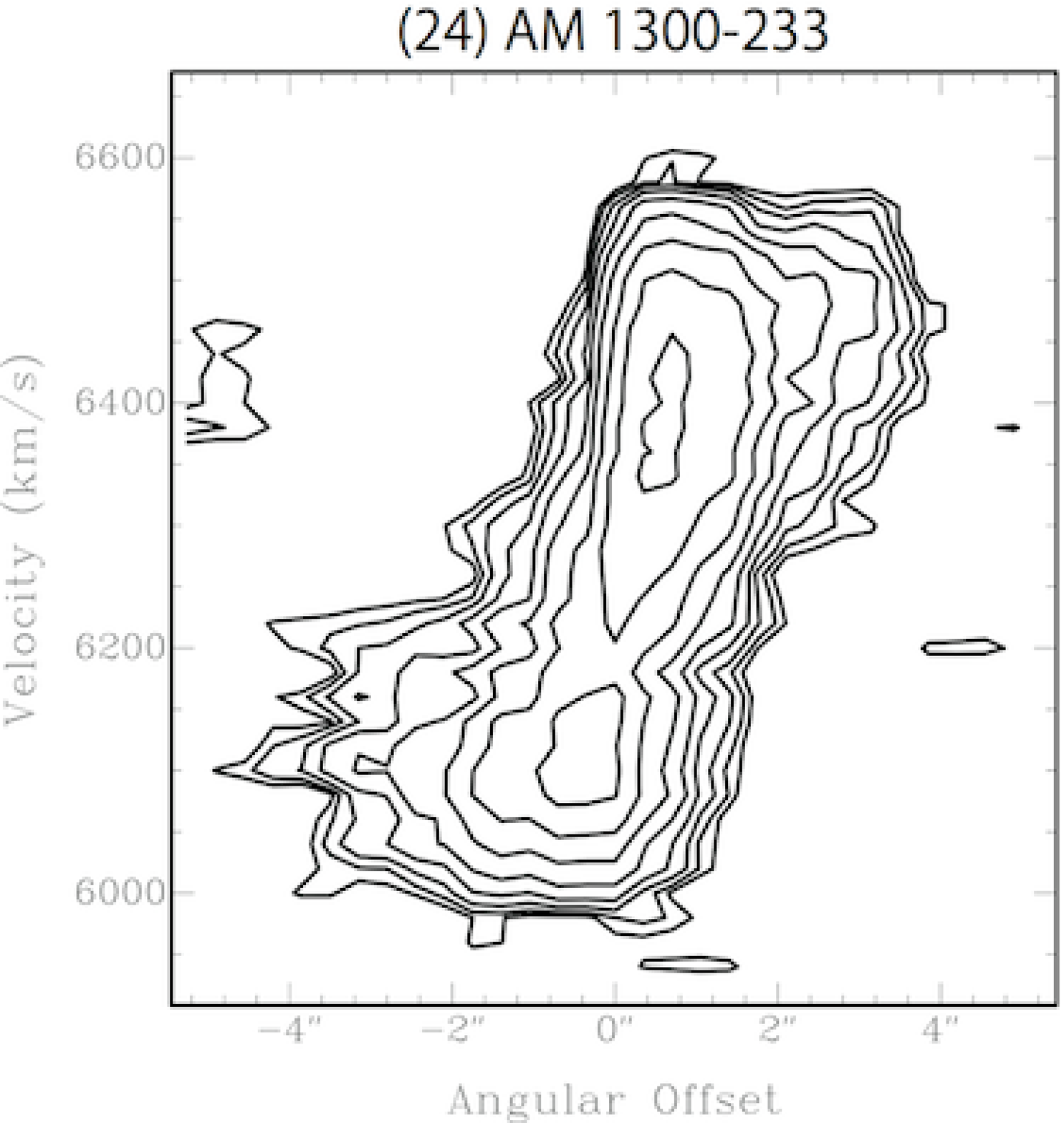}
  \end{center}
 \end{minipage}
 \begin{minipage}{0.24\hsize}
  \begin{center}
   \includegraphics[scale=0.22]{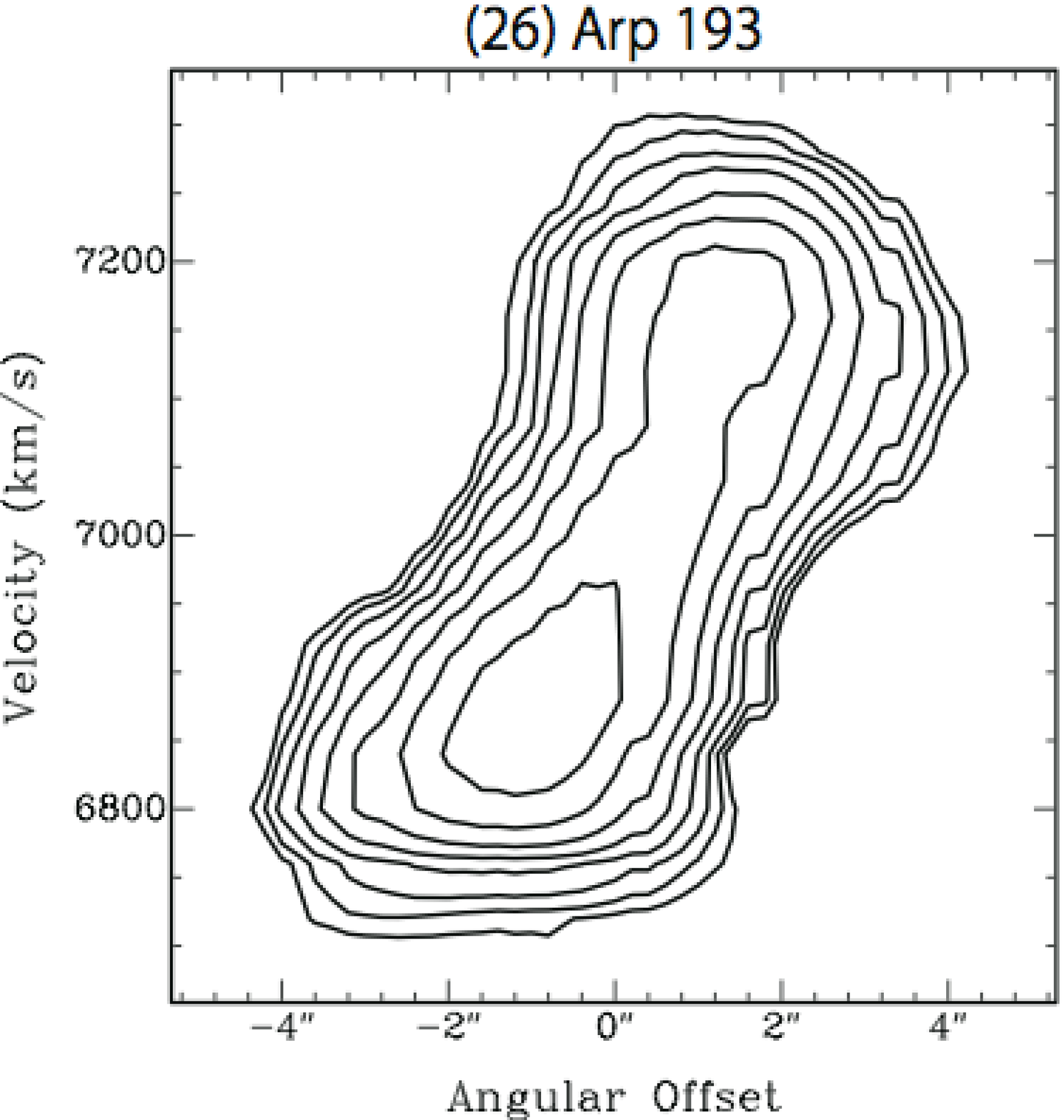}
  \end{center}
 \end{minipage}
 \begin{minipage}{0.24\hsize}
  \begin{center}
   \includegraphics[scale=0.22]{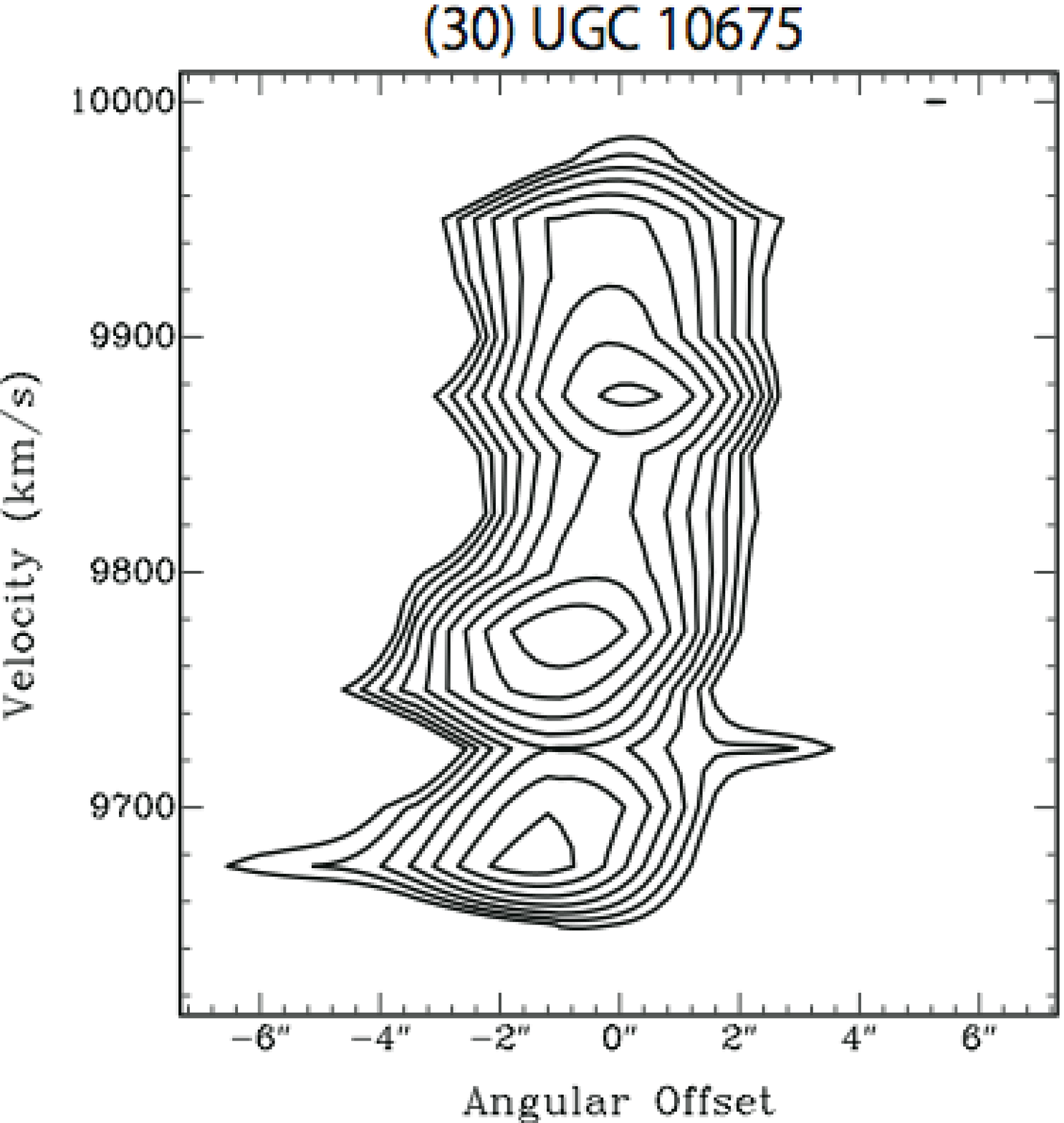}
  \end{center}
 \end{minipage}
  \begin{minipage}{0.24\hsize}
  \begin{center}
   \includegraphics[scale=0.22]{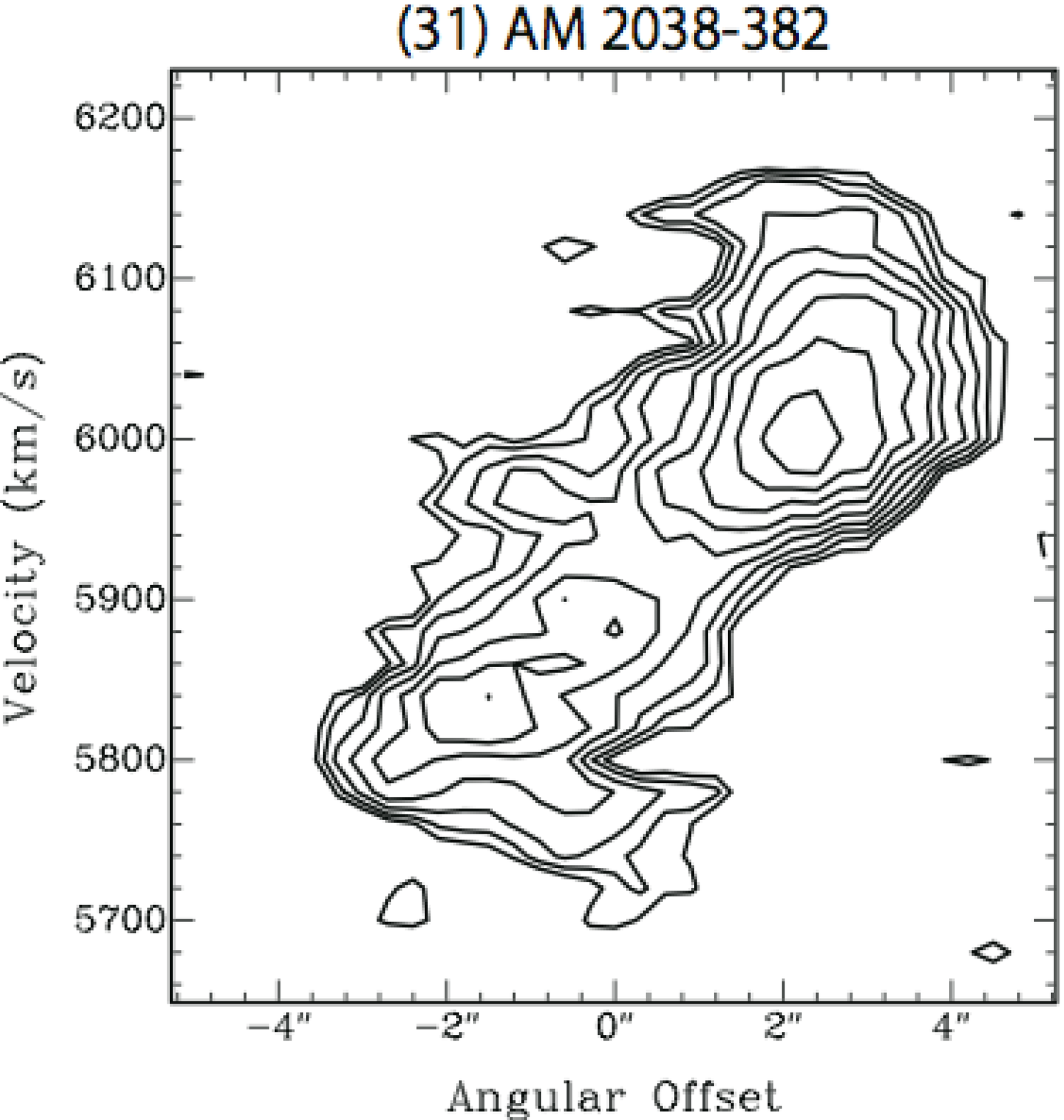}
  \end{center}
 \end{minipage}

 \begin{minipage}{0.24\hsize}
  \begin{center}
   \includegraphics[scale=0.22]{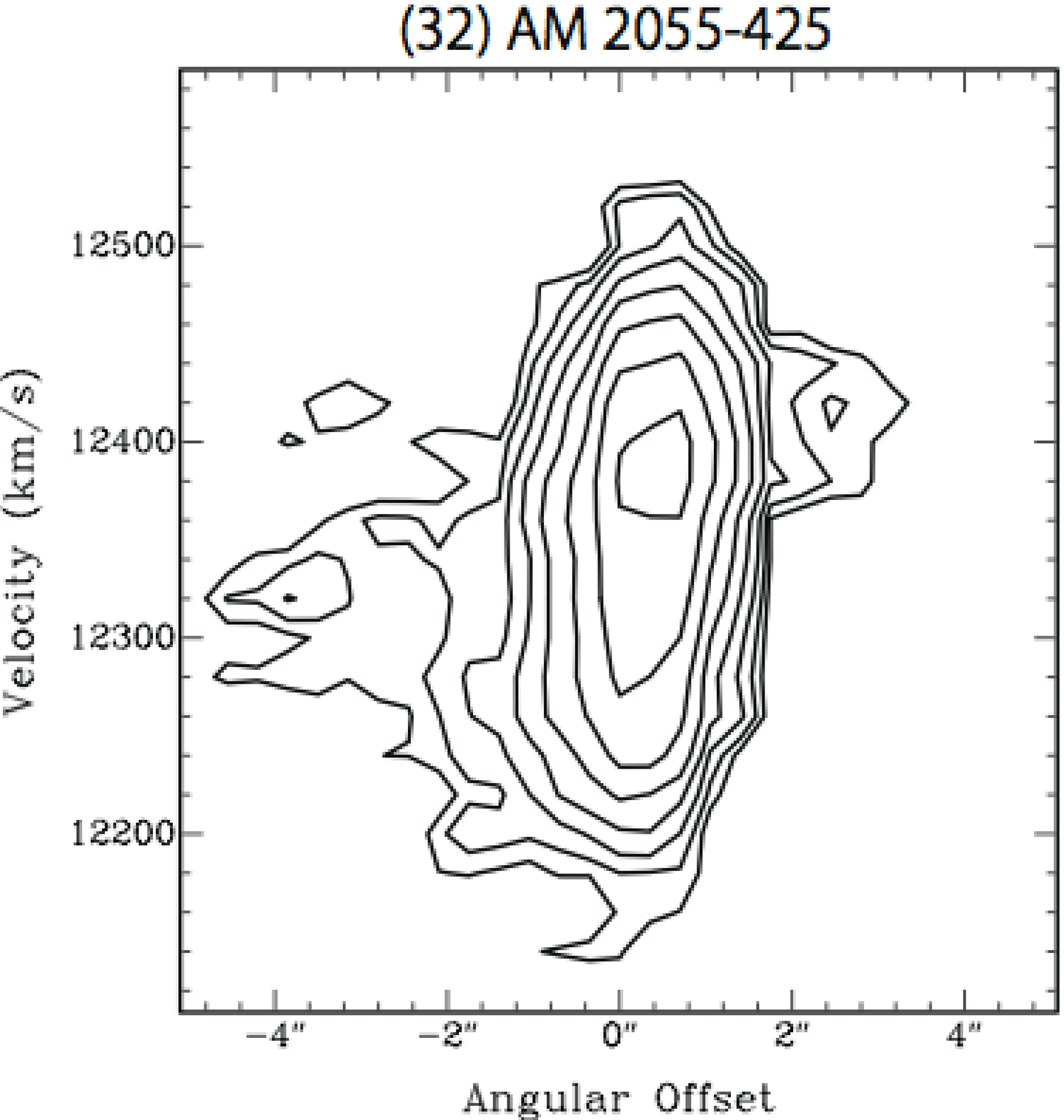}
  \end{center}
 \end{minipage}
 \begin{minipage}{0.24\hsize}
  \begin{center}
   \includegraphics[scale=0.22]{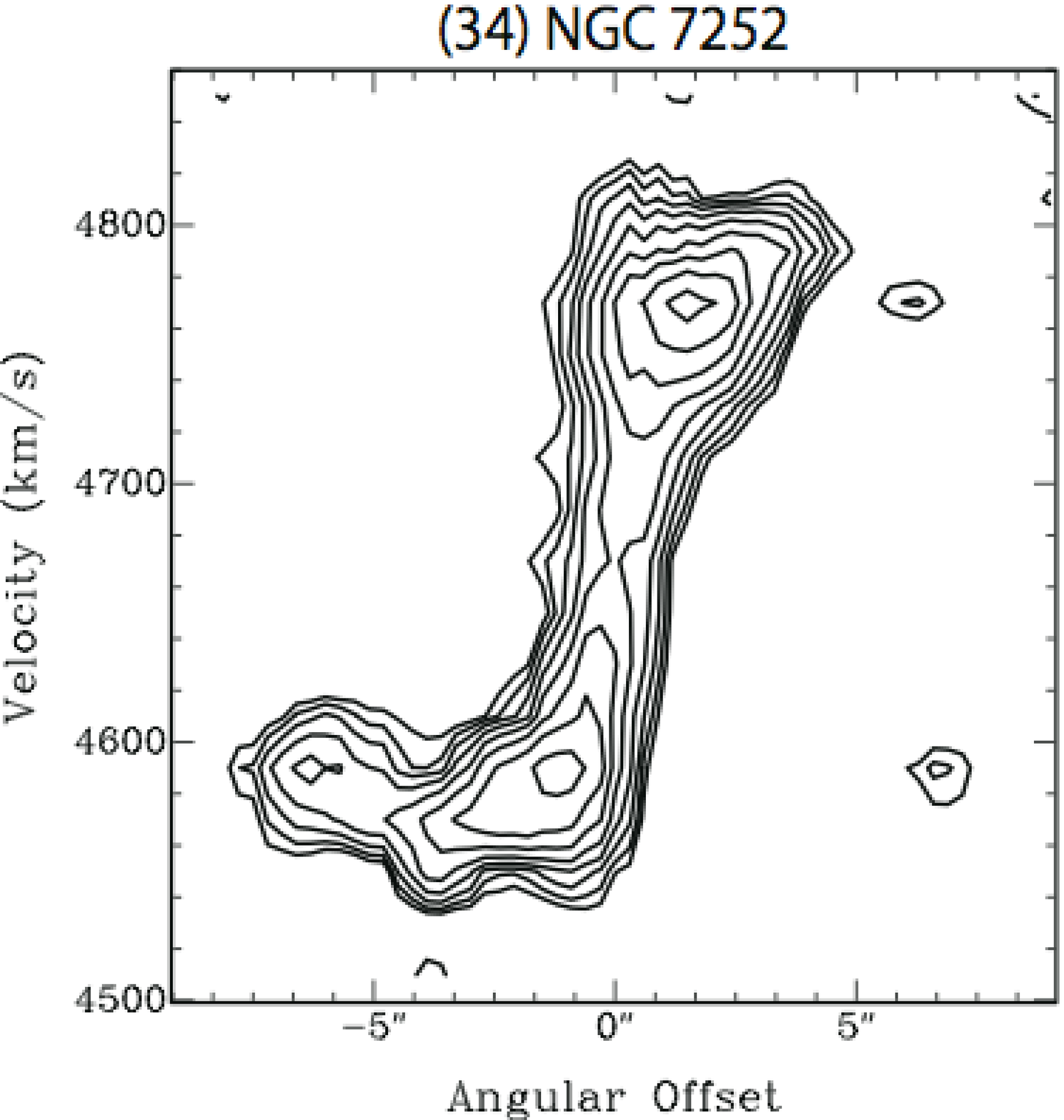}
  \end{center}
 \end{minipage}
 \begin{minipage}{0.24\hsize}
  \begin{center}
   \includegraphics[scale=0.22]{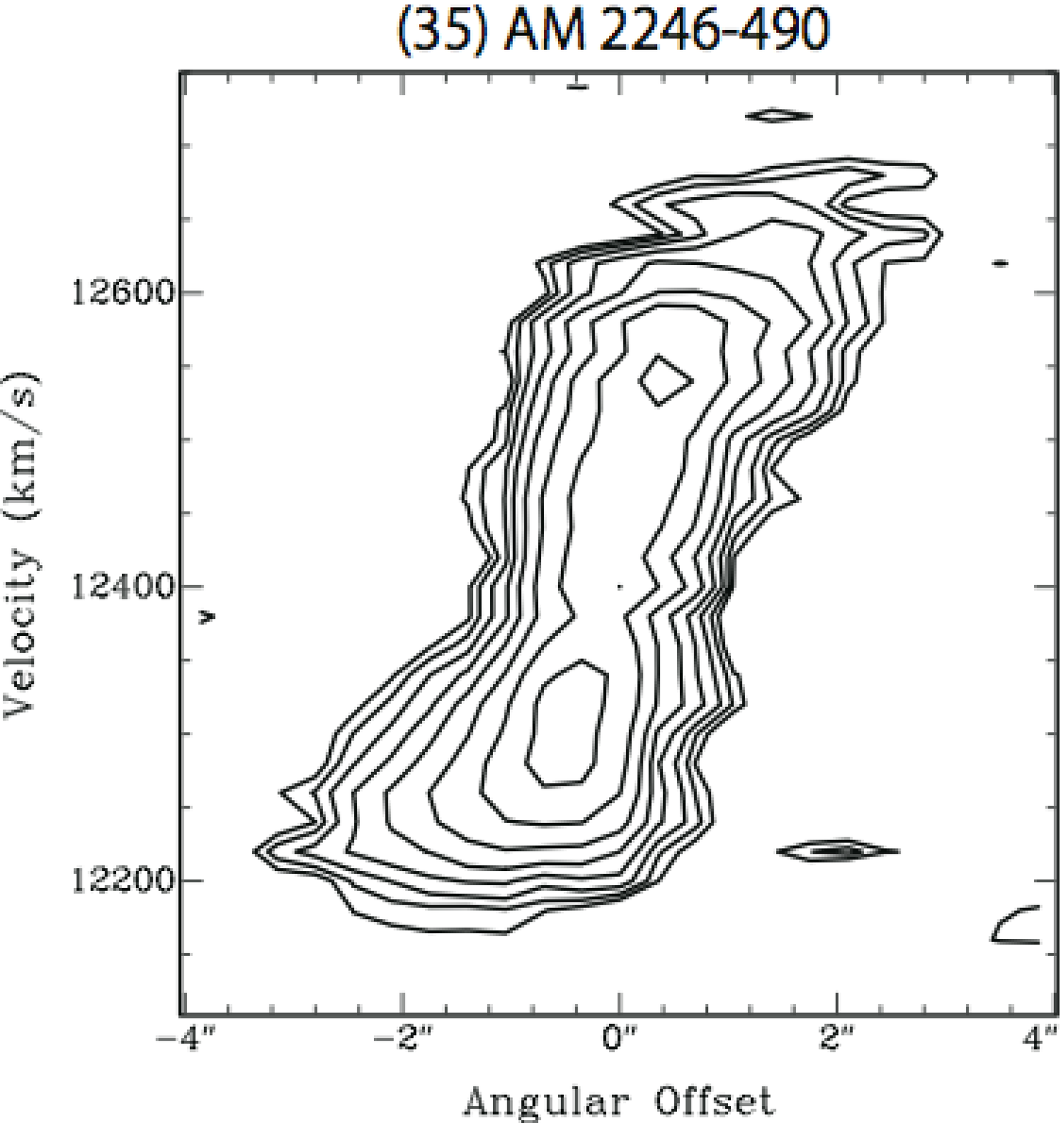}
  \end{center}
 \end{minipage}
  \begin{minipage}{0.24\hsize}
  \begin{center}
   \includegraphics[scale=0.22]{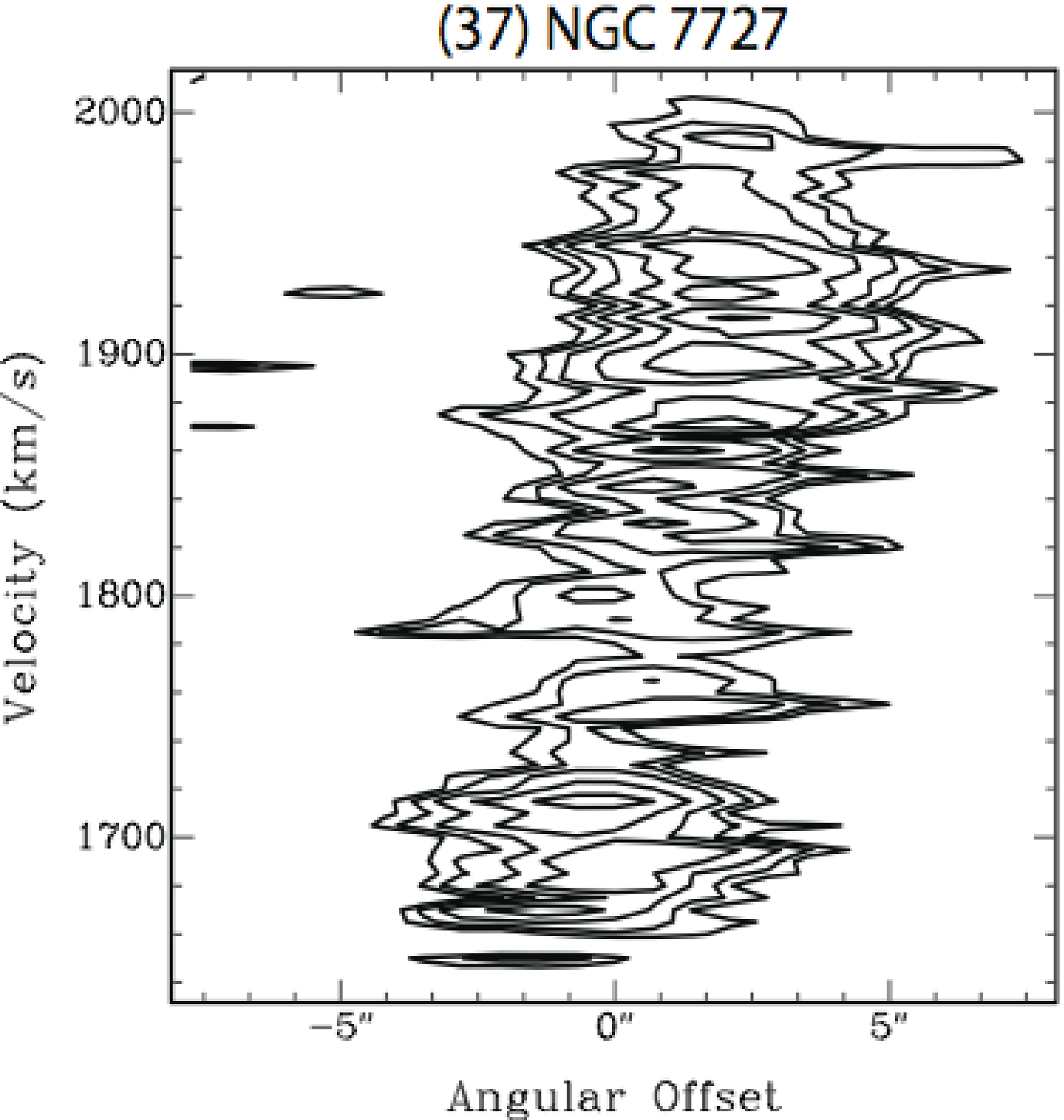}
  \end{center}
 \end{minipage}
 \caption{
 (Continued) The same as the previous figure.  
 The $n$th contours are defined by $cp^{n}$ Jy~beam$^{-1}$, 
 where $c$ is the 2~$\sigma$ level.  The parameters are: ($c$, $p$) = 
 (4.12, 1.5), (52.6, 1.6), (36.0, 1.2), (5.96, 1.3), 
 (4.68, 1.5), (6.72, 1.4), (4.48, 1.4), (8.72, 1.4), 
 respectively, from top-left to bottom-right.
 }
 \label{fig:f4}
\end{figure}

\begin{figure}[htbp]
 \begin{minipage}{0.5\hsize}
  \begin{center}
   \includegraphics{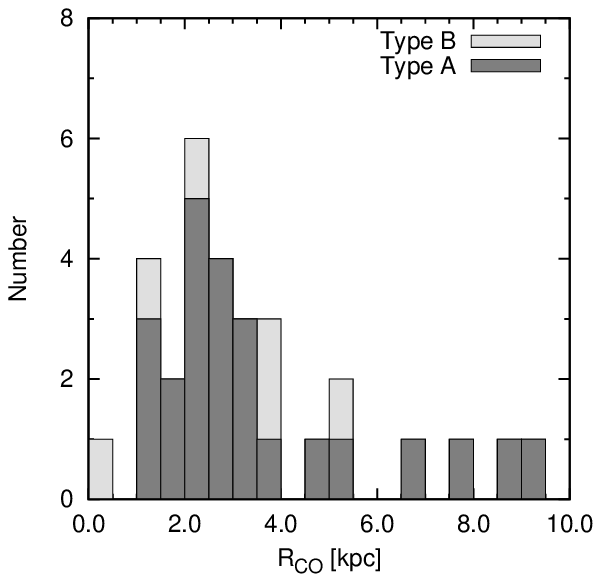}
  \end{center}
 \end{minipage}
 \begin{minipage}{0.5\hsize}
  \begin{center}
   \includegraphics{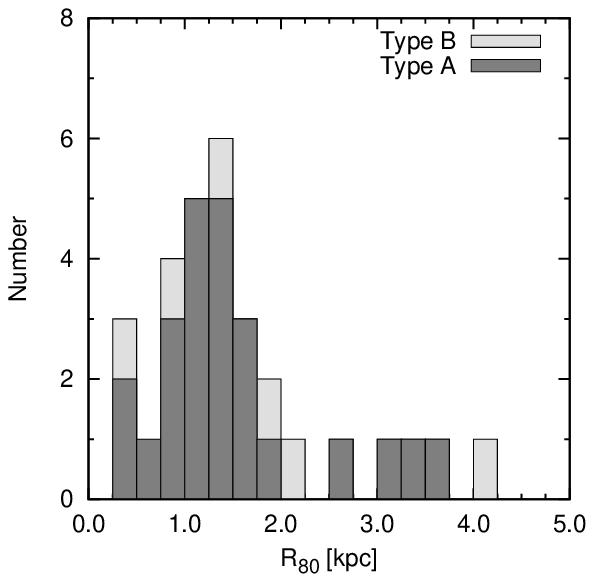}
  \end{center}
 \end{minipage}
 \caption{
 Histogram of the molecular gas extent.  
 $R_{\rm CO}$ is the maximum size of the molecular gas extent (left) and 
 $R_{80}$ is the radius which contains 80~\% of the total CO flux (right).  
 The dark-gray bars show the number of sources with the molecular gas disk 
 (Type~A).  The light-gray bars show the number of sources 
 in which the CO velocity field cannot be modeled by circular motion (Type~B).  
}
 \label{fig:f5}
\end{figure}

\begin{figure}[htbp]
 \begin{center}
  \includegraphics{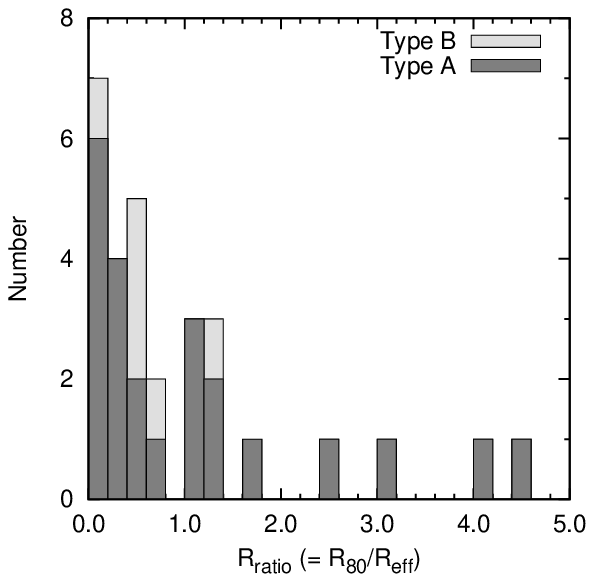}
 \end{center}
 \caption{
 Histogram of the relative size of the molecular gas extent 
 to the stellar component in 29 merger remnants.  
 $R_{80}$ is the radius which contains 80~\% of the total CO flux 
 and $R_{\rm eff}$ is the radius of the isophote 
 containing half of the total $K$-band luminosity.    
 The dark-gray and light-gray bars show the number of 
 Type~A and Type~B, respectively.
 }
 \label{fig:f6}
\end{figure}

\begin{figure}[htbp]
 \begin{center}
  \includegraphics{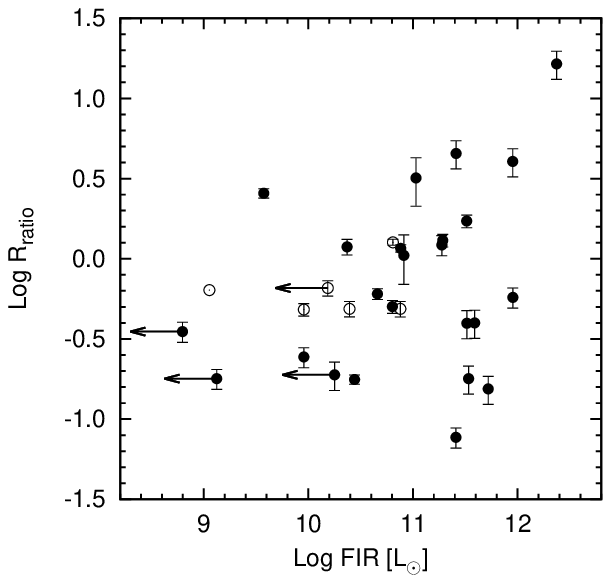}
 \end{center}
 \caption{
 Plot of the FIR luminosity vs. the relative size of the molecular gas extent 
 to the stellar component.  
 The filled circles show Type~A, and the open circles show Type~B.  
 }
 \label{fig:f7}
\end{figure}

\clearpage
\begin{deluxetable}{rlrrrrrrrrr}
\tabletypesize{\scriptsize}
\tablewidth{0pt}
\tablecaption{Merger Remnant Sample \label{tab:t1}}
\tablehead{
&&\multicolumn{1}{c}{R.A.}&\multicolumn{1}{c}{Decl.}&\multicolumn{1}{c}{V$_{\rm sys}$\footnotemark[1]}&\multicolumn{1}{c}{$D_{\rm L}$}&\multicolumn{1}{c}{Scale}&\multicolumn{1}{c}{log~$L_{\rm FIR}$}&\multicolumn{1}{c}{$M_{\rm  K}$\footnotemark[1]}&\multicolumn{1}{c}{$R_{\rm eff}$\footnotemark[2]}&\\
\multicolumn{1}{c}{ID}&\multicolumn{1}{c}{Name}&\multicolumn{1}{c}{(J2000)}&\multicolumn{1}{c}{(J2000)}&\multicolumn{1}{c}{[km s$^{-1}$]}&\multicolumn{1}{c}{[Mpc]}&\multicolumn{1}{c}{[pc/1\arcsec]}&\multicolumn{1}{c}{[$L_{\sun}$]}&\multicolumn{1}{c}{[mag]}&\multicolumn{1}{c}{[kpc]}&\multicolumn{1}{c}{S\'{e}rsic $n$\footnotemark[1]}\\
\multicolumn{1}{c}{(1)}&\multicolumn{1}{c}{(2)}&\multicolumn{1}{c}{(3)}&\multicolumn{1}{c}{(4)}&\multicolumn{1}{c}{(5)}&\multicolumn{1}{c}{(6)}&\multicolumn{1}{c}{(7)}&\multicolumn{1}{c}{(8)}&\multicolumn{1}{c}{(9)}&\multicolumn{1}{c}{(10)}&\multicolumn{1}{c}{(11)}
}
\startdata
  1& UGC~6 & 00 03 09 & 21 57 37 & 6582 & 91.5 & 425 & 10.91 & -24.01 & 0.51 & 10.00\\
  2& NGC~34 & 00 11 06 & -12 06 26 & 5931 & 82.6 & 385 & 11.41 & -24.61 & 0.38 & 10.00\\ 
  3& Arp~230  & 00 46 24 & -13 26 32 & 1742 & 23.9 & 115 & 9.57 & -21.75 & 0.69 & 1.56\\
  4& NGC~455 & 01 15 57 & 05 10 43 & 5269 & 73.3 & 343 & $<$~\hspace{1ex}9.71 & -24.64 & 5.96 & 6.21\\
  5& NGC~828 & 02 10 09 & 39 11 25 & 5374 & 74.6 & 349 & 11.29 & -25.36 & 2.67 & 2.99\\ 
  6& UGC~2238 & 02 46 17 & 13 05 44 & 6436 & 89.8 & 417 & 11.28 & -24.58 & 2.13 & 1.46\\
  7& NGC~1210 & 03 06 45 & -25 42 59 & 3928 & 54.4 & 257 & $<$~\hspace{1ex}9.15 & -23.72 & 2.90 &4.08 \\
  8& AM~0318-230 & 03 20 40 & -22 55 53 & 10699 & 150.7 & 681 & $<$~10.34 & -25.09 & 2.76 & 5.01\\
  9& NGC~1614 & 04 33 59 & -08 03 44 & 4778 & 66.1 & 311 & 11.51 & -24.74 & 3.44 & 10.00\\
10& Arp~187 & 05 04 53 & -10 14 51 & 12291 & 173.8 & 778 & 10.80 & -25.25 & 6.98 & 4.10\\
11& AM~0612-373 & 06 13 47 & -37 40 37 & 9734 & 136.5 & 621 & $<$~10.25 & -25.65 & 7.00 & 3.44\\
12& UGC~4079 & 07 55 06 & 55 42 13 & 6108 & 85.1 & 396 & 10.43 & -23.78 & 7.66 & 3.08\\
13& NGC~2623 & 08 38 24 & 25 45 17 & 5535 & 77.1 & 360 & 11.53 & -24.22 & 2.78 & 10.00\\
14& NGC~2782 & 09 14 05 & 40 06 49 & 2562 & 35.1 & 168 & 10.44 & -23.83 & 6.35 & 6.68\\
15& UGC~5101 & 09 35 51 & 61 21 11 & 11809 & 166.8 & 749 & 11.95 & -25.50 & 0.25 & 10.00\\
16& AM~0956-282 & 09 58 46 & -28 37 19 & 980 & 13.6 & 65 & $<$~\hspace{1ex}9.12 & -20.50 & 2.19 & 2.45\\
17& NGC~3256 & 10 27 51 & -43 54 13 & 2738 & 37.6 & 179 & 11.51 & -24.72 & 1.84 & 2.17\\
18& NGC~3597 & 11 14 42 & -23 43 40 & 3504 & 48.5 & 230 & 10.88 & -23.72 & 1.28 & 1.87\\
19& AM~1158-333 & 12 01 20 & -33 52 36 & 3027 & 41.8 & 199 & 9.96 & -22.61 & 1.59 & 3.75\\ 
20& NGC~4194 & 12 14 09 & 54 31 37 & 2506 & 34.7 & 166 & 10.81 & -23.21 & 1.01 & 4.59\\
21& NGC~4441 & 12 27 20 & 64 48 05 & 2674 & 36.8 & 175 & 9.96 & -22.98 & 3.69 & 6.83\\
22& UGC~8058 & 12 56 14 & 56 52 25 & 12642 & 178.6 & 797 & 12.37 & -27.55 & 0.05 & 10.00\\
23& AM~1255-430 & 12 58 08 & -43 19 47 & 9026 & 126.6 & 578 & $<$~10.18 & -24.93 & 3.01 & 1.87\\
24& AM~1300-233 & 13 02 52 & -23 55 18 & 6446 & 89.8 & 417 & 11.41 & -24.65 & 14.6 & 4.99\\
25& NGC~5018 & 13 13 01 & -19 31 05 & 2794 & 38.5 & 183 & 9.61 & -25.15 & 4.75 & 4.39\\ 
26& Arp~193 & 13 20 35 & 34 08 22 & 7000 & 97.5 & 451 & 11.59 & -24.40 & 2.98& 2.74\\
27& AM~1419-263 & 14 22 06 & -26 51 27 & 6709 & 93.6 & 434 & $<$~10.19 & -24.94 & 5.51 & 4.12\\
28& UGC~9829 & 15 23 01 & -01 20 50 & 8492 & 118.8 & 545 & 10.39 & -24.96 & 8.56 & 2.34\\ 
29& NGC~6052 & 16 05 13 & 20 32 32 & 4716 & 65.3 & 307 & 10.88 & -23.55 & 4.14 & 1.13\\
30& UGC~10675 & 17 03 15 & 31 27 29 & 10134 & 142.5 & 646 & 11.03 & -24.80 & 0.50 & 10.00\\
31& AM~2038-382 & 20 41 13 & -38 11 36 & 6057 & 84.3 & 393 & 10.37 & -24.70 & 0.97 & 10.00\\
32& AM~2055-425 & 20 58 26 & -42 39 00 & 12840 & 181.7 & 810 & 11.95 & -25.08 & 2.96 & 2.34\\
33& NGC~7135 & 21 49 46 & -34 52 35 & 2640 & 36.4 & 173 & 9.05 & -23.95 & 27.1 & 10.00\\
34& NGC~7252 & 22 20 44 & -24 40 42 & 4688 & 64.9 & 305 & 10.66 & -24.84 & 2.43 & 3.32\\
35& AM~2246-490 & 22 49 39 & -48 50 58 & 12884 & 182.1 & 812 & 11.72 & -25.52 & 8.32 & 10.00\\
36& NGC~7585 & 23 18 01 & -04 39 01 & 3447 & 47.7 & 226 & $<$~\hspace{1ex}9.34 & -24.98 & 3.55 & 3.53\\
37& NGC~7727 & 23 39 53 & -12 17 34 & 1855 & 25.6 & 123 & $<$~\hspace{1ex}8.80 & -24.23 & 2.40 & 3.41\\
\hline
\multicolumn{11}{l}{Additional Sources For Single-dish Observations}\\
\hline
38& NGC~2655 & 08 55 37 & 78 13 23 & 1404 & 19.4 & 93 & 9.86 & -23.70 & 1.43 & 3.01\\
39& Arp~156 & 10 42 38 & 77 29 41 & 10778 & 151.6 & 685 & 10.90 & -25.81 & 24.2 & 9.64 \\
40& NGC~3656 & 11 23 38 & 55 50 32 & 2869 & 39.7 & 189 & 10.06 & -23.70 & 5.61 & 4.47\\
41& NGC~3921 & 11 51 06 & 55 04 43 & 5838 & 81.3 & 379 & 10.20 & -25.13 & 1.92 & 5.03\\
42& UGC~11905 & 22 05 54 & 20 38 22 & 7522 & 105.1 & 485 & $<$~10.02 & -24.51 & 2.04 & 5.24 \\
43& IC~5298 & 23 16 00 & 25 33 24 & 8197 & 114.5 & 526 & 11.48 & -24.92 & 2.77 & 3.93
\enddata
\tablecomments{
Col.(1): ID number.  Col.(2): Source name.  Col.(3) \& Col.(4): Right ascension and declination.  
Col.(5): The systemic velocity.  Col.(6): The luminosity distance.  Col(7): The spatial scale.  
Col.(8): The far-infrared luminosity.  The luminosity is estimated using the $IRAS$ catalogs.  
Col.(9): The $K$-band magnitude.  Col.(10): The $K$-band effective radius.  
Col.(11): The S\'{e}rsic index.  The upper limit of the S\'{e}rsic fits is $n$ = 10, 
thus a fit resulting in $n$ = 10 may signify inaccurate fits.  
}
\tablenotetext{a}{Reference: RJ04}
\tablenotetext{b}{The effective radius from RJ04 is corrected to account for the differences 
between RJ04 and this work in the assumed distances to the source.} 
\end{deluxetable}

\begin{deluxetable}{rllcccccc}
\tabletypesize{\scriptsize}
\tablewidth{0pt}
\tablecaption{Properties of Interferometric CO Observations \label{tab:t2}}
\tablehead{
&&&&\multicolumn{1}{c}{Spectral Res.}&\multicolumn{1}{c}{rms}&\multicolumn{1}{c}{Norm. rms}&\multicolumn{1}{c}{Beam Size}&\\
\multicolumn{1}{c}{ID}&\multicolumn{1}{c}{Name}&\multicolumn{1}{c}{Transition}&\multicolumn{1}{c}{Telescope}&\multicolumn{1}{c}{[km s$^{-1}$]}&\multicolumn{1}{c}{[mJy Beam$^{-1}$]}&\multicolumn{1}{c}{[mJy Beam$^{-1}$]}&\multicolumn{1}{c}{[arcsec]}&\multicolumn{1}{c}{Date}\\
\multicolumn{1}{c}{(1)}&\multicolumn{1}{c}{(2)}&\multicolumn{1}{c}{(3)}&\multicolumn{1}{c}{(4)}&\multicolumn{1}{c}{(5)}&\multicolumn{1}{c}{(6)}&\multicolumn{1}{c}{(7)}&\multicolumn{1}{c}{(8)}&\multicolumn{1}{c}{(9)}
}
\startdata
 1& UGC~6 & CO~(1--0) & CARMA & 100 & 3.72 & 8.32 & 1.93 $\times$ 1.44 & Jan./Feb. 2012\\
 3& Arp~230 & CO~(1--0) & ALMA & 5 & 5.83 & 2.92 & 4.45 $\times$ 3.77 & Nov. 2011\\
 4& NGC~455 & CO~(1--0) & ALMA & 20 & 2.07 & 2.07 & 1.58 $\times$ 1.23 & July 2012\\
 5& NGC~828 & CO~(2--1) & SMA & 25 & 21.8 & 24.4 & 4.23 $\times$ 3.10 & Oct. 2011\\ 
 6& UGC~2238 & CO~(2--1) & SMA & 25 & 17.7 & 19.8 & 4.25 $\times$ 2.97 & Oct. 2011\\
 7& NGC~1210 & CO~(1--0) & ALMA & 20 & 2.14 & 2.14 & 1.41 $\times$ 1.11 & July 2012\\
 8& AM~0318-230 & CO~(1--0) & ALMA & 20 & 3.02 & 3.02 & 1.99 $\times$ 1.25 & April 2012\\
10& Arp~187 & CO~(1--0) & ALMA & 20 & 2.52 & 2.52 & 2.19 $\times$ 1.23 & July 2012\\\
11& AM~0612-373 & CO~(1--0) & ALMA & 20 & 2.75 & 2.75 & 1.93 $\times$ 1.51 & July 2012\\\
12& UGC~4079 & CO~(1--0) & CARMA & 100 & 2.05 & 5.23 & 1.74 $\times$ 1.53 & Jan./Feb. 2012 \\
16& AM~0956-282 & CO~(1--0) & ALMA & 5 & 5.71 & 2.86 & 4.21 $\times$ 2.80 & Nov. 2011\\
18& NGC~3597 & CO~(1--0) & ALMA & 20 & 3.98 & 3.98 & 1.56 $\times$ 1.17 & April 2012\\
19& AM~1158-333 & CO~(1--0) & ALMA & 20 & 4.59 & 4.59 & 1.70 $\times$ 1.14 & April 2012\\ 
23& AM~1255-430 & CO~(1--0) & ALMA & 20 & 1.78 & 1.78 & 1.95 $\times$ 1.17 & July 2012\\
24& AM~1300-233 & CO~(1--0) & ALMA & 20 & 2.06 & 2.06 &  2.24 $\times$ 1.07 & July 2012\\
25& NGC~5018 & CO~(1--0) & ALMA & 20 & 5.26 & 5.26 & 1.98 $\times$ 1.21 & April 2012\\ 
27& AM~1419-263 & CO~(1--0) & ALMA & 20 & 1.80 & 1.80 & 1.87 $\times$ 1.11 & July 2012\\ 
28& UGC~9829 & CO~(2--1) & SMA & 25 & 24.4 & 27.3 & 4.11 $\times$ 2.82 & June 2011\\ 
29& NGC~6052 & CO~(2--1) & SMA & 15 & 21.0 & 18.2 & 3.23 $\times$ 2.63 & June 2011\\ 
30& UGC~10675 & CO~(2--1) & SMA & 25 & 18.0 & 20.1 & 3.59 $\times$ 3.04 & June 2011\\
31& AM~2038-382 & CO~(1--0) & ALMA & 20 & 2.98 & 2.98 & 1.47 $\times$ 1.16 & May 2012\\
32& AM~2055-425 & CO~(1--0) & ALMA & 20 & 2.34 & 2.34 & 1.35 $\times$ 1.08 & May 2012\\
33& NGC~7135 & CO~(1--0) & ALMA & 20 & 3.98 & 3.98 & 1.70 $\times$ 1.13 & May 2012\\
34& NGC 7252 & CO~(1--0) & ALMA & 20 & 3.36 & 3.36 & 1.87 $\times$ 1.16 & May 2012\\
35& AM~2246-490 & CO~(1--0) & ALMA & 20 & 2.24 & 2.24 & 1.34 $\times$ 1.17 & May 2012\\
36& NGC~7585 & CO~(1--0) & ALMA & 20 & 4.46 & 4.46 & 2.26 $\times$ 1.21 & May 2012\\
37& NGC~7727 & CO~(1--0) & ALMA & 5 & 4.36 & 2.18 & 4.10 $\times$ 3.74 & Nov. 2011
\enddata
\tablecomments{
Col.(1): ID number.  Col.(2): Source name.  Col.(3): The CO transition.  
Col.(4): Telescope name.  Col.(5): The spectral resolution of the channel map.  
Col.(6): The noise level in the velocity resolution shown in Col. (5).  
Col.(7): The normalized noise level in the velocity resolution of 20~km~s$^{-1}$.  
Col.(8): The beam size.  Col(9): The observation data.
}
\end{deluxetable}

\begin{deluxetable}{rllcccccc}
\tabletypesize{\scriptsize}
\tablewidth{0pt}
\tablecaption{Properties of Archival Interferometric CO Data \label{tab:t3}}
\tablehead{
&&&&\multicolumn{1}{c}{Spectral Res.}&\multicolumn{1}{c}{rms}&\multicolumn{1}{c}{Norm. rms}&\multicolumn{1}{c}{Beam Size}&\\
\multicolumn{1}{c}{ID}&\multicolumn{1}{c}{Name}&\multicolumn{1}{c}{Transition}&\multicolumn{1}{c}{Telescope}&\multicolumn{1}{c}{[km s$^{-1}$]}&\multicolumn{1}{c}{[mJy Beam$^{-1}$]}&\multicolumn{1}{c}{[mJy Beam$^{-1}$]}&\multicolumn{1}{c}{[arcsec]}&Reference\\
\multicolumn{1}{c}{(1)}&\multicolumn{1}{c}{(2)}&\multicolumn{1}{c}{(3)}&\multicolumn{1}{c}{(4)}&\multicolumn{1}{c}{(5)}&\multicolumn{1}{c}{(6)}&\multicolumn{1}{c}{(7)}&\multicolumn{1}{c}{(8)}&\multicolumn{1}{c}{(9)}
 }
\startdata
  2& NGC~34 & CO~(2--1) & SMA & 25 & 13.8 & 15.4 & 4.13 $\times$ 3.11 & Archive (PI: Z. Wang)\\ 
  9& NGC~1614 & CO~(2--1) & SMA & 40 & 11.2 & 15.8 & 3.67 $\times$ 3.33 & \citet{Wilson08}\\
13& NGC~2623 & CO~(2--1) & SMA & 40 & 11.2 & 15.8 & 1.21 $\times$ 0.99 & \citet{Wilson08}\\
14& NGC~2782 & CO~(1--0) & PdBI & 10 & 1.94 & 1.37 & 2.09 $\times$ 1.51 & \citet{Hunt08}\\
15& UGC~5101 & CO~(2--1) & SMA & 40 & 11.9 & 16.8 & 1.20 $\times$ 0.96 & \citet{Wilson08} \\
17& NGC~3256 & CO~(1--0) & ALMA & 40.64 & 1.29 & 1.84 & 7.57 $\times$ 5.43 & Science Verification\\
20& NGC~4194 & CO~(2--1) & SMA & 20 & 6.88 & 6.88 & 1.40 $\times$ 1.30 & Archive (PI: S. Aalto)\\
21& NGC~4441 & CO~(1--0) & PdBI & 6.56 & 2.92 & 1.67 & 3.25 $\times$ 2.65 & \citet{Jutte10}\\
22& UGC~8058 & CO~(3--2) & SMA & 40 & 20.2 & 28.6 & 0.93 $\times$ 0.72 & \citet{Wilson08}\\
26& Arp~193 & CO~(3--2) & SMA & 40 & 26.3 & 37.2 & 2.22 $\times$ 1.98 & \citet{Wilson08} 
\enddata
\tablecomments{
Col.(1): ID number.  Col.(2): Source name.  Col.(3): The CO transition.  
Col.(4): Telescope name.  Col.(5): The spectral resolution of the channel map.  
Col.(6): The noise level in the velocity resolution shown in Col. (5).  
Col.(7): The normalized noise level in the velocity resolution of 20~km~s$^{-1}$.  
Col.(8): The beam size.  Col(9): Reference.
}
\end{deluxetable}

\begin{deluxetable}{rlcrrrr}
\tabletypesize{\scriptsize}
\tablewidth{0pt}
\tablecaption{The CO and 3~$\rm mm$ Continuum Properties \label{tab:t4}}
\tablehead{
&&&\multicolumn{1}{c}{$S_{\nu, \rm co}dv$}&\multicolumn{1}{c}{3~mm}&\multicolumn{1}{c}{Recovered flux}&\multicolumn{1}{c}{log $M_{\rm H_{2}, INT}$}\\
\multicolumn{1}{c}{ID}&\multicolumn{1}{c}{Name}&\multicolumn{1}{c}{Transition}&\multicolumn{1}{c}{[Jy km s$^{-1}$]}&\multicolumn{1}{c}{[mJy]}&\multicolumn{1}{c}{[\%]}&\multicolumn{1}{c}{[$M_{\sun}$]}\\
\multicolumn{1}{c}{(1)}&\multicolumn{1}{c}{(2)}&\multicolumn{1}{c}{(3)}&\multicolumn{1}{c}{(4)}&\multicolumn{1}{c}{(5)}&\multicolumn{1}{c}{(6)}&\multicolumn{1}{c}{(7)}
 }
\startdata
  1& UGC~6 & CO~(1--0) & 15$\pm$3 & \nodata & 32 (56\arcsec, 2) & 9.17\\
  2& NGC~34 & CO~(2--1) & 386$\pm$77 & \nodata & 122 (11\arcsec, 3) & 9.40\\
  3& Arp~230  & CO~(1--0) & 25$\pm$1 & 1.88$\pm$0.09 & 93 (15\arcsec, 1) & 8.23\\
  4& NGC~455 & CO~(1--0) & $<$~0.53 & $<$~0.16 & \nodata & $<$~7.51\\
  5& NGC~828 & CO~(2--1) & 1001$\pm$200 & \nodata & \nodata & 9.73\\
  6& UGC~2238 & CO~(2--1) & 492$\pm$98 & \nodata & \nodata & 9.58\\
  7& NGC~1210 & CO~(1--0) & $<$~0.50 & 0.54$\pm$0.03 & \nodata & $<$~7.23\\
  8& AM~0318-230 & CO~(1--0) & $<$~0.84 & $<$~0.26 & \nodata & $<$~8.33\\
  9& NGC~1614 & CO~(2--1) & 722$\pm$144 & \nodata & 59 (22\arcsec, 3) & 9.48\\
10& Arp~187 & CO~(1--0) & 63$\pm$3 & 24.3$\pm$1.2 & 116 (57\arcsec, 4) & 10.34\\
11& AM~0612-373 & CO~(1--0) & 24$\pm$1 & $<$0.27 & \nodata & 9.71\\
12& UGC~4079 & CO~(1--0) & $<$~1.12 & \nodata & \nodata & $<$ 7.97\\
13& NGC~2623 & CO~(2--1) & 256$\pm$51 & \nodata & \nodata & 9.16\\
14& NGC~2782 & CO~(1--0) & 124$\pm$25 & \nodata & 54 (45\arcsec, 5) & 9.24\\
15& UGC~5101 & CO~(2--1) & 246$\pm$49 & \nodata & 71 (23\arcsec, 6) & 9.81\\
16& AM~0956-282 & CO~(1--0) & 25$\pm$1 & 2.27$\pm$0.11 & \nodata & 7.73\\
17& NGC~3256 & CO~(1--0) & 1664$\pm$83 & 27.5$\pm$1.4 & 80 (44\arcsec, 7) & 9.66\\
18& NGC~3597 & CO~(1--0) & 141$\pm$7 & 5.78$\pm$0.29 & 76 (44\arcsec, 8) & 9.58\\
19& AM~1158-333 & CO~(1--0) & 3$\pm$1 & $<$~0.32 & \nodata & 7.71\\
20& NGC~4194 & CO~(2--1) & 84$\pm$17 & \nodata & 23 (12\arcsec, 3) & 8.77\\
21& NGC~4441 & CO~(1--0) & 43$\pm$9 & \nodata & 68 (15\arcsec, 1) & 8.83\\
22& UGC~8058 & CO~(3--2) & 392$\pm$78 & \nodata & 97 (23\arcsec, 9) & 9.72\\
23& AM~1255-430 & CO~(1--0) & 33$\pm$2 & $<$~0.11 & \nodata & 9.78\\
24& AM~1300-233 & CO~(1--0) & 96$\pm$5 & 8.40$\pm$0.42 & 43 (45\arcsec, 10) & 9.17\\
25& NGC~5018 & CO~(1--0) & $<$~1.31 & 0.72$\pm$0.04 & \nodata & $<$~7.36\\
26& Arp~193 & CO~(3--2) & 910$\pm$182 & \nodata & 89 (22\arcsec, 9) & 9.57\\
27& AM~1419-263 & CO~(1--0) & $<$~4.38 & 1.93$\pm$0.10 & \nodata & $<$~8.64\\
28& UGC~9829 & CO~(2--1) & 63$\pm$13 & \nodata & \nodata & 9.70\\
29& NGC~6052 & CO~(2--1) & 162$\pm$32 &\nodata & 58 (11\arcsec, 3) & 9.60\\
30& UGC~10675 & CO~(2--1) & 44$\pm$9 & \nodata & \nodata & 8.93\\
31& AM~2038-382 & CO~(1--0) & 33$\pm$2 & $<$~0.24 & 53 (45\arcsec, 11) & 9.43\\
32& AM~2055-425 & CO~(1--0) & 51$\pm$3 & 3.85$\pm$0.19 & 43 (46\arcsec, 10) & 9.50\\
33& NGC~7135 & CO~(1--0) & 4$\pm$1 & 7.03$\pm$0.35 & \nodata & 7.73\\
34& NGC~7252 & CO~(1--0) & 87$\pm$4 & 1.99$\pm$0.10 & 55 (45\arcsec, 12) & 9.63\\
35& AM~2246-490 & CO~(1--0) & 39$\pm$2 & 1.91$\pm$0.10 & 46 (46\arcsec, 10) & 9.38\\
36& NGC~7585 & CO~(1--0) & $<$~1.19 & $<$~0.28 & \nodata & $<$ 7.50\\
37& NGC~7727 & CO~(1--0) & 14$\pm$1 & 0.58$\pm$0.03 & \nodata & 8.02
\enddata
\tablecomments{
Col.(1): ID number.  Col.(2): Source name.  Col.(3): The CO transition.  
Col.(4): The CO integrated intensity.  Col.(5): The peak flux of  the 3~mm continuum emission.  
Col.(6): The recovered flux derived by comparing the CO flux in the same circular region 
using the interferometric and single-dish measurements.  The diameter of the circular region 
and reference of the single-dish measurements are noted in brackets.  Reference;  
1: this work, 2: \citet{Maiolino97}, 3: \citet{Albrecht07}, 4: \citet{Evans05}, 5:\citet{Young95}, 
6: \citet{Papadopoulos12b}, 7: \citet{Casoli92}, 8: \citet{Wiklind95}, 9: \citet{Mao10}, 
10: \citet{Mirabel90}, 11: \citet{Horellou97}, 12: \citet{Andreani95}
Col.(7): The molecular gas mass estimated using the interferometric map.  
}
\end{deluxetable}

\begin{deluxetable}{rllcrrrrc}
\tabletypesize{\scriptsize}
\tablewidth{0pt}
\tablecaption{Properties of Single-dish Observations and Archival Data \label{tab:t5}}
\tablehead{
&&&\multicolumn{1}{c}{Beam Size}&\multicolumn{1}{c}{rms}&\multicolumn{1}{c}{$T_{\rm mb}$}&\multicolumn{1}{c}{$I_{\rm CO~(1-0)}$}&\multicolumn{1}{c}{log $M_{\rm H_{2}, SD}$}&\\
\multicolumn{1}{c}{ID}&\multicolumn{1}{c}{Name}&\multicolumn{1}{c}{Telescope}&\multicolumn{1}{c}{[arcse]}&\multicolumn{1}{c}{[mK]}&\multicolumn{1}{c}{[mK]}&\multicolumn{1}{c}{[K km s$^{-1}$]}&\multicolumn{1}{c}{[$M_{\sun}$]}&\multicolumn{1}{c}{Ref.}\\
\multicolumn{1}{c}{(1)}&\multicolumn{1}{c}{(2)}&\multicolumn{1}{c}{(3)}&\multicolumn{1}{c}{(4)}&\multicolumn{1}{c}{(5)}&\multicolumn{1}{c}{(6)}&\multicolumn{1}{c}{(7)}&\multicolumn{1}{c}{(8)}&\multicolumn{1}{c}{(9)}
}
\startdata
  1& UGC~6 & NRAO 12~m & 56 & \nodata &  \nodata & 1.6$\pm$0.2 & 9.66 & 2\\ 
  2& NGC~34 & NRAO 12~m & 56 & \nodata &  \nodata & 3.9$\pm$0.3 & 9.19 & 2\\
  3& Arp~230 & NRO 45~m & 15 & 4.2 (20) & 29.5$\pm$10.6 & 4.1$\pm$0.8 & 7.82 & 1\\ 
  5& NGC~828 & IRAM 30~m & 22 & \nodata &  \nodata & 71.1$\pm$0.7 & 9.58 & 3\\ 
  6& UGC~2238 & NRAO 12~m & 56 & \nodata &  \nodata & 6.0$\pm$1.2: & 9.45 & 4\\ 
  9& NGC~1614 & FCRAO 14~m & 46 & \nodata &  \nodata & 5.8$\pm$0.8 & 9.05 & 5\\
10& Arp~187 &  NRAO 12~m & 57 & \nodata &  \nodata & 1.8$\pm$0.1 & 10.27 & 6\\
12& UGC~4079 & NRO 45~m & 15 & 1.2 (30) & 19.7$\pm$3.1 & 4.0$\pm$0.8 & 8.90 & 1\\ 
13& NGC~2623 & NRO 45~m & 15 & 5.5 (30) & 126.2$\pm$13.7 & 37.4$\pm$7.5 & 9.01 & 1\\
14& NGC~2782 & NRO 45~m & 15 & 7.6 (30) & 121.1$\pm$19.0 & 23.5$\pm$4.7 & 8.91 & 1\\
15& UGC~5101 & NRO 45~m & 16 & 2.8 (30) & 37.5$\pm$7.1 & 17.4$\pm$3.5 & 9.34 & 1 \\
17& NGC~3256 & SEST 15~m & 44 & \nodata &  \nodata & 71$\pm$2 & 9.72 & 7\\
18& NGC~3597 & SEST 15~m & 44 & \nodata &  \nodata & 6.9$\pm$1.4 & 9.71 & 8\\ 
20& NGC~4194 & NRO 45~m & 15 & 7.7 (30) & 262.6$\pm$19.2 & 47.7$\pm$9.5 & 9.21 & 1\\
21& NGC~4441 & NRO 45~m & 15 & 8.8 (30) & 131.6$\pm$21.9 & 22.0$\pm$4.4 & 8.92 & 1\\
22& UGC~8058 & NRO 45~m & 16 & 3.1 (30) & 60.6$\pm$7.7 & 13.0$\pm$2.6 & 9.27 & 1\\
24& AM~1300-233 & SEST 15~m & 43 & \nodata &  \nodata & 8.2$\pm$1.6: & 9.53 & 9\\
25& NGC~5018 & SEST 15~m & 44 & \nodata &  \nodata & $<$~1.4$\pm$0.3: & $<$~8.80 & 10\\
26& Arp~193 & NRAO 12~m & 56 & \nodata &  \nodata & 7.0$\pm$0.3 & 9.59 & 6\\ 
28& UGC~9829 &NRO 45~m & 15 & 1.3 (20) & 46.1$\pm$3.2 & 11.0$\pm$2.2 & 9.63 & 1\\
29& NGC~6052 & NRO 45~m & 15 & 2.9 (10) & 160.2$\pm$7.1 & 16.9$\pm$3.4 & 9.30 & 1\\
30& UGC~10675 & IRAM 30~m & 23 & \nodata & \nodata & 3.7$\pm$0.2 & 8.99 & 11\\
31& AM~2038-382 & SEST 15~m & 43 & \nodata &  \nodata & 2.3$\pm$0.4 & 9.71 & 12\\
32& AM~2055-425 & SEST 15~m & 42 & \nodata &  \nodata & 4.4$\pm$0.9: & 9.87 & 9\\
34& NGC~7252 & SEST 15~m & 43 & \nodata &  \nodata & 5.8$\pm$0.5 & 9.88 & 13\\
35& AM~2246-490 & SEST 15~m & 42 & \nodata &  \nodata & 3.1$\pm$0.6: & 9.72 & 9\\
38& NGC~2655 & NRO 45~m & 15 & 6.3 (30) & 53.7$\pm$15.8 & 12.3$\pm$2.5 & 8.11 & 1\\
39& Arp~156 & NRO 45~m & 15 & 2.9 (30) & 23.5$\pm$7.1 & 10.2$\pm$2.0 & 9.80 & 1\\
40& NGC~3656 & NRO 45~m & 15 & 12.5 (30) & 165.5$\pm$31.2 & 51.6$\pm$10.3 & 9.36 & 1\\
41& NGC~3921 & NRO 45~m & 15 & 1.0 (40) & 25.7$\pm$2.6 & 7.7$\pm$1.5 & 9.30 & 1\\
42& UGC~11905 & NRO 45~m & 15 & 3.8 (30) & 21.7$\pm$9.5 & 10.1$\pm$2.0 &9.49 & 1\\
43& IC~5298 & NRO 45~m & 15 & 5.4 (30) & 48.0$\pm$13.5 & 11.3$\pm$2.3 & 8.83 & 1
\enddata
\tablecomments{
Col.(1): ID number.  Col.(2): Source name.  Col.(3) Telescope.  
Col.(4): The noise level in the velocity resolution shown in ().  
Col.(5): The main beam temperature corrected for atmospheric attenuation.  
Col.(6): The CO~(1--0) line integrated intensity.  Double colon means that 
the flux error is unknown.  The error is assumed to be 20~\% in this paper.  
Col.(8): The molecular gas mass estimated using the single-dish spectra.  
Col.(7): Reference: 
1: this work, 2: \citet{Maiolino97}, 3: \citet{Bertram06}, 4: \citet{Sanders91}, 
5: \citet{Young95}, 6: \citet{Evans05}, 7: \citet{Casoli92}, 8: \citet{Wiklind95}, 
9: \citet{Mirabel90}, 10: \citet{Huchtmeier92}, 11: \citet{Zhu99}, 
12: \citet{Horellou97}, 13: \citet{Andreani95}
}
\end{deluxetable}

\begin{deluxetable}{rlrrrllrc}
\tabletypesize{\scriptsize}
\tablewidth{0pt}
\tablecaption{Properties of the Molecular Gas Disk \label{tab:t6}}
\tablehead{
&&\multicolumn{1}{c}{Position angle}&\multicolumn{1}{c}{Inclination}&\multicolumn{1}{c}{$V_{0}$}&\multicolumn{1}{c}{$R_{\rm CO}$}&\multicolumn{1}{c}{$R_{80}$}&&\\
\multicolumn{1}{c}{ID}&\multicolumn{1}{c}{Name}&\multicolumn{1}{c}{[deg]}&\multicolumn{1}{c}{[deg]}&\multicolumn{1}{c}{[km~s$^{-1}$]}&\multicolumn{1}{c}{[kpc]}&\multicolumn{1}{c}{[kpc]}&\multicolumn{1}{c}{$\frac{R_{80}}{R_{\rm eff}}$}&\multicolumn{1}{c}{Ring}\\
\multicolumn{1}{c}{(1)}&\multicolumn{1}{c}{(2)}&\multicolumn{1}{c}{(3)}&\multicolumn{1}{c}{(4)}&\multicolumn{1}{c}{(5)}&\multicolumn{1}{c}{(6)}&\multicolumn{1}{c}{(7)}&\multicolumn{1}{c}{(8)}&\multicolumn{1}{c}{(9)}
}
\startdata
  1& UGC~6 & 306$\pm$1 & 21$\pm$9 & 6473$\pm$1 & 1.2$\pm$0.2 & 0.5$\pm$0.2 & 1.1$\pm$0.4 & Y\\
  2& NGC~34 & 346$\pm$1 & 26$\pm$1 & 5701$\pm$1 & 3.1$\pm$0.3 & 1.7$\pm$0.3 & 4.5$\pm$0.9 & Y\\
  3& Arp~230 & 113$\pm$1 & 65$\pm$1 & 1694$\pm$1 & 3.2$\pm$0.1 & 1.8$\pm$0.1 & 2.6$\pm$0.2 & Y\\
  5& NGC~828 & 306$\pm$1 & 60$\pm$1 & 5229$\pm$1 & 7.9$\pm$0.3 & 3.5$\pm$0.3 & 1.3$\pm$0.1 & Y\\
  6& UGC~2238 & 138$\pm$1 & 61$\pm$1 & 6361$\pm$1 & 8.5$\pm$0.4 & 2.6$\pm$0.4 & 1.2$\pm$0.2 & Y\\
  9& NGC~1614 & 352$\pm$1 & 36$\pm$1 & 4744$\pm$1 & 4.6$\pm$0.3 & 1.4$\pm$0.3 & 0.40$\pm$0.08 & Y\\
10& Arp~187 & 165$\pm$2 & 53$\pm$5 & 11441$\pm$4 & 9.3$\pm$0.3 & 3.5$\pm$0.3 & 0.50$\pm$0.05 & N\\
11& AM~0612-373 & 123$\pm$1 & 45$\pm$1 & 9465$\pm$1 & 5.0$\pm$0.3 & 1.3$\pm$0.3 & 0.19$\pm$0.04 & Y\\
13& NGC~2623 & 76$\pm$1 & 18$\pm$4 & 5532$\pm$1 & 1.1$\pm$0.1 & 0.5$\pm$0.1 & 0.18$\pm$0.04 & Y\\
14& NGC~2782 & 260$\pm$1 & 35$\pm$1 & 2573$\pm$1 & 2.3$\pm$0.1\footnotemark[1] & 1.1$\pm$0.1\footnotemark[1] & 0.18$\pm$0.01 & Y\\
15& UGC~5101 & 256$\pm$1 & 25$\pm$2 & 11799$\pm$6 & 3.0$\pm$0.2 & 1.0$\pm$0.2 & 4.0$\pm$0.8 & Y\\
16& AM~0956-282 & 353$\pm$2 & 34$\pm$7 & 986$\pm$1 & 1.6$\pm$0.1 & 0.4$\pm$0.1 & 0.18$\pm$0.03 & N\\
17& NGC~3256 & 84$\pm$1 & 43$\pm$1 & 2760$\pm$1 & 6.6$\pm$0.3\footnotemark[1] & 3.2$\pm$0.3\footnotemark[1] & 1.7$\pm$0.2 & N\\
18& NGC~3597 & 264$\pm$1 & 55$\pm$1 & 3475$\pm$1 & 2.3$\pm$0.1 & 1.5$\pm$0.1 & 1.2$\pm$0.1 & Y\\
19& AM~1158-333 & \nodata & \nodata & \nodata & 1.0$\pm$0.1\footnotemark[2] & 0.8$\pm$0.1\footnotemark[2] & 0.48$\pm$0.04 & \nodata\\
20& NGC~4194 &\nodata & \nodata & \nodata & 2.2$\pm$0.1\footnotemark[2] & 1.3$\pm$0.1\footnotemark[2] & 1.3$\pm$0.1 & \nodata\\
21& NGC~4441 & 30$\pm$1 & 31$\pm$3 & 2716$\pm$1 & 2.2$\pm$0.1\footnotemark[1]  & 0.9$\pm$0.1\footnotemark[1]  & 0.24$\pm$0.03 & Y\\
22& UGC~8058 & 98$\pm$1 & 17$\pm$9 & 12641$\pm$5 & 1.5$\pm$0.2 & 0.8$\pm$0.2 & 16$\pm$3 & N\\
23& AM~1255-430 & \nodata & \nodata & \nodata &  3.7$\pm$0.2\footnotemark[2] & 2.0$\pm$0.2\footnotemark[2] & 0.66$\pm$0.07 & N\\
24& AM~1300-233 & 233$\pm$1 & 23$\pm$3 & 6298$\pm$1 & 2.4$\pm$0.2 & 1.1$\pm$0.2 & 0.08$\pm$0.01 & Y\\
26& Arp~193 & 331$\pm$1 & 26$\pm$2 & 6993$\pm$1 & 2.6$\pm$0.2 & 1.2$\pm$0.2 & 0.40$\pm$0.08 & Y\\
28& UGC~9829 & \nodata &\nodata &\nodata & 5.1$\pm$0.5\footnotemark[2] & 4.2$\pm$0.5\footnotemark[2] & 0.49$\pm$0.05 & \nodata\\
29& NGC~6052 & \nodata & \nodata & \nodata & 3.8$\pm$0.2\footnotemark[2] & 2.0$\pm$0.2\footnotemark[2] & 0.49$\pm$0.05 & \nodata\\
30& UGC~10675 & 175$\pm$1 & 31$\pm$4 & 9810$\pm$1 & 2.6$\pm$0.5 & 1.6$\pm$0.5 & 3$\pm$1 & Y\\
31& AM~2038-382 & 330$\pm$1 & 23$\pm$5 & 5935$\pm$1 & 2.2$\pm$0.1 & 1.2$\pm$0.1 & 1.2$\pm$0.1 & Y\\
32& AM~2055-425 & 41$\pm$2 & 24$\pm$10 & 12357$\pm$7 & 3.6$\pm$0.2 & 1.7$\pm$0.2 & 0.57$\pm$0.08 & N\\
33& NGC~7135 & \nodata & \nodata & \nodata & 0.4$\pm$0.1\footnotemark[2] & 0.3$\pm$0.1\footnotemark[2] & 0.01$\pm$0.01 & \nodata\\
34& NGC~7252 & 118$\pm$1 & 23$\pm$3 & 4684$\pm$1 & 2.6$\pm$0.1 & 1.5$\pm$0.1 & 0.60$\pm$0.05 & Y\\
35& AM~2246-490 & 156$\pm$1 & 30$\pm$3 & 12439$\pm$8 & 2.8$\pm$0.3 & 1.3$\pm$0.3 & 0.15$\pm$0.03 & Y\\
37& NGC~7727 & 113$\pm$1 & 62$\pm$2 & 1815$\pm$1 & 1.6$\pm$0.1 & 0.8$\pm$0.1 & 0.35$\pm$0.05 & N
\enddata
\tablecomments{
Col.(1): ID number.  Col.(2): Source name.  
Col.(3) -- (5): The position angle, inclination, and systemic velocity of the molecular gas disk.  
These parameters are estimated by fitting the CO velocity field.  
Col.(6): The radius enclosing the maximum extent of the molecular gas disk and the CO distribution.  
Col.(7): The radius enclosing 80\% of the total CO flux.  
Col.(8): The ratio of $R_{80}$ to the $K$-band effective radius ($R_{\rm eff}$; RJ04).  
Col.(9): ``Y'' means that a ring-like structure is seen in the PV diagram along the kinematical major axis.
}
\tablenotetext{a}{The radii are estimated after excluding the CO extensions which clearly exist outside the molecular gas disk.}
\tablenotetext{b}{The radii are estimated without correcting for the geometry of the galaxy.}
\end{deluxetable}

\clearpage
\makeatletter
    \renewcommand{\thetable}{
    \thesection.\arabic{table}}
   \@addtoreset{table}{section}
  \makeatother
\makeatletter
    \renewcommand{\thefigure}{
    \thesection.\arabic{figure}}
    \@addtoreset{figure}{section}
  \makeatother


\begin{thebibliography}{fun}
\bibitem[Aalto et al.(2012)]{Aalto12} Aalto, S., Garcia-Burillo, S., Muller, S., et al.\ 2012, \aap, 537, A44 
\bibitem[Aalto \& H\"{u}ttemeister(2000)]{Aalto00} Aalto, S., H\"{u}ttemeister, S.\ 2000, \aap, 362, 42 
\bibitem[Alatalo et al.(2013)]{Alatalo13} Alatalo, K., Davis, T.~A., Bureau, M., et al.\ 2013, \mnras, 432, 1796 
\bibitem[Albrecht et al.(2004)]{Albrecht04} Albrecht, M., Chini, R., Kr{\"u}gel, E., M{\"u}ller, S.~A.~H., \& Lemke, R.\ 2004, \aap, 414, 141 
\bibitem[Albrecht et al.(2007)]{Albrecht07} Albrecht, M., Kr{\"u}gel, E., \& Chini, R.\ 2007, \aap, 462, 575 
\bibitem[Alloin \& Duflot(1979)]{Alloin79} Alloin, D., \& Duflot, R.\ 1979, \aap, 78, L5 
\bibitem[Alonso-Herrero et al.(2001)]{Alonso-Herrero01} Alonso-Herrero, A., Engelbracht, C.~W., Rieke, M.~J., Rieke, G.~H., \& Quillen, A.~C.\ 2001, \apj, 546, 952 
\bibitem[Andreani et al.(1995)]{Andreani95} Andreani, P., Casoli, F., \& Gerin, M.\ 1995, \aap, 300, 43 
\bibitem[Armus et al.(2004)]{Armus04} Armus, L., Charmandaris, V., Spoon, H.~W.~W., et al.\ 2004, \apjs, 154, 178 
\bibitem[Arp \& Madore(1987)]{Arp87} Arp, H.~C., \& Madore, B.\ 1987, Cambridge ; New York : Cambridge University Press, 1987.,
\bibitem[Balick et al.(1976)]{Balick76} Balick, B., Faber, S.~M., \& Gallagher, J.~S.\ 1976, \apj, 209, 710 
\bibitem[Barnes(2002)]{Barnes02} Barnes, J.~E.\ 2002, \mnras, 333, 481 
\bibitem[Barnes \& Hernquist(1992)]{Barnes92} Barnes, J.~E., \& Hernquist, L.\ 1992, \araa, 30, 705 
\bibitem[Beck et al.(2014)]{Beck14} Beck, S.~C., Lacy, J., Turner, J., Greathouse, T., \& Neff, S.\ 2014, \apj, 787, 85 
\bibitem[Beichman et al.(1988)]{Beichman88} Beichman, C.~A., Neugebauer, G., Habing, H.~J., Clegg, P.~E., \& Chester, T.~J.\ 1988, Infrared astronomical satellite (IRAS) catalogs and atlases.~Volume 1: Explanatory supplement, 1,  
\bibitem[Bertram et al.(2006)]{Bertram06} Bertram, T., Eckart, A., Krips, M., Staguhn, J.~G., \& Hackenberg, W.\ 2006, \aap, 448, 29 
\bibitem[Biermann et al.(1978)]{Biermann78} Biermann, P., Clarke, J.~N., \& Fricke, K.~J.\ 1978, \aap, 70, L41 
\bibitem[Boer et al.(1992)]{Boer92} Boer, B., Schulz, H., \& Keel, W.~C.\ 1992, \aap, 260, 67 
\bibitem[Borne \& Richstone(1991)]{Borne91} Borne, K.~D., \& Richstone, D.~O.\ 1991, \apj, 369, 111 
\bibitem[Bottinelli et al.(1980)]{Bottinelli80} Bottinelli, L., Gouguenheim, L., \& Paturel, G.\ 1980, \aaps, 40, 355 
\bibitem[Bournaud \& Combes(2002)]{Bournaud02} Bournaud, F., \& Combes, F.\ 2002, \aap, 392, 83 
\bibitem[Braatz et al.(2004)]{Braatz04} Braatz, J.~A., Henkel, C., Greenhill, L.~J., Moran, J.~M., \& Wilson, A.~S.\ 2004, \apjl, 617, L29 
\bibitem[Brassington et al.(2007)]{Brassington07} Brassington, N.~J., Ponman, T.~J., \& Read, A.~M.\ 2007, \mnras, 377, 1439 
\bibitem[Bridge et al.(2010)]{Bridge10} Bridge, C.~R., Carlberg, R.~G., \& Sullivan, M.\ 2010, \apj, 709, 1067 
\bibitem[Brightman \& Nandra(2011)]{Brightman11} Brightman, M., \& Nandra, K.\ 2011, \mnras, 414, 3084 
\bibitem[Bryant \& Scoville(1999)]{Bryant99} Bryant, P.~M., \& Scoville, N.~Z.\ 1999, \aj, 117, 2632 
\bibitem[Cappellari et al.(2011)]{Cappellari11} Cappellari, M., Emsellem, E., Krajnovi{\'c}, D., et al.\ 2011, \mnras, 413, 813 
\bibitem[Carilli et al.(1998)]{Carilli98} Carilli, C.~L., Wrobel, J.~M., \& Ulvestad, J.~S.\ 1998, \aj, 115, 928 
\bibitem[Carilli \& Walter(2013)]{Carilli13} Carilli, C.~L., \& Walter, F.\ 2013, \araa, 51, 105 
\bibitem[Carlson et al.(1999)]{Carlson99} Carlson, M.~N., Holtzman, J.~A., Grillmair, C.~J., et al.\ 1999, \aj, 117, 1700 
\bibitem[Casey et al.(2014)]{Casey14} Casey, C.~M., Narayanan, D., \& Cooray, A.\ 2014, arXiv:1402.1456 
\bibitem[Casoli et al.(1992)]{Casoli92} Casoli, F., Dupraz, C., \& Combes, F.\ 1992, \aap, 264, 49 
\bibitem[Chien \& Barnes(2010)]{Chien10} Chien, L.-H., \& Barnes, J.~E.\ 2010, \mnras, 407, 43 
\bibitem[Chitre \& Jog(2002)]{Chitre02} Chitre, A., \& Jog, C.~J.\ 2002, \aap, 388, 407 
\bibitem[Combes et al.(2013)]{Combes13} Combes, F., Garc{\'{\i}}a-Burillo, S., Braine, J., et al.\ 2013, \aap, 550, A41 
\bibitem[Condon et al.(1991)]{Condon91} Condon, J.~J., Anderson, M.~L., \& Helou, G.\ 1991, \apj, 376, 95
\bibitem[Condon et al.(1998)]{Condon98} Condon, J.~J., Cotton, W.~D., Greisen, E.~W., et al.\ 1998, \aj, 115, 1693 
\bibitem[Condon et al.(1990)]{Condon90} Condon, J.~J., Helou, G., Sanders, D.~B., \& Soifer, B.~T.\ 1990, \apjs, 73, 359 
\bibitem[Corbett et al.(2003)]{Corbett03} Corbett, E.~A., Kewley, L., Appleton, P.~N., et al.\ 2003, \apj, 583, 670 
\bibitem[Costagliola et al.(2011)]{Costagliola11} Costagliola, F., Aalto, S., Rodriguez, M.~I., et al.\ 2011, \aap, 528, A30 
\bibitem[Cox \& Sparke(2004)]{Cox04} Cox, A.~L., \& Sparke, L.~S.\ 2004, \aj, 128, 2013 
\bibitem[Crocker et al.(2011)]{Crocker11} Crocker, A.~F., Bureau, M., Young, L.~M., \& Combes, F.\ 2011, \mnras, 410, 1197 
\bibitem[Daddi et al.(2010)]{Daddi10} Daddi, E., Bournaud, F., Walter, F., et al.\ 2010, \apj, 713, 686 
\bibitem[Dahari \& De Robertis(1988)]{Dahari88} Dahari, O., \& De Robertis, M.~M.\ 1988, \apjs, 67, 249 
\bibitem[Davis et al.(2013)]{Davis13} Davis, T.~A., Alatalo, K., Bureau, M., et al.\ 2013, \mnras, 429, 534 \
\bibitem[Dasyra et al.(2006)]{Dasyra06} Dasyra, K.~M., Tacconi, L.~J., Davies, R.~I., et al.\ 2006, \apj, 651, 835 
\bibitem[Deeg et al.(1993)]{Deeg93} Deeg, H.-J., Brinks, E., Duric, N., Klein, U., \& Skillman, E.\ 1993, \apj, 410, 626 
\bibitem[Dekel et al.(2009)]{Dekel09} Dekel, A., Birnboim, Y., Engel, G., et al.\ 2009, \nat, 457, 451 
\bibitem[Denisiuk et al.(1976)]{Denisyuk76} Denisiuk, E.~K., Lipovetskii, V.~A., \& Afanasev, V.~L.\ 1976, Astrofizika, 12, 665 
\bibitem[Downes \& Solomon(1998)]{Downes98} Downes, D., \& Solomon, P.~M.\ 1998, \apj, 507, 615 
\bibitem[English et al.(2003)]{English03} English, J., Norris, R.~P., Freeman, K.~C., \& Booth, R.~S.\ 2003, \aj, 125, 1134 
\bibitem[Evans et al.(2005)]{Evans05} Evans, A.~S., Mazzarella, J.~M., Surace, J.~A., et al.\ 2005, \apjs, 159, 197 
\bibitem[Evans et al.(2008)]{Evans08} Evans, A.~S., Vavilkin, T., Pizagno, J., et al.\ 2008, \apjl, 675, L69 
\bibitem[Feruglio et al.(2010)]{Feruglio10} Feruglio, C., Maiolino, R., Piconcelli, E., et al.\ 2010, \aap, 518, L155 
\bibitem[Fern{\'a}ndez et al.(2010)]{Fernandez10} Fern{\'a}ndez, X., van Gorkom, J.~H., Schweizer, F., \& Barnes, J.~E.\ 2010, \aj, 140, 1965 
\bibitem[F{\"o}rster Schreiber et al.(2006)]{Forster-Schreiber06} F{\"o}rster Schreiber, N.~M., Genzel, R., Lehnert, M.~D., et al.\ 2006, \apj, 645, 1062 
\bibitem[Fort et al.(1986)]{Fort86} Fort, B.~P., Prieur, J.-L., Carter, D., Meatheringham, S.~J., \& Vigroux, L.\ 1986, \apj, 306, 110 
\bibitem[Franceschini et al.(2003)]{Franceschini03} Franceschini, A., Braito, V., Persic, M., et al.\ 2003, \mnras, 343, 1181 
\bibitem[Farrah et al.(2003)]{Farrah03} Farrah, D., Afonso, J., Efstathiou, A., et al.\ 2003, \mnras, 343, 585 
\bibitem[Gallego et al.(1996)]{Gallego96} Gallego, J., Zamorano, J., Rego, M., Alonso, O., \& Vitores, A.~G.\ 1996, \aaps, 120, 323 
\bibitem[Galletta et al.(1997)]{Galletta97} Galletta, G., Sage, L.~J., \& Sparke, L.~S.\ 1997, \mnras, 284, 773 
\bibitem[Garc{\'{\i}}a-Lorenzo et al.(2008)]{Garcia-Lorenzo08} Garc{\'{\i}}a-Lorenzo, B., Cair{\'o}s, L.~M., Caon, N., Monreal-Ibero, A., \& Kehrig, C.\ 2008, \apj, 677, 201 
\bibitem[Garland et al.(2007)]{Garland07} Garland, C.~A., Pisano, D.~J., Williams, J.~P., et al.\ 2007, \apj, 671, 310 
\bibitem[Genzel et al.(1998)]{Genzel98} Genzel, R., Lutz, D., Sturm, E., et al.\ 1998, \apj, 498, 579 
\bibitem[Genzel et al.(2001)]{Genzel01} Genzel, R., Tacconi, L.~J., Rigopoulou, D., Lutz, D., \& Tecza, M.\ 2001, \apj, 563, 527 
\bibitem[Georgakakis et al.(2001)]{Georgakakis01} Georgakakis, A., Hopkins, A.~M., Caulton, A., et al.\ 2001, \mnras, 326, 1431 
\bibitem[Gon{\c c}alves et al.(1999)]{Goncalves99} Gon{\c c}alves, A.~C., V{\'e}ron-Cetty, M.-P., \& V{\'e}ron, P.\ 1999, \aaps, 135, 437 
\bibitem[Goudfrooij et al.(1994)]{Goudfrooij94} Goudfrooij, P., Hansen, L., Jorgensen, H.~E., \& Norgaard-Nielsen, H.~U.\ 1994, \aaps, 105, 341 
\bibitem[Haan et al.(2011)]{Haan11} Haan, S., Surace, J.~A., Armus, L., et al.\ 2011, \aj, 141, 100 
\bibitem[Hattori et al.(2004)]{Hattori04} Hattori, T., Yoshida, M., Ohtani, H., et al.\ 2004, \aj, 127, 736  
\bibitem[Helfer et al.(2003)]{Helfer03} Helfer, T.~T., Thornley, M.~D., Regan, M.~W., et al.\ 2003, \apjs, 145, 259 
\bibitem[Helou et al.(1985)]{Helou85} Helou, G., Soifer, B.~T., \& Rowan-Robinson, M.\ 1985, \apjl, 298, L7 
\bibitem[Hibbard et al.(1994)]{Hibbard94} Hibbard, J.~E., Guhathakurta, P., van Gorkom, J.~H., \& Schweizer, F.\ 1994, \aj, 107, 67 
\bibitem[Hibbard \& Mihos(1995)]{Hibbard95} Hibbard, J.~E., \& Mihos, J.~C.\ 1995, \aj, 110, 140 
\bibitem[Hibbard \& Yun(1996)]{Hibbard96} Hibbard, J.~E., \& Yun, M.~S.\ 1996, Cold Gas at High Redshift, 206, 47 
\bibitem[Holtzman et al.(1996)]{Holtzman96} Holtzman, J.~A., Watson, A.~M., Mould, J.~R., et al.\ 1996, \aj, 112, 416 
\bibitem[Hopkins et al.(2009)]{Hopkins09} Hopkins, P.~F., Cox, T.~J., Younger, J.~D., \& Hernquist, L.\ 2009, \apj, 691, 1168 
\bibitem[Horellou \& Booth(1997)]{Horellou97} Horellou, C., \& Booth, R.\ 1997, \aaps, 126, 3 
\bibitem[Huchtmeier \& Tammann(1992)]{Huchtmeier92} Huchtmeier, W.~K., \& Tammann, G.~A.\ 1992, \aap, 257, 455 
\bibitem[Hunt et al.(2008)]{Hunt08} Hunt, L.~K., Combes, F., Garc{\'{\i}}a-Burillo, S., et al.\ 2008, \aap, 482, 133 
\bibitem[Imanishi(2006)]{Imanishi06} Imanishi, M.\ 2006, \aj, 131, 2406 
\bibitem[Imanishi et al.(2010)]{Imanishi10} Imanishi, M., Nakagawa, T., Shirahata, M., Ohyama, Y., \& Onaka, T.\ 2010, \apj, 721, 1233 
\bibitem[Imanishi \& Nakanishi(2013)]{Imanishi13} Imanishi, M., \& Nakanishi, K.\ 2013, \aj, 146, 47 
\bibitem[Imanishi et al.(2009)]{Imanishi09} Imanishi, M., Nakanishi, K., Tamura, Y., \& Peng, C.-H.\ 2009, \aj, 137, 3581 
\bibitem[Imanishi et al.(2003)]{Imanishi03} Imanishi, M., Terashima, Y., Anabuki, N., \& Nakagawa, T.\ 2003, \apjl, 596, L167 
\bibitem[Iono et al.(2009)]{Iono09} Iono, D., Wilson, C.~D., Yun, M.~S., et al.\ 2009, \apj, 695, 1537 
\bibitem[Iwasawa et al.(2011)]{Iwasawa11} Iwasawa, K., Sanders, D.~B., Teng, S.~H., et al.\ 2011, \aap, 529, A106 
\bibitem[James et al.(1999)]{James99} James, P., Bate, C., Wells, M., Wright, G., \& Doyon, R.\ 1999, \mnras, 309, 585 
\bibitem[Ji et al.(2000)]{Ji00} Ji, L., Chen, Y., Huang, J.~H., Gu, Q.~S., \& Lei, S.~J.\ 2000, \aap, 355, 922 
\bibitem[Jogee et al.(1999)]{Jogee99} Jogee, S., Kenney, J.~D.~P., \& Smith, B.~J.\ 1999, \apj, 526, 665 
\bibitem[J\"{u}tte et al.(2010)]{Jutte10} J\"{u}tte, E., Aalto, S., H\"{u}ttemeister, S.\ 2010, \aap, 509, A19 
\bibitem[Kere{\v s} et al.(2005)]{Keres05} Kere{\v s}, D., Katz, N., Weinberg, D.~H., \& Dav{\'e}, R.\ 2005, \mnras, 363, 2 
\bibitem[Kewley et al.(2000)]{Kewley00} Kewley, L.~J., Heisler, C.~A., Dopita, M.~A., et al.\ 2000, \apj, 530, 704 
\bibitem[Kim et al.(1988)]{Kim88} Kim, D.-W., Jura, M., Guhathakurta, P., Knapp, G.~R., \& van Gorkom, J.~H.\ 1988, \apj, 330, 684 
\bibitem[Kinney et al.(1984)]{Kinney84} Kinney, A.~L., Bregman, J.~N., Huggins, P.~J., Glassgold, A.~E., \& Cohen, R.~D.\ 1984, \pasp, 96, 398 
\bibitem[Knapen(2005)]{Knapen05} Knapen, J.~H.\ 2005, \aap, 429, 141 
\bibitem[Koda \& Subaru Cosmos Team(2009)]{Koda09} Koda, J., \& Subaru Cosmos Team 2009, The Starburst-AGN Connection, 408, 22 
\bibitem[K{\"o}nig et al.(2013)]{Konig13} K{\"o}nig, S., Aalto, S., Muller, S., Beswick, R.~J., \& Gallagher, J.~S.\ 2013, \aap, 553, A72 
\bibitem[Kotilainen et al.(1996)]{Kotilainen96} Kotilainen, J.~K., Moorwood, A.~F.~M., Ward, M.~J., \& Forbes, D.~A.\ 1996, \aap, 305, 107 
\bibitem[Krips et al.(2007)]{Krips07} Krips, M., Eckart, A., Krichbaum, T.~P., et al.\ 2007, \aap, 464, 553 
\bibitem[Laine et al.(2003)]{Laine03} Laine, S., van der Marel, R.~P., Rossa, J., et al.\ 2003, \aj, 126, 2717 
\bibitem[Lake \& Dressler(1986)]{Lake86} Lake, G., \& Dressler, A.\ 1986, \apj, 310, 605 
\bibitem[Lees et al.(1991)]{Lees91} Lees, J.~F., Knapp, G.~R., Rupen, M.~P., \& Phillips, T.~G.\ 1991, \apj, 379, 177 
\bibitem[Jenkins et al.(2004)]{Jenkins04} Jenkins, L.~P., Roberts, T.~P., Ward, M.~J., \& Zezas, A.\ 2004, \mnras, 352, 1335 
\bibitem[Lin et al.(2004)]{Lin04} Lin, L., Koo, D.~C., Willmer, C.~N.~A., et al.\ 2004, \apjl, 617, L9 
\bibitem[L{\'{\i}}pari et al.(2000)]{Lipari00} L{\'{\i}}pari, S., D{\'{\i}}az, R., Taniguchi, Y., et al.\ 2000, \aj, 120, 645 
\bibitem[Lira et al.(2002)]{Lira02} Lira, P., Ward, M., Zezas, A., Alonso-Herrero, A., \& Ueno, S.\ 2002, \mnras, 330, 259 
\bibitem[Longhetti et al.(1998)]{Longhetti98} Longhetti, M., Rampazzo, R., Bressan, A., \& Chiosi, C.\ 1998, \aaps, 130, 267 
\bibitem[Lonsdale et al.(2003)]{Lonsdale03} Lonsdale, C.~J., Lonsdale, C.~J., Smith, H.~E., \& Diamond, P.~J.\ 2003, \apj, 592, 804 
\bibitem[Lutz(1991)]{Lutz91} Lutz, D.\ 1991, \aap, 245, 31 
\bibitem[Maiolino et al.(2003)]{Maiolino03} Maiolino, R., Comastri, A., Gilli, R., et al.\ 2003, \mnras, 344, L59 
\bibitem[Maiolino et al.(1997)]{Maiolino97} Maiolino, R., Ruiz, M., Rieke, G.~H., \& Papadopoulos, P.\ 1997, \apj, 485, 552 
\bibitem[Malin \& Carter(1983)]{Malin83} Malin, D.~F., \& Carter, D.\ 1983, \apj, 274, 534 
\bibitem[Manthey et al.(2008)]{Manthey08} Manthey, E., Aalto, S., H{\"u}ttemeister, S., \& Oosterloo, T.~A.\ 2008, \aap, 484, 693 
\bibitem[Mao et al.(2010)]{Mao10} Mao, R.-Q., Schulz, A., Henkel, C., Mauersberger, R., Muders, D., \& Dinh-V-Trung 2010, \apj, 724, 1336 
\bibitem[Marino et al.(2009)]{Marino09} Marino, A., Iodice, E., Tantalo, R., et al.\ 2009, \aap, 508, 1235 
\bibitem[M{\'a}rquez et al.(2002)]{Marquez02} M{\'a}rquez, I., Masegosa, J., Moles, M., et al.\ 2002, \aap, 393, 389 
\bibitem[Martini et al.(2003)]{Martini03} Martini, P., Regan, M.~W., Mulchaey, J.~S., \& Pogge, R.~W.\ 2003, \apjs, 146, 353 
\bibitem[McGaugh \& Bothun(1990)]{McGaugh90} McGaugh, S.~S., \& Bothun, G.~D.\ 1990, \aj, 100, 1073 
\bibitem[Mihos \& Bothun(1998)]{Mihos98} Mihos, J.~C., \& Bothun, G.~D.\ 1998, \apj, 500, 619 
\bibitem[Mihos et al.(1993)]{Mihos93} Mihos, J.~C., Bothun, G.~D., \& Richstone, D.~O.\ 1993, \apj, 418, 82 
\bibitem[Miller et al.(1997)]{Miller97} Miller, B.~W., Whitmore, B.~C., Schweizer, F., \& Fall, S.~M.\ 1997, \aj, 114, 2381 
\bibitem[Mirabel et al.(1990)]{Mirabel90} Mirabel, I.~F., Booth, R.~S., Johansson, L.~E.~B., Garay, G., \& Sanders, D.~B.\ 1990, \aap, 236, 327 
\bibitem[Mirabel \& Sanders(1988)]{Mirabel88} Mirabel, I.~F., \& Sanders, D.~B.\ 1988, \apj, 335, 104 
\bibitem[Miralles-Caballero et al.(2011)]{Miralles-Caballero11} Miralles-Caballero, D., Colina, L., Arribas, S., \& Duc, P.-A.\ 2011, \aj, 142, 79
\bibitem[Monreal-Ibero et al.(2010)]{Monreal-Ibero10} Monreal-Ibero, A., Arribas, S., Colina, L., et al.\ 2010, \aap, 517, A28 
\bibitem[Moshir et al.(2008)]{Moshir08} Moshir, M., Kopan, G., Conrow, T., et al.\ 2008, VizieR Online Data Catalog, 2275, 0 
\bibitem[Moshir et al.(1992)]{Moshir92} Moshir, M., Kopman, G., \& Conrow, T.~A.~O.\ 1992, Pasadena: Infrared Processing and Analysis Center, California Institute of Technology, 1992, edited by Moshir, M.; Kopman, G.; Conrow, T.~a.o.,  
\bibitem[Mulchaey et al.(1996)]{Mulchaey96} Mulchaey, J.~S., Wilson, A.~S., \& Tsvetanov, Z.\ 1996, \apjs, 102, 309 
\bibitem[Mullan et al.(2011)]{Mullan11} Mullan, B., Konstantopoulos, I.~S., Kepley, A.~A., et al.\ 2011, \apj, 731, 93 
\bibitem[Naab \& Burkert(2003)]{Naab03} Naab, T., \& Burkert, A.\ 2003, \apj, 597, 893 
\bibitem[Nakajima et al.(2008)]{Nakajima08} Nakajima, T., Sakai, T., Asayama, S., et al.\ 2008, \pasj, 60, 435
\bibitem[Nakanishi \& Sofue(2006)]{Nakanishi06} Nakanishi, H., \& Sofue, Y.\ 2006, \pasj, 58, 847 
\bibitem[Narayanan et al.(2005)]{Narayanan05} Narayanan, D., Groppi, C.~E., Kulesa, C.~A., \& Walker, C.~K.\ 2005, \apj, 630, 269 
\bibitem[Narayanan et al.(2012)]{Narayanan12} Narayanan, D., Krumholz, M.~R., Ostriker, E.~C., \& Hernquist, L.\ 2012, \mnras, 421, 3127 
\bibitem[Neff et al.(1990)]{Neff90} Neff, S.~G., Hutchings, J.~B., Standord, S.~A., \& Unger, S.~W.\ 1990, \aj, 99, 1088 
\bibitem[Neff et al.(2003)]{Neff03} Neff, S.~G., Ulvestad, J.~S., \& Campion, S.~D.\ 2003, \apj, 599, 1043 
\bibitem[Norris \& Forbes(1995)]{Norris95} Norris, R.~P., \& Forbes, D.~A.\ 1995, \apj, 446, 594 
\bibitem[Olsson et al.(2010)]{Olsson10} Olsson, E., Aalto, S., Thomasson, M., \& Beswick, R.\ 2010, \aap, 513, A11 
\bibitem[Osterbrock \& Martel(1993)]{Osterbrock93} Osterbrock, D.~E., \& Martel, A.\ 1993, \apj, 414, 552 
\bibitem[Papadopoulos et al.(2012a)]{Papadopoulos12a} Papadopoulos, P.~P., van der Werf, P., Xilouris, E., Isaak, K.~G., \& Gao, Y.\ 2012, \apj, 751, 10 
\bibitem[Papadopoulos et al.(2012b)]{Papadopoulos12b} Papadopoulos, P.~P., van der Werf, P.~P., Xilouris, E.~M., et al.\ 2012, \mnras, 426, 2601 
\bibitem[Peeples \& Martini(2006)]{Peeples06} Peeples, M.~S., \& Martini, P.\ 2006, \apj, 652, 1097 
\bibitem[Rampazzo et al.(2007)]{Rampazzo07} Rampazzo, R., Marino, A., Tantalo, R., et al.\ 2007, \mnras, 381, 245 
\bibitem[Rampazzo et al.(2003)]{Rampazzo03} Rampazzo, R., Plana, H., Longhetti, M., et al.\ 2003, \mnras, 343, 819 
\bibitem[Regan et al.(2001)]{Regan01} Regan, M.~W., Thornley, M.~D., Helfer, T.~T., et al.\ 2001, \apj, 561, 218 
\bibitem[Richter et al.(1994)]{Richter94} Richter, O.-G., Sackett, P.~D., \& Sparke, L.~S.\ 1994, \aj, 107, 99 
\bibitem[Risaliti et al.(2000)]{Risaliti00} Risaliti, G., Gilli, R., Maiolino, R., \& Salvati, M.\ 2000, \aap, 357, 13 
\bibitem[Risaliti et al.(2006)]{Risaliti06} Risaliti, G., Maiolino, R., Marconi, A., et al.\ 2006, \mnras, 365, 303 
\bibitem[Robertson \& Bullock(2008)]{Robertson08} Robertson, B.~E., \& Bullock, J.~S.\ 2008, \apjl, 685, L27 
\bibitem[Rodr{\'{\i}}guez-Zaur{\'{\i}}n et al.(2011)]{Rodriguez-Zaurin11} Rodr{\'{\i}}guez-Zaur{\'{\i}}n, J., Arribas, S., Monreal-Ibero, A., et al.\ 2011, \aap, 527, A60 
\bibitem[Rothberg \& Fischer(2010)]{Rothberg10} Rothberg, B., \& Fischer, J.\ 2010, \apj, 712, 318 
\bibitem[Rothberg \& Joseph(2004)]{Rothberg04} Rothberg, B., \& Joseph, R.~D.\ 2004, \aj, 128, 2098 
\bibitem[Rothberg \& Joseph(2006a)]{Rothberg06a} Rothberg, B., \& Joseph, R.~D.\ 2006a, \aj, 131, 185 
\bibitem[Rothberg \& Joseph(2006b)]{Rothberg06b} Rothberg, B., \& Joseph, R.~D.\ 2006b, \aj, 132, 976 
\bibitem[Sakamoto et al.(2014)]{Sakamoto14} Sakamoto, K., Aalto, S., Combes, F., Evans, A., \& Peck, A.\ 2014, arXiv:1403.7117 
\bibitem[Sakamoto et al.(2006)]{Sakamoto06} Sakamoto, K., Ho, P.~T.~P., \& Peck, A.~B.\ 2006, \apj, 644, 862 
\bibitem[Sanders \& Mirabel(1996)]{Sanders96} Sanders, D.~B., \& Mirabel, I.~F.\ 1996, \araa, 34, 749 
\bibitem[Sanders et al.(1991)]{Sanders91} Sanders, D.~B., Scoville, N.~Z., \& Soifer, B.~T.\ 1991, \apj, 370, 158 
\bibitem[Schiminovich et al.(2001)]{Schiminovich01} Schiminovich, D., van Gorkom, J.~H., Dijkstra, M., et al.\ 2001, Gas and Galaxy Evolution, 240, 864 
\bibitem[Schiminovich et al.(2013)]{Schiminovich13} Schiminovich, D., van Gorkom, J.~H., \& van der Hulst, J.~M.\ 2013, \aj, 145, 34 
\bibitem[Schombert et al.(1990)]{Schombert90} Schombert, J.~M., Wallin, J.~F., \& Struck-Marcell, C.\ 1990, \aj, 99, 497 
\bibitem[Schweizer(1982)]{Schweizer82} Schweizer, F.\ 1982, \apj, 252, 455 
\bibitem[Schweizer(1996)]{Schweizer96} Schweizer, F.\ 1996, \aj, 111, 109 
\bibitem[Schweizer \& Seitzer(1992)]{Schweizer92} Schweizer, F., \& Seitzer, P.\ 1992, \aj, 104, 1039 
\bibitem[Schweizer \& Seitzer(2007)]{Schweizer07} Schweizer, F., \& Seitzer, P.\ 2007, \aj, 133, 2132 
\bibitem[Schweizer et al.(2013)]{Schweizer13} Schweizer, F., Seitzer, P., Kelson, D.~D., Villanueva, E.~V., \& Walth, G.~L.\ 2013, \apj, 773, 148
\bibitem[Scoville et al.(2000)]{Scoville00} Scoville, N.~Z., Evans, A.~S., Thompson, R., et al.\ 2000, \aj, 119, 991 
\bibitem[Sekiguchi \& Wolstencroft(1993)]{Sekiguchi93} Sekiguchi, K., \& Wolstencroft, R.~D.\ 1993, \mnras, 263, 349 
\bibitem[Shier \& Fischer(1998)]{Shier98} Shier, L.~M., \& Fischer, J.\ 1998, \apj, 497, 163 
\bibitem[Simien \& Prugniel(1997)]{Simien97} Simien, F., \& Prugniel, P.\ 1997, \aaps, 122, 521 
\bibitem[Sliwa et al.(2012)]{Sliwa12} Sliwa, K., Wilson, C.~D., Petitpas, G.~R., et al.\ 2012, \apj, 753, 46 
\bibitem[Smirnova \& Moiseev(2010)]{Smirnova10} Smirnova, A., \& Moiseev, A.\ 2010, \mnras, 401, 307 
\bibitem[Smith(1991)]{Smith91a} Smith, B.~J.\ 1991, \apj, 378, 39 
\bibitem[Smith(1994)]{Smith94} Smith, B.~J.\ 1994, \aj, 107, 1695 
\bibitem[Smith et al.(1996)]{Smith96} Smith, D.~A., Herter, T., Haynes, M.~P., Beichman, C.~A., \& Gautier, T.~N., III 1996, \apjs, 104, 217 
\bibitem[Smith \& Hintzen(1991)]{Smith91b} Smith, E.~P., \& Hintzen, P.\ 1991, \aj, 101, 410 
\bibitem[Soifer et al.(2001)]{Soifer01} Soifer, B.~T., Neugebauer, G., Matthews, K., et al.\ 2001, \aj, 122, 1213 
\bibitem[Solomon \& Barrett(1991)]{Solomon91} Solomon, P.~M., \& Barrett, J.~W.\ 1991, Dynamics of Galaxies and Their Molecular Cloud Distributions, 146, 235 
\bibitem[Solomon et al.(1997)]{Solomon97} Solomon, P.~M., Downes, D., Radford, S.~J.~E., \& Barrett, J.~W.\ 1997, \apj, 478, 144 
\bibitem[Springel \& Hernquist(2005)]{Springel05} Springel, V., \& Hernquist, L.\ 2005, \apjl, 622, L9 
\bibitem[Springob et al.(2005)]{Springob05} Springob, C.~M., Haynes, M.~P., Giovanelli, R., \& Kent, B.~R.\ 2005, \apjs, 160, 149 
\bibitem[Stanford \& Bushouse(1991)]{Stanford91} Stanford, S.~A., \& Bushouse, H.~A.\ 1991, \apj, 371, 92 
\bibitem[Surace et al.(2000)]{Surace00} Surace, J.~A., Sanders, D.~B., \& Evans, A.~S.\ 2000, \apj, 529, 170 
\bibitem[Tacconi et al.(2013)]{Tacconi13} Tacconi, L.~J., Neri, R., Genzel, R., et al.\ 2013, \apj, 768, 74 
\bibitem[Takeuchi et al.(2010)]{Takeuchi10} Takeuchi, T.~T., Buat, V., Heinis, S., et al.\ 2010, \aap, 514, A4 
\bibitem[Taniguchi \& Noguchi(1991)]{Taniguchi91} Taniguchi, Y., \& Noguchi, M.\ 1991, \aj, 101, 1601 
\bibitem[Taylor et al.(1999)]{Taylor99} Taylor, G.~B., Silver, C.~S., Ulvestad, J.~S., \& Carilli, C.~L.\ 1999, \apj, 519, 185 
\bibitem[Theureau et al.(1998)]{Theureau98} Theureau, G., Bottinelli, L., Coudreau-Durand, N., et al.\ 1998, \aaps, 130, 333 
\bibitem[Toomre (1977)]{Toomre77} Toomre, A. 1977, in The Evolution of Galaxies and Stellar Populations, ed. B.M. Tinsley \& R.B. Larson (New Haven: Yale Univ.), 401
\bibitem[Tzanavaris \& Georgantopoulos(2007)]{Tzanavaris07} Tzanavaris, P., \& Georgantopoulos, I.\ 2007, \aap, 468, 129 
\bibitem[van den Broek et al.(1991)]{vandenBroek91} van den Broek, A.~C., van Driel, W., de Jong, T., et al.\ 1991, \aaps, 91, 61 
\bibitem[van Driel et al.(1991)]{vanDriel91} van Driel, W., van den Broek, A.~C., \& de Jong, T.\ 1991, \aaps, 90, 55 
\bibitem[Veilleux et al.(1995)]{Veilleux95} Veilleux, S., Kim, D.-C., Sanders, D.~B., Mazzarella, J.~M., \& Soifer, B.~T.\ 1995, \apjs, 98, 171 
\bibitem[Veilleux et al.(2009)]{Veilleux09} Veilleux, S., Rupke, D.~S.~N., Kim, D.-C., et al.\ 2009, \apjs, 182, 628 
\bibitem[Vlahakis et al.(2013)]{Vlahakis13} Vlahakis, C., van der Werf, P., Israel, F.~P., \& Tilanus, R.~P.~J.\ 2013, \mnras, 433, 1837 
\bibitem[Wang et al.(1992)]{Wang92} Wang, Z., Schweizer, F., \& Scoville, N.~Z.\ 1992, \apj, 396, 510 
\bibitem[Wang et al.(1991)]{Wang91} Wang, Z., Scoville, N.~Z., \& Sanders, D.~B.\ 1991, \apj, 368, 112 
\bibitem[Weistrop et al.(2004)]{Weistrop04} Weistrop, D., Eggers, D., Hancock, M., et al.\ 2004, \aj, 127, 1360 
\bibitem[Whitmore et al.(1990)]{Whitmore90} Whitmore, B.~C., Lucas, R.~A., McElroy, D.~B., et al.\ 1990, \aj, 100, 1489 
\bibitem[Whitmore et al.(1997)]{Whitmore97} Whitmore, B.~C., Miller, B.~W., Schweizer, F., \& Fall, S.~M.\ 1997, \aj, 114, 1797 
\bibitem[Whitmore et al.(1993)]{Whitmore93} Whitmore, B.~C., Schweizer, F., Leitherer, C., Borne, K., \& Robert, C.\ 1993, \aj, 106, 1354 
\bibitem[Wiklind et al.(1995)]{Wiklind95} Wiklind, T., Combes, F., \& Henkel, C.\ 1995, \aap, 297, 643 
\bibitem[Wilson et al.(2008)]{Wilson08} Wilson, C.~D., Petitpas, G.~R., Iono, D., et al.\ 2008, \apjs, 178, 189 
\bibitem[Wright \& Otrupcek(1992)]{Wright92} Wright, A., \& Otrupcek, R.\ 1992, Bulletin d'Information du Centre de Donnees Stellaires, 41, 47 
\bibitem[Windhorst et al.(2002)]{Windhorst02} Windhorst, R.~A., Taylor, V.~A., Jansen, R.~A., et al.\ 2002, \apjs, 143, 113 
\bibitem[Xu et al.(2014)]{Xu14} Xu, C.~K., Cao, C., Lu, N., et al.\ 2014, \apj, 787, 48 
\bibitem[Xu et al.(2010)]{Xu10} Xu, X., Narayanan, D., \& Walker, C.\ 2010, \apjl, 721, L112 
\bibitem[Yamamura et al.(2010)]{Yamamura10} Yamamura, I., Makiuti, S., Ikeda, N., et al.\ 2010, VizieR Online Data Catalog, 2298, 0 
\bibitem[Young et al.(1995)]{Young95} Young, J.~S., Xie, S., Tacconi, L., et al.\ 1995, \apjs, 98, 219 
\bibitem[Yuan et al.(2010)]{Yuan10} Yuan, T.-T., Kewley, L.~J., \& Sanders, D.~B.\ 2010, \apj, 709, 884 
\bibitem[Zepf et al.(1999)]{Zepf99} Zepf, S.~E., Ashman, K.~M., English, J., Freeman, K.~C., \& Sharples, R.~M.\ 1999, \aj, 118, 752 
\bibitem[Zhang et al.(2006)]{Zhang06} Zhang, J.~S., Henkel, C., Kadler, M., et al.\ 2006, \aap, 450, 933 
\bibitem[Zhu et al.(1999)]{Zhu99} Zhu, M., Seaquist, E.~R., Davoust, E., Frayer, D.~T., \& Bushouse, H.~A.\ 1999, \aj, 118, 145 
\bibitem[Zhu et al.(2003)]{Zhu03} Zhu, M., Seaquist, E.~R., \& Kuno, N.\ 2003, \apj, 588, 243 
\end{thebibliography}
\end{document}